\documentclass{article}
\usepackage[default]{cantarell} 
\usepackage[T1]{fontenc}

\usepackage[utf8]{inputenc}
\usepackage[polutonikogreek,english]{babel}

\usepackage{bm}
\usepackage{epubtk}
\usepackage{booktabs}
\usepackage{graphicx}
\usepackage{rotating}
\usepackage{amssymb}
\usepackage{amsmath}
\usepackage{pstricks}
\usepackage{pifont}
\def\msol{M_{\odot}}
\def\Msun{M_{\odot}}
\def\mbh{{\cal M}_{\bullet}}

\def\MBH{{\cal M}_{\bullet}}

\newcommand {\PN}{{\rm PN}}

\newcommand {\kms}{\ensuremath{{\rm km\,s}^{-1}}}

\newcommand{\trlx}{\ensuremath{t_\mathrm{rlx}}}
\newcommand{\RS}{\ensuremath{R_{\rm Schw}}}
\newcommand{\Nstar}{\ensuremath{N_{\star}}}
\newcommand{\tcoll}{\ensuremath{t_\mathrm{coll}}}
\newcommand{\tGW}{\ensuremath{\tau_\mathrm{GW}}}

\newcommand{\partd}[2]{\frac{\partial{#1}}{\partial{#2}}}

\newcommand {\Mpthree}{\ensuremath{M_{\odot}\,\mathrm{pc}^{-3}}}

\newenvironment{itemize_estret}{        %
\begin{itemize}                         %
  \setlength{\itemsep}{1pt}             %
  \setlength{\parskip}{0pt}             %
  \setlength{\parsep}{0pt}              %
}{\end{itemize}}                        %
\newenvironment{enumerate_estret}{      %
\begin{enumerate}                       %
  \setlength{\itemsep}{1pt}             %
  \setlength{\parskip}{0pt}             %
  \setlength{\parsep}{0pt}              %
}{\end{enumerate}}                      %
\begin{document}

\title{Relativistic Dynamics and Extreme Mass Ratio Inspirals}

\author{\epubtkAuthorData{Pau Amaro Seoane}
        {
\\
Institute of Space Sciences (ICE, CSIC) \&\\
Institut d'Estudis Espacials de Catalunya (IEEC)\\
Campus UAB, Carrer de Can Magrans s/n, 08193\\
Barcelona, Spain$^{*}$
        }
       {pau@ice.cat}
       {http://astro-gr.org/}}

\date{}
\maketitle

\begin{abstract}

It is now well-established that a dark, compact object (DCO), very likely a
massive black hole (MBH) of around four million solar masses is lurking at the
centre of the Milky Way.  While a consensus is emerging about the origin and
growth of supermassive black holes (with masses larger than a billion solar
masses), MBHs with smaller masses, such as the one in our galactic centre,
remain understudied and enigmatic.  The key to understanding these holes - how
some of them grow by orders of magnitude in mass - lies in understanding the
dynamics of the stars in the galactic neighbourhood.  Stars interact with the
central MBH primarily through their gradual inspiral due to the emission of
gravitational radiation. Also stars produce gases which will subsequently be
accreted by the MBH through collisions and disruptions brought about by the
strong central tidal field. Such processes can contribute significantly to the
mass of the MBH and progress in understanding them requires theoretical work in
preparation for future gravitational radiation millihertz missions and X-ray
observatories.  In particular, a unique probe of these regions is the
gravitational radiation that is emitted by some compact stars very close to the
black holes and which could be surveyed by a millihertz gravitational wave
interferometer scrutinizing the range of masses fundamental to understanding
the origin and growth of supermassive black holes. By extracting the
information carried by the gravitational radiation, we can determine the mass
and spin of the central MBH with unprecedented precision and we can determine
how the holes ``eat'' stars that happen to be near them.

\end{abstract}

\epubtkKeywords{}

\vfill
\footnotesize
\noindent
and\\
Institute of Applied Mathematics, Academy of Mathematics and Systems Science, CAS, Beijing 100190, China\\
Kavli Institute for Astronomy and Astrophysics, Beijing 100871, China\\
Zentrum f{\"u}r Astronomie und Astrophysik, TU Berlin, Hardenbergstra{\ss}e 36, 10623 Berlin, Germany
\normalsize

\newpage

\subsection*{Foreword}

The volume where capture orbits are produced is so small in comparison to other
typical lengthscales of interest in astrodynamics that it has usually been seen
as unimportant and irrelevant to the global dynamical evolution of the system.
The only exception has been the tidal disruption of stars by massive black
holes. Only when it transpired that the slow, adiabatic inspiral of compact
objects onto massive black holes provides us with valuable information, did
astrophysicists start to address the question in more detail.  Since the
problem of EMRIs (Extreme-Mass Ratio Inspiral) started to draw our attention,
there has been a notable progress in answering fundamental questions of
stellar dynamics. The discoveries have been numerous and some of them remain
puzzling.  The field is developing very quickly and we are making important
breakthroughs even before a millihertz mission flies.

When I was approached and asked to write this review, I was glad to accept it
without realising the dimensions of the task. I was told that it should be \emph{
similar to a plenary talk for a wide audience}.  I have a personal problem with
instructions like this.  I remember that when I was nine years old, our Spanish
teacher asked us to summarise a story we had read together in class.  I asked
her to define ``summarise'', because I could easily produce a summary of one,
two or fifty pages, depending on what she was actually expecting from us. She
was confused and I never got a clear answer.  She replied that ``A summary is a
summary and that's it''. On this occasion I am afraid that I have run into the
same snag and I have gone for the many-pages approach, to be sure that any
newcomer will have a good overview of the subject, with relevant references, in
a single document. If the document is too long, please address your complains
to her, because she is solely responsible.

However, I would like to note that I have \emph{not} focused on gathering as
much information as possible from different sources.  I think it is more
interesting for the reader, though harder for the writer, to have a consistent
document. This can be done by introducing the subject step by step, rather than
working out a compendium of citations of the related literature. For instance,
I present results that I have not previously published that will, I hope,
enlighten the reader. Figures that I prepared myself and are not published
elsewhere do not have a reference.

From the point of view of millihertz gravitational wave (GW) missions, as the
reader probably knows, the Laser Interferometer Space Antenna (LISA), see
\cite{Amaro-SeoaneEtAl2017}, is now the official ESA L3 mission, already
entering the phase A.

\newpage

\section*{Glossary}
\footnotesize
\bgroup
\def\arraystretch{1.5}
\begin{tabular}{ c | l }
Acronym/Symbol       & Meaning\\
\hline
$1\,M_\odot$  & 1 Solar Mass = $1.99\times10^{30}\,{\rm kg}$\\
${\cal M}_{\bullet}$ & Mass of super- or massive black hole\\
$1\,{\rm pc}$ & 1 parsec $ \approx 3.09\times 10^{16}\,{\rm m}$\\
1 Myr/Gyr     & One million/billion years\\
AGN           & Active Galactic Nucleus\\
BH            & Black Hole\\
CO            & Compact Object (a white dwarf or a neutron star),\\
              & or a stellar-mass black hole. In general, a collapsed star\\
              & with a mass $\in [1.4,\,10]\,M_{\odot}$ in this work\\
DCO           & Dark Compact Object\\
DF            & Dynamical Friction\\
EMRI          & Extreme Mass Ratio Inspiral\\
GC            & Galactic Centre \\
GPU           & Graphics Processing Unit\\
GW/GWs        & Gravitational Wave/s \\
HB            & Giant stars in the horizontal branch\\
HST           & Hubble Space Telescope\\
IMBH          & Intermediate-Mass Black Hole ($M \in\,[10^2,\,10^5]\,M_\odot$)\\
IMF           & Initial Mass Function\\
IMRI          & Intermediate Mass Ratio Inspiral\\
LISA          & Laser Interferometer Space Antenna\\
LSO           & Last Stable Orbit\\
MBH           & Massive Black Hole ($M \approx 10^6 M_\odot$)\\
MC            & Monte Carlo\\
MW            & Milky Way\\
NB6           & Direct-summation $N-$body6\\
NS            & Neutron Star\\
PN            & post-Newtonian\\
RG            & Red giant\\
RMS           & root mean square\\
SMBH          & Super Massive Black Hole ($M > 10^6 M_\odot$)\\
SNR           & Signal-To-Noise Ratio\\
SPH           & Smoothed Particle Hydronamics\\
TDE           & Tidal Disruption Event\\
UCD           & Ultra-Compact Dwarf Galaxy\\
z             & redshift\\
\end{tabular}
\egroup

\normalsize

\newpage
\renewcommand{\contentsname}{Contents of the Review}
\tableofcontents

\newpage

\listoffigures

\newpage

\section{Massive dark objects in galactic nuclei}

Massive objects allowing no light to escape from them is a concept that goes
back to the 18th century, when John Michell (1724\,--\,1793), an English
natural philosopher and geologist overtook Laplace by 12 years (see the article
\cite{MontgomeryEtAl2009}) with the idea that a very massive object could be
able to stop light escaping from it thanks to its overwhelming gravity. Such an
object would be \emph{black}, that is, invisible, precisely because of the lack
of light \cite{Michell1784,Schaffer79}. I.e., a \emph{dark star}. He wrote:

\begin{quote}
\em{
``If the semi-diameter of a sphere of the same density as the sun is
in the proportion of five hundred to one, and by supposing light to be
attracted by the same force in proportion to its mass with other
bodies, all light emitted from such a body would be made to return
towards it, by its own proper gravity.''}
\end{quote}

That {dark star} would hence not be directly observable, but if it is in a
binary system, one could use the kinematics of a companion star. He even
derived the corresponding radius, which corresponds to exactly the
Schwarzschild radius.

A ``black hole''\epubtkFootnote{This term was first employed by John Archibald
Wheeler (b. 1911)} means the observation of phenomena which are associated
with matter accretion on to it, for we are not able to directly observe it electromagnetically.
Emission of electromagnetic radiation, accretion discs and emerging jets are
some, among many, kinds of evidence we have for the existence of such massive dark
objects, lurking at the centre of galaxies.

On the other hand, spectroscopic and photometric studies of the stellar and gas
dynamics in the inner regions of local spheroidal galaxies and prominent bulges
suggest that nearly all galaxies harbour a central massive
dark object, with a tight relationship between its mass and the mass or the
velocity dispersion of the host galaxy spheroidal component (as we will see
below). Nonetheless, even though we do not have
any direct evidence that such massive dark objects are black holes, alternative
explanations are sorely constrained (see, for instance, \cite{Kormendy03} and also
\cite{Amaro-SeoaneBarrancoBernalRezzolla10} for an exercise on constraining the properties of
scalar fields with the observations in the galactic centre, although the authors conclude that
one needs a mixed configuration with a black hole at the centre).

Super-massive black holes are ensconced at the centre of \emph{active
galaxies}.  What we understand by \emph{active} is a galaxy in which we can
find an important amount of emitted energy which cannot be attributed to its
``normal'' components.  These active galactic nuclei (AGNs) are powered by a
compact region in their centres.

We will embark in the next sections of this review on a study of the dynamics
of stellar systems harbouring a central massive object in order to extract the
dominant physical processes and their parameter dependences, for instance,
dynamical friction and mass segregation, as a precursor to the astrophysics of
extreme-mass ratio inspirals.

\subsection{Active galactic nuclei}

In this section, and to motivate the introduction of the concept of massive
black holes, I give a succinct introduction to active galactic nuclei, but I
refer the reader to the book of Julian H. Krolik on this topic,
\cite{Krolik1999}.

The expression ``active galactic nucleus'' of a galaxy (AGN
henceforth) is referring to the \emph{energetic phenomena occurring at the
central regions of galaxies which cannot be explained in terms of
stars, dust or interstellar gas}. The released energy is emitted across
most of the electromagnetic spectrum, UV, X-rays, as infrared, radio
waves and gamma rays. Such objects have large luminosities ($10^4$ times
that of a typical galaxy) coming from tiny volumes ($\ll 1\, {\rm
pc}^3$); in the case of a typical Seyfert galaxy the luminosity is
about $\sim 10^{11} ~L_{\odot}$ (where $L_{\odot} :=3.83 \cdot 10^{33}$ erg/s
is the luminosity of the sun), whilst for a typical quasar it is
brighter by a factor $100$ or even more; actually they can emit as
much as some thousand galaxies like our Milky-Way. They are therefore the
most powerful objects in the universe. There is a connection between
young galaxies and the creation of active nuclei, because the
luminosity can strongly vary with the red-shift.

In anticipation of something that I will elaborate on later, nowadays one
explains the generation of energy as a product of matter accreting on to a
super-massive black hole in the range of mass $\mbh \sim 10^{\,6-10}\,\msol$
(where $\mbh$ is the black hole mass). In this process, angular momentum
flattens the structure of the in-falling material to a so-called \emph{accretion
disc}.

For some alternative and interesting schemes to that of MBHs, see \cite{GO64}
spinars, \cite{AKO75} for clusters of stellar mass BHs or neutron stars and
\cite{Terlevich89} for \emph{warmers}: massive stars with strong mass-loss
spend a significant amount of their He-burning phase to the left of the ZAMS on
the HR diagram. The ionisation spectrum of a young cluster of massive stars
will be strongly influenced by extremely hot and luminous stars.

It is frequent to observe jets, which may arise from the
accretion disc, although we do not
dispose of direct observations that corroborate this. Accretion
is a very efficient channel for turning matter into energy. Whilst
nuclear fusion reaches only a few percent, accretion can transfer
almost 50\% of the mass-energy of a star into energy.

Being a bit more punctilious, we should say that hallmark for AGNs
is the frequency range of their electro-magnetic emission,
observed from $\lesssim 100$ MHz (as low frequency radio sources) to
$\gtrsim 100$ MeV (which corresponds to $\sim 2 \cdot 10^{22}$ Hz
gamma ray sources). Giant jets give the upper size of manifest
activity $\lesssim 6~{\rm Mpc} \sim 2
\cdot10^{25}$ cm\epubtkFootnote{If we do not take
into account the ionising radiation on intergalactic medium}, and the
lower limit is given by the distance covered by light in the shortest
X-ray variability times, which is $\sim 2\cdot 10^{12}$ cm.

With regard to the size, we can envisage this as a radial distance from the very
centre of the AGN where, ostensibly, a supermassive black hole (SMBH) is
harboured along with the different observed features of the nucleus.
From the centre outwards, we have first a UV ionising
source amidst the optical continuum region. This, in turn, is enclosed by the
emission line clouds and
the compact radio sources and these between another emitting region.

The radiated power at a certain frequency per \emph{dex}\epubtkFootnote{The
number of orders of magnitude between two numbers. This means that if we have
two numbers within one dex, the ratio between the larger and the smaller number
is less than one order of magnitude.} frequency ranges from $\sim 10^{39}$ erg/s
(radio power of the MW) to $\sim 10^{48}$ erg/s, the emitted UV power of the
most powerful, high-redshifted quasars. Such broad frequency and radius ranges
for emission causes us to duly note that they are far out of thermal
equilibrium. This manifests in two ways: first, smaller regions are hotter;
second, components of utterly different temperature can exist together, even
though components differ by one or two orders of magnitude in size.

\subsection{Massive black holes and their possible progenitors}

The quest for the source of the luminosities of $L \approx 10^{12}\, {\rm
L}_{\odot }$ produced on such small scales, jets and other properties of quasars
and other types of active galactic nuclei led in the 1960s and 1970s to thorough
research that pointed to the inkling of ``super-massive central objects''
or ``dark compact objects'' (DCO) harboured at their centres.

These objects were suggested to be the main source of such characteristics
\cite{LyndenBell67,LR71,Hills75}. \cite{LB69} showed that the release of
gravitational binding energy by stellar accretion on to a MBH could be the
primary powerhouse of an AGN \cite{LB69}.  Following the same argument, 13 years
later So{\l}tan related the quasars luminosity to the accretion rate of mass on
to MBHs, so that if we use the number of observed quasars at different
redshifts, we can obtain an integrated energy density \cite{Soltan82}. This
argument strengthened the thought that MBHs are found at the centre of galaxies
and acted in the past as the engines that powered ultraluminous quasars.

In the last decade, observational evidence has been accumulating that strongly
suggests that MBHs are indeed present at the centre of most galaxies with a
significant spheroidal component.  Mostly thanks to the Hubble Space Telescope
(HST), the kinematics of gas or stars in the present-day universe has been
measured in the central parts of tens of nearby galaxies. In almost all cases
\epubtkFootnote{With the possible exception of M33
\cite{GebhardtEtAl01,MFJ01} and M31, see e.g. \cite{BenderKormendyEtAl2005}},
proper modelling of the measured motions requires the presence of a central
compact dark object with a mass of a few $10^{6}$ to $10^{9}\,\msol$, see
\cite{FPPMWJ01,GRH02,PinkneyEtAl03,Kormendy03,GenzelEtAl10} and references therein.
Note,
however, that the conclusion that such an object is indeed a MBH rather than a
cluster of smaller dark objects (like neutron stars, brown dwarfs etc) has only
been reached for a two galaxies. The first one is the Milky Way itself at the
centre of which the case for a $3\mbox{\,--\,}4\times 10^{6}\,\msol$ MBH has
been clinched, mostly through ground-based IR observations of the fast orbital
motions of a few stars (\cite{GhezEtAl05,SchoedelEtAl03} and see
\cite{GenzelEtAl10} for a review).  The second case is NGC4258, which possesses
a central Keplerian gaseous disc with $\rm{H_2O}$ MASER strong sources allowing
high resolution VLBI observations down to 0.16\,pc of the centre
\cite{Miyoshi95,Herrnstein99,MoranEtAl99}.

It is hence largely accepted that the central dark object required to explain
kinematics data in local active and non-active galaxies should be a MBH. The
large number of galaxies surveyed has allowed us to study the demographics of
the MBHs and, in particular, to look for correlations with properties of the
host galaxy. Indeed, a deep link exists between the central MBH and its host
galaxy \cite{KormendyHo2013}, illuminated by the discovery of correlations
between the mass of the MBH, $M_{\bullet}$, and global properties of the
surrounding stellar system, e.g.  the velocity dispersion $\sigma$ of the
spheroid of the galaxy, known as the $M-\sigma$ relation.  In spite of some
progress in recent decades, many fundamental questions remain open.  There is
still no clear evidence of MBH feedback in galaxies, and the low mass end of the
$M-\sigma$ relation is very uncertain.  These facts certainly strike a close
link between the formation of the galaxy and the massive object harboured at its
centre.

It is also important to note that claims of detection of ``intermediate-mass''
black holes (IMBHs) at the center of globular clusters raise the possibility
that these correlations could extend to much smaller systems, see e.g.
\cite{GRH02,GerssenEtAl02}.  The origin of these (I)MBH is still shrouded in
mystery, and many aspects of their interplay with the surrounding stellar
cluster remain to be elucidated.

\subsection{Tidal disruptions}

The centre-most part of a galaxy, its \emph{nucleus} consists of a cluster of a
few $10^7$ to a few $10^8$ stars surrounding the DCO, assumed from now onward to
be a MBH, with a size of a few pc. The nucleus is naturally expected to play a
major role in the interaction between the DCO and the host galaxy, as we
mentioned before. In the nucleus, stellar densities in excess of $10^6\,{\rm
pc}^{-3}$ and relative velocities of order a few 100 to a few $1000\,\kms$ are
reached. In these exceptional conditions, unlike anywhere else in the bulk of
the galaxy, collisional effects come into play.  These include 2-body
relaxation, i.e., mutual gravitational deflections, and genuine contact
collisions between stars.

This means that, if a star happens to pass very close to the MBH, some part of
it or all of it may be torn apart because of the tidal gravity of the central
object. The difference in gravitational forces on points diametrically separated
on the star alter its shape, from its initial approximately spherical
architecture to an ellipsoidal one and, in the end, the star is disrupted. This
radius can be easily calculated as follows. The star gets disrupted whenever the
work exerted over it by the tidal force exceeds its own binding energy, (all
energies are per unit mass). We can hence derive the radius where this happens
easily. The binding energy of the star is

\begin{equation}
  E_{\rm bind} =\alpha\,\frac{Gm_{\star}}{r_{\star}},~\alpha=\frac{3}{5-n},
\label{eq.Ebind}
\end{equation}
In the equation $r_{\star}$
and $m_{\star}$ are the radius and mass of the star, respectively, $G$ the
gravitational constant and $n$ the polytropic index \cite{Chandra42}.

\epubtkImage{.png}{
\begin{figure}[htbp]
\centerline{\includegraphics[width=\textwidth]{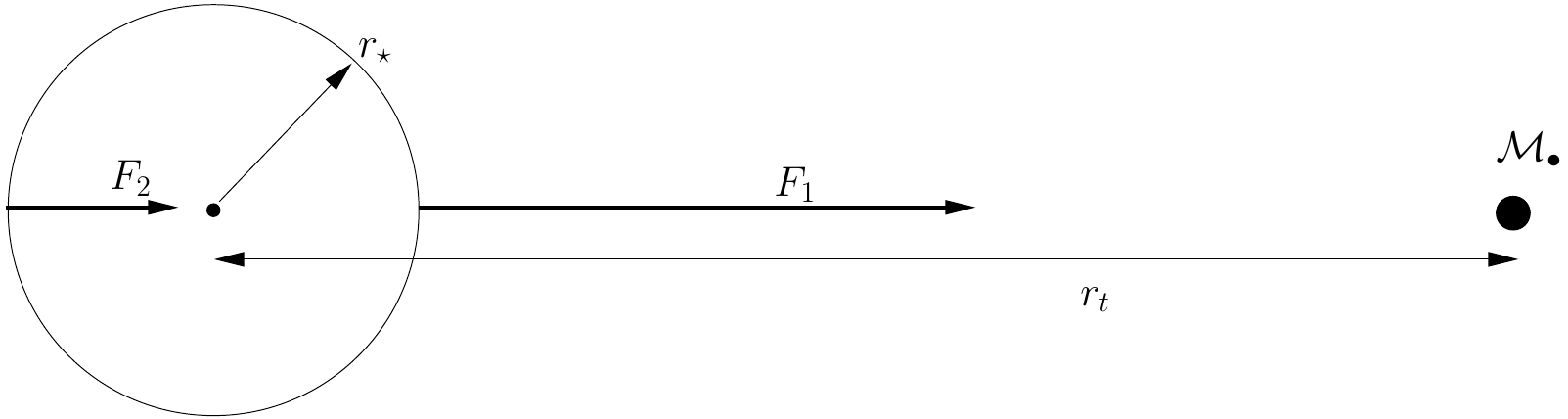}}
\caption{Decomposition of the tidal forces over a star. The tidal radius is
$r_{\rm t}$, ${\cal M}_{\bullet}$ the mass of the MBH and $F_{1}$, $F_{2}$ the
forces exerted on two points of the star which are diametrically separated.}
\label{fig.bind_energy}
\end{figure}
}

We now equate the binding energy of the star to the work exerted over it on two
points diametrically separated,

\begin{equation}
(F_{1}-F_{2})\,2r_{\star}=\alpha\frac{Gm_{\star}}{r_{\star}},
\label{eq.F1_F2}
\end{equation}

\noindent
with

\begin{align}
  F_{1}&=\frac{G{\cal M}_{\bullet}}{(r_{\rm t}-r_{\star})^2},\nonumber \\
  F_{2}&=\frac{G{\cal M}_{\bullet}}{(r_{\rm t}+r_{\star})^2}.
\label{eq.F1_F2_b}
\end{align}

Considering $r_{\star}\ll r_{\rm t}$, we can approximate the expressions:
\begin{align}
\frac{1}{(r_{\rm t}-r_{\star})^2} &\approx  \frac{1}{r_{\rm t}^2}+ \frac{2r_{\star}}{r_{\rm t}^3} \nonumber \\
\frac{1}{(r_{\rm t}+r_{\star})^2} &\approx  \frac{1}{r_{\rm t}^2}- \frac{2r_{\star}}{r_{\rm t}^3};
\label{eq.F1_F2_c}
\end{align}
then,

\begin{equation}
r_{\rm t}=\Bigg[\frac{2}{3} (5-n)
\frac{{\cal M}_{\bullet}}{m_{\star}}\Bigg]^{1/3} r_{\star}.
\label{eq.r_tid_bind}
\end{equation}
For solar-type stars it is (considering a $n=3$ polytrope)

\begin{equation}
r_{\rm t} \simeq 1.4\times 10^{11} \left( \frac {{\cal M}_{\bullet}}{M _{\odot}} \right)^{1/3}~{\rm cm}.
\label{eq.r_tib_bind_b}
\end{equation}

In Figure~\ref{fig.BUUUM} I show the simulation of the tidal disruption of a
star.  The initial spherical architecture of the star is altered after the
passage through periapsis, as we can see in the second snapshot. The third and
fourth panels show the star at much later times. We can see the core of the star
in the last one, idenfitied as a bright, spherical condensate of SPH particles.

In Figure~\ref{fig.rxj1242}, on the left I show a Chandra X-ray image of
J1242-11 with a scale of 40~arcsec on a side. This figure pinpoints one of the
most extreme variability events ever detected in a galaxy.  One plausible
explanation for the extreme brightness of the ROSAT source could be accretion of
stars on to a super-massive black hole. On the right we have its optical
companion piece, obtained with the 1.5~m Danish telescope at ESO/La Silla. The
right circle indicates the position of the Chandra source in the centre of the
brighter galaxy.

\epubtkImage{.png}{
\begin{figure}[htbp]
\centerline{\includegraphics[scale=0.5,clip]{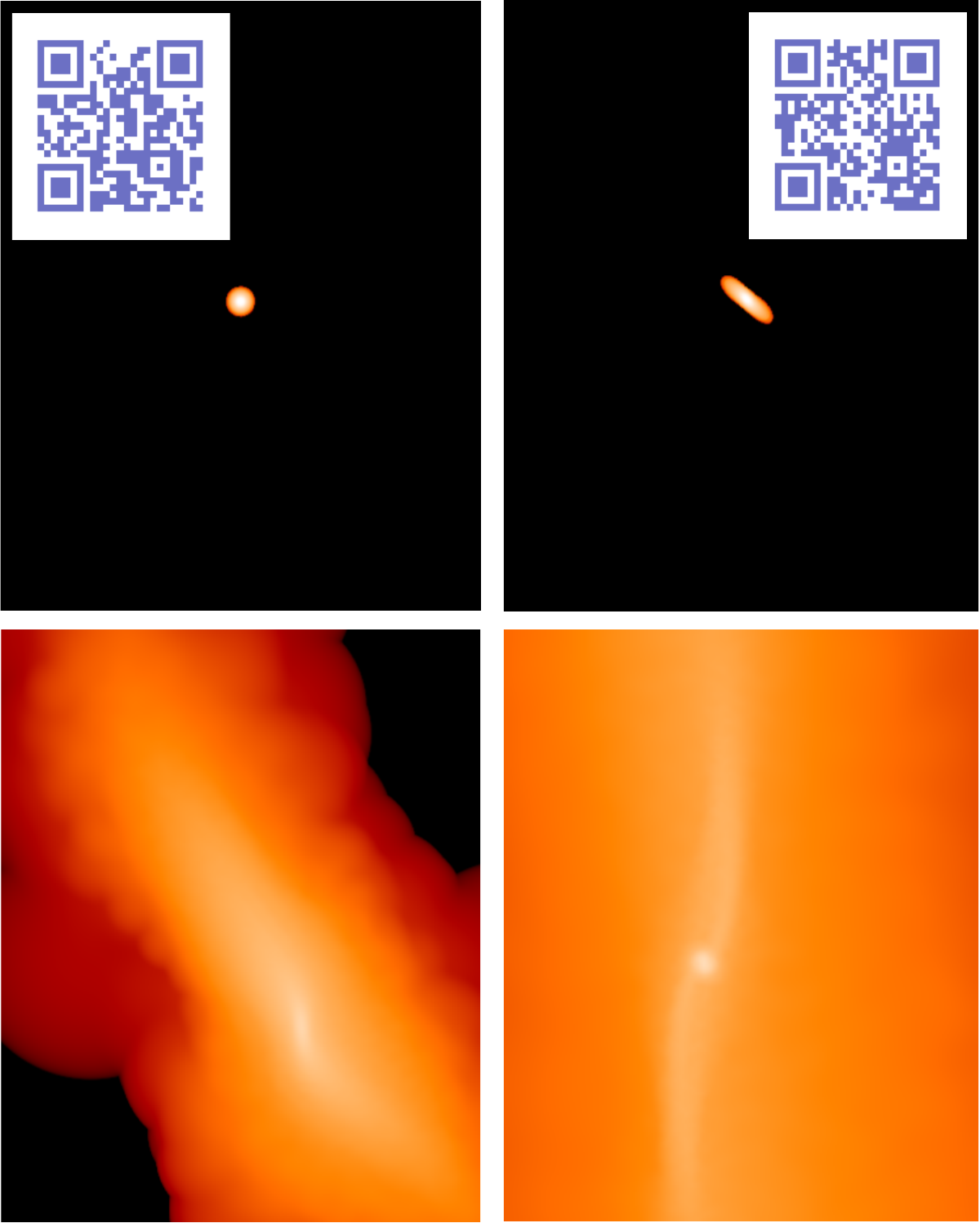}}
\caption{
Four snapshots in the evolution of a tidal disruption of a star. In this
simulation, which I have done with GADGET-2 \cite{Springel2005}, the star is
modelled as a polytrope using $5\cdot 10^4$ particles. The penetration factor, which
is defined to be the ratio between the tidal radius and the distance of
periapsis, has been set to 9. The mass of the MBH is $10^6\,M_{\odot}$ and of
the star $1\,M_{\odot}$. The snapshots correspond to the initial time, and three
later moments in the evolution. The left and right quick response codes link to
two movies in the frame of the star and the general one, which point to
the URLs \url{https://youtu.be/Ryc44v4Eb7I} and
\url{https://youtu.be/uZqXBD8R9Dw}, respectively.
}
\label{fig.BUUUM}
\end{figure}
}


These processes may contribute significantly to the mass of the MBH, see e.g.
\cite{MCD91,FB02b}. Tidal disruptions trigger phases of bright accretion that
may reveal the presence of a MBH in an otherwise quiescent, possibly very
distant, galaxy \cite{Hills75,GezariEtAl03}.

\subsection{Extreme-mass ratio inspirals}

On the other hand, stars can be swallowed whole if they are kicked directly
through the horizon of the MBH (the so-called \emph{direct plunges}) or
gradually inspiral due to the emission of GWs The latter process, known as an
``\emph{Extreme Mass Ratio Inspiral}'' (EMRI) is one of the main objects of
interest for LISA, see
\cite{Amaro-SeoaneEtAl2017,Amaro-SeoaneEtAl2013,Amaro-SeoaneEtAl2012,Amaro-SeoaneEtAl2012b}.
A compact object, such as a star so dense that it will not be disrupted by the
tidal forces of the MBH, (say, a neutron star, a white dwarf or a small
stellar-mass black hole), is able to approach very close to the central MBH.
When the compact object comes very close to the MBH, a large amount of orbital
energy is radiated away, causing the semi-major axis shrink.  This phenomenon
will be repeated thousand of times as the object inspirals until is swallowed
by the central MBH.

The ``doomed'' object spends many orbits around the MBH before it is swallowed.
When doing so, it radiates energy which can be conceptualised as a snapshot
containing detailed information about space-time and \emph{all the physical
parameters} that characterise the binary, the MBH and the stellar-mass black
hole: their masses, spins, inclination and their sky position.  The emitted GWs
encode a map of the space-time. If we can record and decode it, then we will be
able to test the theory that massive dark objects are indeed Kerr black holes as
the theory of general relativity predicts, and not exotic objects such as boson
stars. This would be the ultimate test of general relativity.

The detection of such an EMRI will allow us to do very exciting science: EMRIs
will give us measurements of the masses and spins of BHs to an accuracy which
is beyond that of any other astrophysical technique. Such information will tell
us about cosmic evolution, about the history and growth of MBHs in the nearby
universe, with unprecedent accuracy.

\epubtkImage{.png}{
\begin{figure}[htbp]
\centerline{\includegraphics[width=\textwidth]{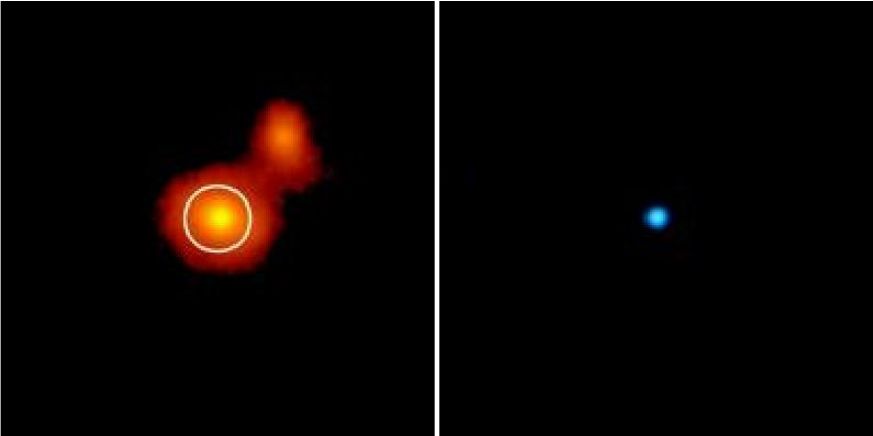}}
\caption{Optical and X-ray images of RX J1242-11.
Credits: (left) ESO/MPE/S.Komossa and (right)
NASA/CXC/MPE, \cite{KomossaEtAl2004}.}
\label{fig.rxj1242}
\end{figure}
}

%

The theoretical study of the structure and evolution of a stellar cluster
(galactic nucleus or globular cluster) harbouring a central MBH started a few
decades years ago. However, due to the complex nature of the problem which
includes many physical processes and span a huge range of time and length
scales, our understanding of such systems is still incomplete and, probably,
subjected to revision. As in many fields of astrophysics, analytical
computations can only been applied to highly idealised situations and only a
very limited variety of numerical methods have been developed so far that can
tackle this problem.  In the next sections I will address the most relevant
astrophysical phenomena for EMRIs and in the last section I give a description
of a few different approaches to study these scenarios with numerical schemes.

\section{GWs as a probe to stellar dynamics and the cosmic growth of SMBHs}

\subsection{GWs and stellar dynamics}

The challenge of detection and characterization of gravitational waves is
strongly coupled with the dynamics of dense stellar systems. This is especially
true in the case of the capture of a compact object by a MBH.

In order to estimate how many events one can expect and what we can assess
about the distribution of parameters of the system, we need to have a very
detailed comprehension of the physics. In this regard, the potential detection
of GWs is an incentive to dive into a singular realm otherwise irrelevant for
the global dynamics of the system.

As mentioned, a harbinger in this respect has been the tidal disruption of stars
as a way to feed the central MBH. About 50\% of the star is bound
to the MBH and accreted on to it, producing an electromagnetic flare which tops
out in the UV/X-rays, emitting a luminosity close to Eddington. Nonetheless, the
complications of accretion are particularly intricate, tight on many different
timescales to the microphysics of gasous processes. Even on local, galactic
accreting objects the complications of accretion are convoluted. It is thus
extremely difficult to understand how to extract very detailed information about
extragalactic MBHs from the flare.  The question of feeding a MBH is a
statistical one. We do not care about individual events to understand the growth
in mass of the hole, but about the statistics of the rates on cosmological
timescales. Obviously, if we tried to understand the individual processes, we
would fail.

As for the fate a compact object which approaches the central MBH, this was
never addressed before we had the incentive of direct detection of gravitational
radiation. Astrophysical objects such as a black hole binary, generate
perturbations in space and time that spread like ripples on a pond.  Such
ripples, known as ``gravitational waves'' or ``gravitational radiation'', travel
at the speed of light, outward from their source. These gravitational waves are
predicted by general relativity, first proposed by Einstein. Measurement of
these gravitational waves give astrophysicists a totally new and different way
of studying the Universe: instead of analysing the propagation and
transformation of particles such as photons, we have direct information from the
fabric of spacetime itself. The information carried by the gravitational
radiation will tell us in exquisite detail about the history, behaviour and
structure of the universe: from the Big Bang to black holes.

When we started to look into this problem, we realised that there were many
questions of stellar dynamics that either did not have an answer or that had not
even been addressed at all.  In this review I will discuss the relaxation
processes that we know to play a major role in the dynamics of this particular
regime. This involves two-body as well as many-body- coherent or non-coherent
relaxation, and relativity.  The list of processes is most likely incomplete,
for there can still be additional, even more complicated processes unknown to
us.  We now have more questions than answers.

\subsection{The mystery of the growth of MBHs}
\label{sec.GrowthMBHs}

One of the most exciting results of modern astronomy is the discovery, mostly
through high-resolution observations of the kinematics of stars and gas, that
most, if not all, nearby bright galaxies harbor a dark, massive, compact object
at their centre, see \cite{FF04,Kormendy04}, \cite{GuelketinEtAl09}, from which
we reproduce their figure in Fig.({\ref{fig.GuelketinEtAl2009_f1}}), and
\cite{KormendyHo2013}. The most spectacular case is our own galaxy, the Milky
Way, see \cite{GenzelEtAl10} for a review.  By tracking and interpreting the
stellar dynamics at the centre of our galaxy, we have the best evidence for the
existence of a massive dark object, very probably a MBH.

\epubtkImage{.png}{
\begin{figure}[htbp]
\centerline{\includegraphics[width=0.6\textwidth]{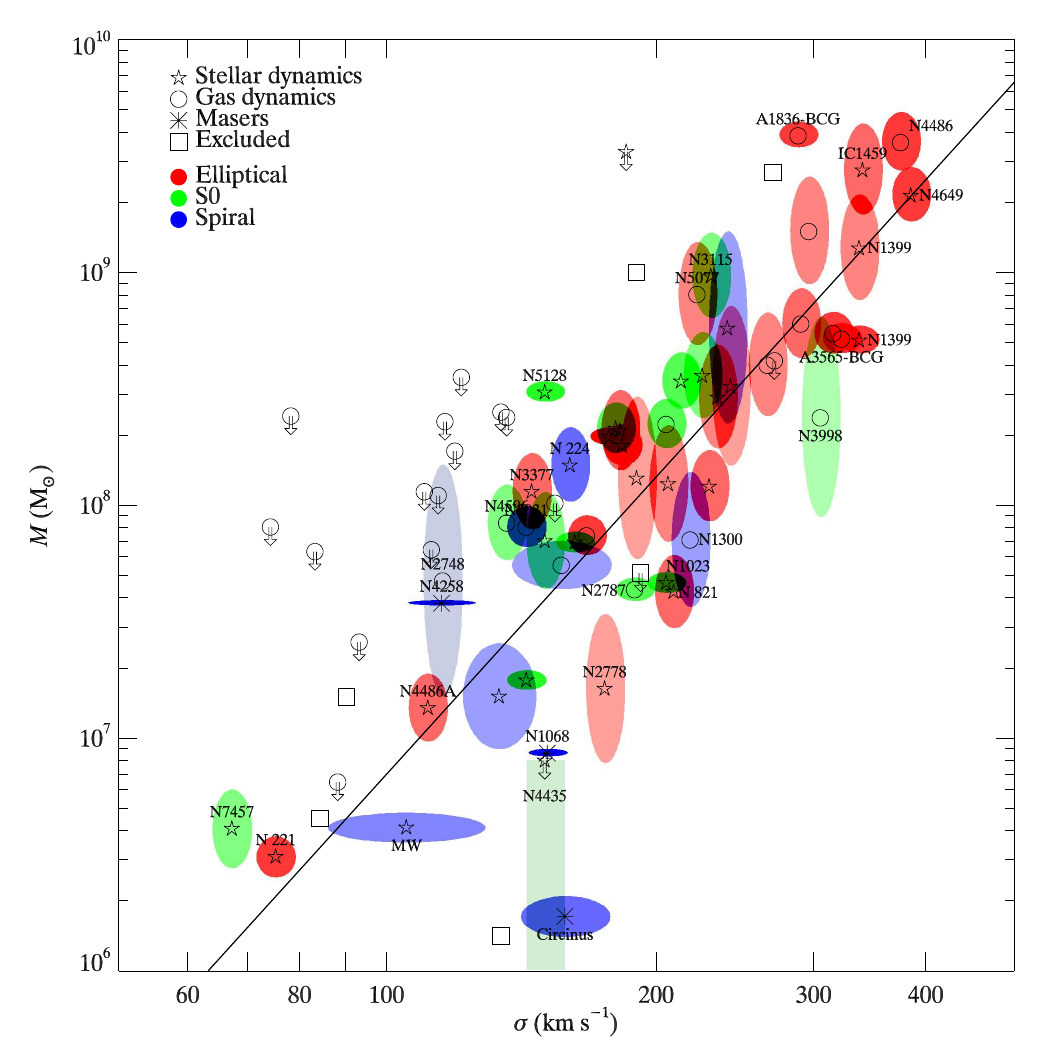}}
\caption{Correlation between the mass of supermassive black holes and the
velocity dispersion of their host galaxies, taken from \cite{GuelketinEtAl09}.}
\label{fig.GuelketinEtAl2009_f1}
\end{figure}
}

The close examination of the Keplerian orbits of the so-called ``S-stars''
(also called S0-stars, where the letter S stands simply for source) has
revealed the nature of the central dark object located at the Galactic Center.
By following one of them, S2 (S02), the mass enclosed by the orbit, a volume
with radius no larger than 6.25 light-hours, was estimated to be about
$3.7\times 10^6\,M_{\odot}$ \cite{SchoedelEtAl03,GhezEtAl03b}.  More recent
data based on many years of observations set the mass of the central MBHs to
$\sim 4 \times 10^{6} \, M_{\odot}$.

Observations of other galaxies indicate that the masses of SMBH can reach a few
billion solar masses ($\Msun$), they correlate tightly with the stellar
properties of the host galaxies (e.g. the velocity dispersion $\sigma$ of galaxy
bulge). The existence of such a SMBH population in the present-day universe is
strongly supported by So{\l}tan's argument that the average mass density of
these SMBHs agrees with expectations from integrated luminosity of quasars
\cite{Soltan82,YT02}.  Claims of detection of ``intermediate-mass'' black holes
(IMBHs, with masses ranging between $100-10^4\,\Msun$) at the center of globular
clusters \cite{GRH02,GerssenEtAl02} raise the possibility that these
correlations extend to much smaller systems, but so far the strongest, although
not conclusive, observational support for the existence of IMBHs are
ultra-luminous X-ray sources \cite{MC03,KongEtAl09}.

Although there is an emerging consensus regarding the growth of large-mass MBHs
thanks to So{\l}tan's argument, MBHs with masses up to $10^7\,M_{\odot}$, such
as our own MBH in the Galactic Centre (with a mass of $\sim
4\times10^6\,M_{\odot}$), are enigmatic. There are many different explains of
their masses: accretion of multiple stars from arbitrary directions, see
\cite{Phinney1989,MT99,SU99,Hills75,Rees1988}, mergers of compact objects such
as stellar-mass black holes and neutron stars, see \cite{QS90}, or IMBHs falling on
to the MBH, \cite{PortegiesZwartEtAl06}. Other more peculiar means are
accretion of dark matter \cite{Ostriker00} or collapse of supermassive stars
\cite{Hara78,ST79,Rees84,Begelman10}. The origin of these low-mass MBHs and,
therefore, the early growth of {\em all} MBHs, remains a conundrum.

The centre-most part of a galaxy, its {\em nucleus}, consists of a
nuclear star cluster of a few millions of stars surrounding the MBH, see
\cite{SchoedelEtAl2014}.  The nucleus is naturally expected to play a major role
in the interaction between the MBH and the host galaxy. In the nucleus, stellar
densities in excess of a million stars per cubic parsec and relative velocities
of the order $\sim$ $100-1000\,\kms$ can be reached. In these conditions, as
mentioned before, collisional effects are important come into play. This is true
except in globular clusters, but one important difference is that the SMBH gives
the central part of the cluster almost a Keplerian potential, and thus very
tricky resonance characteristics. This is one reason it has been difficult to
analyse the stars here.

\subsection{A magnifying glass}

The Laser Interferometer Space Antenna, see in particular the document prepared
in response to the call for missions for the L3 slot in the Cosmic Vision
Programme, \cite{Amaro-SeoaneEtAl2017}, but also
\cite{Danzmann00,Amaro-SeoaneEtAl2012,Amaro-SeoaneEtAl2012b}, will be our
reference point throughout my review.  LISA consists of three spacecraft
arranged in an equilateral triangle with sides of length 2.5 million kilometre.
LISA  will scan the entire sky and covers a band from below $10^{-4}\,$Hz to
above $10^{-1}\,$Hz.  In this frequency band, the Universe is populated by
strong sources of GWs such as binaries of supermassive black holes merging in
the centre of galaxies, massive black holes ``swallowing'' entirely small
compact objects like stellar-mass black holes, neutron stars and white dwarfs.
The information is encoded in the gravitational waves: the history of galaxies
and black holes, the physics of dense matter and stellar remnants like
stellar-mass black holes, as well as general relativity and the behaviour of
space and time itself. Chinese mission study options, such as Taiji,
\cite{BenderEtAl05,GongEtAl11,GongEtAl2015,HuangEtAl2017} will also be able to
catch these systems with good signal-to-noise ratios.

In any case, a key property of GW astrophysics is the fact that GWs interact
only very weakly with matter, except for high-z. The observations we will make
with LISA will not suffer any of the usual problems in astrophysics -
absorption, scattering, or obscuration.  This is what makes LISA-like missions
such as LISA or Taiji unique. It is not ``merely'' a test of general
relativity; these missions would be able to corroborate the underlying theory
of the nature of the central dark objects which we now observe in most
galaxies. We will get direct information from the heart of the densest stellar
systems in the Universe: galactic nuclei, nuclear stellar clusters and globular
clusters.  The LISA mission technology has been successfully tested with the
LISA Pathfinder\epubtkFootnote{\url{http://sci.esa.int/lisapf}} mission, an
ESA-led mission with a contribution from NASA, launched in 2015 from Kourou,
French Guiana.  In Fig.(\ref{fig.Pathfinder_Fig1}) I reproduce figure 1 of
\cite{ArmanoEtAl2018}.  This publication has remarkably improved the previous
results of \cite{ArmanoEtAl2016}, which showed that LISA Pathfinder has
satisfied the mission requirements by factors ranging from 10-1000 depending on
the frequency range, achieving a sub-Femto-$g$ in free fall
\cite{ArmanoEtAl2016}. Indeed, the results published in 2018 show that,
actually, LISA Pathfinder has exceeded the requirements for LISA by more than
a factor of two over the whole observation band (down to 20 $\mu$Hz).

\epubtkImage{.png}{
\begin{figure}[htbp]
\centerline{\includegraphics[width=0.8\textwidth]{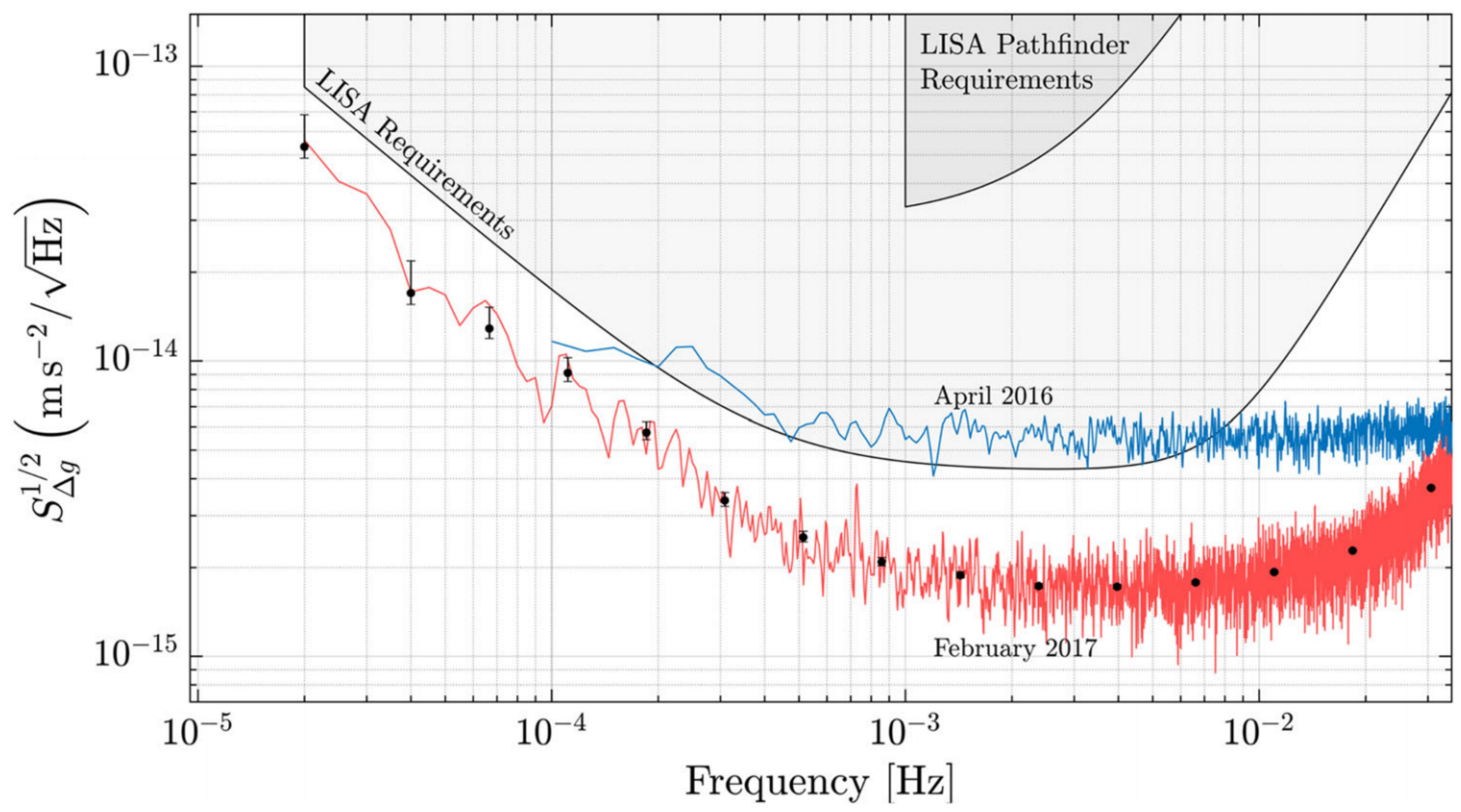}}
\caption{
Amplitude spectral density of LISA Pathfinder as compared to the previous
publication of the LPF group, \cite{ArmanoEtAl2016}, which is the curve in
blue.  The data are compare with LISA requirements, as presented in
\cite{Amaro-SeoaneEtAl2017}.  We can see that LISA Pathfinder exceeds the
requirements for key technologies for LISA over a factor of two over the entire observation
band.
}
\label{fig.Pathfinder_Fig1}
\end{figure}
}

For the full success of a mission such as LISA, it is important that we {\em
understand} the systems that we expect to observe. A deep
theoretical comprehension of the sources which will populate LISA's field of
view is important to achieve its main goals.

Whilst main-sequence stars are tidally disrupted when approaching the central
MBH, compact objects (stellar-mass black holes, neutron stars, and white dwarfs)
slowly spiral into the MBH and are swallowed whole after some $\sim 10^5$
orbits in the LISA band. At the closest approach to the MBH, the system emits a
burst of GWs which contains information about spacetime and the masses and
spins of the system. We can envisage each such burst as a snapshot of the
system.  This is what makes EMRIs so appealing: a set of $\sim 10^5$ bursts of
GWs radiated by \emph{one} system will tell us with the utmost accuracy about
the system itself, it will test general relativity, it will tell us about the
distribution of dark objects in galactic nuclei and globular clusters and,
thus, we will have a new understanding of the physics of the process.  New
phenomena, unknown and unanticipated, are likely to be discovered.

If the central MBH has a mass larger than $10^7\,M_{\odot}$, then the signal of
an inspiraling stellar-mass black hole, even in its last stable orbit (LSO) will
have a frequency too low for detection. On the other hand, if it is less
massive than $10^4\,M_{\odot}$, the signal will also be quite weak unless the
source is very close. This is why one usually assumes that the mass range of
MBHs of interest in the search of EMRIs for LISA is between
$[10^7,\,10^4]\,M_{\odot}$. Nonetheless, if the MBH is rotating rapidly, then
even if it has a mass larger than $10^7\,M_{\odot}$, the LSO will be closer to
the MBH and thus, even at a higher frequency, the system should be detectable.
This would push the total mass to a few $\sim 10^7\,M_{\odot}$.

For a binary of a MBH and a stellar-mass black hole to be in the LISA band, it has
to have a frequency of between roughly $10^{-5}$ and 1 Hz. The emission of
GWs is more efficient as they approach the LSO, so that LISA will detect the
sources when they are close to the LSO. The total mass required to observe
systems with frequencies between $0.1$ Hz and $10^{-4}$ is of $10^4 -
10^7\,M_{\odot}$. For masses larger than $10^7\,M_{\odot}$ the frequencies
close to the LSO will be too low, so that their detection will be very
difficult. On the other hand, for a total mass of less than $10^3\,M_{\odot}$
we could in principal detect them at an early stage, but then the amplitude of
the GW would be rather low.

On top of this, the measurement of the emitted GWs will give us very
detailed information about the spin of the central MBH. With current
techniques, we can only hope to measure MBH spin through X-ray observations of
Fe~K$\alpha$ profiles, but the numerous uncertainties of this technique may
disguise the real value. Moreover, such observations can only rarely be made.

This means that LISA will scrutinize exactly the range of masses fundamental to
the understanding of the origin and growth of supermassive black holes.  By
extracting the information encoded in the GWs of this scenario, we can determine
the redshifted mass and spin of the central MBH with an astonishing relative
precision. Additionally, the mass of the compact object which falls into the MBH
and the eccentricity of the orbit will be recovered from the gravitational
radiation with a tiny fractional accuracy. All this means that LISA will not be
``just'' the ultimate test of general relativity, but an exquisite probe of the
spins and range of masses of interest for theoretical and observational
astrophysics and cosmology.

\subsubsection{A problem of $\sim$ 10 orders of magnitude}

For the particular problem of how does a compact object end up being an extreme-mass
ratio inspiral, we have to study very different astrophysical regimes, spanning over
many orders of magnitude.

\paragraph{Galactic or Cosmological dynamics}

In Figure~\ref{fig.ThreeRealStellDyn} I depict the three different realms of
stellar dynamics of relevance for the problem of EMRIs.  At the largest scale
exists the galaxy, with a size of a few kiloparsecs. Just as a point of
reference, the gravitational radius of a MBH of $10^6\,M_{\odot} \sim 5\cdot
10^{-8}$ pc. The relaxation time, $t_{\rm rlx}$ which I will introduce with
more detail ahead, is a timescale which can be envisaged as the required time
for the stars to exchange energy and angular momentum between them: it is
the time that the stars need to ``see'' each other individually and not only
the average, background stellar potential of the whole stellar system. For the
galaxy, $t_{\rm rlx}$ is larger than the Hubble time, which means that, on
average, it has no influence on the galaxy at all. A test star will only feel
the mean potential of the rest of the stars and it will never exchange either
energy or angular momentum with any other star. The system is ``collisionless'', meaning
that two-body interactions can be neglected. This defines the realm of stellar
galactic dynamics, the one investigated in Cosmological simulations using, e.g.,
$N$-body integrators. Since we do not have to take into account the strong
interactions between stars, one can easily simulate ten billion particles with
these integrators.

\paragraph{Cluster dynamics}

If we zoom in by typically a factor of $10^3$, we enter the (mostly Newtonian)
stellar dynamics of galactic nuclei. There, $t_{\rm rlx} \sim 10^8 - 10^{10}$
yrs. In this realm stars do feel the graininess of the stellar potential.  The
closer we get to the central MBH, the higher $\sigma$ will be, if the system is
in centrifugal equilibrium; the stars have to orbit around the MBH faster.  In
particular, S2 (or S02), one of the S-stars (S0-stars) for which we have enough
data to reconstruct the orbit to a very high level of confidence -- as we saw in
the previous section -- has been observed to move with a velocity of $15\cdot
10^{3}\,{\rm km\,s}^{-1}$.  Typically, $t_{\rm rlx}$ is (on occasion
\emph{much}) shorter than the age of the system, of a few $\sim 10^{8} -
10^{10}$ yrs.  For these kind of systems one has to take into account
relaxation, exchange of energy and angular momentum between stars. The system is
``collisional''. When we have to take into account this in the numerical
simulations, the result is that we cannot simulate with $N$-body integrators
more than some thousands of stars on a regular computer. To get to more
realistic particle numbers one has to resort to many computers operating in
parallel, special-purpose hardware or the graphic processor units. I will
discuss this later.

\paragraph{Relativistic stellar dynamics}

Last, in the right panel of figure~\ref{fig.ThreeRealStellDyn}, we have the
relativistic regime of stellar dynamics when we enlarge the previous by a factor
of ten million. There the role of relativistic effects is of paramount
importance for the evolution of the system. In this zone, generally, \emph{there
are no stars}. Even at the densities which characterise a galactic nucleus, the
probability of having a star in such a tiny volume is extremely small. Moreover,
even if we had a significantly larger volume, or a much higher density for
the galactic nucleus, so that we had a few stars close to the MBH, these would
quickly merge with the MBH due to the emission of GWs, which is what defines an
EMRI. But they do it \emph{too} fast.  These systems can be collisional or
collisionless, depending on how many stars we have at a given time. If they are
there, they will exchange energy and angular momentum between them.
Nevertheless, relaxation is not well-defined in this regime.

The key point here is how to replenish that area, so that there are other stars
replacing those which merge quickly with the central MBH. On average, there are
\emph{zero} stars. As a matter of fact, and in general, for the general study of
the stellar dynamics of galactic nuclei, the role of this last realm is
negligible. One does not have to bother with the effects of GR; most, if not
all, stars are on a Newtonian regime. The impact on the dynamics of galactic
nuclei is zero. It is impressive that this last region dominated by the effects
of GR has an effect worth studying at all. But, as we will see ahead, the
encoded information that one can recover from the detection of an EMRI about its
surrounding dynamical system is dramatic.  If we want to address this problem,
we need to cope with a range of scales that spans over seven orders of magnitude
when understanding the role of the dynamics of galactic nuclei in relativistic
dynamics, and of ten orders of magnitude in the big picture.

\epubtkImage{.png}{
\begin{figure}[htbp]
\centerline{\includegraphics[width=0.7\textwidth]{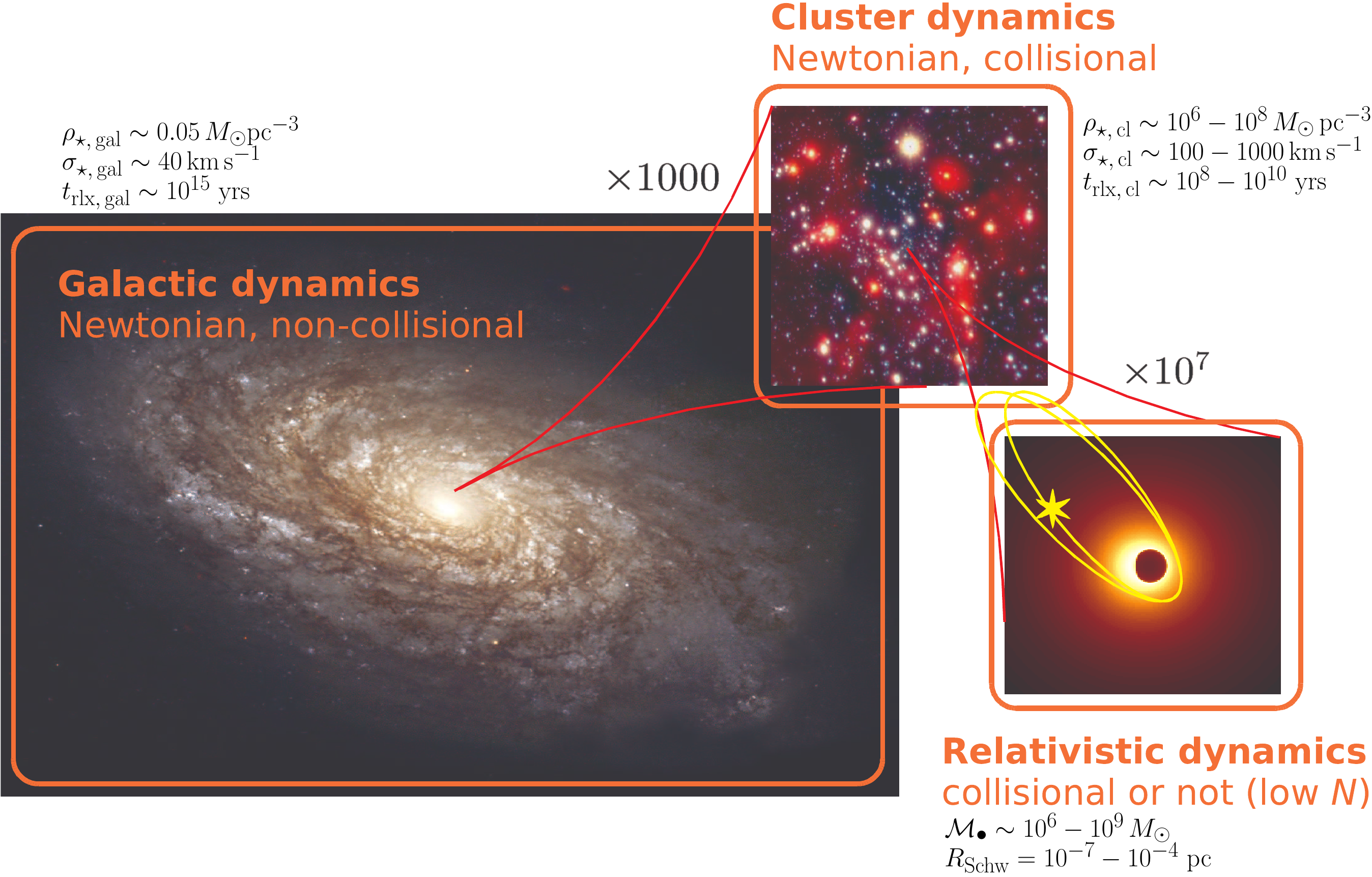}}
\caption{On the left, and with the largest scale, the galaxy has an average density of
stars of about $0.05\,M_{\odot}{\rm pc}^{-3}$.  The velocity dispersion is
$\sim 40\,{\rm km\,s}^{-1}$. From these quantities one can infer that the
relaxation time in the vicinity of our Sun is $t_{\rm rlx} \sim 10^{15}$ yrs.
The upper panel shows the galactic nucleus that such a galaxy has. A typical
size for it is $\sim\,1$ pc, the stellar density ranges between $10^6
-10^8\,M_{\odot}\,{\rm pc}^{-3}$ and the velocity dispersion is of $\sigma \sim
100 - 1000\,{\rm km\,s}^{-1}$. In this region, $t_{\rm rlx} \sim 10^{8} -
10^{10}$ yrs. In the last panel, we have that the dynamics of the system is
dominated by General Relativity.
As a reference point, the Schwarschild radius of a $10^6\,M_{\odot}$
($10^9\,M_{\odot}$) is $10^{-7}$ pc ($10^{-4}$ pc).
}
\label{fig.ThreeRealStellDyn}
\end{figure}
}

\subsection{How stars distribute around MBHs in galactic nuclei}

In Figure~\ref{fig.Merritt06_Fig2} I show data constrainted by electromagnetic
measurements. One of the very first questions one has to address when trying to
understand the stellar dynamics around a MBH is \emph{how many stars are there
and how do they distribute around it}? Unfortunately there are very few
observations for this because we are interested in nuclei that harbour
lower-mass MBHs, i.e. with masses ranging $10^4$ and $10^6\,M_{\odot}$, so that
they therefore have a small radius of influence $r_{\rm inf}$ and, thus, they
are observationally very difficult to resolve. Currently there are only a very
few galaxies that are both in the range of GW frequencies interesting to us and
that have a resolved $r_{\rm inf}$. For these we have information on how bound
stars that can become EMRIs are distributed around the central MBH. Obviously,
the Milky Way (MW) is one of these galaxies. In figure~\ref{fig.Merritt06_Fig2}
the stellar density profile of the MW is displayed. We see that it goes up to at
least $10^8\,M_{\odot}/{\rm pc}^3$ in the inner regions. This number has been
calculated by assuming a population of stars; one has to deproject the
observation, because we are only seeing a few of the total amount of stars, the
brightest ones. One assumes that the observed stars are tracing an underlying
population invisible to us. This requires a considerable amount of modelling to
obtain the final results. These are uncertain by, at most, a factor of ten.  In
the same figure we have another nucleus, M32, which should be harbouring a MBH
with a mass similar to the one located in the GC. The density profile happens to
be similar to the one corresponding to the GC. Whether this is a coincidence or
something deeper is not clear. In any case, and to \emph{first order of
approximation}, we can state that once we know the mass of the MBH, we know the
way stars distribute around it. Later the relevance of this point will be
obvious to the lector.

\epubtkImage{.png}{
\begin{figure}[htbp]
\centerline{\includegraphics[width=0.6\textwidth]{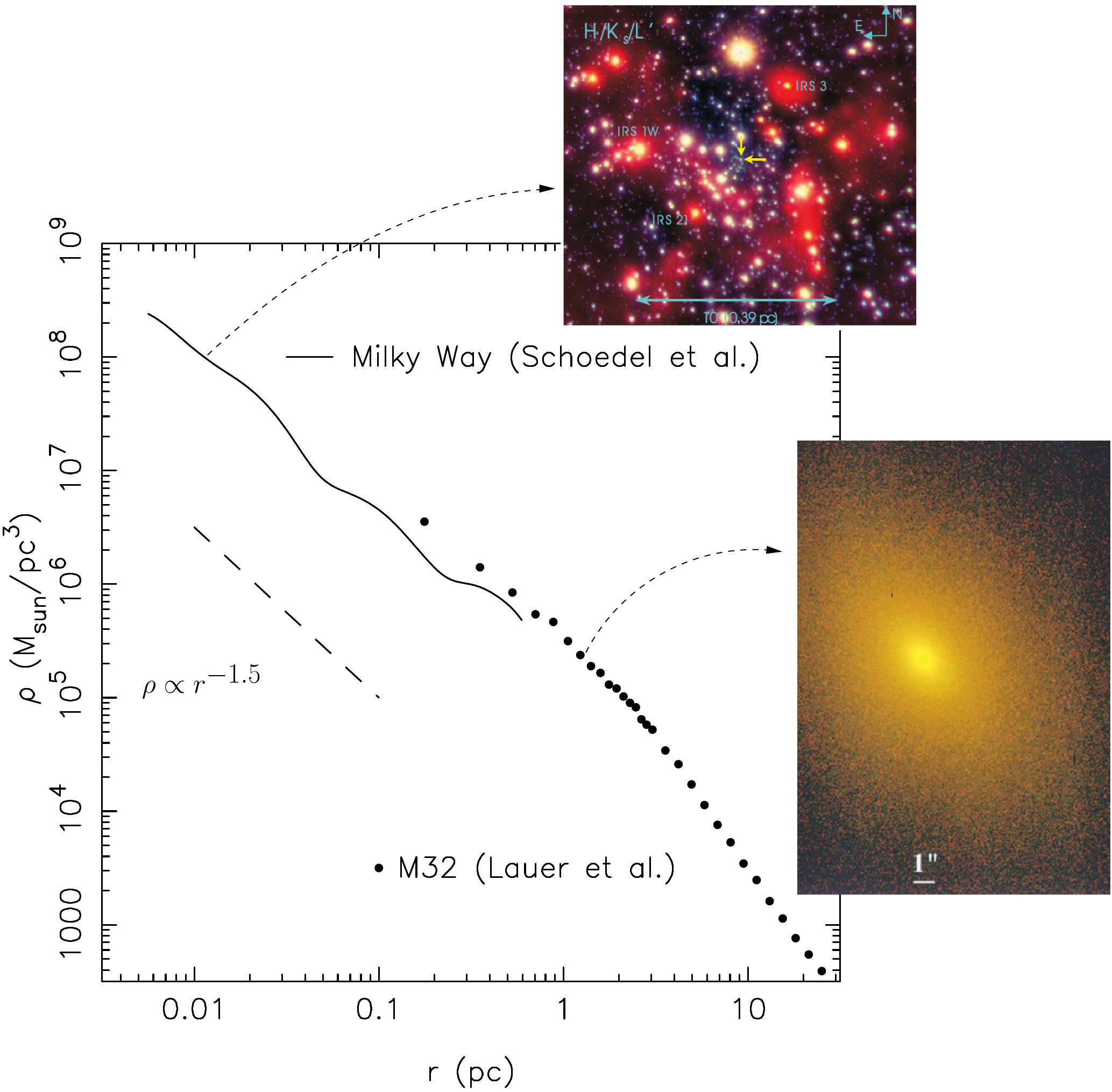}}
\caption{Density profile for the GC of the MW and for M32, both with
MBHs of masses $3\times 10^6\,M_{\odot}$ and influence radii $\sim 3$ pc. The dashed
curve on the very left corresponds to a slope of $\rho \propto r^{-3/2}$,
adapted from \cite{Merritt06,Schoedel02,Lauer98}.}
\label{fig.Merritt06_Fig2}
\end{figure}
}

\section{A taxonomy of orbits in galactic nuclei}
\label{sec.taxonomy}

Before we address the physics and event rate estimates of EMRIs, it is crucial
that we have a good understanding of the kind of orbits that we might expect in
the enviroments natural to COs in dense systems around a MBH.  An important
factor in understanding how a star can become an EMRI is the shape and evolution
of its orbit. In this section I will address these two aspects. First, we will
\emph{not} take into account the role of relaxation. The stellar potential in
which our test star $m_{\bullet}$ is moving is completely smooth.  For any
purpose, the test star will not feel any individual star, but a background
potential.

\subsection{Spherical potentials}

Consider now Figure~\ref{fig.rosettes}; there we have two orbits which differ in
their eccentricity. The rosettes are characterised by their energy and angular
momentum. Since the test stars do not suffer any individual gravitational tug
from the stellar system (at least not on a noticeable timescale), the orbital
elements are kept constant.  The periapsis\epubtkFootnote{In the related
literature there exist other terms to refer to the distance of maximum or
minimum approach to a black hole; namely \emph{peribarathron} and
\emph{apobarathron}, respectively.  There seems to be a confusion and wrong use
of the later. I discuss this in section~\ref{sec.barathron}.} is fixed because
the angular momentum is conserved, so that the test star will never come
arbitrarily close to the central MBH. In order to achieve anything interesting,
one needs to perturbate the system.

A different situation, however, is when the orbit of the test star is
\emph{within} the $R_{\rm infl}$ of the MBH. In this case, the orbits look more
and more like Keplerian ellipses, unless one gets \emph{very} close to the
central MBH, so that we get relativistic precession. In
Figure~\ref{fig.NewtonianPeriShift1} we have an ellipse which precesses with
time. This is neither the relativistic precession nor an advance, but a purely
Newtonian perihelion (periapsis) retard, counterclockwise. The timescale for it
is

\begin{equation}
  T_{\rm New,\,PS} \approx \frac{{\cal M}_{\bullet}}{M_{\star}(a)}\,P_{\rm orb}
                   \approx \frac{R_{\rm infl}}{a}P_{\rm orb}
\end{equation}
In this last equation, $M_{\star}(a)$ is the amount of stellar mass encompassed
within the orbit. The Newtonian periapsis retard is the result of the fact that
we do not have a perfect Keplerian orbit because we do not have a point mass,
but an \emph{extended} mass distribution. As an exercise, we can compare the last
equation to the relativistic periapsis \emph{advance} (in order of magnitude),

\begin{equation}
T_{\rm Rel,\,PS} \approx \frac{R_{\rm peri}}{R_{\rm Schw}}\,P_{\rm orb}
\end{equation}
This equation is only relevant for orbits whose periapsis is very small, whilst the
later one is only important for relatively extended orbits (because $M_{\star}(a)$ is
larger)

\epubtkImage{.png}{
\begin{figure}[htbp]
\centerline{\includegraphics[width=\textwidth]{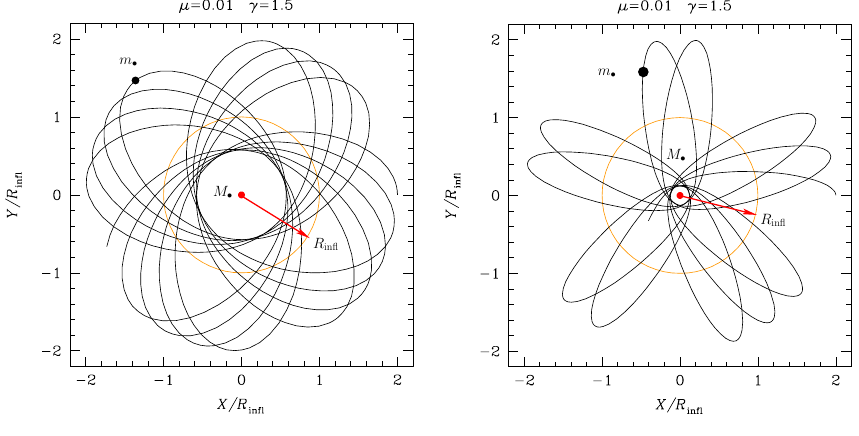}}
\caption{Projection in the X-Y plane of the evolution of two test star orbits in a
stellar system without relaxation. The central, orange point represents the
position of the MBH, the black dots on the orbits the position of the test stars and
the red arrow delimits the influence radius $R_{\rm infl}$ of the MBH. The
right panel represents a case with a larger eccentricity. The orbits extend further
than the $R_{\rm infl}$.}
\label{fig.rosettes}
\end{figure}
}

\epubtkImage{.png}{
\begin{figure}[htbp]
\centerline{\includegraphics[width=\textwidth]{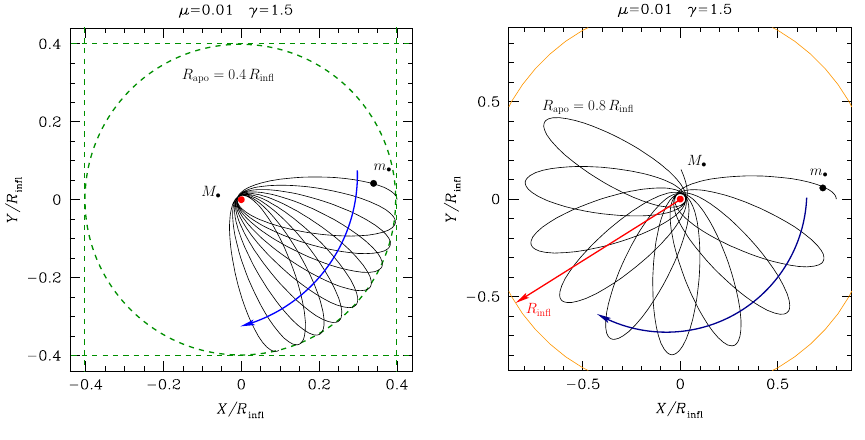}}
   \caption{Same as Figure~\ref{fig.rosettes} for an apoapsis $R_{\rm apo}=0.4,\,0.8\,R_{\rm infl}$ and
     a velocity of the CO of $0.2\,V_{\rm circ}$, with $V_{\rm circ}$ the circular velocity.}
\label{fig.NewtonianPeriShift1}
\end{figure}
}

\epubtkImage{.png}{
\begin{figure}[htbp]
\centerline{\includegraphics[width=\textwidth]{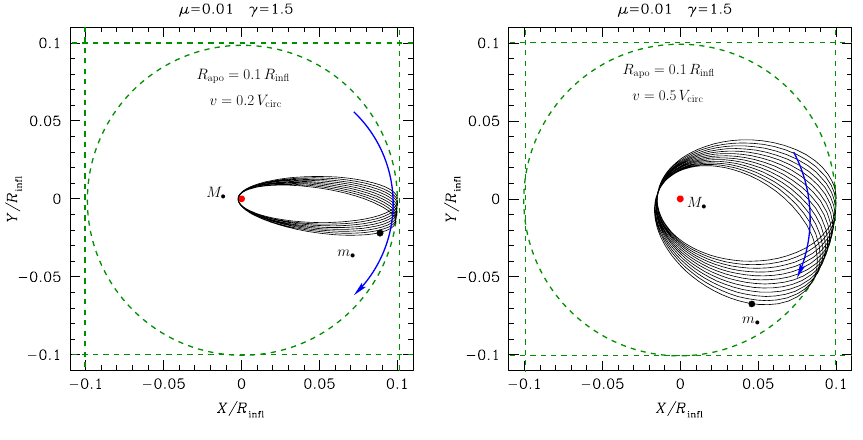}}
   \caption{Same as Figures~\ref{fig.rosettes} and \ref{fig.NewtonianPeriShift1} for different values of the
     apoapsis radius and velocity of the CO.}
\label{fig.NewtonianPeriShift2}
\end{figure}
}

\epubtkImage{.png}{
\begin{figure}[htbp]
\centerline{\includegraphics[width=0.7\textwidth]{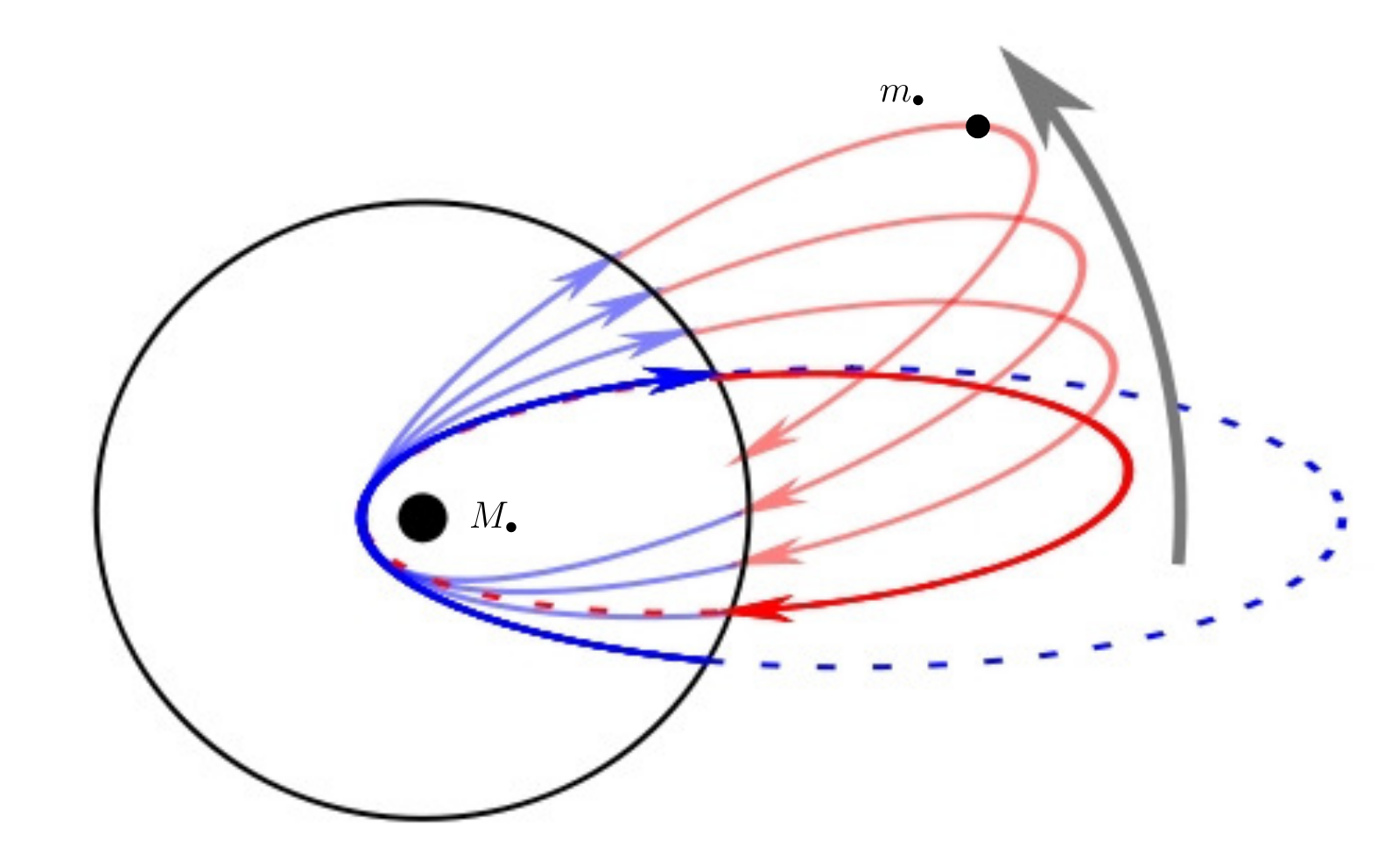}}
\caption{In the Newtonian case we have an
extended mass distribution, so tha the star feels more mass far away than closer to the centre.
When the star traverses the ``sphere'', the trajectory abruptly changes and becomes a smaller ellipse
Thus, the object goes back to the centre faster; the orbit precesses in the opposite direction to the orbital one
In the relativistic case the
kinetic energy of the star increases its gravitational mass when it's close to the centre: The effective attraction is more efficient and the trajectory is more curved towards the centre.}
\label{fig.preces_newt_xfig}
\end{figure}
}

\subsection{Non-spherical potentials}

The most general case is the triaxial potential, in which we still have symmetry
but neither spherical nor axial-symmetry, it is a general ellipsoidal
configuration. The angular momentum has no component conserved. This, obviously,
allows orbits to get ``as close as they want'' to the centre. Not {\em all}
orbits will, but there are specific families of orbits which, if one waits long
enough, will get arbitrarily close to the centre. This is evidently very
relevant for our study.  These orbits are refered to as \emph{centrophilic}
orbits for clear reasons. Studies of models of triaxial galaxies have found that there is a
significant fraction of such orbits even very close to the central MBH. At
distances as short as $r<R_{\rm infl}$, within the sphere of influence, some
models have as many as 20\% of stars that are on centrophilic orbits. One should
nevertheless bear in mind that these are \emph{models}, not corroborated by
direct observations of galaxies. They depend on a number of set-up parameters
which will result in strong fluctuations of the final result: the true number
could be between 0\,--\,20\% according to these models. Therefore,
unfortunately we do not know what the real implications
are for observed nuclei, since it is not well-constrained. Of course,
one can resort to (non-collisional) $N$-body simulations to study the merger of
two galaxies to see in the resulting product how many of these orbits one can
get.

As for the implications of the detection rates of EMRIs, this could have a huge
impact, but the problem should probably be revisited due to
the enormous difficulties that force us to make broad simplifications. For
instance, we should explore the behaviour of the potential \emph{very} close to
the MBH because, by definition, at some point the potential is completely
dominated by the MBH and, thus, spherically symmetric.  The only realistic hope
here are those stars that typically are on orbits with semi-major axis much
larger than the radii of interest to us, so that even if they spend most of
the time very far away from the MBH, they will be set on a centrophilic orbit
due to the triaxiality of the system, but it is unclear whether these can
contribute significantly to the local density around the MBH. As an example of
the kind of orbits one can get in a triaxial galactic nucleus, in
Figure~\ref{fig.PoonMerrittFig1} I show some representative examples of {\em
centrophobic} orbits from \cite{PoonMerritt01} (cases b, c, d, e). This means
that the stars never reach the centre. The lack of conservation of the angular momentum can set
stars on either centrophilic orbits or, alternatively, on centrophobic orbits. These can
be envisaged as a generalisation of rosette orbits. Nevertheless, since we are
interested in EMRIs, we will focus on centrophilic orbits and leave the further
description of centrophobic orbits aside.  I refer the interested reader
to the work by \cite{Holley-BockelmannEtAl2001,Holley-BockelmannEtAl2002}, and also
to the more recent one by \cite{MerrittVasiliev11}.

We have two different kinds of centrophilic orbits: (i) pyramid or box orbits. These
are still regular but a star on such an orbit can reach arbitrarily small distances
in its periapsis; (ii) stochastic orbits, which also come arbitrarily close to the centre.
The probability for an orbit to get within a distance $d$ from the central
MBH, the very centre of the potential, is proportional to $d$.

\epubtkImage{.png}{
\begin{figure}[htbp]
\centerline{\includegraphics[width=\textwidth]{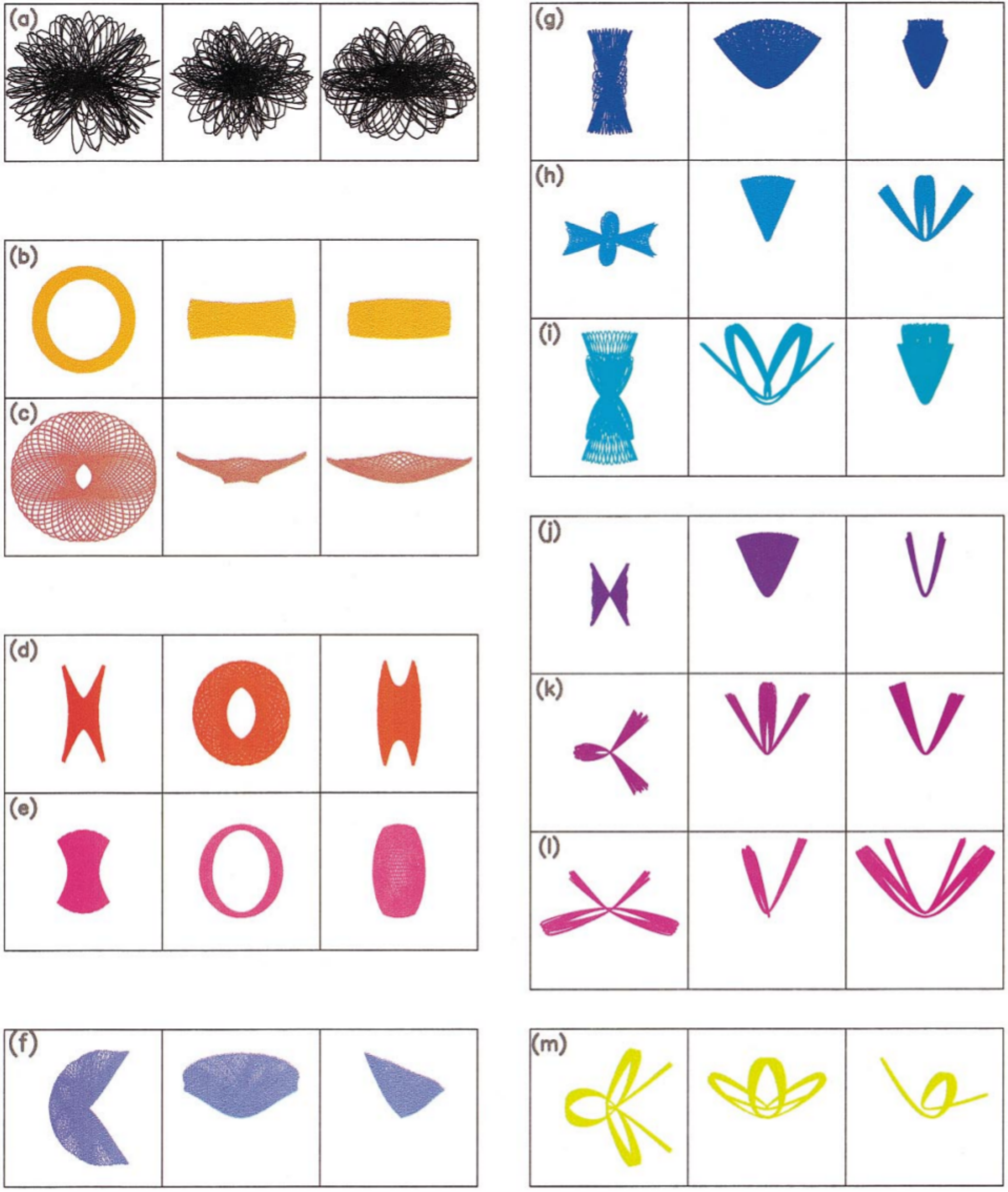}}
\caption{From the left to the right and from the top to the bottom, we
  have stochastic orbits, short-axis tube orbit, saucer orbit, a resonant short-axis tube,
  inner long-axis tube, long-axis tube, resonant, pyramid, resonant pyramid,
  resonant pyramid orbit, banana orbit, 2 : 3 : 4 resonant
        banana, 3 : 4 : 6 resonant banana, and a 6 : 7 : 8 resonant orbit. This figure is
        Figure~1 of \cite{PoonMerritt01}. We note that the projections of
  ``tube'', centrophilic stellar orbits around a MBH in triaxial
nuclei will look aligned in one or another plane depending on the potential.}
\label{fig.PoonMerrittFig1}
\end{figure}
}

This is non-intuitive. If you have a target with a mass and you shoot a
projectile from random directions, the probability of coming within a distance
$d$ of the target $R_{\rm p}<d$ is proportional to $d$ itself and not $d^2$
(which would have been the case for a totally random experiment, without
focusing).  In the case of a star on an orbit towards the MBH, the number of
times you have to ``throw'' it to get to a periapsis distance closer than $d$
is, $N_{\rm pass}\,(R_{\rm p}<d) \propto d$.  The reason for this is that our
target is a particular one and influences the projectile through a process
called gravitational focusing.  The projectile, the star, is attracted by the
target, the MBH.

Something to also bear in mind is that all of these simulations are limited by a
particular resolution, which is still far from being close to reality, so that
we are not in the position of extrapolating these results to the distances where
the star will be captured by the MBH and become and EMRI.

\section{Two-body relaxation in galactic nuclei}

\subsection{Introduction}

We are now back to a spherical system world, in which orbits such as those in the
previous section do not exist. Therefore, one needs an additional
factor to bring stars close to the MBH. As I have already discussed before, a possibility, is to have
a source of exchange of energy and angular momentum. We use and abuse the term
collisional to refer to
any effect not present in a smooth, static potential, including secular effects. Among these, standard two-body
relaxation excells not due to its relevance of contributing to EMRI sources,
but due to the fact that this is the best-studied effect; namely the exchange
of energy and angular momentum between stars due to gravitational interactions.

Another possibility is physical collisions.\epubtkFootnote{The terminology is
somehow, and as forewarned, misleading; whilst in general we refer to
``collisional'' to any effect leading to exchange of energy and angular momentum
among stars, here I mean real collisions between two stars. For a thorough
discussion of the mechanism and an extremely detailed numerical study, I refer
the reader to \cite{FB02b}.}  The stars come so close to each other that they
collide, they have a hydrodynamic interaction; the outcome depends on a number
of factors, but the stars involved in the collision could either merge with each
other or destroy each other completely or partially. Contrary to what one could
expect, the impact of these processes for the global evolution of the dynamics
of galactic nuclei is negligible \cite{FB02b}. In most of the cases, when these
extended stars, such as main-sequence stars (MS) collide, they do not merge due
to the very high velocity dispersion, and they will also not be totally
destroyed, because for that they would need a nearly head-on collision, so that
they have a partial mass-loss and are for our purposes uninteresting. For the
kind of objects of interest to us in this review, stellar-mass black holes, the
probability that they physically collide is negligible.

A third way of altering the angular momentum of stars are secular effects. They
do nevertheless \emph{not} modify the energy. If we assume that the orbits
around the MBH are nearly Keplerian, the shape, an ellipse, does not change, and
the orientation will not change much. If we have another orbit with a different
orientation, both orbits will exert a torque $\mathcal T$ on each other. This
will change angular momentum but not energy.  A Keplerian orbit can be described
in terms of its semi-major axis and eccentricity. The semi-major axis is only
connected to energy and, for a given semi-major axis the eccentricity is
connected to the angular momentum. If one changes the angular momentum but not
the energy, the eccentricity will vary but not the semi-major axis. By
decreasing the angular momentum, one increases the eccentricity.

In this section, however, I introduce the fundamentals of relaxation theory,
focusing on the aspects that will be more relevant for the main interest of
this review.  Further ahead, in Section~\ref{ch.exotica}, I will address resonant
relaxation and other ``exotic'' (in the sense that they are not part of the
traditional two-body relaxation theory) processes.  For a comprehensive
discussion on two-body relaxation, I recommend the text books \cite{Spitzer69}
and \cite{BinneyTremaine08} or, for a shorter but very nice introduction, the
article \cite{FB01a}.

I will first introduce handy timescales in Section~\ref{sec.2body} that will
allow us to pinpoint the relevant physical phenomena that reign the process of
bringing stars (extended or compact) close to the central MBH.  I will then
address a particular case of relaxation, in Section~\ref{sec.DynFric}, dynamical
friction.  Later, in Section~\ref{sec.angles}, I will define more concisely the
region of space-phase in which we expect stars to interact with the central MBH.
Once we have all these concepts, we can cope with the problem of how mass
segregates in galactic nuclei, in Section~\ref{sec.MassSegr}. We will first see
in detail the ``classical'' although academic solution, and later a more recent
and physical result, the so-called strong mass segregation, in
Section~\ref{sec.StrMassSegr}

\subsection{Two-body relaxation}
\label{sec.2body}

I introduce now some useful time-scales to which I will refer often throughout
this review; namely the relaxation time, the crossing time and the dynamical
time. These three time-scales allow us to delimit our physical system.

\paragraph{The relaxation time}

In \cite{Chandra42} a time-scale was defined which stems from the 2-body
small-angle encounters and gives us a typical time for the evolution
of a stellar system.

This relaxation time could be regarded as an analogy of the shock time
of the gas dynamics theory, by telling us when a particle (a star) has
forgotten its initial conditions or, expressed in a different way, when
the local thermodynamical equilibrium has been reached. Then, we can
roughly say that the most general idea is that this is the time over
which the star ``forgets'' its initial orbit due to the series of
gravitational tugs caused by the passing-by stars. After a relaxation
time the system has lost all information about the initial orbits of
all the stars.  This means that the encounters alter the star orbit
from the one it would have followed if the distribution of matter
was smooth. Hence, we can regard {\sl the relaxation time} as the
time interval required for the velocity distribution to reach the
Maxwell-Boltzmann form.

Consider two stars of masses $m_1$ and $m_2$ deflecting each other as in
Figure~\ref{fig.Deflection}. The deflection angle $\theta$ is given by the relation

\epubtkImage{.png}{
\begin{figure}[htbp]
\centerline{\includegraphics[width=\textwidth]{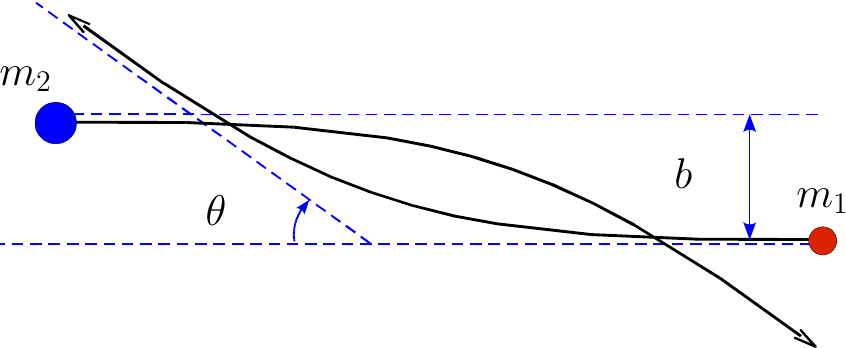}}
\caption{Deflection angle $\theta$ of a ``test'' star of mass $m_1$ with a field star
of mass $m_2$
\label{fig.Deflection}
}
\end{figure}
}

\begin{equation}
\tan\frac{\theta}{2}=\frac{b_0}{b},~{\rm with}~b_0=\frac{G(m_1+m_2)}{v_{\rm rel}^2}
\end{equation}
If the relative velocity $v_{\rm rel}$ is high, $\theta$ is small and the
larger the mass, the stronger the deflection. This simple relation expresses
the kernel of relaxation. One has to integrate it over all possible parameters
to get the relaxation rate. When we do the integration over the impact
parameter $b$ whilst keeping $v_{\rm rel}$ and the masses fixed, we have the
picture of Figure~\ref{fig.Deflection2}.  The test star encounters a lot
of field stars, all of them with the same mass $m_2$ and relative velocity
${\bf v}_{\rm rel}$. After a time $\delta t$, the velocity vector of the test
star has slightly changed direction by an angle $\theta_{\delta t}$. On
average, $\langle\theta_{\delta t}\rangle=0$ but

\begin{equation}
\langle\theta_{\delta t}^2\rangle=\left(\frac{\pi}{2}\right)^2\,\frac{\delta t}{\hat{t}_{\rm rlx}}
\label{eq.avthetasq}
\end{equation}

\epubtkImage{.png}{
\begin{figure}[htbp]
\centerline{\includegraphics[width=\textwidth]{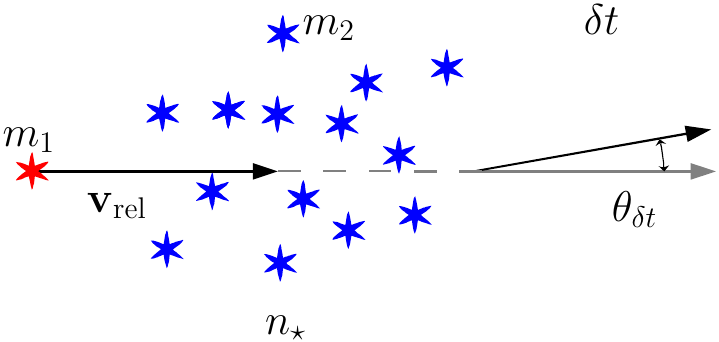}}
\caption{The test stars suffers a change in direction by $\theta_{\delta t}$
due to the accumulation of encounters with field stars.
\label{fig.Deflection2}
}
\end{figure}
}
Therefore it is a \emph{diffusion} process; $\langle\theta_{\delta t}^2\rangle \propto \delta t$,
see e.g. \cite{SH71a,Henon75}. I have introduced the special relaxation time for this situation as

\begin{equation}
\hat{t}_{\rm rlx} = \frac{\pi}{32}\frac{v_{\rm rel}^3}{\ln\Lambda\,G^2\,n_{\star}(m_1+m_2)^2}
\label{eq.hatrlx}
\end{equation}
In this last equation, $\ln\Lambda$, the Coulomb logarithm, has appeared as a result of
integrating for all impact parameters. The information encoded in it is how many orders
of magnitude of $b$ contribute to the relaxation,

\begin{equation}
\ln\Lambda=\ln\,\frac{b_{\rm max}}{b_0}\simeq
           \ln\,\frac{P_{\rm orb}}{b_0/v_{\rm rel}}
\label{eq.Lambda}
\end{equation}
In this last equation $b_0$, which I introduced before, is the effective
minimum impact parameter for relaxation. Our main focus is not a detailed review of stellar dynamics.  For a detailed
description of the Coulomb logarithm, I refer the reader to
\cite{BinneyTremaine08,Spitzer87}. Therefore, I will simply comment that, for
our purposes, $\ln\Lambda \approx 10-15$ always. This is very useful because
the exact calculation can be rather arduous and almost an incubus which to our
knowledge nobody has attempted to implement in any calculation. Therefore we
mention only two special cases for the argument of the logarithm,

\begin{equation}
\Lambda \approx \left\{ \begin{array}{ll}
 0.01\,N_{\star} & \textrm{(a) for a self-gravitating}\\
                 &  \textrm{stellar cluster}\\
 {\cal M}_{\bullet}/m   & \textrm{(b) close to the MBH}
                            \end{array}
                     \right.
\label{eq.Lambda2}
\end{equation}
In case (a) we have a self-gravitating cluster of stars in equilibrium with
itself but lacking a central MBH.  The argument is proportional to the number
of stars in the system. In the situation in which a star is orbiting the MBH,
the previous value is formally no longer valid and one should use the value
(b). Nevertheless, in effect this is neglected because the value turns out to
be $\sim 10$. To define a local average value of the relaxation time we
integrate over the
distribution of relative velocities.

It must nevertheless be noted that the way in which I have introduced the
concept of the relaxation time is a particular one. In equation~\ref{eq.hatrlx}
I have introduced the ``encounter relaxation time'' to stress that it depends
on the characteristics of a peculiar class of encounter: a star of mass $m_1$
with ``field stars'' of mass $m_2$ with a local density $n_{\star}$ and a
relative velocity $v_{\rm rel}$.  It can be envisaged as the required time to
deflect gradually the motion of star $m_1$ due to encounters with field stars
by a root mean square (RMS) angle $\pi/2$. This definition is useful to
understand the fundamentals of relaxation, but it must be noted that it is
subject to this very peculiar type of encounter.

However, in a general case, we define relaxation by simplyfing the problem: (i)
We restrict to the radius of influence for a system in which the distribution
of stars is spherically symmetric, (ii) stars are treated as single objects,
with a two-body relaxation as the only mechanism that can change the angular
momentum, and (iii) we neglect mass segregation.

The influence radius within which the central MBH dominates the gravitational
field is

\begin{equation}
r_{\rm infl}=\frac{G\MBH}{\sigma_0^2}\approx 1~{\rm pc}\left(\frac{\MBH}{10^6\,M_\odot}\right)
\left(\frac{60~{\rm km/s}}{\sigma_0}\right)^2.
\end{equation}

\noindent
Hence, in our approximation, the relaxation time is

\begin{equation}
t_{\rm rlx}(r)=\frac{0.339}{\ln\Lambda}\frac{\sigma^3(r)}{G^2\langle m\rangle
m_{\rm CO}n(r)}
\simeq 1.8\times 10^8\,{\rm yr}\left(\frac{\sigma}{100\,\kms}\right)^3
\left(\frac{10\,\Msun}{m_{\rm CO}}\right)
\left(\frac{10^6\,\Msun{\rm pc}^{-3}}{\langle m\rangle n}\right).
\end{equation}

Here $\sigma(r)$ is the local velocity dispersion. It is approximately equal to the Keplerian
orbital speed $\sqrt{G\MBH r^{-1}}$ for $r<r_{\rm infl}$ and has
a value $\approx\sigma_0$ outside of it. In the expression $n(r)$ is the local number density of
stars, $\langle m\rangle$ is the average stellar mass, $m_{\rm CO}$ is the mass
of the compact object (we take a standard $m_{\rm CO}=10\,M_\odot$ for
stellar-mass black holes).

{For typical density profiles, $t_{\rm rlx}$ decreases slowly with decreasing
$r$ inside $r_{\rm infl}$. It should be noted that the exchange of energy
between stars of different masses ---sometimes referred to as dynamical
friction, as we will see ahead, in section~\ref{sec.DynFric} in the case of one
or a few massive bodies in a field of much lighter objects--- occurs on a
timescale shorter than $t_{\rm rlx}$ by a factor of roughly $M/\langle
m\rangle$, where $M$ is the mass of a heavy body.}

As we will see later, relaxation redistributes orbital energy amongst
stellar-mass objects until the most massive of them (presumably stellar-mass
black holes) form a power-law density cusp around the MBH, $n(r)\propto r^{-\gamma}$ with
$\gamma$ ranging between $\simeq 1.75$\,--\,$2.1$, which depends on the
solution to mass segregation considered, while less massive species arrange
themselves into a shallower profile, with $\alpha \simeq 1.4-1.5$
\cite{BW76,LS77,DS83,FB02b,ASFS04,BME04a,PMS04,FASK06,HA06b,AlexanderHopman09,Merritt2010,PretoAmaroSeoane10,Amaro-SeoanePreto11} (see also
Section~\ref{sec.EMRIStatMethods}). Nuclei likely to host MBHs in the LISA mass
range ($\MBH\lesssim \mbox{few}\times 10^6\,\Msun$) probably have relaxation
times comparable to or less than a Hubble time, so that it is expected that
their heavier stars form a steep cusp.

\paragraph{Collision time}

${t_{\rm coll}}$ is defined as the required mean time for
the number of stars within a volume $V=\Sigma  v_{\rm rel}
 \triangle t$ to be one (see Figure~\ref{fig.coll_time}),
where $v_{\rm rel}$ is the relative velocity at infinity of two
colliding stars.

\epubtkImage{.png}{
\begin{figure}[htbp]
\centerline{\includegraphics[width=0.5\textwidth]{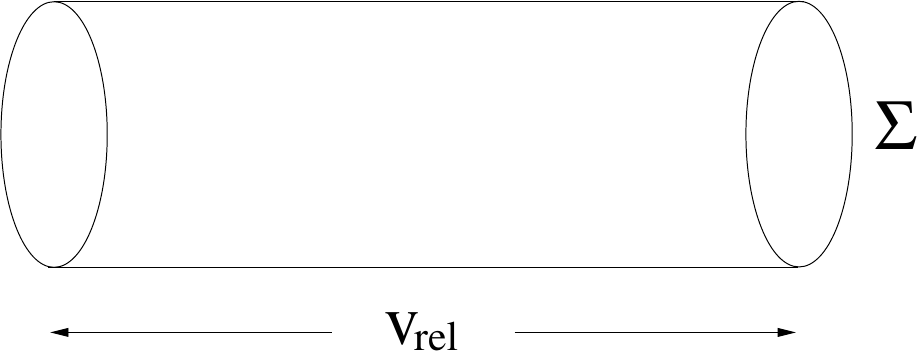}}
\caption{Definition of the {collision} time
\label{fig.coll_time}
}
\end{figure}
}
Computed for an average distance of closest approach $\bar{r}_{\rm min}=
\frac{2}{3}r_{\star}$, this time is given by
\begin{equation}
n_{\star}V(t_{\rm coll})=1= n_{\star}\,\Sigma\, v_{\rm rel}\, t_{\rm coll}.
\end{equation}
And so,
\begin{equation}
t_{\rm coll}=\frac{m_{\star}}{\rho_{\star}\Sigma\sigma_{\rm rel}},
\end{equation}
where
\begin{equation}
\Sigma=\pi \bar{r}_{\rm min}^2 \left( 1+ \frac{2Gm_{\star}}
{\bar{r}_{\rm min}\sigma_{\rm rel}^2}\right),
\end{equation}
$\sigma_{\rm rel}^2=2\sigma_{\star}^2$ is the stellar velocity dispersion and
$\Sigma$ a {collision}al cross section with gravitational focusing.

\paragraph{The crossing time}

As the name suggests, this is the required time for a star to pass
through the system, i.e. to {\sl cross} it. Obviously, this value is given
by the ratio between space and velocity,
\begin{equation}
t_{\rm cross}= \frac {R}{v},
\end{equation}
where $R$ is the size of the physical system and $v$ the velocity of the
star crossing it.

For instance, in a star cluster it would be:

\begin{equation}
t_{\rm cross}= \frac{r_{\rm h}}{\sigma_{\rm h}};
\end{equation}

\noindent
where $r_{\rm h}$ is the radius containing 50 \% of the total mass and
$\sigma_{\rm h}$ is a typical velocity taken at $r_{\rm h}$. One
denominates it \emph{velocity dispersion} and is introduced by the
statistical concept of RMS dispersion; the {\em
variance} $\sigma^2$ gives us a measure of the dispersion, or
scatter, of the measurements within the statistical population, which
in our case is the star sample:

\begin{equation}
\sigma^2= \frac{1}{N} \sum_{i=1}^{N} (x_{i}- \mu_{a})^2. \nonumber
\end{equation}

\noindent
In the last expression $x_{i}$ are the individual stellar velocities and
$\mu_{a}$ is the arithmetic mean,

\begin{equation}
\mu_{a} \equiv  \frac{1}{N} \sum_{i=1}^{N}x_{i}. \nonumber
\end{equation}
If virial equilibrium prevails, we have
$\sigma_{\rm h}
\approx \sqrt{GM_{\rm h}/r_{\rm h}}$, then we get the dynamical time-scale
\begin{equation}
t_{\rm dyn} \approx \sqrt {\frac{r_{\rm h}^3}{GM_{\rm h}}}\approx
\frac {1}{\sqrt{G\rho_{\star}}},
\end{equation}
where $\rho_{\star}$ is the mean stellar density.

Contrary to gas dynamics, the thermodynamical equilibrium time-scale $t_{\rm
rlx}$ in a stellar system is large compared with the crossing time $t_{\rm
cross}$. In a homogeneous, infinite stellar system, we expect some kind of
stationary state to be established in the limit $t\to \infty$. The decisive
feature for such a virial equilibrium is how quickly a perturbation of the
system will be smoothed down.

The dynamical time in virial equilibrium is, cf., e.g., \cite{Spitzer87}:
\begin{equation}
t_{\rm dyn}\propto\frac{\log (\gamma N)}{N} t_{\rm rlx}
\ll t_{\rm rlx}.
\end{equation}

If we have perturbations in the system because of the heat conduction,
star accretion on to the MBH, etc. a new virial equilibrium will be
established within a time $t_{\rm dyn}$, which is short. This means that we
get again a virial-type equilibrium in a short time. This situation
can be considered not far from a virial-type equilibrium. We say that
the system {\sl changes in a quasi-stationary way}.

\subsection{Dynamical friction}
\label{sec.DynFric}

Consider now a star more massive than the average. In this case, relaxation
boils down to dynamical friction (DF). The massive intruder will suffer from
dynamical friction, which is an effect of all encounters with lighter stars.
For this special kind of star, the timescale over which its orbital parameters
change is not the relaxation time. This star will lose kinetic energy in
the following timescale:

\begin{equation}
t_{\rm DF} \sim \frac{\langle m \rangle}{m}\,t_{\rm rlx}.
\label{eq.tDF}
\end{equation}

As we can see, if the object is 10\,--\,20 times more massive than the average,
as in the case of a stellar-mass black hole, this timescale is correspondigly 10\,--\,20
times shorter than the $t_{\rm rlx}$.

\epubtkImage{.png}{
\begin{figure}[htbp]
\centerline{\includegraphics[width=0.7\textwidth]{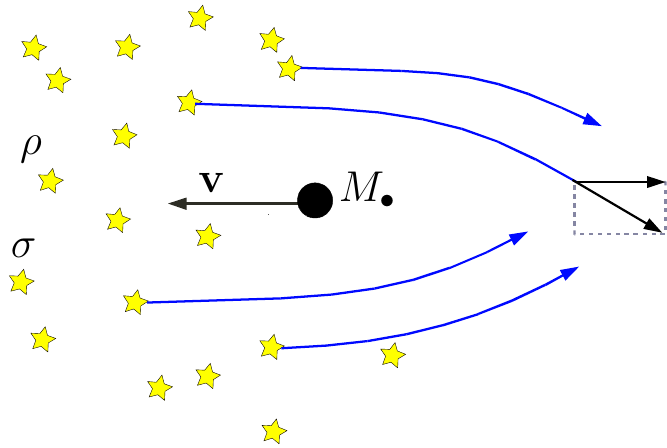}}
\caption{In the reference frame of the encounter I depict a massive interloper, a
stellar-mass black hole, traversing a sea of lighter stars which are deflected by
it. The velocity vector of the stellar-mass black hole is bearely modified (at least
in direction) by the deflections, because they cancel out on average.}
\label{fig.DynFric}
\end{figure}
}
In Figure~\ref{fig.DynFric} we have an illustration for what DF is. A massive
intruder, a stellar-mass black hole, is travelling in a homogeneous sea of stars of
density $\rho$ and velocity dispersion $\sigma$. The velocity vectors of the
stars is rotated after the deflection and the projected component in the
direction of the deflection is shorter. Therefore, the massive object is
accumulating just after it a high-density stellar region. The perturber will feel
a drag from that region from the conservation of angular momentum in the direction of its
velocity vector, just as depicted in Figure~\ref{fig.DynFric2}. The direction does
not change to first-order, but the amplitude decreases. The intruder will feel
a force (acceleration) given by the Chandrasekhar formula,

\begin{equation}
\vec{a}_{\rm DF} = -\frac{\vec{v}}{t_{\rm DF}}
                 -\frac{4\pi\ln\Lambda\,G^2\rho\,M}{v^3}\,\xi(X)\vec{v}.
\label{eq.aDF}
\end{equation}

\noindent
In this last equation,

\begin{align}
\xi(X) & = {\rm erf}(X)-2\pi^{-1/2}Xe^{-X^2},\\\nonumber
    X  & = \frac{v}{\sqrt{2}\sigma}
\end{align}

\epubtkImage{.png}{
\begin{figure}[htbp]
\centerline{\includegraphics[width=0.7\textwidth]{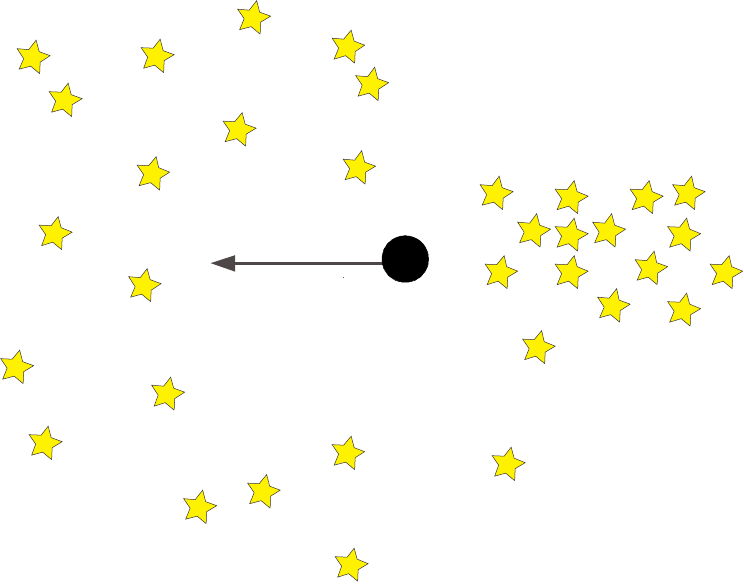}}
\caption{The accumulation of stars right behind the massive perturber
creates a region of stellar overdensity that acts on the perturber,
slowing it down, braking it.}
\label{fig.DynFric2}
\end{figure}
}

The most interesting point is that if we plug into Eq.~\ref{eq.aDF} the velocity
of the perturber which is $v \approx \sigma$, we have that
\begin{equation}
t_{\rm DF} \sim \frac{m}{M}t_{\rm rlx} \ll t_{\rm rlx}
\label{eq.}
\end{equation}

As I have already mentioned before, galactic nuclei in the range of what a
mission like LISA could observe have relaxation times that are shorter than a
Hubble time. In Figure~\ref{fig.TcollTrelLISA}, which is a modified version of
the figure to be found in article \cite{FB05}, we have a schematic
representation of what relaxation times in other observed galaxies could be.
Each dot shows the mass of the central MBH or the upper limit to it (the
arrows). From this mass we can derive what the velocity dispersion would be at
0.1~pc, and from observations of the brightness surface profiles we can estimate
what the stellar density at that distance would be. In many cases this distance
is usually not resolvable, so that one has to extrapolate in order to obtain the
density at 0.1~pc, which is what has been done in
Figure~\ref{fig.TcollTrelLISA}. The blue, dashed lines correspond to $t_{\rm
rlx}(r=0.1\,{\rm pc})$, the relaxation time at that distance. Any system below
$10^{10}$ yrs should be relaxed and is, hence, interesting. For the range of
frequencies we are interested in, MBHs with masses typically less than a few
$10^7\,M_{\odot}$ (the region below the red line) we can see only three
(since M110 is only an upper limit and M33 possibly lacks an MBH). This low
number does not mean that nuclei in the range of frequencies of interest are
rare, it simply means that it is hard to observe MBHs in that range of masses.
In this regard, a GW mission that could observe MBHs in that region would
provide us with very valuable information, since in the electromagnetic domain
we are still far from resolving those nuclei.

\epubtkImage{.png}{
\begin{figure}[htbp]
\centerline{\includegraphics[width=0.7\textwidth]{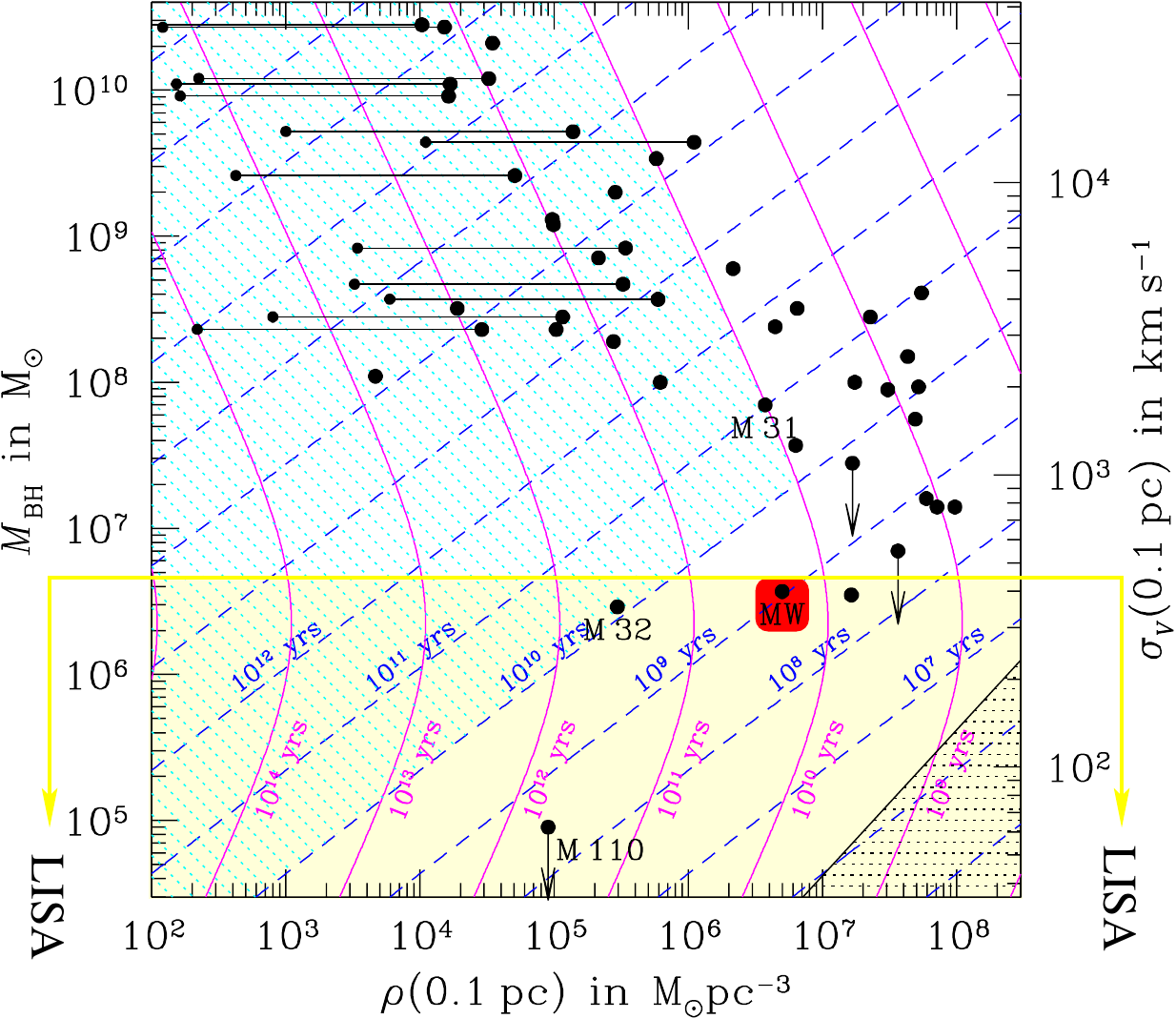}}
\caption{
Plane of the stellar density at 0.1\,pc and the mass of the central MBH,
taken from \cite{FB05} Relaxation (and collision times) at 0.1\,pc from an
MBH in the centre of a galactic nucleus.
\label{fig.TcollTrelLISA}
}
\end{figure}
}

\subsection{The difussion and loss-cone angles}
\label{sec.angles}

As we have seen, the relaxation time is the required time to induce a change in
the perpendicular velocity component of the same order as the perpendicular
velocity component itself, i.e. $\triangle v_{\perp}^2 /v_{\perp}^2 \simeq 1$.
Therefore,

\begin{equation}
\triangle v_{\perp}^2 = n_{\rm rlx} \cdot \delta v_{\perp}^2.
\end{equation}

\noindent
Hence,

\begin{equation}
\triangle  v_{\perp}^2 / v_{\perp}^2 = 1 = \frac{n_{\rm rlx} \cdot \delta v_{\perp}^2}
{ v_{\perp}^2}.
\label{eq.thetaD_a}
\end{equation}

\noindent
And then,

\begin{equation}
t_{\rm rlx}=n_{\rm rlx} \cdot t_{\rm dyn} = \left(
\frac {v_{\perp}^2}{\delta v_{\perp}^2} \right) \cdot t_{\rm dyn},
\label{eq.thetaD_b}
\end{equation}
where $n_{\rm rlx}$ is the numbers of crossings for $\triangle v_{\perp}^2 /
v_{\perp}^2 \simeq 1$. This conforms to the definition of the relaxation time,
$\triangle v_{\perp}^2 / v_{\perp}^2 = t/t_{\rm rlx}$, see \cite{BT87}.

A useful quantity to derive is the diffusion angle, $\theta_{\rm D}$, which is
defined to be the mean deviation of a star orbit in a dynamical time,
i.e. $t_{\rm rlx} \simeq t_{\rm dyn}/{\theta_{\rm D}^2}$. I assume
that this angle must be a very small one, so that

\begin{equation}
\sin{\theta_{\rm D}} \simeq \frac{ \delta v_{\perp}}{v} \simeq \theta_{\rm D}.
\label{eq.thetaD_c}
\end{equation}

\noindent
Therefore,

\begin{equation}
\theta_{\rm D} \simeq \sqrt{\frac {t_{\rm dyn}}{t_{\rm rlx}}}.
\label{eq.thetaD_d}
\end{equation}

I now introduce the loss-cone angle $\theta_{\rm lc}$ as an illustrative
example. Suppose that the central object with mass ${\cal M}_{\bullet}$ has an
influence radius $r_{\rm h}$. To define this radius we say that a star will
interact with the central object only when $r \le r_{\rm h}$. Then, we look for
a condition at a place $r>r_{\rm h}$ for a star to touch or to cross the
influence radius of the central object within a crossing time $t_{\rm cross}= r/
\sigma_{\rm r}$.

\epubtkImage{.png}{
\begin{figure}[htbp]
\centerline{\includegraphics[width=0.7\textwidth]{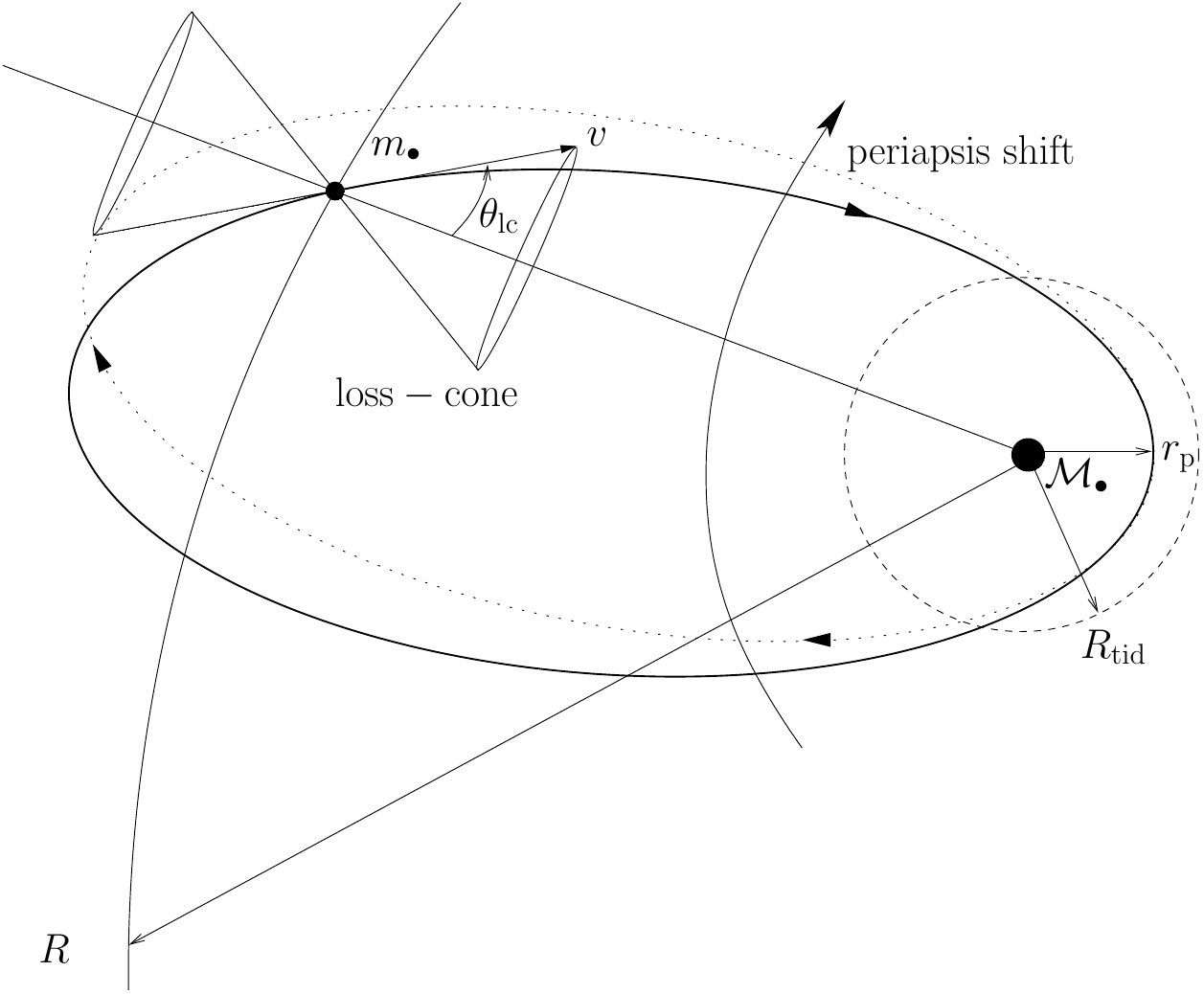}}
\caption{Definition of the loss-cone angle $\theta_{\rm lc}$. The star has a
mass $m_{\bullet}$, the MBH a mass $\cal{M}_{\bullet}$, $r_{\rm p}$ is the
periapsis distance, $R_{\rm tid}$ the tidal radius and $R$ the distance to the
MBH.}
\label{fig.theta_lc}
\end{figure}
}

I depict this in Figure~\ref{fig.theta_lc}: A star on a certain orbit will get
into the tidal disruption radius of the MBH if its velocity vector is such that
the distance of periapsis is within that radius.  The velocity and radial
distance vectors define the angle of the cone in phase-space for this to happen.
Extended stars are torn apart and lost for the system, which is why we refer to
that angle as the loss-cone angle.  If the star is a stellar-mass black hole, it
can withstand the tidal forces. Although I have also illustrated the effect of
periapsis shift in the figure, I do not take it into account for the derivation
of the loss-cone. It is meant to illustrate the complexity of the problem we are
interested in, the gravitational capture of compact objects. As we have seen before,
the condition that defines this angle is the following:

\begin{align}
r_{\rm p}(E,L) & \leq r_{\rm t},\nonumber \\
\theta & \leq \theta_{\rm lc},
\label{eq.thetalc_a}
\end{align}

\begin{align}
\sin{\theta} & = \frac{v_{\rm t}}{v},~{\rm with}~\theta \ll 1 \nonumber \\
\theta & \simeq  \frac{v_{\rm t}}{v}= \frac {L/r}{v}.
\label{eq.thetalc_b}
\end{align}

\noindent
In the last expression I have introduced $L:=r\,v_{\rm t}$ as the specific
angular momentum. Now, I derive an expression for this angle in terms of the
influence radius. Within the region {$r \le r_{\rm h}$}, the star moves under
the MBH potential influence, then

\begin{align}
\sigma(r) & \approx \sqrt{ \frac {G{\cal M}_{\bullet}}{r}}
= \sqrt { \frac {G{\cal M}_{\bullet}}{R_{\rm h}}} \sqrt { \frac {R_{\rm h}}{r}} \nonumber \\
&         = \sigma (R_{\rm h})  \sqrt {R_{\rm h}/r} = \sigma _{c}  \sqrt {R_{\rm h}/r},
\label{eq.thetalc_c}
\end{align}
since $ \sigma_{c}^2 \equiv G{\cal M}_{\bullet} / R_{\rm h}.$
The typical velocity of the orbit is $\langle v^2 \rangle \simeq 3 \sigma^3$, where the factor
three accounts for the three directions in the space. Since $\sigma$ means the one-dimensional
dispersion,  we have to take into account the dispersion of the
velocity in each direction. Then,

\begin{equation}
\langle v \rangle \simeq \sqrt{3} \sigma_{c} \sqrt {r_{\rm h}/r}.
\label{eq.thetalc_d}
\end{equation}

Finally, we obtain the loss-cone angle,

\begin{equation}
\theta _{\rm lc} = \sqrt { \frac {2}{3} \frac {r_{\rm t}}{r}}.
\label{eq.thetalc_e}
\end{equation}

In the region in which {$r \ge r_{\rm h}$} we can consider that the velocity
dispersion is more or less constant from this $r_{\rm h}$ onwards, $v \approx
\sqrt{3} \sigma_{c}$,

\begin{align}
\theta_{\rm lc} & = \frac { \sqrt {2G{\cal M}_{\bullet}r_{\rm t}}}{ \sqrt {3} r \sigma _{c}};\nonumber \\
\sigma _{c} & = \sqrt {G{\cal M}_{\bullet}/r_{\rm h}}.
\label{thetalc_f}
\end{align}

The angle  is
\begin{equation}
\theta_{\rm lc} \approx \frac{1}{r}\,\sqrt{\frac {2\,r_{\rm t}\, r_{\rm h}}{3}}
\label{eq.thetalc_g}
\end{equation}

I have derived the loss-cone velocity $v_{\rm lc}(r)$ using angular momentum and
energy conservation arguments. We just have to evaluate it at a general radius
$r$ and at the tidal radius $r_{\rm t}$, where the tangential velocity is
maximal and the radial velocity cancels (see Figure~\ref{fig.peribarath}).

\epubtkImage{.png}{
\begin{figure}[htbp]
\centerline{\includegraphics[width=\textwidth]{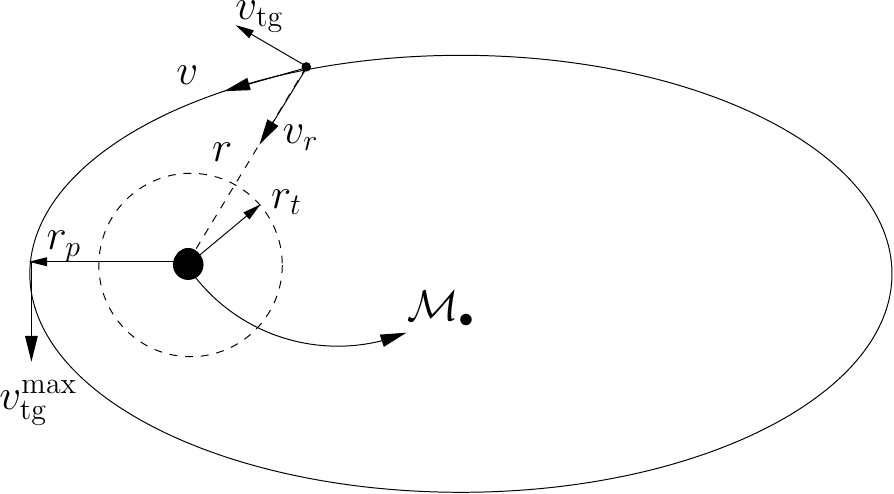}}
\caption{Definition of the tidal radius ``$r_{\rm t}$'',
and depiction of the distance of closest approximation of the star in its orbit to the MBH. In this
point the radial component of the velocity of the star cancels and
the tangential component is maximum. In the figure ``$r_{\rm p}$'' stands
for the periapsis radius.}
\label{fig.peribarath}
\end{figure}
}

\noindent
For a general radius we have that

\begin{align}
E(r)=& \phi(r) - \frac{v_{\rm tg}(r)^2}{2} - \frac{v_{\rm r}(r)^2}{2} \nonumber \\
L(r)=&r v_{tg}(r)
\label{eq.lc_vel_a}
\end{align}

\noindent
For the tidal radius:

\begin{align}
E(r_{\rm t})=& \phi(r_{\rm t}) - \frac{v_{tg}(r_{\rm t})^2}{2} \nonumber \\
L(r_{\rm t})=&r_{\rm t} v_{tg}(r_{\rm t}),
\label{eq.lc_vel_a}
\end{align}

\noindent
Hence, from momentum conservation and the fact that $v_{\rm r}(r_{\rm t})=0$,
we get

\begin{equation}
v_{\rm tg}(r_{\rm t})= \frac {r}{r_{\rm t}} v_{\rm tg}(r).
\label{eq.lc_vel_b}
\end{equation}

\noindent
Using energy conservation and the last result,

\begin{align}
\phi(r) & - \frac {v_{tg}(r)^2}{2} - \frac {v_{\rm r}(r)^2}{2} = \nonumber \\
&\phi (r_{\rm t}) -
\frac{r^2}{2r_{\rm t}^2}  v_{tg}(r)^2.
\label{eq.lc_vel_c}
\end{align}
Then we get the tangential velocity of the stars in terms of r; namely, the
loss-cone velocity:

\begin{align}
v_{\rm lc}(r) &= \frac {r_{\rm t}} {\sqrt {r^2-r_{\rm t}^2}} \times  \nonumber \\
&  \sqrt {2[ \phi (r_{\rm t}) -
\phi (r)] + v_{\rm r}(r)^2}.
\label{eq.lc_vel_d}
\end{align}

The angular momentum is

\begin{align}
L(r_{\rm t})=&r_{\rm t} v_{\rm tg}(r)|_{\rm max}=r_{\rm t}
\frac{r}{r_{\rm t}}v_{\rm tg}(r)= \nonumber \\
r v_{\rm tg}(r)&=r  \frac {r_{\rm t}}{\sqrt {r^2-r_{\rm t}^2}}
\sqrt {2\triangle \phi + v_{\rm r}(r)^2},
\label{eq.lc_vel_d2}
\end{align}
where

\begin{align}
\triangle \phi & \equiv \phi (r_{\rm t}) - \phi (r)= \nonumber \\
& \frac{G{\cal M}_{\bullet}}{r_{\rm t}} + \phi_{\star}(r_{\rm t})-
\frac {G{\cal M}_{\bullet}}{r}- \phi_{\star}(r)
\label{eq.lc_vel_e}
\end{align}

If we use the fact that $r \gg r_{\rm t}$, then

\begin{equation}
\frac{G{\cal M}_{\bullet}}{r_{\rm t}}
\gg \left( \frac{G{\cal M}_{\bullet}}{r} +
\phi_{\star}(r) \right)= \phi(r)
\label{eq.lc_vel_f}
\end{equation}
Also, since ${\cal M}_{\bullet}\gg {\cal M}_{\star}(r_{\rm t})$,

\begin{equation}
\frac {G{\cal M}_{\bullet}}{r_{\rm t}} \gg \phi_{\star}(r_{\rm t}).
\label{eq.lc_vel_g}
\end{equation}
Thus,

\begin{equation}
v_{\rm lc}(r) \approx \frac{r_{\rm t}}{r}
\sqrt{ \frac{2G{\cal M}_{\bullet}}{r_{\rm t}}}.
\label{eq.lc_vel_h}
\end{equation}
If we use now the fact that

\begin{align}
\sigma_{\rm r}(r)& =\sigma_{\rm r}(r_{\rm t})
\left( \frac{r}{r_{\rm t}} \right)^{-1/2}=\nonumber \\
& \sqrt{ \frac{G{\cal M}_{\bullet}}{r_{\rm t}}}
 \left( \frac{r}{r_{\rm t}} \right)^{-1/2},
\label{eq.lc_vel_i}
\end{align}
we have that

\begin{equation}
\sqrt{ \frac{G{\cal M}_{\bullet}}{r_{\rm t}}}
= \sigma_{\rm r}(r) \left( \frac{r}{r_{\rm t}}
\right)^{-1/2}
\label{eq.lc_vel_j}
\end{equation}
And so,

\begin{equation}
v_{\rm lc}(r) \approx \frac {r_{\rm t}}{r}
\sqrt{ \frac{2G{\cal M}_{\bullet}}{r_{\rm t}}}
\approx \sigma_{\rm r}(r)
 \left( \frac{r_{\rm t}}{r} \right)^{1/2}.
\label{eq.lc_vel_k}
\end{equation}

\section{``Standard'' mass segregation}
\label{sec.MassSegr}

\subsection{Introduction}

In order to address the question of how many objects a year get close
enough to the central MBH to be tidally destroyed, in the case of an extended
star, or captured, if a compact object, the zero-th order problem we must solve
is how stars distribute around MBHs.

In a system with a spectrum of masses initially distributed uniformely, the more
massive ones have a higher kinetic energy than the lighter ones, simply due to
the fact that they have the same velocity dispersion but a higher mass. The
heavy stars exchange energy with each other and with the light stars through
relaxation. The exchange of energy goes in the direction of equipartition,
because the system searches the equilibrium. The heavy stars will lose energy
to the light ones. When they do so, since they feel their own potential or the
potential well of the MBH, their semi-major axis shrinks and they segregate to
the centre of the system. When doing so, their kinetic energy will become
higher.  The system tries to re-equilibrate itself; the velocity dispersion is
larger as it was when the massive stars were at larger distances from the
centre. As they approach the MBH, their kinetic energy will be higher as
compared to the light stars, which are pushed to the
outskirts of the system.

In Figure~\ref{fig.PowerLawReconstr} we have a density profile which shows us
the evolution of a single-mass galactic nucleus with a MBH while letting
relaxation play a role (i.e., the simulations were run for at least a $T_{\rm
rlx}$). The initial density profile is depicted in red and shows already a cusp
because the authors were using a King model \cite{FAK06a,FAK06b}, so that it
diverges at the centre. When we let it evolve, the profile obtains a much
steeper cusp, the blue curve, reaching later a power-law cusp of $\rho \propto
R^{-1.75}$. This cusp is kept as the system continues to evolve and the cluster
expands. The time units are expressed in Fokker--Plank units
\epubtkFootnote
{
We can relate standard $N$-body time units $T_{\rm NB}$ as defined in e.g.
\cite{HeggieHut03} to Fokker-Planck time units $T_{\rm FP}$ as follows: $T_{\rm
FP} = T_{\rm NB} \cdot {N}_{\star} / \ln(\gamma \cdot {N}_{\star})$, with
${N}_{\star}$ the number of stars in the system.
}.

This is not intuitive. This phenomenon occurs because at the centre we have a
sink, the MBH is removing stars, either through tidal disruptions or EMRIs. The
stars removed from the system must have a very negative energy, they are very
close to the centre, and stars also physically collide with each other and they
are partially or totally destroyed in the process, which also represents a loss
of stellar mass in the system. For the rest of the system, this represents
actually \emph{a source of heat}. The total energy in the system has increased.
We can also envisage the picture as follows: the stars that will be removed
have to give energy to the rest of the stellar system in order to approach the
central sink. When they do so, they heat up the system.

\epubtkImage{.png}{
\begin{figure}[htbp]
\centerline{\includegraphics[width=0.7\textwidth]{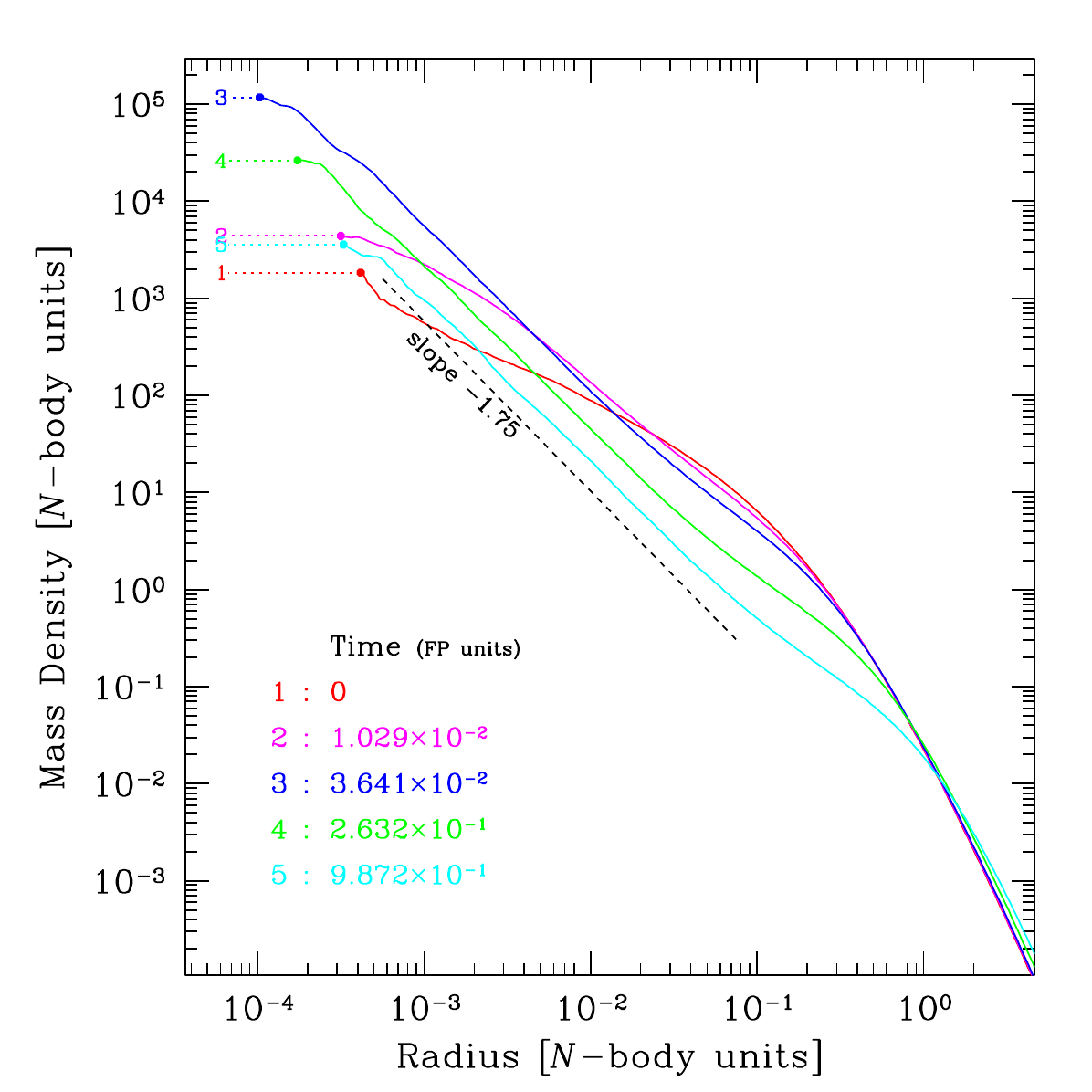}}
\caption{Density profile for a galactic nucleus with a single stellar population
in different moments of the evolution of the system, taken from \cite{FAK06a},
in Fokker--Plank units (FP), as defined in the footnote.}
\label{fig.PowerLawReconstr}
\end{figure}
}

In Figure~\ref{fig.DensProfilesMultimass} we have a somehow more realistic
situation. In this figure the authors depict the mass density distribution for a system
that has different stellar components and not only single-mass stars. After
some $10^{10}$ yrs the total density has not changed much but in the centre,
within $\sim 0.1$ pc, the stellar-mass black holes overwhelmingly defeat the rest of
the stellar components. Therefore, within a radius of $\sim 0.1$ pc around a
MBH such as the one in our GC, the mass density will be dominated by the
stellar-mass black holes. This does not apply to the number density of stellar black
holes.  They are less numerous as compared to MS stars, but more massive. The
important point here is that we expect to have about $2\cdot 10^3$ stellar
black holes within 0.1~pc, or $2\cdot 10^4$ within 1~pc of Sgr~A*
\cite{FAK06a,FAK06b}.

\epubtkImage{.png}{
\begin{figure}[htbp]
\centerline{\includegraphics[width=\textwidth]{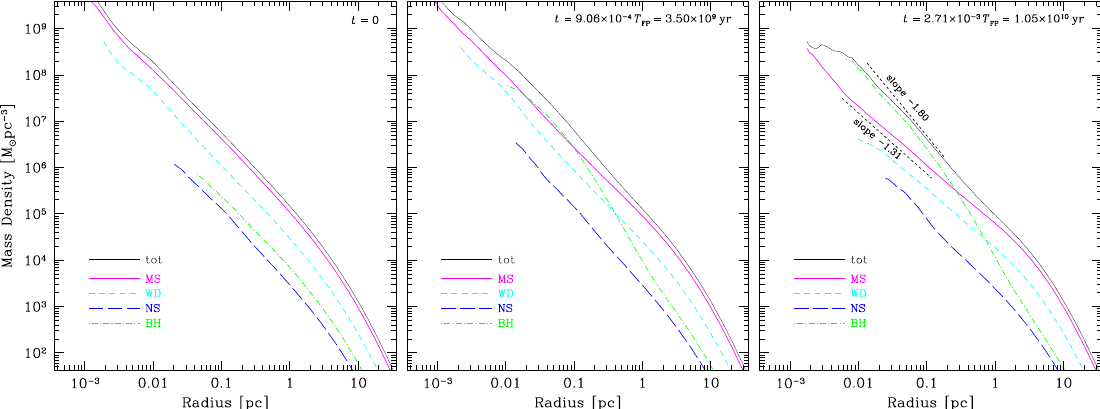}}
\caption{Evolution of a multi-mass system corresponding to Sgr~A*. The model, which is
taken from \cite{FAK06a}, their Figure~10, contains stellar-mass black holes, with masses
between 10\,--\,30 times larger than MS stars on average. On the left panel we have the initial
conditions, at $t=0$ yr. When we leave the system evolve, the components separate and roughly after
a Hubble time we obtain the situation corresponding to the right panel.}
\label{fig.DensProfilesMultimass}
\end{figure}
}

Before we further analyse realistic models with a mass spectrum and address the
potential implications for EMRI production, we will start assuming that all
stars have the same mass. As we mentioned in the foreword, the main goal of this
document is to give a self-consistent starting point to
understand the complexity of the different astrophysical phenomena associated with EMRIs.
Thus, the first kind of systems I will address will contain only one kind of
star.  As Donald Lynden Bell puts it in page 515, Sec. 4.5 of
\cite{Lynden-BellWood1968},

\begin{quotation}
  \emph{``Our other excuse for leaving out high order correlations
is that \textbf{only a fool tries the harder problem when he does
not understand the simplest special case.}''}
\end{quotation}

In this section I will illustrate the different phenomena with numerical
simulations published for the first time in this review.

\subsection{Single-mass clusters}

The work of Peebles in \cite{Peebles72} was the first to realise\epubtkFootnote{We note that eight
years ago the article \cite{Gurevich64} had an interesting first idea of this concept: The
authors obtained a similar solution for how electrons distribute around a
positively charged Coulomb centre.} that the statistical thermal equilibrium
in a stellar cluster, i.e., the fact that the distribution of energy in the
cluster is $f(E) \propto e^{-E/\sigma^2}$, with $\sigma$ the velocity
dispersion, must be violated when we are close to the MBH, because we have
three characteristic radii within which stars are lost for the system. These
are the tidal radius, $R_{\rm t}$, the ``Schwarzschild radius'' $R_{\rm Schw}$
(i.e., the capture radius via gravitational loss), and the collisional radius
$R_{\rm coll}$. Peebles found that there should be a steady state with a net
inward flux of stars and energy in the stellar system.  Nevertheless, well
within the influence radius $R_{h}$ of the MBH but far from $R_{\rm t}$, the
stars should have       nearly-isotropic velocities. Peebles derived a solution
in the form of a power-law for a system in which all stars have the same mass.
The quasi-steady solution takes the form (for an isotropic distribution
function)  $f(E)\sim E^p$, $\rho(r)\sim r^{-\gamma}$, with $\gamma = 3/2 + p$.
Nevertheless, Peebles derived the wrong exponent. A few years later, \cite{BW76}
did an exhaustive kinematic treatment for single-mass systems and found that
the exponent should be $\gamma=7/4$ and $p=\gamma-3/2=1/4$. This solution has been
corroborated in a number of semi-/analytical approaches, and approximative numerical schemes,
see e.g. \cite{SM78,MS79,MS80,ST85,FB01a,ASEtAl04}, as well as direct-summation
$N$-body simulations, of which the work of \cite{PretoMerrittSpurzem04} was the first one.

This is one of the most important phenomena in the production of EMRIs, since
the galactic nuclei of interest for us, the ones which are thought to be
harbouring EMRIs in their cores and are in the range of frequencies of
interest, are relaxed. These nuclei are relatively small and are
likely to have at least gone through at least one full relaxation time. In
general, nuclei in the range of interest for LISA are relaxed (see the rule of
thumb introduced in the work of \cite{Preto2010}).

\subsection{Mass segregation in two mass-component clusters}

As we have just seen, the processes that one-component clusters bring about are
nowadays relatively well understood and has been plentifully studied by
different authors to check for the quality of their approaches.  Nonetheless,
the properties of multi-mass systems are only very poorly represented by
idealised models in which all stars have a single mass.
New features of these systems' behaviour arise when we consider a
stellar system in which masses are divided into two groups.  Hence, since the
idealised situation in which all stars in a stellar cluster have the same mass
has been arduously examined in literature, we have the right to extend
the analysis a further step.  Here I address more realistic
configurations in which the stellar system is split into various components.
The second integer immediately after one is two, so we will first
extend, cautious and wary as we are, our models to two-component star
clusters.

Initial mass functions (IMFs), introduced with more detail in
Section~\ref{ref.MassSpecNoMBH}, ranging between $[0.1,\, \sim 120] M_\odot$ can
be approximated to first order by two well-separated mass scales : one with a
mass of the order of ${\cal O}(1 M_\odot)$ (which could represent main-sequence
stars, MS, white dwarfs, WD, or neutron stars NS) and ${\cal O}(10 M_\odot)$
(stellar-mass black holes).  Depending on how the system taken into
consideration is configured we will exclude \emph{dynamical equilibrium}
(meaning that the system is not stable on dynamical time-scales) or
equipartition of different components kinetic energies is not allowed
(\emph{thermal equilibrium}).

The work of \cite{Spitzer69} was in this respect pioneering. For some clusters it seemed
impossible to find a configuration in which they enjoy dynamical and thermal
equilibrium together. The heavy components sink into the centre because they
cede kinetic energy to the light ones when reaching equipartition. The process
will carry on until equipartition is fully gained. In most of the cases, equipartition
happens to be impossible, because the subsystem of massive stars will undergo
core collapse before equipartiton is reached.  Anon, a \emph{gravothermal
collapse} will happen in this component and, as a result, a small dense core
of heavy stars is formed \cite{Spitzer69,LF78}.  This gravothermal contraction is
a product of negative heat capacity, a typical property of gravitationally
bound systems \cite{ElsonEtAl87}.

Different authors have addressed the problem of thermal and dynamical
equilibrium in such systems, using techniques such as direct $N$-body
\cite{PZMM00,KhalEtAl07} and Monte Carlo simulations \cite{WJR00} to direct
integration of the Fokker--Planck equation \cite{IW84,KLG98} or moments of it
\cite{ASEtAl04}, including Monte Carlo approaches to the numerical integration
of this equation \cite{SH71b}. For a general and complete overview of the
historical evolution of two-stars stellar components, see \cite{WJR00,ASEtAl04}
and references therein.

If we do not have any energy source in the cluster and stars do not collide
(physically), the contraction carries on self-similarly indefinitely; in such a
case, one says that the system undergoes \emph{core-collapse}. This phenomenon
has been observed in a large number of works using different methods
\cite{Henon73,Henon75,SS75a,Cohn80,MS80,Stodol82,
Takahashi93,GH94b,Takahashi95,SA96,Makino96,Quinlan96,DCLY99,JRPZ00}. Core
collapse is not just a characteristic of multi-mass systems, but has been also
observed in single mass analysis.

\epubtkImage{.png}{
\begin{figure}[htbp]
\begin{center}
\centerline{\includegraphics[width=0.6\textwidth]{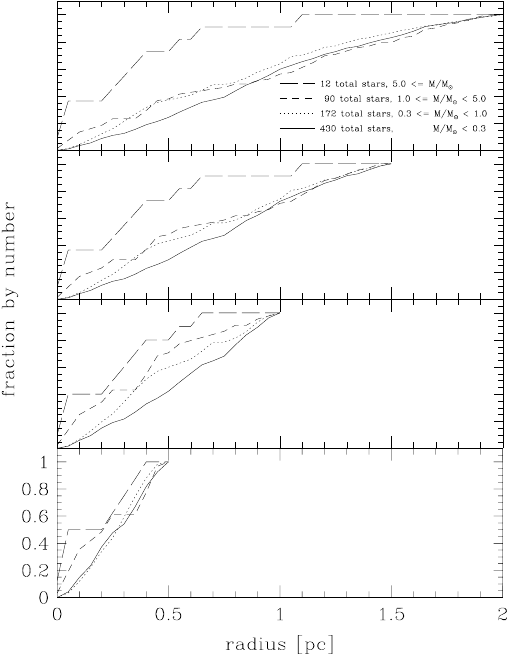}}
\caption{In this plot taken from \cite{HH98}, we see
mass-segregation of stars more massive than $5\,\msol$ (long-dashed
lines) toward the cluster centre and some evidence for general mass
segregation persisting down to $1\mbox{\,--\,}2\,\msol$ in the Orion Nebula
cluster. The cumulative radial distributions of source counts over
different mass intervals are shown. To clarify the sensitivity of the
cumulative plots to the outer radius they have shown here four panels
with four different limiting radii
\label{fig.hillen_mass_seg} }
\end{center}
\end{figure}
}

The work of \cite{Spitzer69} gives the analytical criterion to determine whether a
two-component system has achieved energy equipartition. According to this
analysis, energy equipartition between the light and heavy component exists if
the following inequality holds

\begin{equation}
{S}:= \left(\frac{{\cal M}_{\rm h}}{{\cal M}_{\rm l}}\right)
\left(\frac{m_{\rm h}}{m_{\rm l}}\right)^{3/2} < 0.16.
\label{eq.spitzer_stab}
\end{equation}

\noindent
Where ${\cal M}_{\rm l}$ and ${\cal M}_{\rm h}$ are the total numbers of light
and heavy components, respectively (i.e., the total stellar mass in light stars
and heavy stars in the system). More numerical calculations \cite{WJR00} have
settled this criterion to

\begin{equation}
{\Lambda}:= \left(\frac{{\cal M}_{\rm h}}{{\cal M}_{\rm l}}\right)
\left(\frac{m_{\rm h}}{m_{\rm l}}\right)^{2.4} < 0.16
\label{eq.spitzer_stab_new}
\end{equation}

When we modify the ratio ${\cal M}_{\rm max}/{\cal M}$, the time required to
reach core-collapse is different. In a cluster with, for instance, a broad
Salpeter initial mass function (IMF) between $[0.2\,\msol,\,120\,\msol]$,
core-collapse takes place after a time $\lesssim 0.1\,t_{\rm rh}(0)$, while for a
single-mass Plummer model it occurs after a time $\gtrsim 10\,t_{\rm rh}(0)$ (this
specific example was taken from the Monte Carlo-based calculations of
\cite{GFR04}).

There is an ample evidence for mass segregation in observed clusters.
\cite{McMS94} and \cite{HH98} provided deep infrared observations of the
Trapezium cluster in Orion that clearly show the mass segregation in the system,
with the highest mass stars segregated into the centre of the cluster.  To test
whether there is evidence for more general mass segregation, they showed in a
plot reproduced in Figure~\ref{fig.hillen_mass_seg} cumulative distributions
with radius of stars contained within different mass intervals. They include in
the figure four different panels in order to make clear the sensitivity to the
limiting radius.  They find that, inside 1.0 pc, general mass segregation
appears to be established in the cluster, with stars of masses less than 0.3,
0.3-1.0, 1.0-5.0 $M_{\odot}$, and greater than 5 $\msol$ progressively more centrally
concentrated with increasing mass.

\epubtkImage{.png}{
\begin{figure}[htbp]
\begin{center}
\centerline{\includegraphics[width=0.4\textwidth]{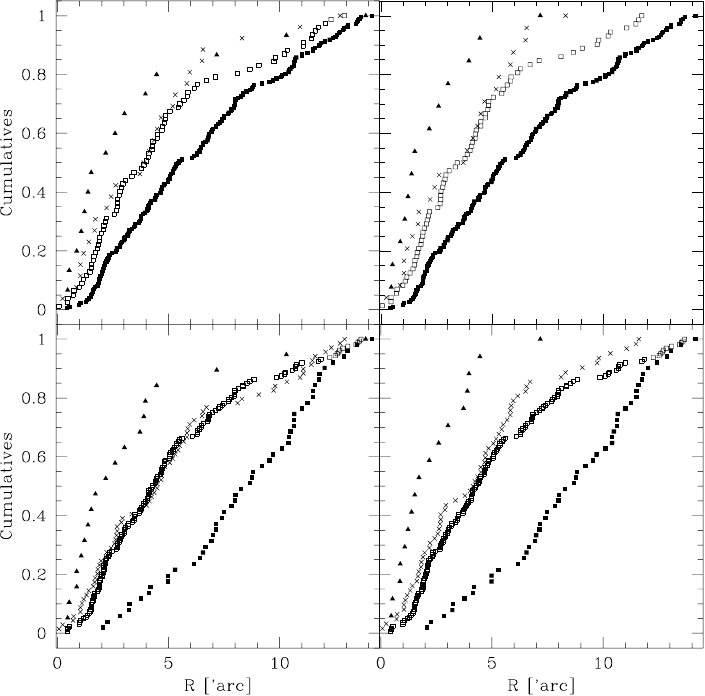}}
\caption{Mass segregation in NGC~623 for two mass interval sets, taken from
\cite{RM98}. The two
left panels include all sample stars, while the right ones do not
include the 9 bright stars of the cluster corona. For the two top
figures $M<5\,\msol$ (filled squares), $M \in \,[5,\,10[\,\msol$ (open
squares), $M \in \,[10,\,20[\,\msol$ (crosses) and $M \geq 20\,\msol$
(triangles).  For the two bottom figures, $M<2.5\,\msol$ (filled
squares), $M \in \,[2.5,\,6.3[\,\msol$ (open squares), $M \in
\,[6.3,\,15.8[\,\msol$ (crosses) and $M \geq 15.8\, \msol$ (triangles)
\label{fig.raboud_mass_seg} }
\end{center}
\end{figure}
}

At this point, the question looms up of whether for very young clusters mass
segregation is due to relaxation, like in our models, or rather reflects the
fact that massive stars are formed preferentially around the centre of the
cluster, as some models predict.

The work of \cite{RM98} addressed the radial structure of Praesepe and of the
very young open cluster NGC~6231. There they find evidence for mass segregation
among the cluster members and between binaries and single stars. They put it
down to the greater average mass of the multiple systems. In
Figure~\ref{fig.raboud_mass_seg} I reproduce a plot of \cite{RM98}, where again
we have clear evidence for mass segregation in NGC~6231. In the two first panels
the mass intervals are set in a different way to those in the bottom.

\epubtkImage{.png}{
\begin{figure}[htbp]
\begin{center}
\centerline{\includegraphics[width=0.7\textwidth]{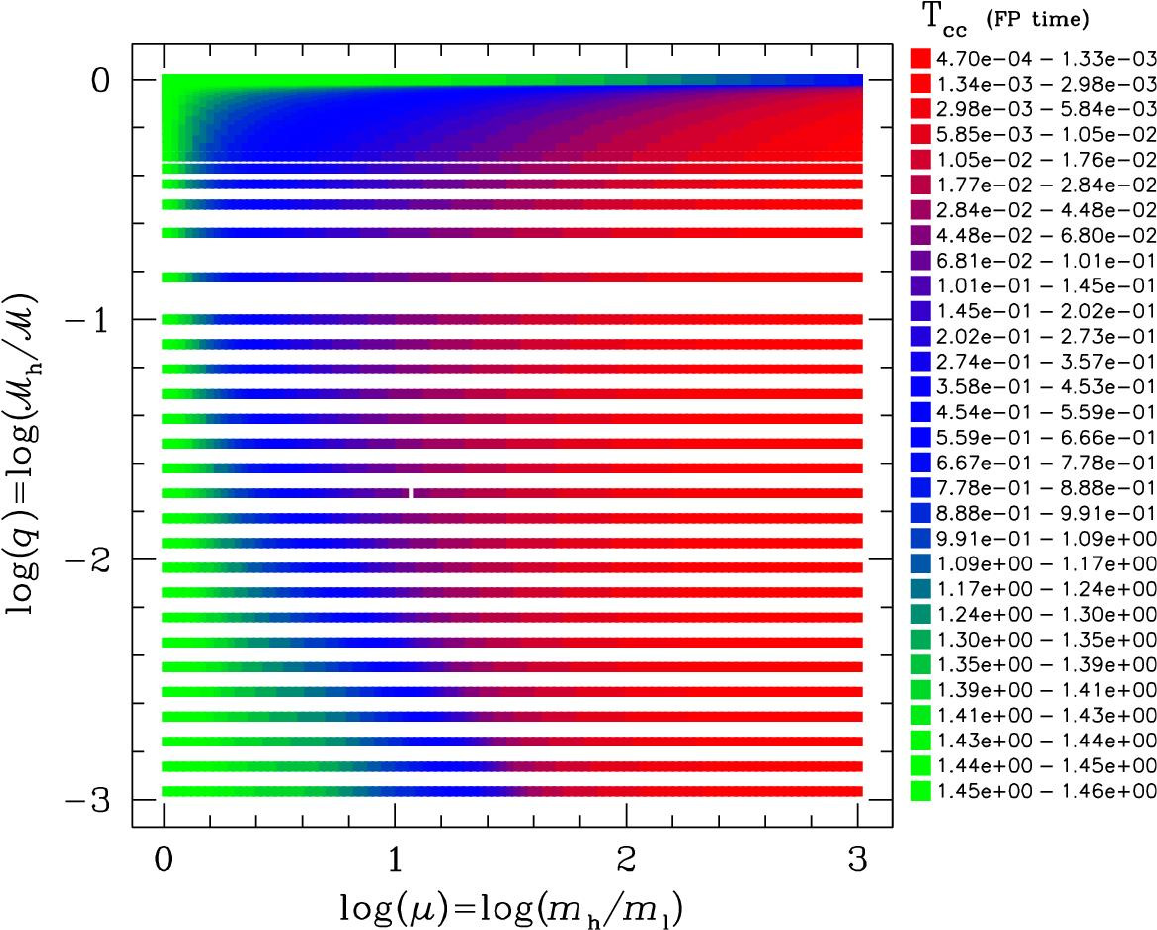}}
\caption{Parameter space for the set of $10^4$ simulations.
Here $t_{\rm end}$ stands for the core collapse time and is expressed
in FP units (see text); time at which the simulation ended. $q$ and
$\mu$ are plotted logarithmically.
\label{fig.mass_seg_10_4}}
\end{center}
\end{figure}
}

The two left-hand panels of Figure~\ref{fig.raboud_mass_seg} include the 9
bright stars of the cluster Corona, while on the right do not.
The manifestation of mass segregation for massive stars (triangles) is clearly
displayed, while stars with masses between $[5,\,20]\,\msol$ are spatially
well mixed (open squares and crosses); i.e., mass segregation is not yet
established over a rather large mass interval. This population is more
concentrated than the lower-mass population (here shown with filled squares).
They derive from Figure~\ref{fig.raboud_mass_seg} that only a dozen,
bright, massive, mainly binary stars are well concentrated toward the cluster
centre.

It therefore seems interesting to set up multi-mass models with two-components
as a starting point, since they are well-studied and we have robust
observational evidence of this phenomenon. On the other hand, observations do
not tell us whether mass segregation is due to relaxation. I now show the
results from a set of $10^4$ simulations for two-component models using the
``Gaseous Model'' programme to illustrate this (see
Section~\ref{ch.Integration}). I define two parameters now that describe the
physics of the system,

\begin{align}
  q &:= {{\cal M}_{\rm h}}/{\cal M},\nonumber \\
\mu &:= {m_{\rm h}}/{m_{\rm l}}
\end{align}
In this definition, ${\cal M}$ is the total mass of the system,
${\cal M}_{\rm h}$ the total mass in heavy stars and ${m_{\rm h,\,l}}$
the mass of one heavy (light) star. In the expression, $q$ is the total
stellar mass in heavy stars normalised to the total mass of the system, and
$\mu$ the mass ratio between heavy and light stars.

Now I introduce the quantity $\zeta \equiv 1-q$, and we let $\zeta$ vary from
$10^{-4}$ to $9.99 \cdot 10^{-1}$.  For each $\zeta$ value, we let $\mu$ vary
between $1.03$ and $10^3$.  The values for $q$ are regularly distributed in
$\log{(\zeta)}$. For $\zeta \approx 1$ we have added a series of values in
$\log{(\zeta-1)}$. The mean particle mass is $1,\msol$ and the total mass $10^6
\,\msol$, but this is not important for our study, because the physics of the
system is driven by relaxation and therefore the only relevant concept is the
relaxation time. We can always extend the physics to any other system containing
more particles, with the proviso that only relaxation is at play.  The mean mass
is therefore just a normalisation. What really determines the dynamics of the
system are the mass ratios, $q$ and $\mu$, which is the reason why I use them
to explore the system.

In Fig.\,(\ref{fig.mass_seg_10_4}) I show the whole $(q,\,\mu)$-parameter space
in a plot where the time at which the core-collapse begins is included. The
green zone corresponds to the quasi single-mass case. In the red zone we have
the largest difference between masses and blue is an intermediate case.

In Figure~\ref{fig.Tcc_q_diff_mu} I show collapse times for cluster models
with two mass components normalised to the single-mass core-collapse time
for different values of $\mu$. The initial clusters
are Plummer spheres without segregation.  The collapse times are displayed as a
function of the mass fraction of the heavy component in the cluster. When
compared to single-mass component systems, we see that the core-collapse time
is accelerated notably for a wide range of the heavy component ${\cal M}_{\rm
h}$ ($M_2$). Even a small number of heavy stars accelerate the core-collapse time.

\epubtkImage{.png}{
\begin{figure}[htbp]
\begin{center}
\centerline{\includegraphics[width=0.6\textwidth]{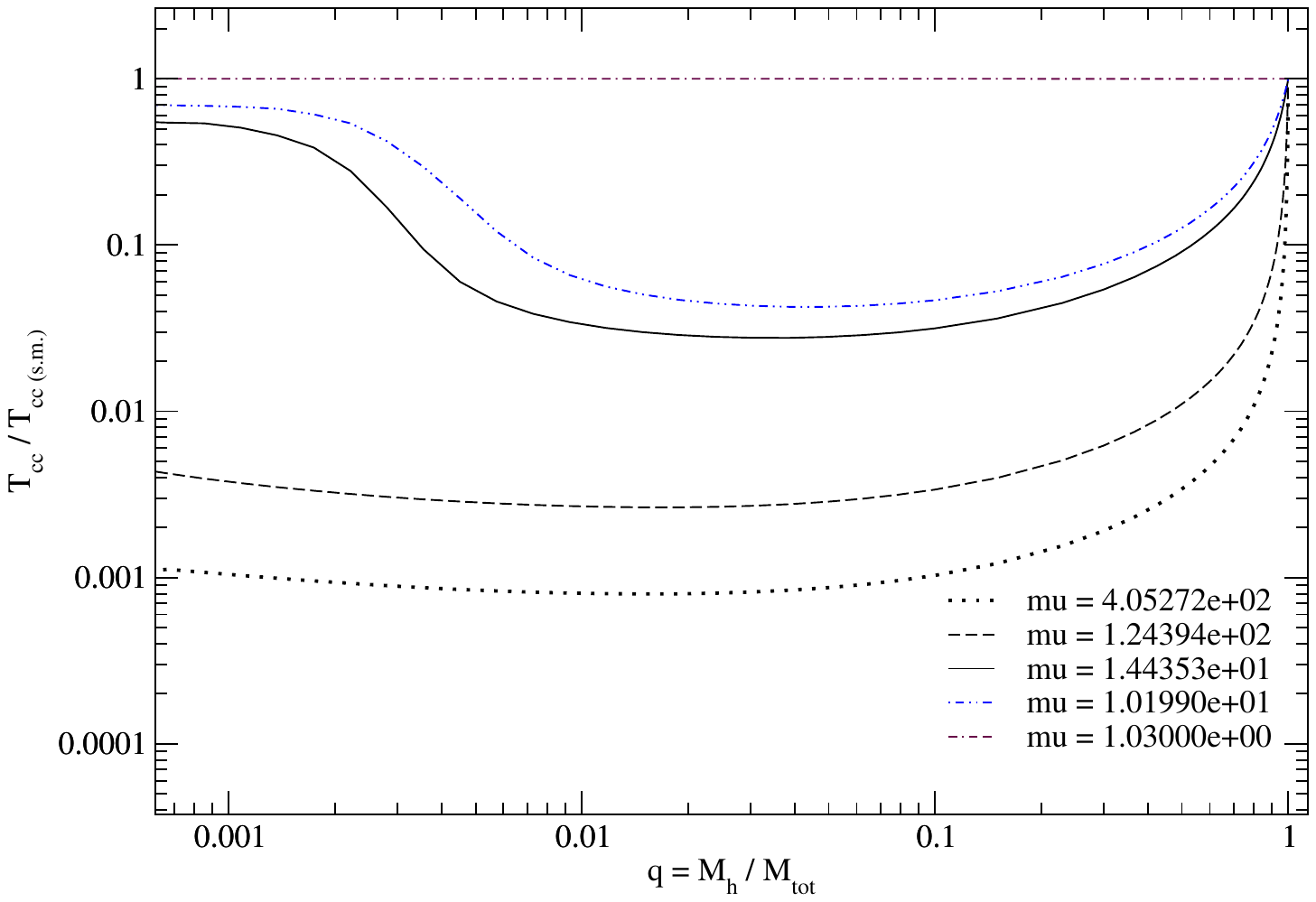}}
\caption{Core-collapse time for different values of $q$ and $\mu$
\label{fig.Tcc_q_diff_mu}}
\end{center}
\end{figure}
}

It is really interesting to compare the capacity of our approach by comparing
the results of this set of simulations to the $N$-body calculations of star
clusters with two-mass components performed by \cite{KhalEtAl07} with
direct-summation techniques.  For this aim, I plot the evolution of the
average mass in Lagrangian shells of the cluster from the averaged mass in
Lagrangian \emph{spheres} containing the following mass percentages
$[0-1],\,[2-5],\,[10-20],\,[40-50],\,[75-95]~\%$, among others, to be able to
compare with the results of \cite{KhalEtAl07}.  These are the comprised volume
between two Lagrangian radii, which contain a fixed mass fraction of the bound
stars in the system.

We have calculated the average mass as follows: If $M_{\rm r}^{({\rm i})}$ is
the total mass for the component ${\rm i}$ comprised at the radius r and
$\bar{m}_{\star}^{({\rm i})}$ is the average mass for this component within
that radius, we can find out what is the value of $\bar{m}_{\star}^{({\rm
i};\,{\rm i}+1)}$ (the average mass between $\bar{m}_{\star}^{({\rm i})}$ and
$\bar{m}_{\star}^{({\rm i}+1)}$) knowing $M_{\rm r}^{({\rm i})}$, $M_{\rm
r}^{({\rm i}+1)}$, $\bar{m}_{\star}^{({\rm i})}$ and $\bar{m}_{\star}^{({\rm
i}+1)}$. This is schematically shown in Figure~\ref{fig.lagr_av_mass}. Indeed,

\epubtkImage{.png}{
\begin{figure}[htbp]
\centerline{\includegraphics[width=0.3\textwidth]{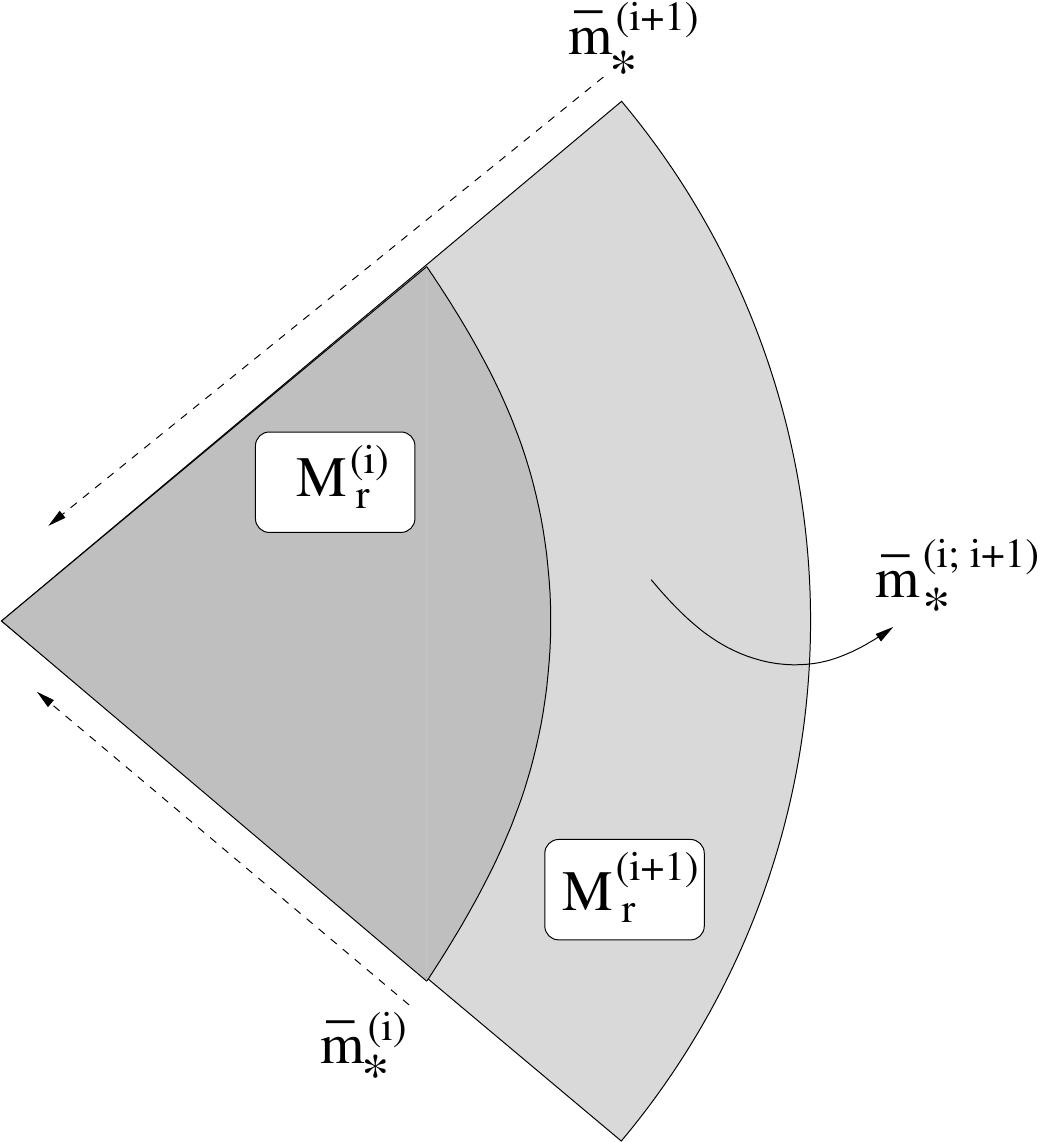}}
\caption{Average mass in Lagrangian shells from
averaged mass in Lagrangian spheres.}
\label{fig.lagr_av_mass}
\end{figure}
}

\begin{equation}
M_{\rm r}^{({\rm i}+1)} =N_{\rm r}^{({\rm i})}\cdot
\bar{m}_{\star}^{({\rm i})} + N_{\rm r}^{({\rm i};\,{\rm i}+1)} \cdot
\bar{m}_{\star}^{({\rm i};\,{\rm i}+1)}
 = N_{\rm r}^{({\rm i}+1)} \cdot
\bar{m}_{\star}^{({\rm i}+1)}.
\label{eq.aver_mass_Lag1}
\end{equation}
Since

\begin{equation}
N_{\rm r}^{({\rm i}+1)}=N_{\rm r}^{({\rm i})}+N_{\rm r}^{({\rm
i};\,{\rm i}+1)},
\end{equation}
where

\begin{equation}
N_{\rm r}^{({\rm i})}=\frac{M_{\rm r}^{({\rm
i})}}{\bar{m}_{\star}^{({\rm i})}}
\end{equation}
we have that, from Eq.~(\ref{eq.aver_mass_Lag1}),

\begin{equation}
\bar{m}_{\star}^{({\rm i};\,{\rm i}+1)}=
\frac{M_{\rm r}^{({\rm i}+1)}-M_{\rm r}^{({\rm i})}}
{\frac{M_{\rm r}^{({\rm i}+1)}}{\bar{m}_{\star}^{({\rm i}+1)}} -
\frac{M_{\rm r}^{({\rm i})}}{\bar{m}_{\star}^{({\rm i})}}}
\end{equation}

I show in Figures~\ref{fig.nbody_spedi_2comp} and
\ref{fig.spedi_spedi_2comp} the curves corresponding to the values
shown in Table~\ref{tab.mu}.

\epubtkImage{.png}{
\begin{figure}[htbp]
\centerline{\includegraphics[width=0.8\textwidth]{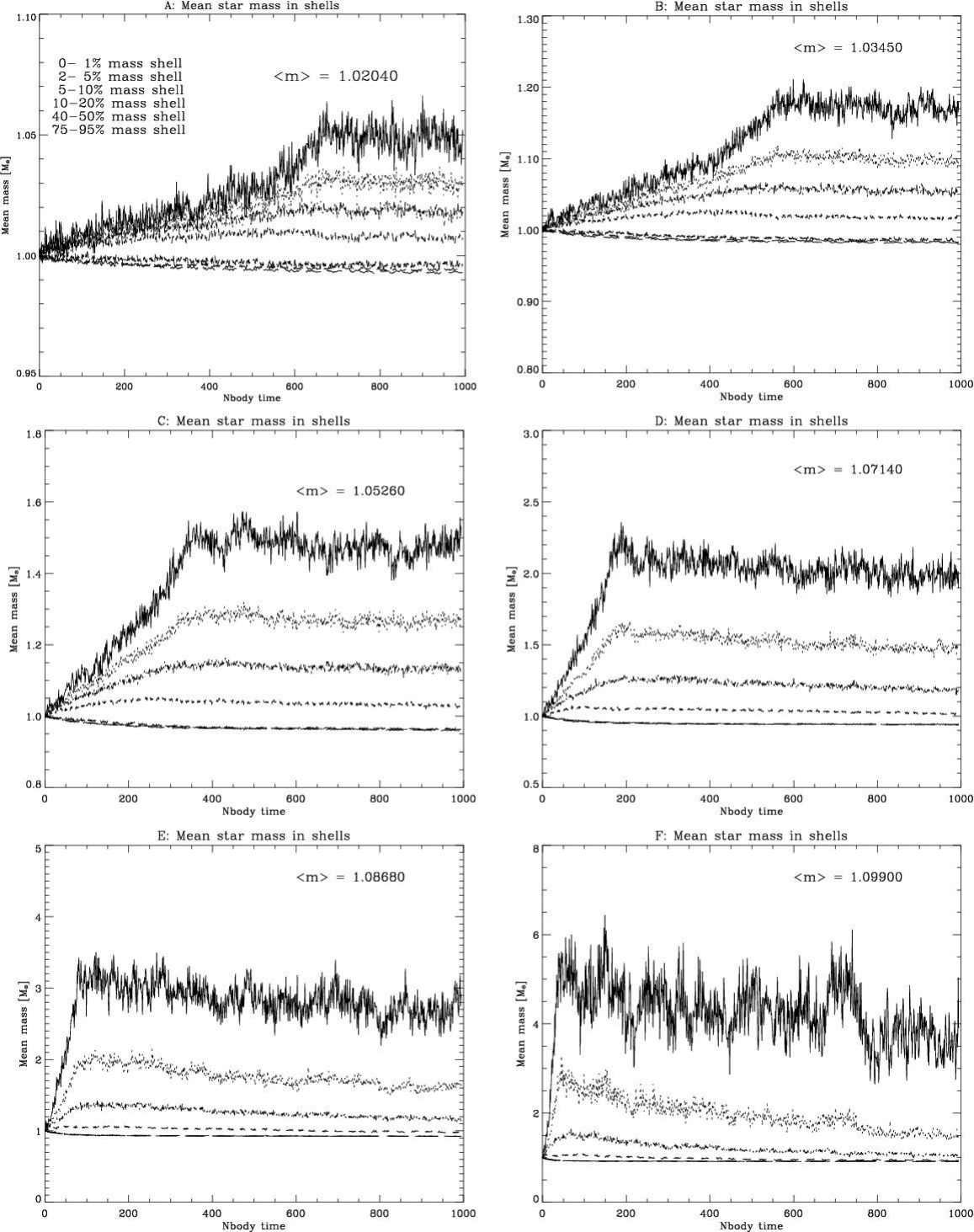}}
\caption{Average Lagrangian radii shells for the
$N$-body models of \cite{KhalEtAl07} (see the text for further explanation).}
\label{fig.nbody_spedi_2comp}
\end{figure}
}

\epubtkImage{.png}{
\begin{figure}[htbp]
\centerline{\includegraphics[]{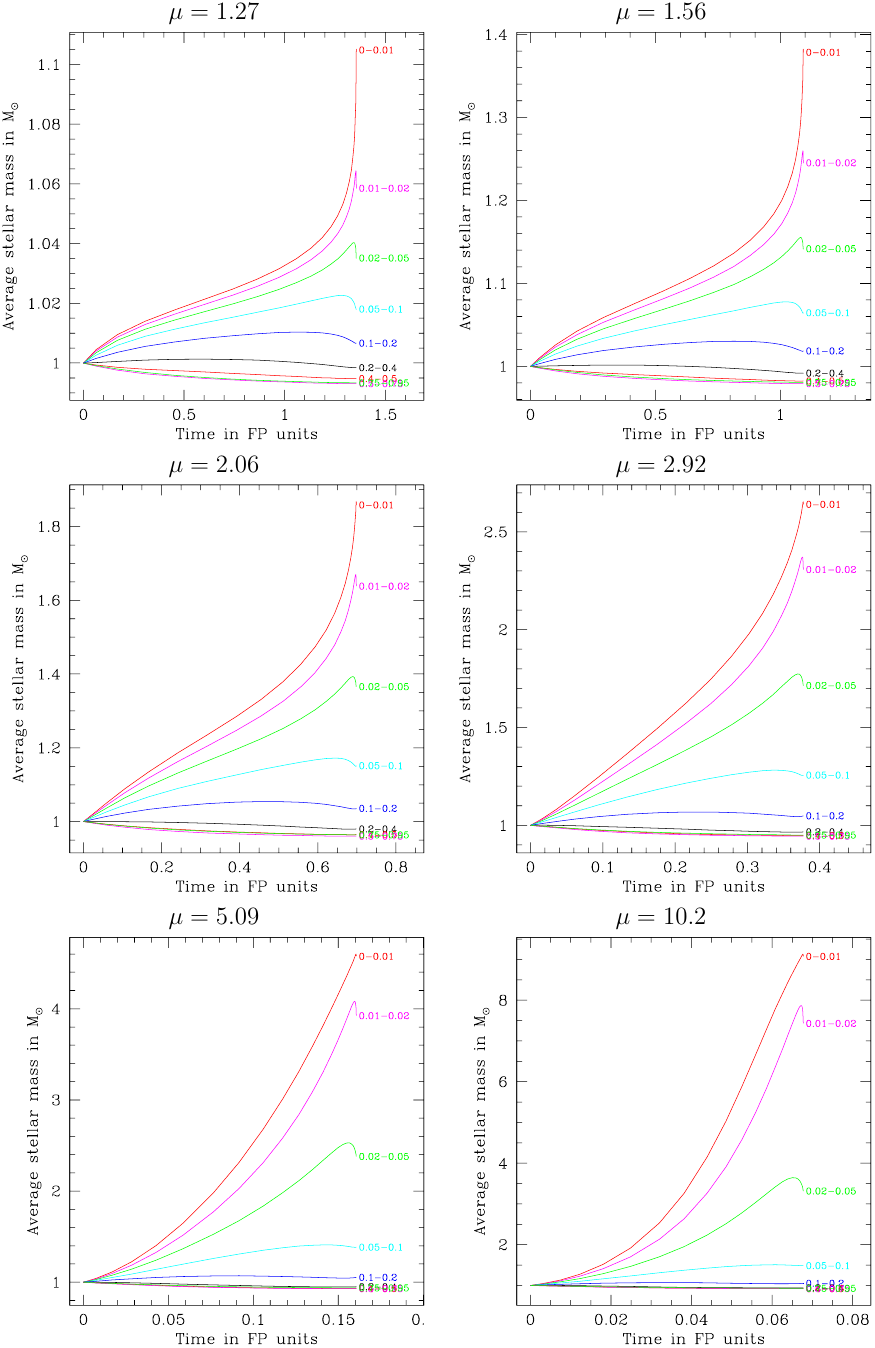}}
\caption{Average Lagrangian radii shells for our models, equivalents
to those of Figure~\ref{fig.nbody_spedi_2comp}.}
\label{fig.spedi_spedi_2comp}
\end{figure}
}

\begin{table}[htbp]
\caption{Different $\mu$ values used in the $N$-body calculations
and in our gaseous model results of Figure~\ref{fig.nbody_spedi_2comp}.}
\label{tab.mu}
\centering
\begin{tabular}{c |c c}
$\mu$ in \cite{KhalEtAl07} &
\multicolumn{2}{c}{$\mu$ in this work} \\
\hline \hline
1.25 & 1.27 \\
1.5 & 1.56 \\
2 & 2.06 \\
3 & 2.92 \\
5 & 5.09 \\
10 & 10.2 \\
\hline
\end{tabular}
\end{table}

We have followed in the curves the evolution of the system until a deep
collapse of the system. These figures show the evolution until the most massive
component dominates the centre.

In order to compare our plots with those of \cite{KhalEtAl07}, one should look
in the diagrams from the work of these authors in the region \emph{during} core
contraction. At this point, we can observe in
Fig.\,(\ref{fig.nbody_spedi_2comp}) a self-similarity after core-collapse
\cite{GH96}. Binaries are responsible for interrupting core-collapse and driving
core re-expansion in the $N$-body simulations. The flattening in the $N$-body
plots at the moment of core-collapse is due to the binary energy generation.
This means that we can only compare the steep rise, but not the saturation.

For instance, in the second plot of the $N$-body set (second column on the top
of Figure~\ref{fig.nbody_spedi_2comp}), we have to look at the point at which
the average mass of the $N$-body system is about 1.20 in the 0\,--\,1\% shell.
This establishes the limit until which we can really compare the behaviour as
given by both methods. Our simulations yield a very similar evolution until that
point. The gaseous model behaves (it clearly shows the tendency) like the
$N$-body result.

By converting the Fokker-Planck units, we find that the conversion factor is the
same; namely, for $\gamma =0.11$, $\ln (\gamma \cdot {\cal N}_{\star})/ {\cal
N}_{\star}= 0.0022$. On the other hand, the value of $\gamma$ is not so well
defined and depends on the mass spectrum \cite{Henon75}. This means that
potentially it is not the same for the different models. For a broader mass
spectrum, $\gamma$ is about 0.01 and, unfortunately, in the case of having a
small particle number, it will definitively make an important difference despite
the ``smoothing'' effect of the logarithm, viz $\ln (\gamma \cdot {\cal
N}_{\star})/ {\cal N}_{\star} = 0.0013$. Thus, in order to be able to compare
the different models, one should consider $\gamma$ as a free parameter ranging
between 0.01 to 0.2 and look for the best fit for the majority of cases. On the
other hand, we must bear in mind that the $N$-body simulations of
\cite{KhalEtAl07} do not go into deep core collapse and so, the moment at which
the core radius reaches a minimum is not the same as for our model. To sum up,
although we cannot say exactly to what point we can compare the two methods (the
Gas Model and direct-summation simulations), because the core collapse time will
be different, the physics of the system is the same in the two cases.  This
should provide the reader with a good understanding of the phenomena in play, as
well as a proof that they are independent of the details of the algorithm used.

\subsection{Clusters with a broader mass spectrum with no MBH}
\label{ref.MassSpecNoMBH}

In order to understand the phenomena that I will describe later, which is
crucial for EMRI formation, it is of relative relevance to understand first the
physics behind cluster dynamics \emph{without} a central MBH.  This section is
also interested in interpreting observations of young stellar clusters extending
to a larger number of mass components. In clusters with realistic IMFs,
equipartition cannot be reached, because the most massive stars build a
subsystem in the cluster's centre as the process of segregation goes on thanks
to the kinetic energy transfer to the light mass components until the cluster
undergoes core collapse \cite{Spitzer69,IW84,IS85}.  Although the case in which
the MBH is lurking at the centre of the host cluster is more attractive
for EMRI production and from a dynamical point of view, one should study, in a
first step, more simple models.

In this section we want, thus, to go a step further and evaluate stellar
clusters with a broad mass function (MF hereafter).  For this, I will again be
using the Gas Model, because it is a good compromise between accuracy and
integration time for this review.

We study those clusters for which the relaxation time is relatively short,
because the most massive stars will sink to the centre of the system due to mass
segregation before they have time to leave the main sequence (MS). In this
scenario we can consider, as an approximation, that stellar evolution plays no
role; stars did not have time to start evolving. The configuration is similar to
that of \cite{GFR04}, but they employ a rather different approach based on a
Monte Carlo code (MC), using the ideas of \cite{Henon73} that allow one to study
various aspects of the stellar dynamics of a dense stellar cluster with or
without a central MBH. Our scheme, although being more approximate than MC codes
(and direct-summation $N$-body ones) and unable, in its present version, to
account for collision has the advantage, as we will see in the
section~\ref{ch.Integration}, of being much faster to run, and of providing data
that has no numerical noise. It captures the essential features of the physical
systems considered in our analysis and is an interesting, powerful tool for
illustrating the different scenarios in this review.

One of the first questions we should address is the maximum number of components one
should take into consideration when performing our calculations. Since the
computational time becomes larger and larger when adding more and more
components to the system -even for an approximative scheme such as the Gas
Model-, we should first find out what is a realistic number of components in
our case. For this end I have performed different computations with different
number of stellar components.

For the simulations shown here, the initial cluster models are Plummer models
with a Salpeter IMF \cite{Salpeter55},

\begin{equation}
\frac{dN_{\star}}{dM_{\star}} \propto M_{\star}^{-\alpha}
\label{eq.salpeter}
\end{equation}
between $0.2$ and $120\,\msol$. In this equation $\alpha=2.35$.
There is no initial mass segregation.
The discretisation of the mass components follows this recipe:

\begin{equation}
\log (M_{\rm comp}|_i) =\log (M_{\rm min})+
  \log \left(\frac{M_{\rm max}}{M_{\rm min}}\right) \cdot
\left(  \frac{i}{N_{\rm comp +1}} \right)^{\delta}
\label{eq.discIMF}
\end{equation}

In this equation $\delta$ is the discretisation exponent. If $\delta >1$ we
have more bins at low mass; for $\delta <1$, we have more bins at high mass.
I.e., $\delta$ allows one to put more discretised mass components at low masses
($\delta>1$) or at high masses ($\delta <1$), $\delta$=1 gives the
logarithmical equal spacing. $M_{\rm max,\,min}$ are, respectively, the maximum
and minimum individual stellar masses for the components. For all simulations
that I present, the number of mass bins has been typically set to 15. I
have chosen a Plummer model by default and the stellar clusters have $10^6$
stars. The model radius by default is $R_{\rm Pl}=1$ pc. The default initial
mass function is Salpeter.

In Figure~\ref{fig.how_many_comp} we see the Lagrangian radii for ten
different models and look for the main dynamical characteristics of the system: the
core collapse time and the Lagrangian radii containing 90, 70, 50, 20, 10, 3,
1, 0.3, 0.1, $3\cdot 10^{-2}$, $10^{-2}$, $3\cdot 10^{-3}$ and $10^{-3}\%$ of
the stellar mass.
In this plot, $N_{\rm comp}$ stands for the mass spectrum different components
number.  For $N_{\rm comp}=6$ I have performed three simulations varying the
$\delta$ parameter between 1.0 (equal logarithmic spacing of components), 0.75
(more massive components) and 0.5 (even more). For $N_{\rm comp}=12$ I have
performed only one simulation (with $\delta=1$, by default); for the $N_{\rm
comp}=20$ case I have repeated the same procedure as with $N_{\rm comp}=6$, the
penultimate one that I have chosen is $N_{\rm comp}=20$ and, in this case, we
studied two grid resolutions, $N_{\rm sh}=200$ (the default value) and 400 grid
points, in order to check whether this could influence the results.
To finish with, a last simulation with $N_{\rm comp}=50$ was
performed and included in the analysis. Whilst we can see an important
difference between models of 6 and 12 components, we see that the global
behaviour from 12 components onwards is very similar. Therefore, unless
indicated, I choose 15 components in our study in this section, since a higher
number would not contribute anything essential.

\epubtkImage{.png}{
\begin{figure}[htbp]
\centerline{\includegraphics[width=0.8\textwidth]{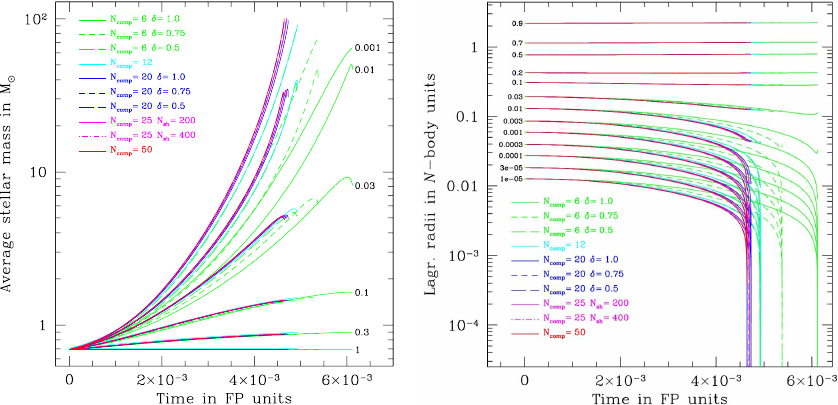}}
\caption{Lagrangian radii and average stellar mass for 10 models
with different mass spectrum (see text for details).}
\label{fig.how_many_comp}
\end{figure}
}

To see this in more detail, in Figure~\ref{fig.how_many_2figs} I show the
Lagrangian radii for each stellar mass $m_i$ and the corresponding mass
fraction $f_m$ for the 25 and 15 components simulations. Again, we cannot see
any substantial difference between the 25 and 15 cases.

\epubtkImage{.png}{
\begin{figure}[htbp]
\centerline{\includegraphics[]{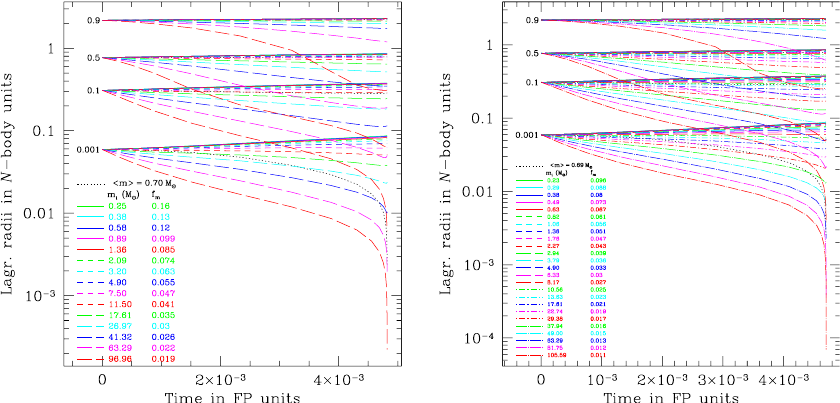}}
\caption{Lagrangian radii for each stellar mass $m_i$ for the cases of 25
and 15 mass components.}
\label{fig.how_many_2figs}
\end{figure}
}

Taking the last arguments into account, I have done an analysis of mass
segregation in multi-mass models with more than two stellar components without
MBH. In Figures~\ref{fig.mass_seg_multi_noBH_1} and
\ref{fig.mass_seg_multi_noBH_2}, I show the evolution of a stellar
cluster of 15 components (in colours); $m$ is the mass (in $\msol$) of the
stars in each component and $f_m$ the corresponding fraction of the total mass.
In the upper box we have the density profile, where the solid black line
represents the total density; below, we have the average total mass for the
system. I show different snapshots of the system. At $T=0$ we have the initial
model, which duly shows no mass segregation. As time passes, at $T=5.30\cdot
10^{-2}\,T_{\rm rh}(0)$, with $T_{\rm rh}(0)$ the value of $T_{\rm rh}$ at the
beginning of the simulation, we observe how mass segregation has fragmented the
initial configuration; the heavy components have sunk into the central
regions of the stellar cluster and, thus, increased the mean average mass. The
outer parts of the system start losing their heavy stars quickly and,
consequently, their density profile decreases. This becomes more acute for
later times at $T=6.75\cdot 10^{-2}\,T_{\rm rh}(0)$, as the plots on the right
in Figure~\ref{fig.mass_seg_multi_noBH_2} show. In these plots and, more clearly
in the right panel of density profile, we can observe a depletion at
intermediate radii.

\epubtkImage{.png}{
\begin{figure}[htbp]
\centerline{\includegraphics[width=0.7\textwidth]{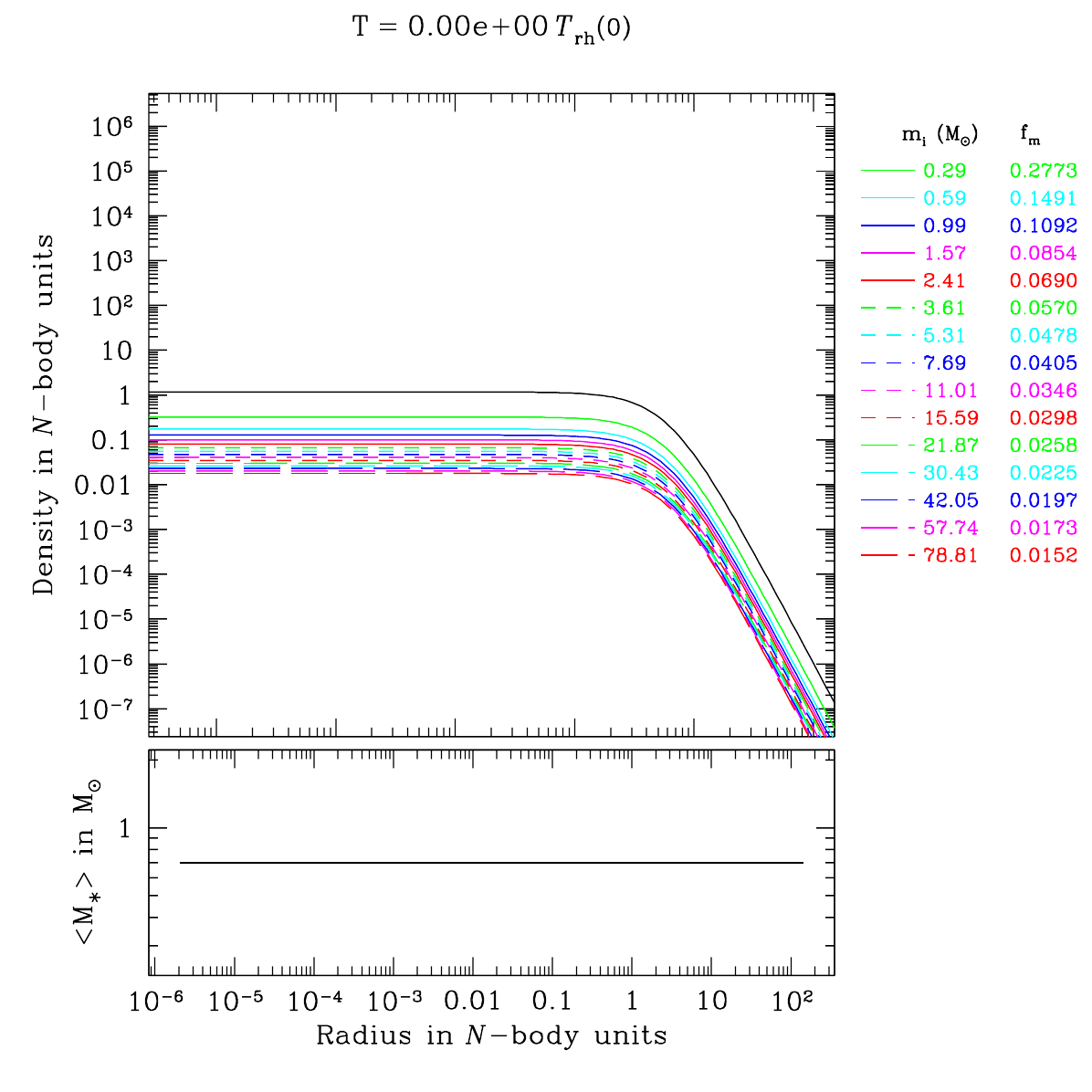}}
\caption{Initial density profile for a stellar cluster with 15
components (upper panel) in $N$-body units and average total
stellar mass in $\msol$ (lower panel).}
\label{fig.mass_seg_multi_noBH_1}
\end{figure}
}

\epubtkImage{.png}{
\begin{figure}[htbp]
\centerline{\includegraphics[width=\textwidth]{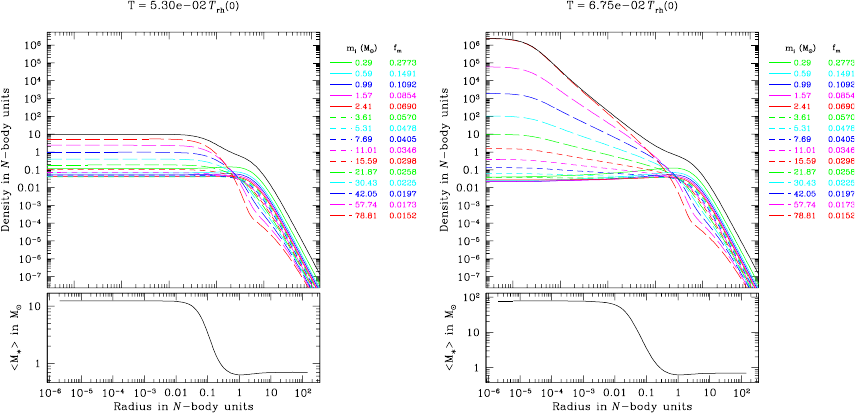}}
\caption{Same situation as in Figure~\ref{fig.mass_seg_multi_noBH_1} but
at later times. See text for explanations.}
\label{fig.mass_seg_multi_noBH_2}
\end{figure}
}

\subsection{Core-collapse evolution}

\cite{GFR04} show that for a broad MF -- either Salpeter or
Kroupa --, mass segregation produces a core-collapse of the system that happens
very fast. For clusters of moderate initial concentration, they find that this
happens in about 10\% of the $T_{\rm rh}(0)$, the initial half-mass relaxation
time (i.e., the half-relaxation time that the cluster had when time started, at
$t=0$). A good and clear illustration of this is Figure~\ref{fig.msegr} and
Figure~\ref{fig.core_coll_spedi_king}. In the former, on the left panel we
have the initial configuration of the system. On the right one, we have the
cluster at the moment of core-collapse. In the figure, all stars within a slice
containing the centre have been depicted. On the other hand, this does not
represent a real physical system, because all radii have been magnified
(see the bottom of each panel). The dashed circles represent spheres containing 1,
3 and 10\,\% of the total cluster mass (from the centre). We can clearly see
how the massive, large stars are segregated towards the centre. In
Figure~\ref{fig.core_coll_spedi_king}, I show the core-collapse evolution of
a multi-mass stellar cluster simulated with the gaseous model. As usual, $m$ is
the mass (in $\msol$) of the stars in each component and $f_m$ the
corresponding fraction of the total mass. On the left panel I display the time
evolution of the central density for a model in which I have employed 15
individual mass components. The total density is given by the dotted line. On
the right panel we have the evolution of the central velocity dispersions. The
dotted black line shows the mass-averaged value

\begin{equation}
\bar{\sigma}^2=\sqrt{\frac{\sum_{i=1}^{15}m_i\,\sigma_i^2}{\bar{m}}},
\end{equation}
and I use $N$-body units for the $y$-axes.

\epubtkImage{.png}{
\begin{figure}[htbp]
\centerline{\includegraphics[]{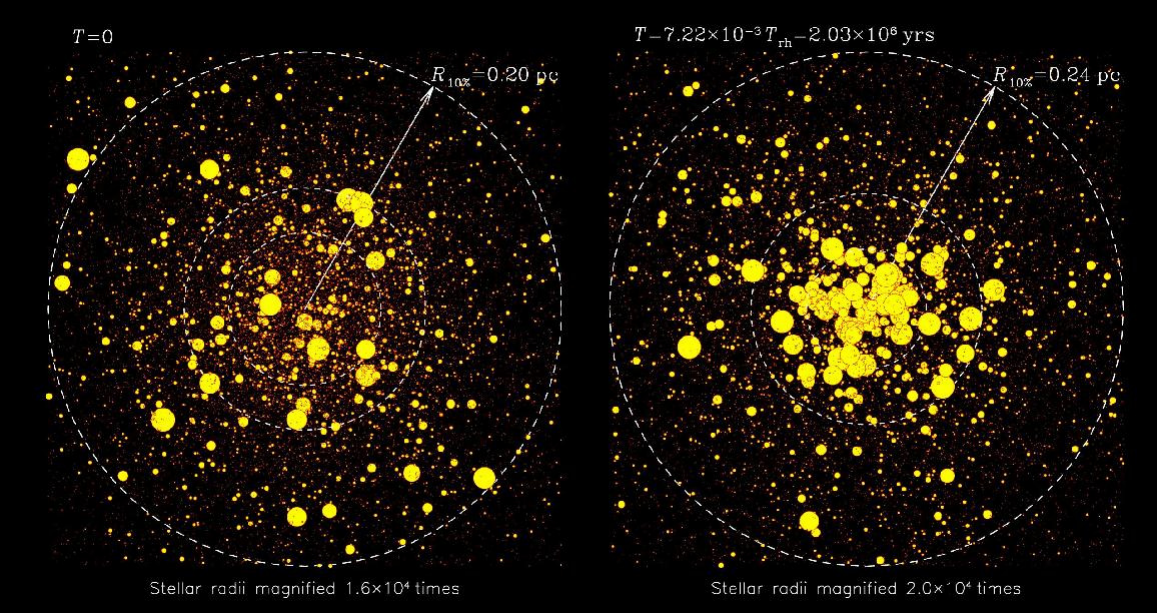}}
\caption{Illustration of core-collapse in multi-mass systems treated with
a Monte Carlo approach (courtesy of M.\ Freitag).}
\label{fig.msegr}
\end{figure}
}

\epubtkImage{.png}{
\begin{figure}[htbp]
\centerline{\includegraphics[width=0.9\textwidth]{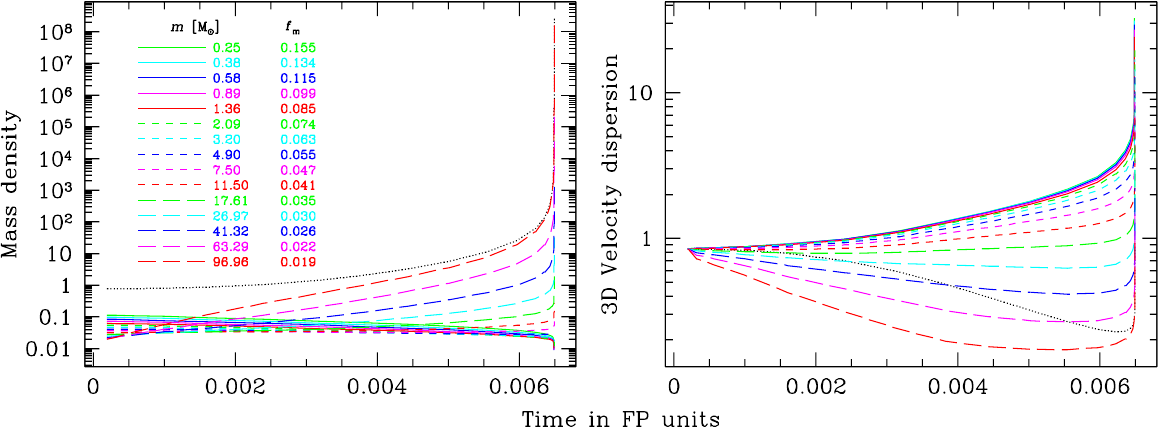}}
\caption{Evolution of the central density and $3D$-velocity dispersion
in a model with 15 components (see text for further explanations).}
\label{fig.core_coll_spedi_king}
\end{figure}
}

One notes that, during core collapse, the central regions of the cluster become
completely dominated by the most massive stars. But, contrary to the case of
single-mass clusters, the central velocity dispersion \emph{decreases} (see
Figure~\ref{fig.core_coll_spedi_king}).

\subsection{Clusters with a broader mass spectrum with a MBH}
\label{sec.BroadMassMBH}

Afer having addressed the systems studied in previous sections we now look into
the dynamical problem of a multi-mass component cluster harbouring a central
seed MBH that grows due to stellar accretion.

In this section I extend our analysis to systems for which I use an
evolved mass function of an age of about 10~Gyr. We consider a mass spectrum
with stellar remnants. We employ a Kroupa IMF \cite{KroupaEtAl93,Kroupa01}
with ZAMS mass\epubtkFootnote{The zero age main sequence (ZAMS) corresponds to the
position of stars in the Hertzsprung--Russell diagram where stars begin hydrogen
fusion.} from $0.1$ to $120\,\msol$ with the turn-off mass of $1\,\msol$.  I
have chosen the following values for the exponent according to the mass
interval,

\begin{equation}
 \alpha = \left\{ \begin{array}{lll}
 1.3, & 0.008 \leq m_{\star}/\msol < 0.5  \\
 2.2, & 0.5 \leq m_{\star}/\msol < 1 \\
 2.7, & 1 \leq m_{\star}/\msol \leq 120.
                              \end{array}
                              \right.
\end{equation}
And with the following distribution of components,

\begin{itemize_estret}

\item[(i)]  Main sequence stars of $0.1\mbox{\,--\,}1\,\msol$ ($\sim$ 7
components)
\item[(ii)] White dwarfs of $\sim \, 0.6\,\msol$ (1 component)
\item[(iii)] Neutron stars of $\sim \,1.4\,\msol$ (1 component)
\item[(iv)] Stellar black holes of $\sim\,10\msol$ (1 component)

\end{itemize_estret}

The defined IMF evolves and provides an evolved population with compact
remnants. This means that main sequence stars can be transformed into white
dwarfs, neutron stars or stellar-mass black holes according to their masses. If
$m_{\rm MS}$ is the mass of a MS star, I have defined the following mass ranges
for the evolution into compact remnants:

\begin{itemize_estret}

\item[(a)] White dwarfs in the range of $1 \leq m_{\rm MS}/\msol < 8$
\item[(b)] Neutron stars for masses $8 \leq m_{\rm MS}/\msol < 30$
\item[(c)] Stellar black holes for bigger masses, $\geq 30 \msol$

\end{itemize_estret}

As I have already mentioned, I place at the centre a seed BH whose initial mass is
$50\,\msol$. The initial model for the cluster is a Plummer sphere with a
Plummer radius $R_{\rm Pl}=1$ pc. The total number of stars in the system is
${\cal N}_{\rm cl}=10^6$.

The presence of a small fraction of stellar remnants may greatly affect the
evolution of the cluster and growth of the MBH because they segregate to the
centre, and in doing so, they expel MS stars from it but, being compact, they
cannot be tidally disrupted. This kind of evolution is shown in
Figures~\ref{fig.norm10comp_raylag} and \ref{fig.before_afterCC_multiBH}.

Figure~\ref{fig.norm10comp_raylag} shows us the time evolution of different
Lagrange radii with 0.1, 10, 50, 80\% of the mass of each component.
Here the core collapse happens at about $T=0.18\,T_{\rm rh}(0)$.  The
later re-opening out is due MBH accretion.

In Figure~\ref{fig.before_afterCC_multiBH}, I plot the density profiles of the
system before and after the post-collapse phase. We can also see that the slope
of $\rho\propto R^{-7/4}$ on account of the cusp of stellar-mass black holes
that has formed around the central MBH. We can see how the different components
redistribute in the process, as I mentioned at the beginning of this section.
We can see how the MBH dominates the dynamics at the centre.

\epubtkImage{.png}{
\begin{figure}[htbp]
\centerline{\includegraphics[width=0.6\textwidth]{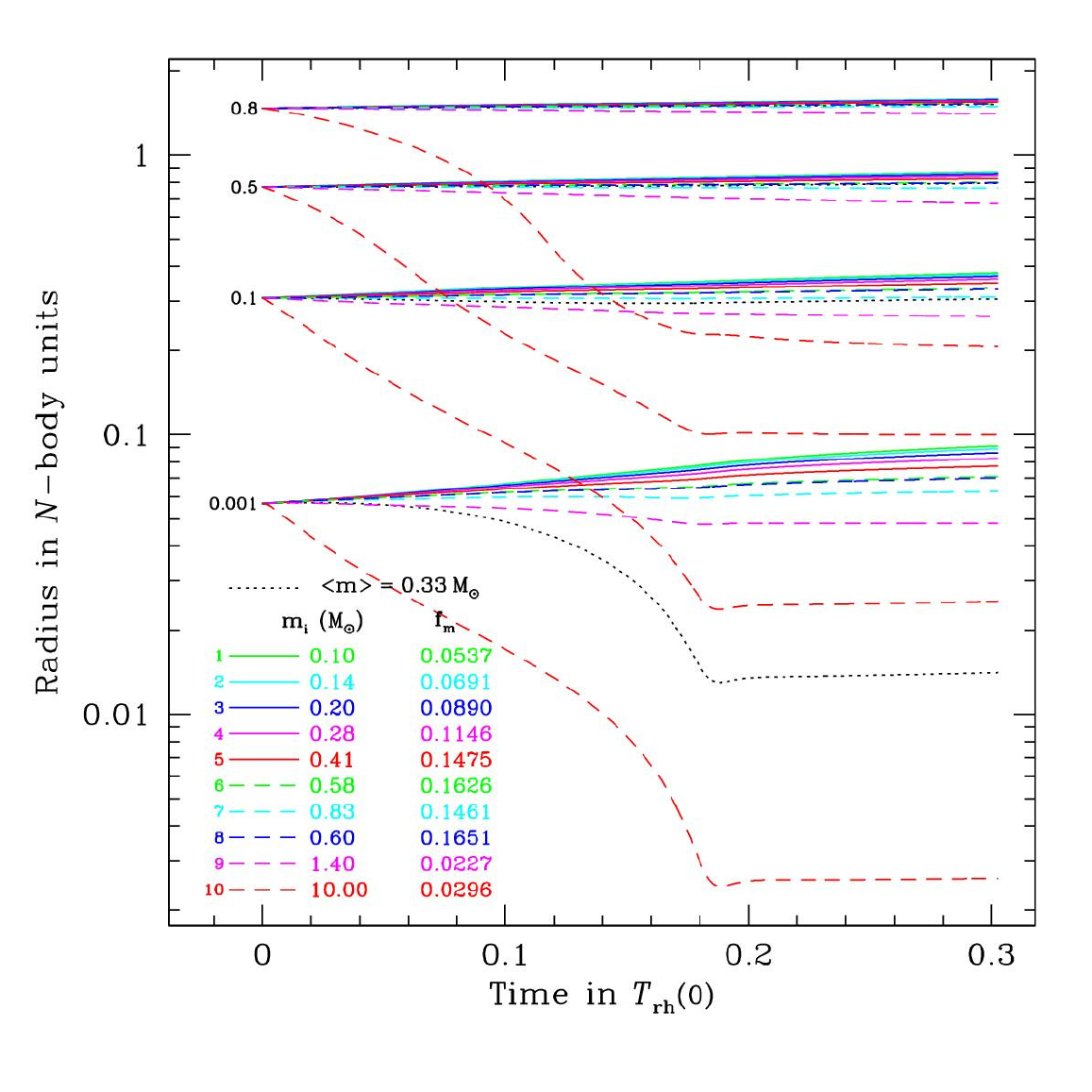}}
\caption{Lagrange radii evolution for a 10 components
calculation with a seed BH and stellar remnants.}
\label{fig.norm10comp_raylag}
\end{figure}
}

\epubtkImage{.png}{
\begin{figure}[htbp]
\centerline{\includegraphics[]{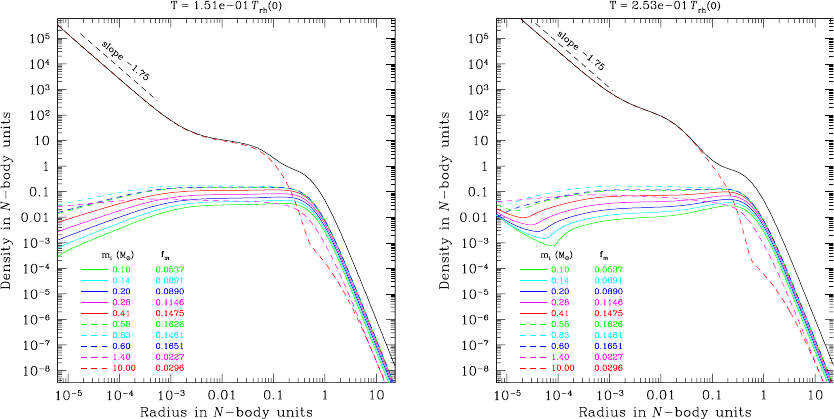}}
\caption{Density profiles in a multi-mass system with seed
BH before and after core-collapse (see text).}
\label{fig.before_afterCC_multiBH}
\end{figure}
}

We can study how the system evolves from the point of view of the distribution
of kinetic energies between the different components of the clusters during the
process of mass segregation.

In Figure~\ref{fig.temp_10comp_BH} I show the evolution of the
``temperature'' of the system, defined as the mean kinetic energy per star
divided by the global mean mass (in order to have a ``temperature'' expressed
in square velocity units).
In this plot I show the core collapse situation corresponding to
Figures~\ref{fig.norm10comp_raylag} and \ref{fig.before_afterCC_multiBH}. I
consider a 10 component cluster with the characteristics explained before. The
mean temperature is defined as

\begin{equation}
\langle T \rangle = \frac{\sum n_{\rm i}\,T_{\rm i}}{\sum n_{i}},
\end{equation}

\noindent
where $n_{\rm i}$ is the numerical local density for component $i$. This
corresponds to the mean kinetic energy per star. We can see
in Figure~\ref{fig.temp_10comp_BH} that it
is about the same as the heaviest component in the inner regions, even though
one could think that segregation should not have set in the beginning. This
is due to the fact that the moment does not correspond to exactly the initial
moment, $T=0$. We can already see how the mean central
temperature moves back as time passes (solid black line) and the most massive component
(dashed red line) increases. For later times, the kinetic energies of the
different components rise at the inner part of the cluster and the most massive
one approaches the sum of all of them. This is even more evident in the last
plot, where the temperatures of all components sink except for that
corresponding to the most massive one.

\epubtkImage{.png}{
\begin{figure}[htbp]
\centerline{\includegraphics[width=\textwidth]{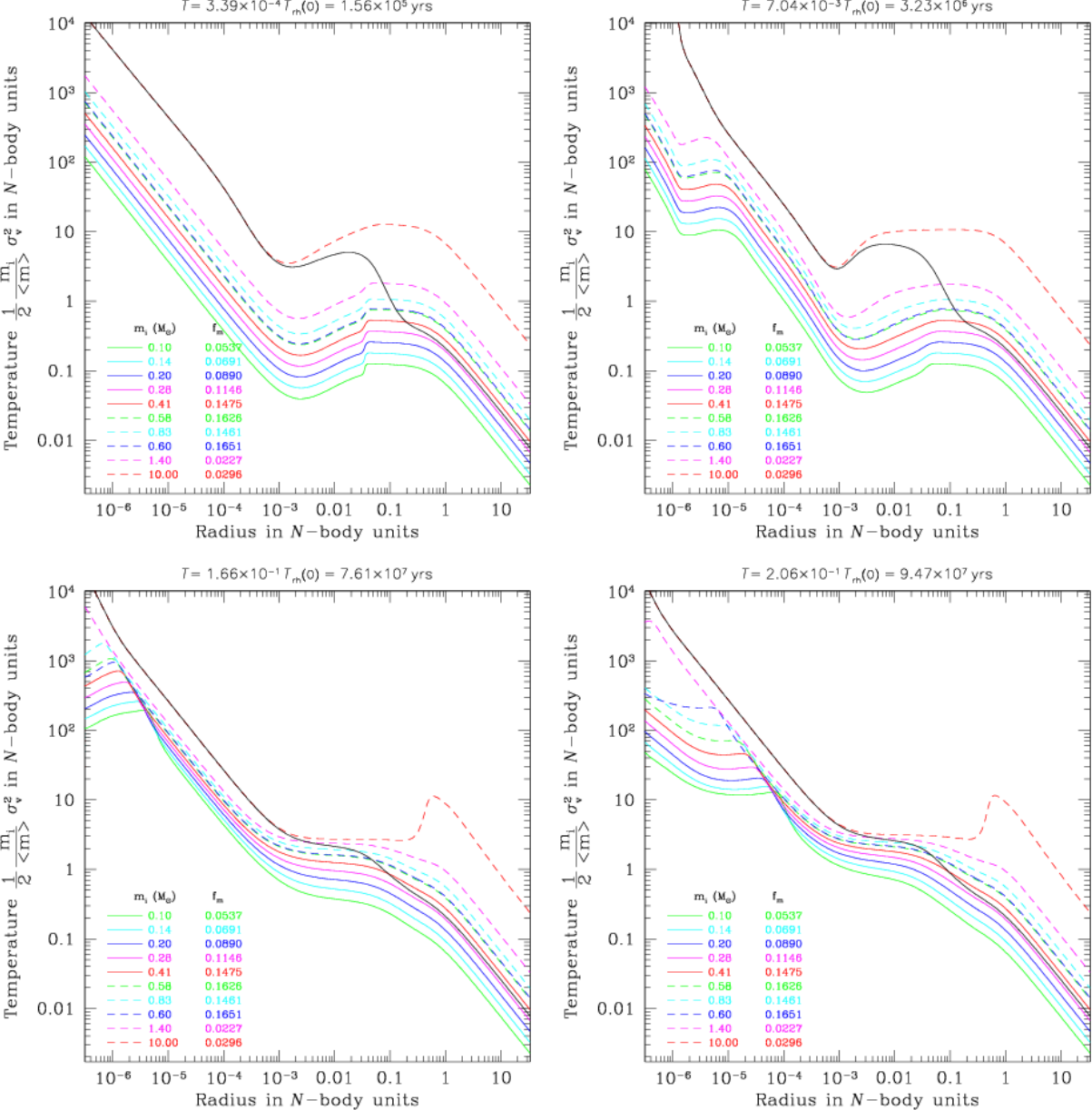}}
\caption{Different moments in the evolution of the cluster
temperature for a 10 stellar components system with a seed MBH.}
\label{fig.temp_10comp_BH}
\end{figure}
}

\section{Two-body Extreme-mass ratio inspirals}
\label{ch.TwoBodyEMRIs}

After the first sections we have a good understanding of the fundamentals of
two-body relaxation in dense stellar systems, including mass segregation and
dynamical friction, which could be roughly described as ``relaxation when we
have a large mass ratio''. In this section I address the subject of capture of
compact objects by a massive black hole considering that the driving mechanism
in the production is two-body relaxation.

\subsection{A hidden stellar population in galactic nuclei}

The question about the distribution and capture of stellar-mass black holes at
the Galactic Centre has been addressed a number of times by different authors,
from both a semi- or analytical and numerical standpoint, see e.g.
\cite{SR97,MEG00,Freitag01,Freitag03,FAK06a,FAK06b,HopmanAlexander06,Amaro-SeoaneEtAl07,PretoAmaroSeoane10,Amaro-SeoanePreto11}.
Addressing this problem has implications for a variety of astrophysical
questions, including of course inspirals of compact objects onto the central
MBH, but also on the distribution of X-ray binaries at the Galactic Centre,
tidal disruptions of main sequence stars, and the behaviours of the so-called
``source'' stars, which were introduced in Section~\ref{sec.GrowthMBHs}. Even if
we only consider single stellar-mass black holes, the impact they can have on
the S-stars is not negligible; a distribution of non-luminous matter around the
Galactic Centre would have a clear fingerprint on their orbits. Current data are
insufficient to detect such an extended non-luminous cusp which typically would
induce a slight Newtonian retrograde precession \cite{MouawadEtAl05}, so that we
will have to wait for future telescopes before we can hope to see such
trajectory deflections. The study of \cite{WeinbergEtAl05} estimated that proposed 30 to 100
meter aperture telescopes will allow us to observe about three trajectory
deflections per year between any of the monitored ``source'' stars and a
stellar-mass black hole.

The centermost part of the stellar spheroid, the \emph{galactic nucleus,}
constitutes an extreme environment for stellar dynamics.  With stellar densities
highter than $10^6\,\Mpthree$, relative velocities in excess of 100\,{\kms} the
nucleus (unlike most of the rest of the galaxy) is the site of a variety of
``collisional processes'' -- both close encounters and actual collisions between
stars, as we have seen in the previous sections.  The central MBH and the
surrounding stellar environment interact through various mechanisms: some are
global, like the accretion of gases liberated by stellar evolution or the
adiabatic adaptation of stellar orbits as the mass of the MBH increases; others,
which involve the close interaction between a star and the MBH -- EMRIs and
stellar disruptions -- are local in nature.  As we have seen in
Section~\ref{sec.angles}, to interact closely with the central MBH, stars have
to find themselves on ``loss-cone'' orbits, which are orbits elongated enough to
have a very close-in periapsis \cite{FR76,LS77,AS01}.

The rate of tidal disruptions can be established (semi-)analytically if the
phase space distribution of stars around the MBH is known, see
\cite{MT99,SU99,WM04} for estimates in models of observed nearby nuclei.
However, in order to account for the complex influence of mass segregation,
collisions and the evolution of the nucleus over billions of years, detailed
numerical simulations are required, as in the work of
\cite{DDC87a,DDC87b,MCD91,FB02b,BME04b,FAK06a,KhalEtAl07,PretoAmaroSeoane10,Amaro-SeoanePreto11}.

In the case of a gradual inspiral following the ``capture'' of a compact object
(i.e., an EMRI), the situation becomes even more complex, even in the idealized case
of a spherical nucleus with stars all of the same mass. As the star spirals
down towards the MBH, it has many opportunities to be deflected back by
two-body encounters on to a ``safer orbit'', i.e., an orbit which does not
lead to gravitational capture,  \cite{AH03}, hence even the
definition of a loss-cone is not straightforward. Once again, the problem is a
compound of the effects of mass segregation, general relavity and resonant
relaxation, to mention three main complications. As as result, considerable
uncertainties are attached to the (semi-)analytical predictions of capture
rates and orbital parameters of EMRIs.

Only self-consistent stellar dynamical modeling of galactic nuclei will provide
us with a better understanding of these questions. Some steps in that direction
have been made by\cite{Freitag01,Freitag03,Freitag03b} using Monte Carlo
simulations. Later, \cite{FAK06a,FAK06b} improved upon these results. Yet these
studies neglected a direct estimation of EMRIs or ``direct plunges'', due in part
to the fact that, to follow stars on very eccentric orbits, one needs the
combined effects of GW emission and relaxation on timescales much shorter than
the capabilities of the numerical Monte-Carlo code.  Much work remains to be
done to confirm these results and improve on them with a more accurate treatment
of the physics, to extend them to a larger domain of the parameter space and to
more general situations, including non-spherical nuclei.

Classical studies based on approximate stellar dynamics methods that neglect,
in particular, the motion of the central MBH and strong 2-body interactions,
indicate that, in dense enough clusters, a ``seed'' MBH (in the IMBH mass range)
could swallow a significant fraction of the cluster mass, and thus become a MBH
over the span of a few Gyrs \cite{MCD91,FB02b,ASFS04}.  More detailed, higher
fidelity $N$-body simulations of relatively small clusters
\cite{BME04a,BME04b} have not confirmed this classical result, calling for a
critical re-examination and improvement of approximation techniques, the only
ones that can cope with the high particle numbers found in massive clusters
such as galactic nuclei.  It has also been suggested that some processes, such
as the effects of chaotic orbits in a slightly non-spherical potential, may
effectively keep the loss-cone orbits populated. In this case disruptions and
captures can efficiently feed the central MBH and produce the $M-\sigma$
relation \cite{ZHR02,MP04}.

Understanding the astrophysical processes within galacto-centric clusters that
give rise to EMRI events has significant bearing on LISA's applicability to this
science. Accurate predictions of the event rate are important for preparing LISA
data analysis and design --- many events lead to source-confusion, which must be
dealt with, while a few events necessitate identifying weak sources in the
presence of instrumental noise \cite{Amaro-SeoaneEtAl07}. More importantly, LISA
observations alone cannot decouple the mass distribution of the galactic black
hole population from the mass-dependence of the EMRI rate within a single
system.  If we can improve our understanding of the latter, we improve LISA's
utility as a probe of the former.  In this section I elaborate in detail on the
``standard'' physics leading to sources of gravitational radiation in the
milliHertz regime --i.e.  in the bandwidth of a LISA-like detector-- originating
in two-body relaxation processes.

\subsection{Fundamentals of EMRIs}
\label{sec.fundamentalEMRIs}

In the simplest idealization, an EMRI consists of a binary of two compact
objects, a massive black hole (MBH) and a -- typically -- stellar black hole
(SBH) describing a large number of cycles around the MBH as it approaches the
LSO, emitting important, coherent amounts of GWs at every
periapsis passage\epubtkFootnote{The systems emits gravitational radiation all
the time, but the most important bursts of energy occur at periapsis}. After
every $2\pi$ around the orbit, the semi-major axis decays a fraction
proportional to the energy loss. After typically some $10^{4\mbox{\,--\,}5}$
cycles, the small body, the CO, plunges through the horizon of the MBH and is
lost. The emission of GW finishes. This is what makes this system so attractive.
We can regard it as a camera flying around a MBH taking extremely detailed
pictures of the space and time around it. With one EMRI we are provided with a
set of $\sim 10^{4\mbox{\,--\,}5}$ pictures from a binary, and the information
contained in them will allow us also to know with an unprecedent accuracy in the
history of astronomy about the mass of the system, the inclination, the
semi-major axis, the spin, to mention some, and it will also be an accurate test
of the general theory of relativity.

At first glance the task seems simple and, of course, worth doing; we just have
to analyse a binary which decays slowly in time proportionally to $a^4$, where
$a$ is the semi-major axis. The work seems to be easy for such a big gain. The
only problem is that it is not as easy as it seems, because we need to
understand how a star can become an EMRI in such a dynamically complex system as
a galactic nucleus. Also, the EMRI might suffer perturbations either from gas
or from the stellar system
\cite{KocsisEtAl11,Amaro-SeoaneBremCuadraArmitage2012,BarausseEtAl2014}.

In Figure~\ref{fig.EMRI_LISAplot_4MBHs} I show what systems would missions such
as LISA be more sensitive to. Obviously, this is only an illustration and the
data analysis of the signal will be much more complicated in reality, but it is
just an indication already that if the central MBH has a mass larger than
$10^7\,M_{\odot}$, then the signal, even at the LSO, will have a frequency too
low for detecting the system. On the other hand, if it is less massive than
$10^4\,M_{\odot}$, the signal will also be quite weak unless the source is very
close. This is why one usually assumes that the mass range of MBHs of interest
in the search of EMRIs for LISA is between $[10^4,\,10^7]\,M_{\odot}$. We note
that this picture is shifted towards lighter masses in the eLISA configuration,
as explained in \cite{Amaro-SeoaneEtAl2012,Amaro-SeoaneEtAl2012b}. Nonetheless,
if the MBH is rotating fast, then even if it has a mass larger than
$10^7\,M_{\odot}$, the LSO will be closer to the MBH and thus, even at a higher
frequency the system should be detectable. This would push to the left the total
mass to a few $\sim 10^7\,M_{\odot}$.  Indeed, in Figure~1 of \cite{Gair2009} we
can see how the sensitivity varies as we vary the spin of the MBH. The
sensitivity limit for non-spinning black holes is about $5\times 10^6\,M_{\odot}$,
but this goes up to a few times $10^7\,M_{\odot}$ for prograde inspirals into
rapidly spinning black holes.  More recently, in Figure~5 of
\cite{BabakEtAl2017} we have sky-average horizons for prograde inspirals into
maximally spinning black holes. The authors show that we can see inspirals out
to $z \sim 1$ even if the MBH has a mass of $10^7\,M_{\odot}$.  From the point
of view of astrophysics, this range of masses corresponds to low-mass SMBHs.
They are not easily detectable and we do not know much about them.

A different way of looking at the same picture is
Figure~\ref{fig.Binary_Porb_Diag_LISA}. I depict, as a function of the total,
non-redshifted mass of the binary $M_1 + M_2$, the semi-major axis of the binary
assuming zero eccentricity.  We note here that, even if for some particular
models, LISA can in principle detect EMRIs out to a redshift of $\sim 4$, see
the work of \cite{BabakEtAl2017}, most EMRIs will very likely originate from
within $z\sim 1$, so that for the rest of this work I neglect it.  In
Figure~\ref{fig.Binary_Porb_Diag_LISA} I show the orbital frequency of the
binary. Obviously, for the binary to be in the LISA band, it has to have a
frequency of roughly --being generous-- between $1$ and $10^{-5}$ Hz. The
emission of GWs is more efficient as they approach the LSO, so that LISA will
detect the sources when they are close to the LSO line. For masses larger than
$10^7\,M_{\odot}$ the frequencies even close to the LSO will be too low, so that
their detection will be very difficult. On the other hand, for a total mass of
less than $10^3\,M_{\odot}$ in principal we could detect them at an early stage
but then the amplitude of the GWs would be rather low. On top of that, the
existence of intermediate-mass black holes is uncertain.

\epubtkImage{.png}{
\begin{figure}[htbp]
\centerline{\includegraphics[width=0.55\textwidth]{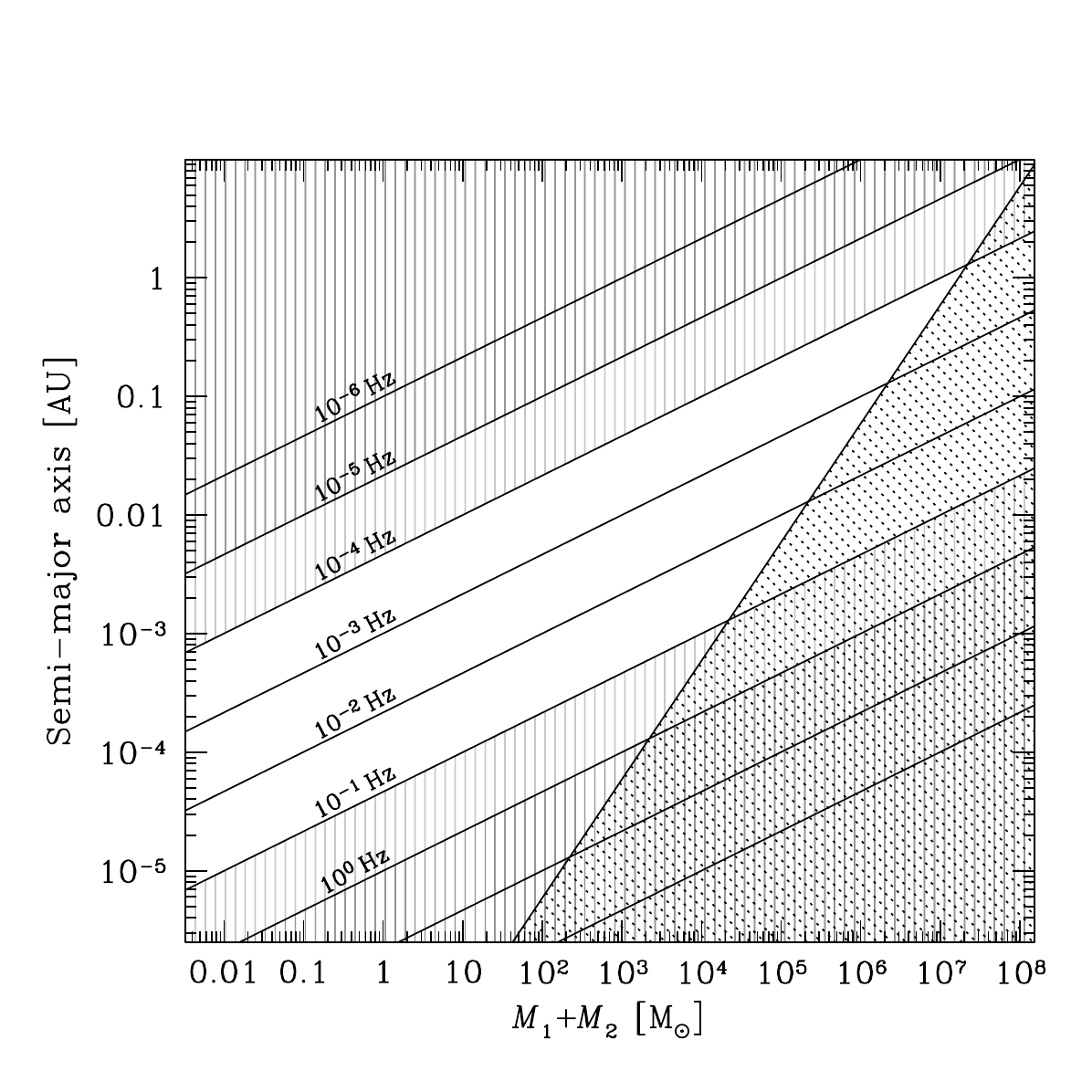}}
\caption{Frequency of a binary of total mass $M_1 + M_2$ against their semi-major
axis and the corresponding frequencies. The solid, dark straigt line delimites the
LSO, so that anything on the right of that line is of no
interest for our purposes.}
\label{fig.Binary_Porb_Diag_LISA}
\end{figure}
}

\epubtkImage{.png}{
\begin{figure}[htbp]
\centerline{\includegraphics[width=0.52\textwidth]{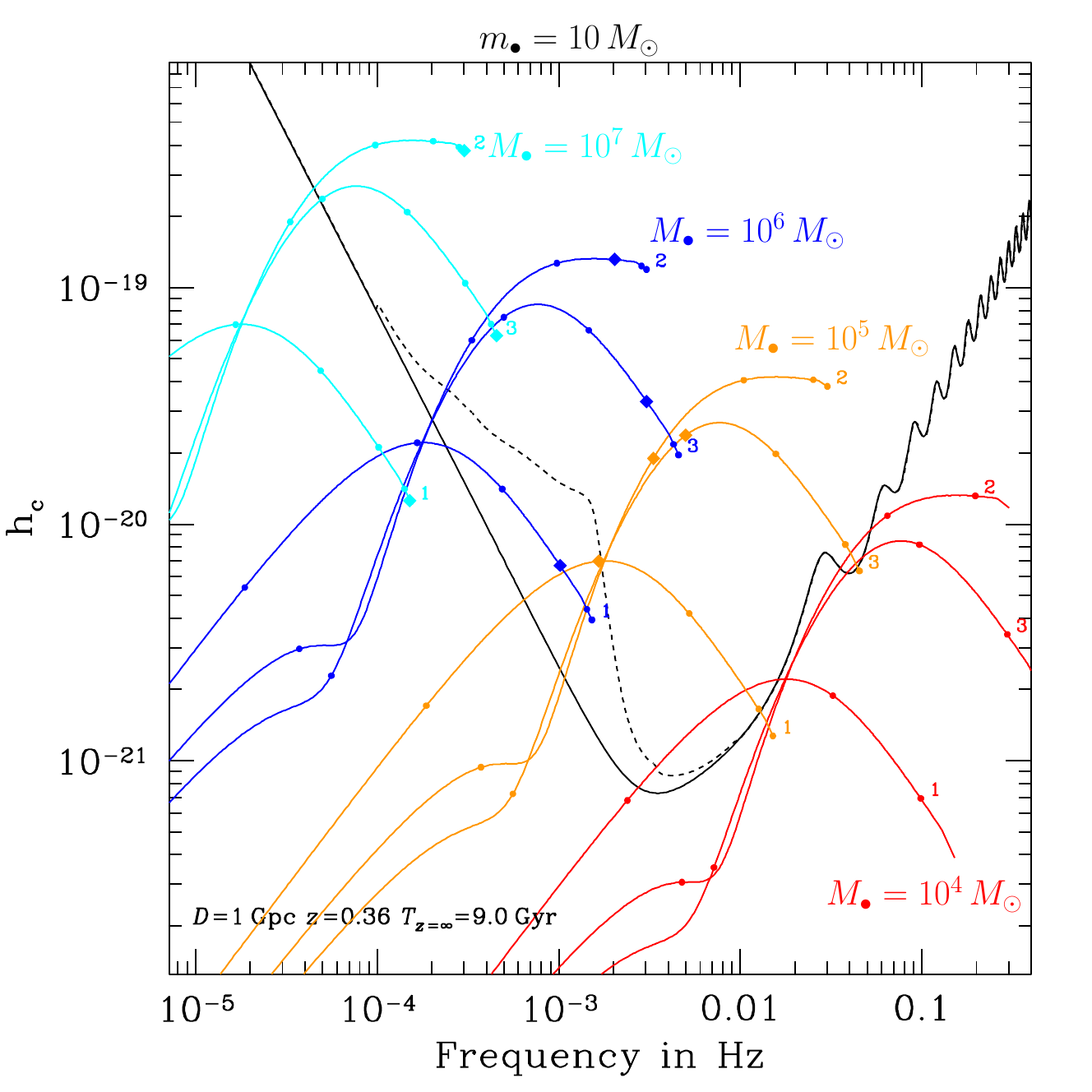}}
\caption{LISA's sensitivity window and four EMRI signals. The groups of colour
correspond to the 1st, 2nd and 3rd harmonic in the quadrupole approximation
of \cite{Peters64} for a SBH of $10\,M_{\odot}$ inspiralling on to a MBH of mass $10^7\,M_{\odot}$
(cyan, left ``cascade'' of harmonics), $10^6\,M_{\odot}$ (blue, second
group from the left), $10^5\,M_{\odot}$ (orange, third cascade) and
and $10^4\,M_{\odot}$ (red cascade, first from the right). In each case, the
distance to the source has been set up to 1~Gpc.}
\label{fig.EMRI_LISAplot_4MBHs}
\end{figure}
}

In a spherical potential, at any given time, the stars and compact objects in
the nucleus simply orbit the MBH with their semi-major axes and eccentricities
changing slowly, owing to 2-body relaxation. For an EMRI to occur, in this
standard picture, 2-body relaxation has to bring a compact remnant on to an
orbit with such a small periapsis distance that dissipation of energy by
emission of GWs becomes significant.

If the object is on a very eccentric orbit but one for which the timescale for
passage through periapsis, $t_{\rm peri}\simeq (1-e)^{3/2}P$, is less than $\sim
10^4$s, the source will generate bursts of gravitational radiation in the LISA
band each time the object passes through periapsis. However, such GW signals
consist of bursts which can probably not benefit from coherent signal
processing even if they repeat with a periodicity shorter than LISA mission
duration. Only if they reside at the Milky Way centre is there a non-vanishing
probability for LISA to detect such sources
\cite{RHBF06,HopmanFreitagLarson07,BerryGair2013}. An extra-galactic
source is only likely to be detectable if it radiates continuously in the LISA
band. As a rough guide, therefore, a detectable EMRI source must have an
orbital frequency higher than about $f_{\rm LISA}=10^{-4}\,$Hz, corresponding
to the condition on the semi-major axis

\begin{equation}
a\lesssim 0.5\,{\rm AU}\, \left(\frac{f_{\rm LISA}}{10^{-4}\,{\rm Hz}}\right)^{-2/3}
\left(\frac{\MBH}{{10^6\,\Msun}}\right)^{1/3}.
\label{eq.}
\end{equation}

As there is no
sharp cut-off in the predicted LISA sensitivity curve at $10^{-4}\,$Hz, a
strong source might be detectable at a lower frequency.

Not all objects with an inspiral time by GW emission shorter than a Hubble time
will end up as EMRIs. This is because, although relaxation can increase the
eccentricity of an object to very high values, it can also perturb the orbit
back to a more circular one for which GW emission is completely negligible.
Typically, neglecting GW emission, it takes a time of the order $t_{\rm rlx}\ln
(1-e)$ for an orbit to reach a (large) eccentricity $e$ through the effects of
2-body relaxation. However, the periapsis distance $R_{\rm p}=a(1-e)$ can be
significantly altered by relaxation on a timescale $t_{\rm rel,p} \simeq
(1-e)\,t_{\rm rlx}$, so the condition for a star to become an EMRI is that it
moves onto an orbit for which the timescale for orbital decay by GW emission,
$\tGW$ (see Eq.~\ref{eq.tGW}) is sufficiently shorter than $(1-e)\,t_{\rm rlx}$.
If the semi-major axis of the orbit is too large, this condition cannot be
obeyed unless the star actually finds itself on an unstable, plunging orbit,
with $e\ge e_{\rm pl}(a) \equiv 1-4\RS/a$ where $\RS$ is the Schwarzschild
radius of the MBH. The very short burst of gravitational radiation emitted
during a plunge through the horizon can only be detected if originating from
the Galactic centre \cite{HopmanFreitagLarson07}. Coherent integration of the GW signal for
$>10^4$ cycles with a frequency in LISA band is required for detection of
extragalactic EMRIs.  Therefore a central concern in the determination of EMRI
rates is to distinguish between plunges and progressive inspirals
\cite{HB95,HA05}.

The situation for EMRI production in the standard picture is more complicated
than that of tidal disruptions by the MBH (e.g., \cite{Rees88,MT99,SU99,WM04})
or GW bursts from stars on very eccentric orbits
\cite{RHBF06,HopmanFreitagLarson07} because these processes require a single
passage within a well-defined distance $R_{\rm enc}$ from the MBH to be
``successful''. In such cases, at any distance from the centre and for any given
modulus of the velocity, as mentioned in Section~\ref{sec.angles} and later,
there exists a ``loss cone'' inside which the velocity vector of a star has to
point for it to pass within $R_{\rm enc}$ of the MBH \cite{FR76,BW77,LS77,AS01}.
In contrast, an EMRI is a progressive process which will only be successful (as
a potential source for LISA) if the stellar object experiences a very large
number of successive dissipative close encounters with the MBHs \cite{AH03}.
There is no well-defined loss cone for such a situation.

As described above, a source becomes an EMRI when the orbital period becomes
shorter than about $10^4\,$s.  Even at those distances, the evolution of such a
tight orbit could in principle be modified by other stars
\cite{Amaro-SeoaneBremCuadraArmitage2012}, but based on our current knowledge of nuclei it is an extreme situation,
because it requires a second star being very close to the EMRI.  It is not so
unlikely at earlier stages of the inspiral as 2-body relaxation, experienced
mostly at apoapsis, can easily induce a change in the periapsis distance large
enough to either render GW emission completely insignificant or, on the
contrary, cause a sudden plunge into the MBH \cite{HB95,HA05}\epubtkFootnote{This is
not strictly true, the spin of the MBH might ``push out'' the LSO and so
Schwarzschild plunges are Kerr EMRIs; see
\cite{Amaro-SeoaneSopuertaFreitag2012}.}. The condition for a successful inspiral
is not that the periapsis distance must be sufficiently small, like for tidal
disruptions or GW bursts, but that the timescale for orbit evolution by
emission of GWs (see Eq.~\ref{eq.tGW}) is sufficiently shorter than the
timescale over which 2-body relaxation can affect the periapsis distance
significantly,

\begin{equation}
\tGW < C_{\rm EMRI}\, (1-e)\,\trlx.
\label{eq.EMRIcond}
\end{equation}

\noindent
What ``sufficiently shorter'' means is the main problem and is encoded
in $C_{\rm EMRI}$, a ``safety'' numerical constant that makes this condition
sufficient ($C_{\rm EMRI}<1$). For a given semi-major axis, one can define
a critical eccentricity $\tilde{e}(a)$ above which GW emission dominates over
orbital evolution due to relaxation and a corresponding time scale

\begin{equation}
\tilde{\tau}(a)\equiv \tGW(\tilde{e},a)\equiv C_{\rm EMRI}(1-\tilde{e})\,\trlx
\label{eq.}
\end{equation}
Plunging orbits (for non-rotating MBH, see section~\ref{sub.SchwBarr} to
understand how this picture changes for Kerr MBH) have

\begin{equation}
e\ge e_{\rm pl}\,(a) \equiv 1-\frac{4\RS}{a},
\end{equation}
so EMRIs (as opposed to direct plunges) can only happen if $e_{\rm
pl}(a)>\tilde{e}(a)$. This defines a critical semi-major axis which is a
typical value for an EMRI at the moment orbital evolution starts being
dominated by GW emission,

\begin{equation}
a_{\rm EMRI}  = 5.3\times 10^{-2}\,{\rm pc}\,C_{\rm EMRI}^{2/3} \times
\nonumber
 \left(\frac{\trlx}{10^9\,{\rm yr}}\right)^{2/3}
 \left(\frac{m}{10\,\Msun}\right)^{2/3}
\left(\frac{\MBH}{10^6\,\Msun}\right)^{-1/3}.
\label{eq.aEMRI}
\end{equation}
The corresponding eccentricity is given by

\begin{equation}
1-e_{\rm EMRI}
= 7.2\times 10^{-6}\,C_{\rm EMRI}^{-2/3} \times
\left(\frac{\trlx}{10^9\,{\rm yr}}\right)^{-2/3} \times
 \left(\frac{m}{10\,\Msun}\right)^{-2/3}
\left(\frac{\MBH}{10^6\,\Msun}\right)^{4/3}.
\end{equation}

The situation is represented in Figure~\ref{fig.Capture_Disruption} in the
semi-major axis -- eccentricity plane. I plot schematically the trajectory for
a typical EMRI evolving according to the standard scenario (labelled ``1-body
inspiral'' to distinguish it from the binary tidal separation scenario discussed
later). Initially the values of semi-major axis and eccentricity perform a
random walk due to 2-body relaxation. As it takes of the order of $t_{\rm rlx}$
to change semi-major axis by a factor of 2 but only $(1-e)t_{\rm rlx}$ to change
the value of $1-e$ (and hence the periapsis), the random walk seems more and
more elongated in the horizontal direction, the smaller the value of $1-e$. It
is much more likely for a star to cross over to the plunging or GW-dominated
region by acquiring a very high eccentricity than by shrinking the semi-major
axis significantly. Typically, an EMRI ``progenitor'' starts with a semi-major
axis slightly lower than $a_{\rm EMRI}$. It takes on average a time of order
$\ln (1-\tilde{e})^{-1} t_{\rm rlx} \simeq 10 t_{\rm rlx}$ for relaxation to
produce an eccentricity such that GW emission becomes dominant. From that point,
the object will follow a path closer and closer to a pure inspiral (as approximated
by Peters equations \cite{Peters64}).  At larger semi-major axis values,
inspirals are practically impossible because GW emission is not significant in
comparison to relaxation even on plunge orbits. Unless they first shrink their
orbit through 2-body relaxation, these objects will be swallowed by the MBH on a
direct plunge. Inspirals staring with $a\ll a_{\rm EMRI}$ are rare because, for
a density cusp $n\propto r^{-\alpha}$ with $\alpha\simeq 1.4-1.8$
\cite{BME04a,BME04b,FASK06,HA06b}, the number of stars per unit $\log(a)$ is
roughly ${\rm d}\Nstar/{\rm d}(\log a)\propto a^{(3-\alpha)}$.  Also, as one
goes inwards, the value of $\alpha$ is lowered by the progressively larger
plunge loss cone \cite{LS77,ASFS04}. In other words, the stellar density is
reduced there (in comparison to a pure power law) because to come and populate
this region a star has to spend several relaxation times drifting down in energy
while avoiding entering the GW-dominated region and inspiraling quickly.

Implementing this basic scenario in various ways (see
Section~\ref{sec.EMRIStatMethods}), several authors have estimated the rate at which
stellar remnants are captured by the central MBH, with results between $\sim
10^{-6}-10^{-8}~{\rm yr}^{-1}$ for a $10^6\,M_\odot$ central black hole
\cite{HB95,SR97,Ivanov02,HA05}.  When combined with the uncertainty in the
number density of massive black holes with $\MBH<{\rm few}\times
10^6\,M_\odot$, the net predicted number of detections that LISA can make spans over
three orders of magnitude, from a few to a few thousand events per year.

We note, incidentally, that even in the LISA band (in the final year of
inspiral), the eccentricity of the typical EMRI in the standard picture is high
enough that a large number of harmonics are likely to contribute to the gravitational waves
\cite{Freitag03,BC04,HA05}.  In addition, the orbital plane of the EMRIs is
unlikely to be significantly correlated with the spin plane of the MBH.  These
characteristics are distinct from those in non-standard scenarios (discussed
below), leading to optimism that some aspects of the nuclear dynamics could be
inferred from just a few events.

The word ``capture'' is sometimes used to refer to EMRIs, but this is misleading as, in the
standard picture, stellar objects are not captured by emission of GWs. They are
already bound to the MBH when they are brought into the GW-dominated regime by
2-body relaxation.  A star originally unbound to the MBH, with energy
$\frac{1}{2}v^2$, will be left bound to it by GW emission if it passes with a
periapsis distance smaller than

\begin{equation}
{r_{\rm capt}} \approx 5\RS\left(\frac{m}{10\,M_\odot}\right)^{2/7}
\left(\frac{\MBH}{10^6\,M_\odot}\right)^{-2/7}
 \left(\frac{v}{100~{\rm km/s}}\right)^{-4/7}.
\end{equation}

\noindent
In order to become an EMRI (rather than experience a direct plunge), the
semi-major axis has to be smaller than a few $10^{-2}\,$pc (see
Figure~\ref{fig.Capture_Disruption} and Eq.~\ref{eq.aEMRI}), requiring a passage
within a distance

\begin{equation}
{r_{\rm capt,\,EMRI}}  \approx 3\RS\left(\frac{m}{10\,M_\odot}\right)^{2/7}
 \left(\frac{\MBH}{10^6\,M_\odot}\right)^{-4/7}
 \left(\frac{a_{\rm capt}}{0.05~{\rm pc}}\right)^{2/7}.
\end{equation}

\noindent
Therefore for masses significantly smaller than $10^6\,\Msun$ there is a
possibility of capturing unbound (or loosely bound) stars directly on to EMRI
orbits. To my knowledge, the contribution of this channel to EMRI rates has not
been estimated in detail but is probably small because it is present only for
the lowest-mass MBHs in the LISA range, although we should note that it would be
on the ``sweet spot'' of the LISA configuration
\cite{Amaro-SeoaneEtAl2012,Amaro-SeoaneEtAl2012b}.

\epubtkImage{.png}{
\begin{figure}[htbp]
\centerline{\includegraphics[width=\textwidth]{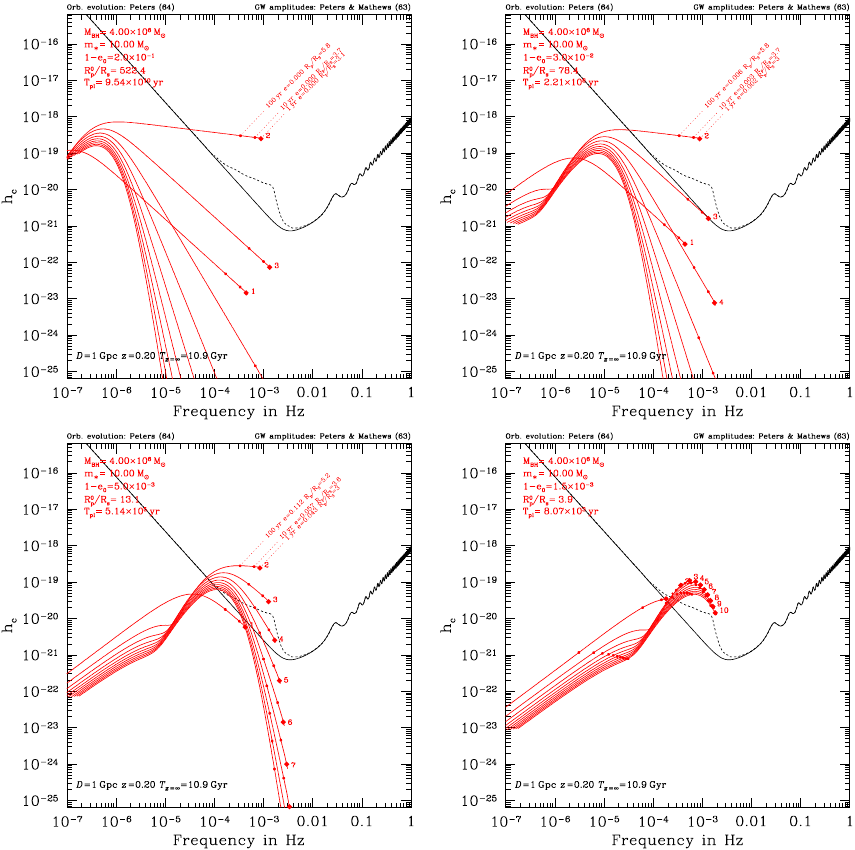}}
\caption{Characteristic amplitude, introduced in Equation~\ref{eq.hc}, of the first harmonics of the quadrupolar
gravitational radiation emitted during the inspiral of a stellar-mass BH of
$m_{\bullet}=10\,M_{\odot}$ ($m_*$ in the plot) into a MBH of mass ${\cal
M}_{\bullet}=4\times10^{6}$ ($\rm M_{BH}$ in the plot). I assume the source is at a distance $D=1\,$Gpc.
I indicate the noise curve $\sqrt{f\,S_h(f)}$ for a LISA-like detector
\cite{LHH00,Larson03}, with the Galactic binary white dwarf confusion
background in dashed line \cite{BH97}. Note that the height of the point for the
amplitude above the curve does \emph{not} represent the SNR (see text). From the
top to the bottom and from left to right, the panels represent a binary which
starts at a semi-major axis of $10^{-3}$ pc and we change the eccentricity,
$e=0.8,\,0.97,\,0.995,\,0.9985$. I show for each panel the ratio $R_{\rm p}^0/R_{\rm s}$,
the initial periapsis distance over the Schwarzschild radius of the system.
For the first three panels I display three moments in the evolution for which the
time to coalescence in the is 100, 10 and 1 yr.} \label{fig.PetersHarmonics}
\end{figure}
}

\subsection{Orbital evolution due to emission of gravitational waves}
\label{sec.OrbEvol}

Consider a binary with component masses $m_1$ and $m_2$, which thus has total
mass $M=m_1+m_2$ and reduced mass $\mu=m_1m_2/M$.  Suppose that its semi-major
axis is $a$ and eccentricity is $e$.  The Peters equations for gravitational
wave emission from a Keplerian orbit~\cite{Peters64} give

\begin{equation}
\left\langle \frac{da}{dt}\right\rangle=-\frac{64}{5}\frac{G^3\mu M^2}{
c^5a^3(1-e^2)^{7/2}}
\left(1+\frac{73}{24}e^2+\frac{37}{96}e^4\right)
\label{aevol}
\end{equation}
and

\begin{equation}
\left\langle \frac{de}{dt}\right\rangle =-\frac{304}{15}e\frac{G^3\mu M^2}
{c^5a^4(1-e^2)^{5/2}}
\left(1+\frac{121}{304}e^2\right)\; .
\label{eccevol}
\end{equation}
We note that the Peters formalism does not
capture the orbital evolution in the strong-field regime, before
plunge. In particular, for EMRIs around a spinning MBH, a slight \emph{
increase} in eccentricity might occur in the late evolution
\cite{GG06}. This does not affect the present discussion. From Eq.~\ref{eccevol},
the characteristic time to change the eccentricity is

\begin{equation}
\tau_{\rm GW}  = \frac{e}{|de/dt|}\approx \frac{15}{304}\frac{c^5a^4(1-e^2)^{5/2}}{G^3\mu M^2}
 \approx 8\times 10^{17}~{\rm yr}\left(\frac{M_\odot}{\mu}\right)\left(\frac{M_\odot}{M}\right)^2
  \left(\frac{a}{1{\rm AU}}\right)^4 \left(1-e^2\right)^{5/2}.
\label{eq.tGW}
\end{equation}
Here I neglect the near-unity factor $(1+121e^2/304)$.

We can rewrite this in terms of gravitational wave frequency.  Let us consider
in particular the frequency emitted at periapsis.  If the orbit is
substantially eccentric, then the orbital frequency at that point will be
approximately $\sqrt{2}$ times the circular frequency at that radius (because
the speed is $\sqrt{2}$ times greater than a circular orbit).  If we dictate
a maximum gravitational wave frequency $f_{\rm max}$ to be double the frequency
at periapsis, then

\begin{equation}
f_{\rm max}\approx {1\over\pi}\left[2GM\over{(a(1-e))^3}\right]^{1/2}\; .
\end{equation}
Therefore

\begin{equation}
a^4= 0.75{\rm AU}^4\left(\frac{M}{10^6\,M_\odot}\right)^{4/3}\left(\frac{f_{\rm max}}{10^{-4}~{\rm Hz}}\right)^{-8/3} (1-e)^{-4},
\end{equation}
and

\begin{align}
\tau_{\rm GW} & \approx 6\times 10^2~{\rm yr}\left(\frac{\mu}{10^3\,M_\odot}\right)^{-1}\left(\frac{M}{10^6\,M_\odot}\right)^{-2/3}
 \left(\frac{f_{\rm max}}{10^{-4}~{\rm Hz}}\right)^{-8/3}(1+e)^{5/2}(1-e)^{-3/2}\nonumber \\
 & \approx 3\times 10^3~{\rm yr}\left(\frac{\mu}{10^3\,M_\odot}\right)^{-1}\left(\frac{M}{10^6\,M_\odot}\right)^{-2/3}
 \left(\frac{f_{\rm max}}{10^{-4}~{\rm Hz}}\right)^{-8/3}(1-e)^{-3/2}
\end{align}
where in the last line I assume a relatively high eccentricity,
so that $1+e\approx 2$.

A classic EMRI, with $M=10^4-10^7\,M_\odot$ and $\mu=1-10\,M_\odot$, could have
a significant eccentricity if (as expected in galactic nuclei) the orbits come
in from large distances, $a>10^{-2}\,$\,pc with $e\gtrsim 0.9999$. Hopman and
Alexander \cite{HA05} made an estimate of the distribution of eccentricities
for one body inspiral and their results showed that it is skewed to high-e
values, with a peak of the distribution at $e \sim 0.7$, at an orbital period
of $10^4$ s. On the other hand, following a binary separation event (and
possibly the tidal capture of giant's core), the compact star is deposited on
an orbit with semi-major axis of order a few tens to a few hundreds of AU. In
this case, the GW-dominated regime is reached with an eccentricity smaller than
0.99 and the orbit should be very close to circular when it has shrunk into the
LISA band.  Such typical orbital evolutions for EMRIs are shown in
Figure~\ref{fig.Capture_Disruption}.

\subsection{Decoupling from dynamics into the relativistic regime}

In the late stage of the inspiral, a binary may become a detectable source of
GWs. The characteristic amplitude of the gravitational radiation from a source
emitting at frequency $f$ is

\begin{equation}
h_{\rm c} = \frac{(2\dot E/\dot f)^{1/2}}{\pi D}
\label{eq.hc}
\end{equation}
where $D$ is the distance to the source, $\dot E$ is the power emitted and
$\dot f$ the time derivative of the frequency \cite{FT00}.  With this
definition, the signal-to-noise ratio (SNR) of an event is obtained, assuming
ideal signal processing, by the integral\epubtkFootnote{This is only meant as a
very general illustrative description. I refer the reader to
\cite{Amaro-SeoaneEtAl07,BabakEtAl10} for a detailed introduction to the
problem of detection and parameter estimation of EMRIs.}

\begin{equation}
({\rm SNR})^2=\int_{f_1}^{f_2}\frac{h^2_{\rm c}(f)}{f\,S_h(f)}\,d(\ln f)
\label{eq.SNR}
\end{equation}
where $f_1$ and $f_2$ are the initial and final frequencies of the source
during the observation and $S_h(f)$ is the instrumental noise of the detector
at frequency $f$ \cite{Phinney02,BC04}.

In Figure~\ref{fig.PetersHarmonics} I follow the signal emitted by a binary
consisting of a Milky Way-like MBH and a stellar BH during their GW-driven
inspiral without taking into account any possible dynamical interaction; i.e.
we only allow the system to evolve via gravitational radiation emission. I
plot the five lowest harmonics of the quadrupolar emission in a rough
approximation \cite{PM63}, only useful for illustrative purposes. In
this figure, I assume a distance of 1~Gpc.

For low-eccentric captures only the $n=2$ harmonic is detectable, during the
last few years of inspiral. However, the small residual eccentricity induces a
difference in the phase evolution of the $n=2$ signal compared to a perfect
circular inspiral \cite{ASF06}.  If the source is followed from a time $\tau_{\rm
LSO}$ before merger until merger, the accumulated phase shift is

\begin{equation}
\Delta \psi_e  \simeq \left(\frac{e_{10^{-4}\rm Hz}}{0.05}\right)^2
\left(\frac{\tau_{\rm LSO}}{1\,\rm yr}\right)^{17/12}
 \left(\frac{{\cal M}_z}{10^3\,\Msun}\right)^{25/36},
\end{equation}
where $e_{10^{-4}\rm Hz}$ is the eccentricity when the $n=2$ signal
has reached a frequency of $10^{-4}$\,Hz and

\begin{equation}
{\cal M}_z \equiv (1+z)\frac{({\cal M}_{\bullet} m_{\bullet})^{3/5}}{({\cal M}_{\bullet}+m_{\bullet})^{1/5}}
\end{equation}
is the redshifted chirp mass of the binary \cite{CH06}. This means that in
principle we can easily distinguish between high-eccentricity captures and low-eccentricity captures. The implications of this
result will become clear in the next sections.

I display in Figure~\ref{fig.Capture_Disruption} the last stable orbit in the
effective Keplerian approximation ($R_{\rm p}\simeq 4\,R_{\rm Schw}$ for $e\ll
0.1$, see \cite{CKP94} with a solid, thick diagonal line. The thin dotted
blue lines are contours of constant time left until the final coalescence,
$T_{\rm GW}$ in the \cite{Peters64} approximation. I show the years on the
right. The thin diagonal green lines are the inspiral, capture orbits due \emph{
only} to the emission of GWs. The upper dash-dotted red line shows
$\tilde{e}(a)$, defined by $t_e=T_{\rm GW}$ (Eq.~\ref{eq.EMRIcond} with $C_{\rm
EMRI}=1$) assuming a constant value $t_{\rm rlx}=1$\,Gyr. The lower dash-dotted
red lines depict the same threshold times a factor 10, 100, 1000, 10,000 and
100,000. On the right hand side of these lines the evolution of the binary is
driven mainly by relaxation, GW emission is totally negligible and vice-versa;
i.e., on the left hand side the evolution is led by the loss of energy in GWs.
An interesting point is the intersection of the first of these red lines (the
uppermost one) with the last stable orbit line. This is the transition between
the so-called direct plunges and the EMRIs.

The thick, dashed black line shows the tidal disruption radius. Any extended
star fording that radius will be torn apart by tidal forces of the MBH, which
we assume to have a mass ${\cal M}_{\bullet} = 4 \times 10^6\,M_{\odot}$
($\rm M_{MBH}$
in the plot). Then, as an illustration, I depict the trajectory of a $10\,\Msun$
stellar BH ($m_{bh}$ in the plot) inspiralling into the MBH. We can separate
two kind of sources according to their astrophysical origin; namely low-eccentricity captures, stars
captured by tidal binary separation, and high-eccentricity captures, stemming from ``simple'' two-body
relaxation.  The latter initially have semi-major axis values of order 100\,--\,1000~AU
[$5\times(10^{-4}-10^{-3})\,$pc] and $e=0.9-0.99$ \cite{MFHL05}.  The evolution
of the eccentricity is a random walk leading to nearly-circular orbits after a timescale of about $T_{\rm
rlx}\ln(1-\tilde{e})^{-1}$.  The latter correspond
to stars on capture orbits due to diffusion form large radii or capture by GW
emission and have initially have a much larger value of semi-major axis and hence a higher
eccentricity. If a star has a semi-major axis $\gtrsim 5\times 10^{-2}$\,pc, it
will not reach small orbital periods, i.e., it will not enter a milliHertz
detector such as LISA unless the semi-major axis is reduced considerably, which in the context
of ``normal'' relaxation theory, takes about a time $t_{\rm rlx}$.

A different way of looking at the same picture is by displaying the energy and
angular momentum of the system. Working in terms of energy and angular momentum has advantages
that can be important to understand some very subtle phenomena that possibly
play a major role in the process of capturing stars. We can see this in
Figure~\ref{fig.EMRI_PhaseSpace} (courtesy of Tal Alexander): To get close to the central MBH, it is faster
to relax angular momentum than to relax energy.  Let us assume that we do not
have any dissipation mechanism. Figure~\ref{fig.EMRI_PhaseSpace} depicts the phase-space of the
system in terms of energy and angular momentum and I use the convention that energy is defined
with a negative sign, so that high positive values of energy mean that the star
is very close to the MBH. The red region represents the zone where the star
cannot exist, i.e., we are closer to the MBH than the LSO. The upper right
diagonal line expresses the fact that for a value of energy you can only have up
to some maximum value of $J_c$, the angular momentum of a circular orbit.  Our test star, a
compact remnant, will suffer gravitational tugs whenever it is far away from
the energy and angular momentum edges. These tugs are random and originate from interactions with
other stars that happen to have a very close position in phase-space and the
scattering rate is very similar in both directions.  This means that the time
spent in one of the horizontal segments is approximately the same as the time
spent in one of the vertical segments in the zig-zag trajectory displayed in
the figure:

\begin{equation}
t_J \sim \left(\frac{J}{J_c(E)}\right)^2\,t_E \sim t_E;
\label{eq.}
\end{equation}

\noindent
i.e., the timescale to change angular momentum, $t_J$ is approximately the same as the
timescale to change energy, $t_E$.  This means that if every zig-zag represents a
change over a fixed amount of time, say $10^9$ yrs, the star will travel
approximately the same distance in one or the other way. If the star gets close
to a very low angular momentum, which is statistically probable, then the picture changes:
the rate at which the star will change angular momentum will be much shorter than the rate at which
it changes energy. The star moves approximately in phase-space in one dimension,
horizontally in the figure. If we wait long enough the star will eventually
enter the loss-cone and ``plunge'' on to the central MBH. I.e., the source of GWs is lost after a few periapsis passages, a
few intense GW bursts and is not as interesting as a gradual, slowly
inspiraling source. This picture corresponds to the general scenario that was
described already a few decades ago, when people were investigating ways of
feeding the MBH \cite{LS77,CK78}.

\epubtkImage{.png}{
\begin{figure}[htbp]
\centerline{\includegraphics[width=\textwidth]{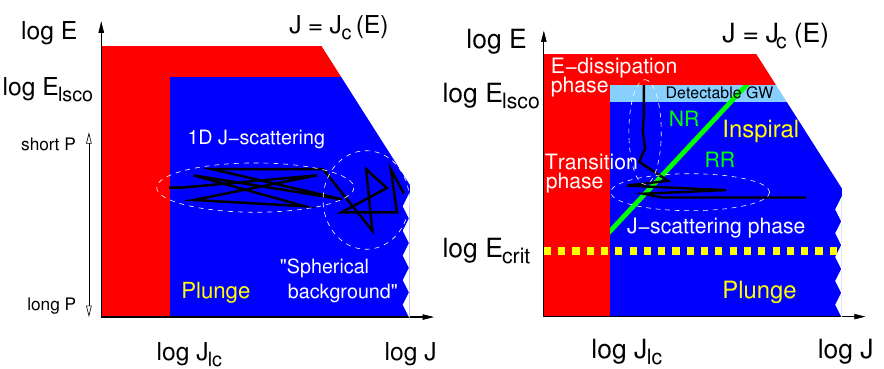}}
\caption{Plunge, inspiral and critical energy of an EMRI in phase-space. \emph{Left
panel:} Direct plunge of a source. \emph{Right panel:} Adiabatic inspiral of a
source subject to dissipation of energy at periapsis and limiting critical energy
(image courtesy of Tal Alexander)}
\label{fig.EMRI_PhaseSpace}
\end{figure}
}

However, if we have a dissipation process acting on to the star, which could be
energy loss in the form of GWs as well as drag forces originating in an
accretion disc or, obviously tidal forces created by the central MBH, the
picture changes significantly.  The process follows the same path and, at some
point, the star reaches the region in which it is on a very radial orbit, i.e.
where the zig-zag stops and we can approximate the curve by a horizontal line.
Nonetheless, in this case, at every periapsis passage, the star will emit an
intense burst of GWs and, thus, shrink its semi-major axis. If this happens
``efficiently enough'', i.e., ``fast enough'' (we will elaborate on this later),
the star is more and more bound to the central MBH and drifts away (goes up in
the energy axis of the figure). The danger of being scattered away from the capture
orbit by other stars decreases more and more and the compact object finds
itself on a safe inspiraling EMRI orbit. The precise details of the dynamics
that lead to this situation determines the distribution of eccentricities that
we can expect. The semi-major axis shrinks to the point that the source enters
the ``Detectable GW'' regime (light blue band in the right panel of
Fig.(\ref{fig.EMRI_PhaseSpace}). As the source advances in that band, the
period becomes shorter and shorter and, hence, the power (emitted energy per unit of
time) grows larger and larger, so that the gravitational radiation can be
measured when it enters the frequency band of the observatory.

The \emph{statistical} orbital properties of the EMRI in the region where GW
emission is
prominent are fully determined by the transition phase between the region
dominated by 2-body scattering processes (the right part of the curve) of the
random walk in phase-space and the \emph{deterministic} dissipation part of the
capture trajectory, i.e., where the energy loss occurs.

As described in \cite{HopmanAlexander05}, in this statistical treatment there
is a critical energy, i.e., a certain distance from the central MBH, of the
order $\sim 10^{-2}$ pc, that can be envisaged as the threshold between the two
regions. This means that stars with energy below the yellow dashed line of the
right panel of Fig.(\ref{fig.EMRI_PhaseSpace}) will have ``longer horizontal
segments'', they will scatter faster in angular momentum than in energy and then
they will end up as
direct plunges.  They approach the central MBH in such a radial orbit that they
are swallowed after one or, at most, a few intense bursts of GWs. This
situation is reverted if the energy of the star is above the line; the star
will spiral in adiabatically and it will not be perturbed out of the EMRI
trajectory, with a significant amount of GW bursts at periapsis before
coalescing with the MBH.

Hence and, again, statistically, at first order, one has to consider only stars
whose energy falls within the critical region and we can ignore all other
stars, even if their energy and angular momentum indicate that they are good
candidates for
EMRIs. Thus, \emph{the event rate will be determined by the ``microphysics''
affecting the innermost volume around the MBH, of radius $\sim 10^{-2}$ pc}.
As the reader will surely have guessed by now, the task is non-trivial.

\epubtkImage{.png}{
\begin{figure}[htbp]
\centerline{\includegraphics[width=\textwidth]{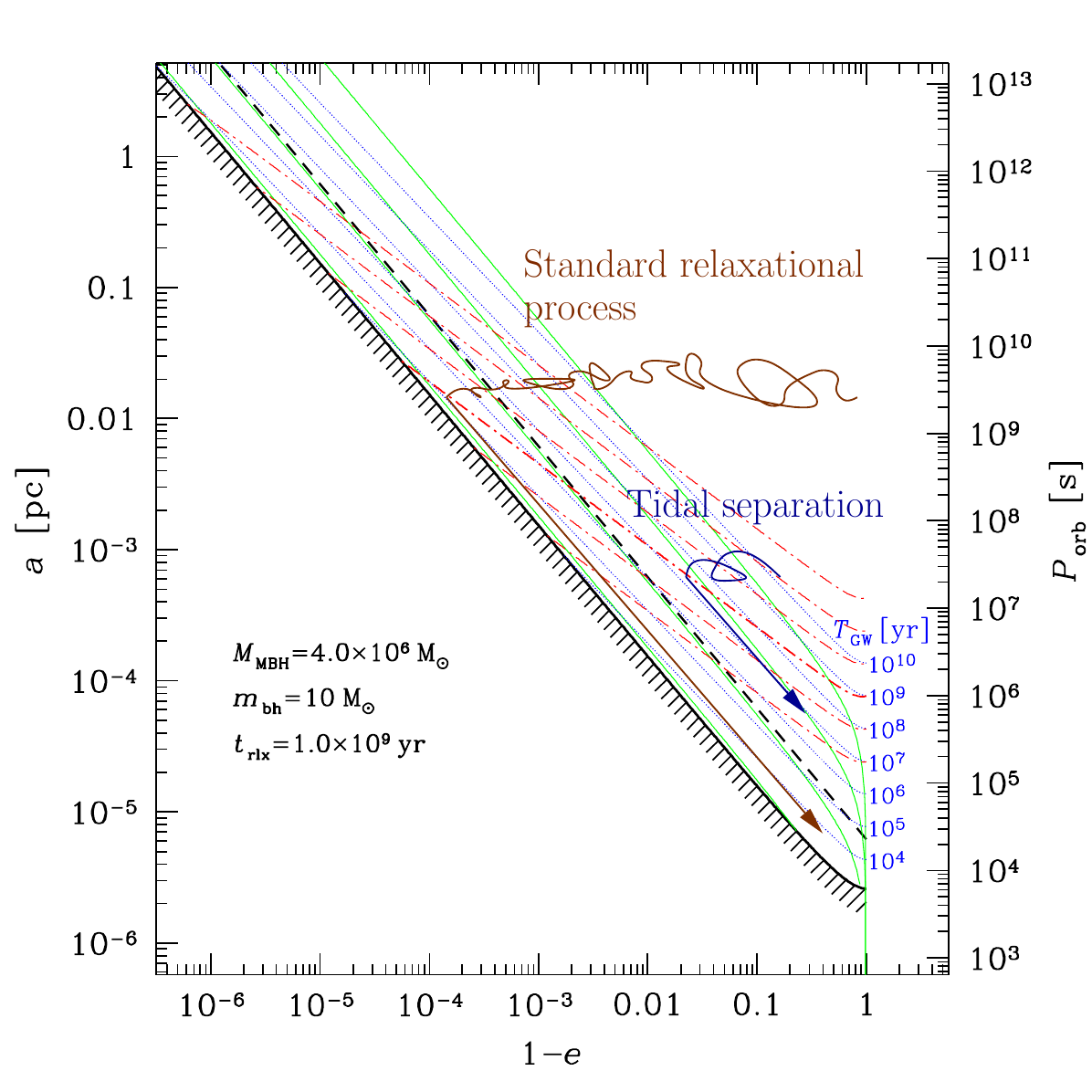}}
\caption{
Capture trajectories in the $a$\,--\,$(1-e)$ plane and tidal disruption limit.
See text for a detailed explanation of the figure.
}
\label{fig.Capture_Disruption}
\end{figure}
}

\section{Beyond the standard model of two-body relaxation}
\label{ch.exotica}

\subsection{The standard picture}

The intelligent reader will very surely have realised that the picture is much
more complex than plain two-body relaxation. Quoting something that Sterling
Colgate said once in Aspen,

\begin{quotation}
\noindent
\emph{``Do you know what the standard (American) model is? : One
  gallon per flush.''}
\end{quotation}

Although Sterling was not directly referring to our standard model, of course.
This means that, illustrating and enlightening as it might be, the standard
model we have been describing so far must be regarded as a (probably very well)
educated guess.

As the interest in a milliHertz mission started to grow and develop,
astrophysicists started to dedicate more and more time to a problem that,
naively, was not very difficult. How do you get a small black hole into a
massive black hole in a galactic nucleus? Now, some decades after the very
first estimates, we have a much better and clear vision of the main phenomena
at play in the process. Well before any space-borne mission is launched, our
understanding of theory related to stellar dynamics has become much broader and
new, unexpected effects have emerged.

\subsection[Coherent or resonant relaxation]{Coherent or resonant relaxation
\epubtkFootnote{The reason for the title of this section is that
  probably the choice of ``resonant'' for this process is not a good
  one. The authors of \cite{RT96} coined this term thinking of the effect of a
  resonance between the radial and azimuthal periods in a Keplerian
  orbit.}
}

As I have discussed previously, in a gravitational potential with a high degree of
symmetry, a test star will receive
gravitational tugs from the rest of the field stars which are not totally
arbitrary and hence do not add up in a random walk way, but \emph{coherently}.
As we have seen in Section~\ref{sec.taxonomy}, the potential will prevent
stellar orbits from evolving in an erratic way. In a two-body Keplerian system,
a SBH will orbit around the MBH in a fixed ellipse. The stellar BH will not
feel random gravitational tugs. It evolves coherently as the result of the
action of the gravitational potential.  When an EMRI approaches the periapsis
of its orbit, we can envisage the situation as a pure two-body problem;
initially Newtonian but later GR effects must be taken into account as the
periapsis grows smaller and smaller.  Nonetheless, as the stellar BH goes back
to the apoapsis, it will feel the surrounding stellar system, distributed in
the shape of a cusp which grows in mass the further away we are from the
periapsis.  The time spent in the region in which we can regard this as a
two-body problem is much shorter than the time in which the stellar BH will
feel the rest of the stellar system. This is particularly true for the kind of
objects of our interest, since the very high eccentricity implies a large
semi-major axis. The time spent on periapsis is negligible as compared with the
time spent on apoapsis, so that the stellar BH can feel the graininess of
the potential. The gravitational tugs from other stars will alter its orbit.
The mean free path in angular momentum-space of that test stellar BH is very large and thus,
it has a \emph{fast} random walk. Both the magnitude and direction of angular momentum of the
stellar black hole are altered.  When the magnitude changes but not the
direction, we talk of ``escalar'' resonant relaxation, and correspondingly when
the direction is changed but not the size, ``vector'' resonant relaxation.

A very radial orbit can become a very eccentric one, so that a compact object
initially set on a potential EMRI orbit can be ``pushed out'' of it.  In a more
general case, a spherical potential that is non-Keplerian, the orbits, as we
have described before, are rosettes and averaged over time they are circular
anuli. In that case we can change the direction of angular momentum but not the modulus. An
eccentric orbit will stay eccentric, but any coherence that was there will be
washed out.

In particular, as illustrated in Figure~\ref{fig.Scal_Coh_Relax}, in the
potential of a point mass, orbits are frozen fixed ellipses that exert a
continuous torque on the test star. A test star does not feel random kicks from
all directions.  When we add up the individual contributions coming from all
the rest of stars on to the test star, there is a residual, non-negligible
torque that will influence its evolution. The mean free path of the star in angular momentum
space is very large. I will refer to this phenomenon as \emph{scalar} coherent
relaxation, because it can change both the magnitude of angular momentum and the inclination of
the orbital plane of the test star. This scenario is a possible way to
alter an initially very circular orbit and modify it in such a way that the
test star will get very close to the MBH after the torques have acted. I.e., we
open a new window for stars to fall into a capture orbit that will lead to an
EMRI.

\epubtkImage{.png}{
\begin{figure}[htbp]
\centerline{\includegraphics[width=0.5\textwidth]{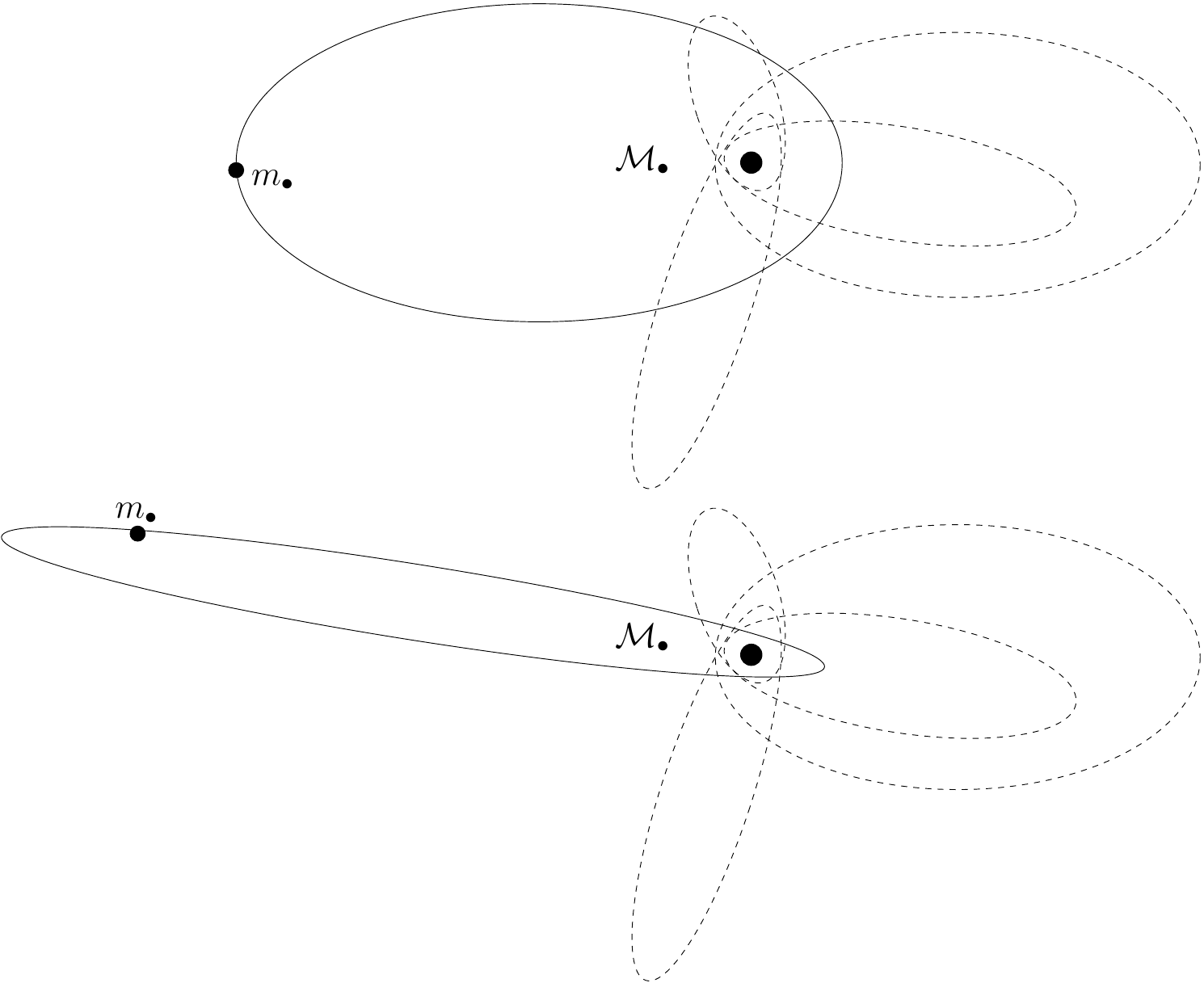}}
\caption{If we have a point-like potential, scalar coherent relaxation can modify the
size of angular momentum and the inclination of the orbital plane of a test star. In dashed
lines I depict the perturbing orbits on the test star, $m_{\bullet}$, whose
orbit is displayed as a solid line in two moments of the evolution.}
\label{fig.Scal_Coh_Relax}
\end{figure}
}

In a more general case, if we have a potential that is simply spherical but not
necessarily Keplerian (a point mass), the field stars, the perturbing orbits to
the test star, describe rosettes -as we have seen- and averaged over time they
can be approximated by a set of anuli that share a centre. From a secular point
of view, the masses are smeared over those anuli which create torques that do
not change the magnitude of angular momentum but they do change the orientation because of
reasons of symmetry \cite{RT96,RI98,HopmanAlexander06}. Hence a circular test
star will keep a negligible eccentricity and it will \emph{not} approach the
central MBH.  Any coherence that was present in the system will nonetheless be
destroyed. I will refer to this as \emph{vectorial} coherent relaxation. From
the standpoint of EMRI production, though, this process is not as relevant and
we will not elaborate on it further, though it can be very relevant for
phenomena related to galactic nuclei, for instance, warping of accretion
discs \cite{BregmanAlexander09}.

However one must note that these illustrations are oversimplifications and
depict perfect symmetries that might be affected or even totally cancelled out
by other effects such as, e.g., the relativistic periapsis shift or Newtonian
precesion. Thus, after a certain time this symmetry is broken and the evolution
is again a random walk, one with very large stepsize.  I refer the reader to
the review of Tal Alexander for a detailed and excellent description of these
processes \cite{Alexander07}.

\epubtkImage{.png}{
\begin{figure}[htbp]
\centerline{\includegraphics[width=0.7\textwidth]{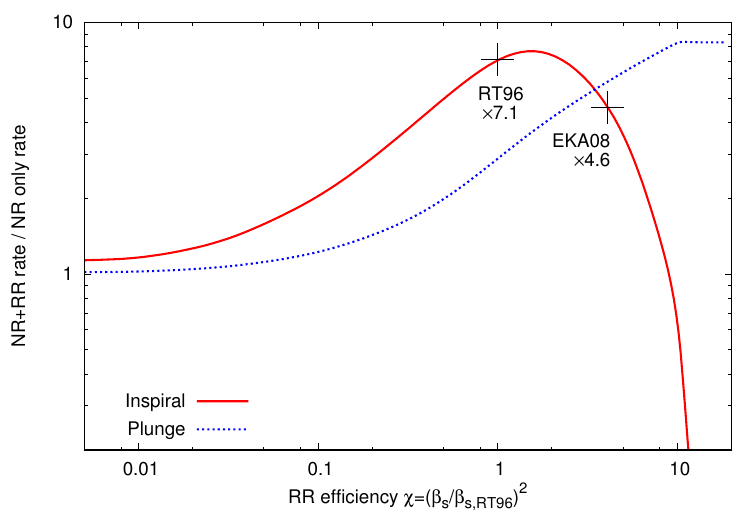}}
\caption{
Total event rates produced by relaxation and coherent relaxation normalised
by the rates obtained by considering only relaxation.
This is Figure~6 of \cite{EilonEtAl09}.}
\label{fig.EilonEtAl09_Fig6}
\end{figure}
}

The impact of coherent relaxation on the production of EMRIs is important.
While the underlying physics of the process is very robust, it is a rather
difficult task to ponder the efficiency of the different parameters involved
in the process. A possible way of evaluating it is as in the work of
\cite{HopmanAlexander06,EilonEtAl09}.  In Figure~\ref{fig.EilonEtAl09_Fig6},
which is Figure~6 of \cite{EilonEtAl09}, we show the rate of EMRIs and plunges
in a system in which we take into account both orthodox or regular relaxation
and coherent relaxation normalised to what one can expect when only taking into
account normal relaxation as function of the $\Xi$ parameter, which gives us
the efficiency of coherent relaxation. The units of $\Xi$ are such that the
value suggested in \cite{RT96} is unity. We note that the work of these authors
was limited to a very low number of particles, but we can consider it as a
reference point to refer to. Thus, if coherent relaxation is more efficient
than what they found, $\Xi>1$ and vice-versa, i.e., we approach the regime in
which there is not coherent relaxation. It is very remarkable to see that by
choosing the value suggested \cite{RT96} we achieve the maximum of the EMRI
rate curve. If the ``real'' value of $\Xi$ happened to be a factor 10 larger,
then we would be drastically dropping the rates and increasing the direct
plunges and, of course, also the tidal disruptions event rate, since these
occur at larger radii.

At first glimpse everything seems to boil down to calculating the precission of
coherent relaxation. One obvious way is to do large-particle number
simulations, since the first attempt of \cite{RT96} was really \emph{very}
limited and difficult to interpret (they were using fewer than 100 particles).
However, the systems we are trying to simulate are much more complicated than
something a simplified approach will be able to investigate. From a numerical point
of view the complications are big and non-negligible. Nevertheless, there has
been an important and impressive advance in this front recently but, before we
address it, the results and interpretations, it is probably better to have a
look at a very familiar system for us, Sgr~A*.  \cite{HopmanAlexander06} have
done this interesting and useful exercise, which is summarised in their figure
6, which I reproduce here in Figure~\ref{fig.Coherent_Relax_SrgAstar}.  In
this figure, the authors display the relevance of different dynamical
components in an attempt to constrain the strength of coherent
relaxation.

On the vertical axes we have the age of different systems found in the GC as
function of the semi-major axis of the stars with the object in Sgr~A*. On
the top of the figure we see a line giving us the timescale for normal
relaxation, $T_{\rm NR}$ to use the same nomenclature as the authors and their plot, which is shorter than the Hubble time but not much
shorter. The following two curves from the top give the timescale for scalar
coherent relaxation for two cases, the first curve from the top corresponds to
a system of $1\,M_{\odot}$ stars and $10\,M_{\odot}$ stars At large values the
effect is quenched by the presence of an extended mass, i.e., Newtonian
precession and at short distances it is periapsis shift that decreases its
strength. The minima displayed in the figure fence in the potential range of
values for the efficiency. The ``real'' value probably lies somewhere in the
middle.

It is nevertheless important to note that the authors did not take into account
the effect of a mass spectrum. In this respect, while it is easier to
understand the fundamentals of the scenario, the system lacks an important
ingredient in realism that could significantly change the narrative.

On the lower right corner of the figure we have vector coherent relaxation,
which is much more efficient with associated timescales shorter than a million
years for a short enough semi-major axis.

In the same figure we display the area from which we believe that
EMRIs originate; i.e., within $\sim 0.01$ pc. These objects are typically
compact remnants and, hence, will be accumulated in the top left corner of the
figure because they are older than the typical time for relaxation. As we can
see, and as shown in the calculations of the authors, they are embedded in the
area which is totally dominated by coherent relaxation.  This is a very
striking result from the standpoint of standard relaxation theory: The dynamics
of EMRIs will be dominated by this new ``exotic'' form of relaxation, coherent
relaxation and not by normal (two-body) relaxation.

As I have already explained previously, there are different populations of
stars in the GC that we can observe. One of these is the disc stars, some
$\sim$ 50\,--\,100 very massive and young stars observed to be orbiting on discs
and almost circularly.  The upper limit on the edge is of a few
$10^6\,M_{\odot}$ and, thus, the strip in the figure is very narrow. These
discs are characterised by having a relatively well-defined and sharp inner
cut-off. It is remarkable to note that the cut-off happens to be exactly at the
place in the figure where the timescale associated with vectorial coherent
relaxation ($T^{\rm V}_{\rm RR}$ in the plot) crosses the strip, without a fit,
as the authors of the work \cite{HopmanAlexander06} claim. On the left side of the line we have the S-stars, which
are \emph{not} on circular orbits, nor aligned with the disc, but randomly
orientated. They are sometimes envisaged as the low-mass members of the disc of
stars.  In any case, it is intriguing that these stars lie exactly on the left
of the curve, where we expect any disc structure to be destroyed by vectorial
coherent relaxation.  This would imply that the values derived by \cite{RT96}
are very close to the real ones.  While it is probably too early to make any
strong statement from this fact, it is encouraging enough to keep us studying and
trying to understand normal as well as coherent relaxation in galactic nuclei.
Another interpretation of Figure~\ref{fig.Coherent_Relax_SrgAstar} is that we can expect some of the S-stars
to have random eccentricities due to the fact that those which are close enough
are affected by scalar coherent relaxation.  Also, we can in principle explain
why late-type giants do not have any particular orientation in their orbits,
since they are in that part of the plot.

\epubtkImage{.png}{
\begin{figure}[htbp]
\centerline{\includegraphics[width=0.7\textwidth]{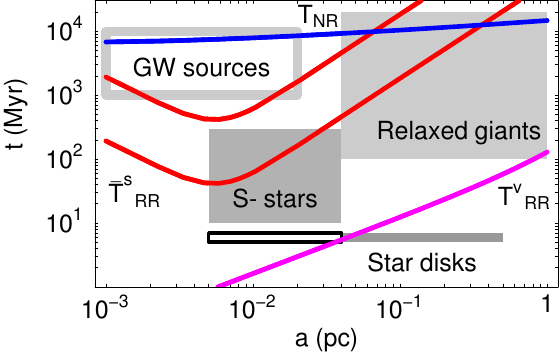}}
\caption{The different timescales dominating stellar dynamics in the
Galactic Centre. This is figure 6 of \cite{HopmanAlexander06}.}
\label{fig.Coherent_Relax_SrgAstar}
\end{figure}
}

The numerical simulations of \cite{EilonEtAl09} show that coherent relaxation can
enhance the EMRI rate by a factor of a few over the rates predicted assuming only
slow stochastic two-body relaxation, as the authors prove.

\subsection{Strong mass segregation}
\label{sec.StrMassSegr}

We have seen in Section~\ref{sec.MassSegr} that stars with different mass get
distributed around a MBH in a galactic nucleus with different density profile.
We devoted a significant part of that section to studying the case of single-mass, which
was described in an analytical way by the work of \cite{BW76}, and previously in
\cite{Peebles72}. The
authors extended the work to stellar systems with two mass components and
argued heuristically for a scaling relation that depends on the star's mass
ratio only, namely $p_L = m_L/m_H \times p_H$ \cite{BW77}. They did not give a
general result on inner slope of the heavy stars (the stellar-mass black holes in our case) and they
did not discuss the dependence of the result on the component's number
fractions.  Fortunately, \cite{AlexanderHopman09} addressed this issue and
found that there exist two branches of solutions, parametrised by

\begin{equation}
  \Delta \approx \frac{N_H m_H^2}{N_Lm_L^2}  \frac{4}{3+m_H/m_L},
\label{eq.DeltaTalClovis}
\end{equation}

\noindent
where capital letters denote total stellar mass and lower-case letters
individual masses of stars.  The quantity $\Delta$ gives us a measure of the
relevance of the SBH self-coupling relative to the other species coupling, the
lighter stars in the system, and the main advantage of it is that it depends
basically on the mass and number ratios. In this respect, the authors of
\cite{AlexanderHopman09} extend the study of \cite{BW77} to an additional,
crucial parameter. For values of $\Delta > 1$ we recover the scaling solutions
of \cite{BW77}. This is the regime that \cite{AlexanderHopman09} refer to as the
``weak branch'' of the solution.  On the other hand, for $\Delta < 1$, we have a
new kind of solution that generalises the solution of \cite{BW77}. This is the
``strong mass segregation'' regime, because the density slopes that one obtains
in this case are steeper.

Inspired by their work, \cite{PretoAmaroSeoane10} and \cite{Amaro-SeoanePreto11}
used direct-summation simulations as a calibration to Fokker--Planck experiments
that allowed them to explore this new solution. This is a priori not obvious,
since we are in a regime in which scattering is dominated by uncorrelated,
2-body, encounters and dense stellar cusps are robust against ejections.
The authors proved that the agreement between both methods is quite good.

\subsection{The cusp at the Galactic Centre}

The implications of these results are interesting and important for EMRI
science, but also particularly timely. This is so because of recent progress in
electromagnetic observations of our Galactic Centre.  Some years ago, two
independent groups have observed that there seems to be a deficit of old stars
based on number counts of spectroscopically identified, old stars in a
sub-parsec region around Sgr$A^*$ (down to magnitude $K=15.5$, see
\cite{BuchholzEtAl09} and \cite{DoEtAl09}).  In
figure~\ref{fig.DoEtAl_2009Fig9_BuchholzEtAl_2009Fig9} we show their main
results. The best fits seem to favor slopes $\gamma < 1$ and the possibility of
a core with the stellar density decreasing, $\gamma < 0$ is not excluded
\cite{BuchholzEtAl09,DoEtAl09,BartkoEtAl10}.  One must take into account that
detectable stars are essentially giants and they represent only a very small
percentage of the underlying population, and the slope of the density profile
is still weakly constrained and such a fit is only marginally better than one
with $\gamma \sim 1/2$.

\epubtkImage{.png}{
\begin{figure}[htbp]
\centerline{\includegraphics[width=\textwidth]{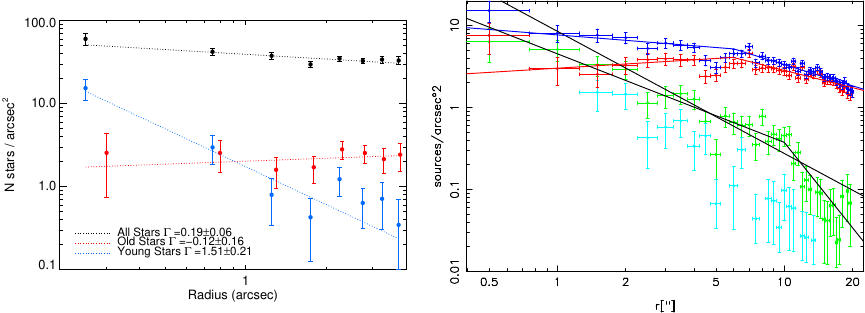}}
\caption{Number of sources per arcsec$^2$ as a function of radius from the Galactic Centre in seconds.
These two figures are from \cite{DoEtAl09} (left panel) and \cite{BuchholzEtAl09} (right panel).}
\label{fig.DoEtAl_2009Fig9_BuchholzEtAl_2009Fig9}
\end{figure}
}

Indeed, the work of 2017 of Gallego, Sch{\"o}del and collaborators
\cite{GallegoEtAl2017,SchoedelEtAl2017} suggests that the observational data of
the Galactic Centre had to be re-analysed. They show that the red- and
brighter giants display a core-like surface density profile within a projected
radius of $R<0.3$ pc of the central MBH, in agreement with previous studies,
but show a cusp-like surface density distribution at larger radii.  The authors
conclude that the observed stellar density at the Galactic Centre is consistent
with the existence of a stellar cusp around the Milky Way's MBH, and that it is
well developed inside its influence radius.  It is remarkable that this
observational study agrees very well with the numerical work of
\cite{BaumgardtEtAl2017}. The authors of the paper ran a series of
direct-summation $N-$body simulations of the Galactic Centre and found that the
distribution of stars is what one might expect from usual two-body relaxation,
without the need of invoking exotic phenomena. The comparison between the numerical
simulation and the observational data is shown in Fig.(\ref{fig.schoedel2017diff}).

The apparent lack of stars at projected distances of $R < 0.3$ pc can be
explained in the theoretical framework of the work
\cite{Amaro-SeoaneChen2014,ChenAmaro-Seoane2014a}: The fragmenting past of the
stellar disc we observe now in our Galactic Centre would have been responsible
for the apparent absence of bright giants. They would have lost their envelopes
by interacting with the high-density clumps that formed in the fragmenting
disc.

\epubtkImage{.png}{
\begin{figure}[htbp]
\centerline{\includegraphics[width=0.5\textwidth]{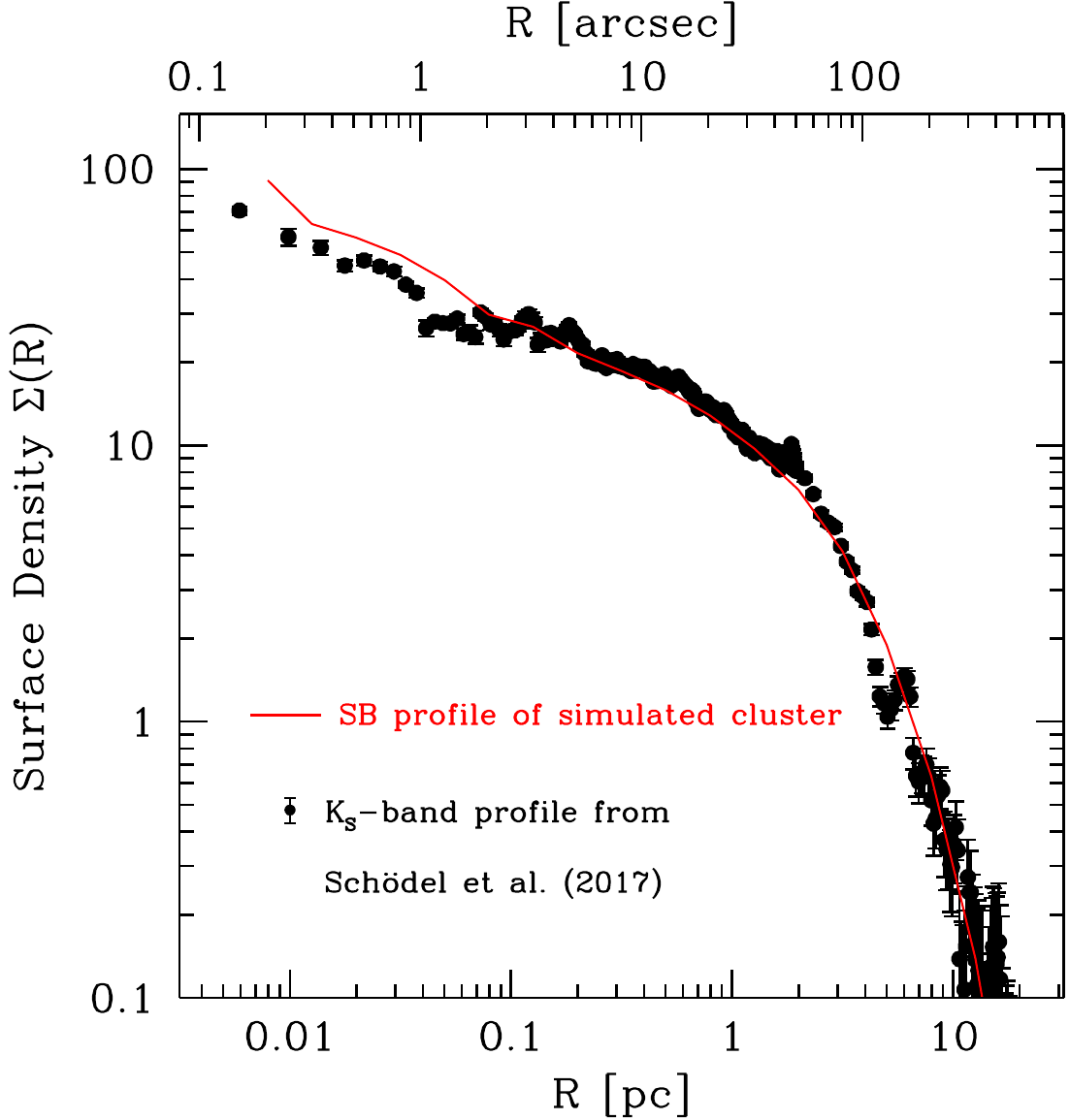}}
\caption{
Comparison of the observational data of the surface luminosity
profile of the diffuse light of the galactic centre of the work
\cite{SchoedelEtAl2017} with a direct-summation $N-$body simulation (red line).
This is Figure 4 of the work \cite{BaumgardtEtAl2017}.}
\label{fig.schoedel2017diff}
\end{figure}
}

Nevertheless, \cite{PretoAmaroSeoane10} and \cite{Amaro-SeoanePreto11} explored
the following situation: How long would cusp growth take if at some point a
central core is carved in the stellar density in a galactic nucleus similar to
the Milky Way? They choose a model with $\gamma=1/2$ as an initial condition, so
that the isotropization time is $\ll t_{\rm rlx}(r_h)$. The results are shown in
figure~\ref{fig.Fig3_PretoAmaroSeoane2010a}.

\epubtkImage{.png}{
\begin{figure}[htbp]
\centerline{\includegraphics[width=0.8\textwidth]{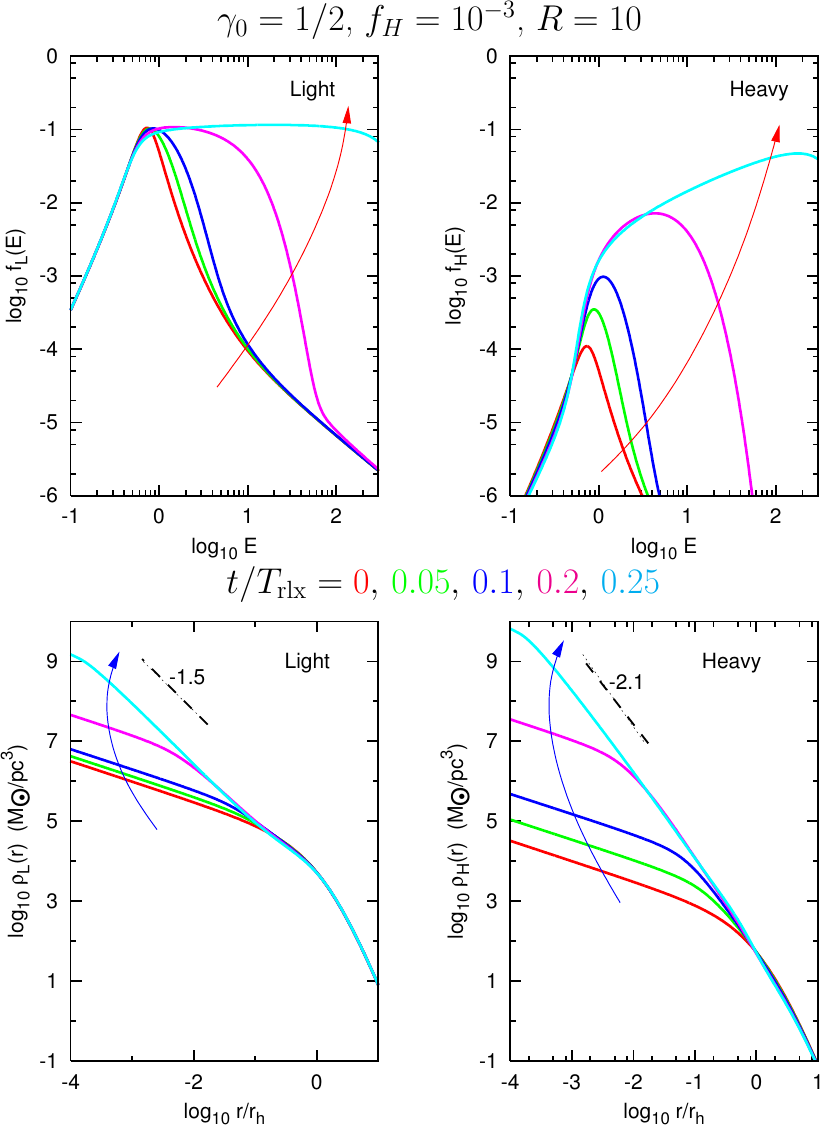}}
\caption{\textit{Upper panels:} Evolution of the phase-space $f(E)$ density, defined
as in \cite{BT87} \textit{Lower panel:} Same for the spatial
density $\rho(r)$ ,
  from the simulations of \cite{PretoAmaroSeoane10} using their
  FP code which is isotropic, orbit-averaged multi-mass in energy space. This
  is adapted from Figure~3 of \cite{PretoAmaroSeoane10}. In different colours we
have different
  moments in the evolution of the system in units of the relaxation time.}
\label{fig.Fig3_PretoAmaroSeoane2010a}
\end{figure}
}

We can see in the figure that by $t \sim 0.25 \ t_{\rm rlx}(r_h)$, cusps with
$\gamma_L\sim1.5$ and $\gamma_H \sim 2$ ($p_L\sim 0.05$ and $p_H\sim 0.5$,
where the subscript ``L'' refers to the light species and ``H'' to the heavy
stars) are \emph{fully developed} ($\sim 0.02$ pc if scaled to a Milky Way-like
nucleus). For masses similar to SgrA$^*$, ${\cal M}_{\bullet} \lesssim 5 \times
10^6 M_\odot$, this is shorter than a Hubble time. Hence, if indeed a carving
event depleted the inner agglomeration of stars around the MBH, as soon as only
$6$ Gyr later a very steep cusp of stellar-mass black holes would have had time to re-grow.

I must note that this result is different to what \cite{Merritt2010}
finds, but this is probably due to the fact that the author only takes into account the
effect of dynamical friction from the light stars over the heavy stars, and he
neglects the scattering of the heavy stars. In this respect, he is limited in
his approach to the early evolution of the system, when the heavy stars only
represent a minor perturbation on the light stars. As a matter of fact, very similar
results were derived later by \cite{GualandrisMerritt2012}.

The impact on EMRI production is the following: If carved nuclei were common in
the range of masses relevant to an observatory like LISA, then we would be
cutting down production of old remnants significantly.  However, even if our
Milky Way \emph{had} a hole in its stellar cusp, LISA EMRI rates peak around
${\cal M}_{\bullet} \sim 4 \times 10^5 - 10^6 M_\odot$ and re-growth times are
$\lesssim 1$ Gyr for ${\cal M}_{\bullet} \lesssim 1.2 \times 10^6 M_\odot$, so
that we still expect that a substantial fraction of EMRI events will originate
from segregated stellar cusps

On the other hand, strong mass segregation not only ``comes to the rescue'' in
the case of carved nuclei. It helps in the production of EMRIs.
The authors of \cite{Amaro-SeoanePreto11} estimate that thanks to strong mass segregation one
might expect EMRI even rates to be $\sim 1-2$ orders of magnitude larger than
one would expect from using the Bahcall and Wolf solution, as they show.

Their solution for the weak branch is physically unrealistic, since it predicts
a too high event rate because it uses unreasonably high number fractions of
stellar-mass black holes $f_\bullet$ ($\ge 0.05$).  In a more realistic case,
when $\Delta \sim 0.03$, ($f_\bullet \sim 10^{-3}$) the Bahcall and Wolf
solution would lead to a strong supression of the EMRI rate to --at best -- a
few tens of events per Gyr.

The new solution of strong mass segregation implies a higher $\rho_{\bullet}$
well inside the influence radius of the MBH, so that we have a boost in the
diffusion of stellar-mass black holes close to the MBH. When going from number
fractions that are based on unrealistic IMF, such as in the work of \cite{BW77}
(say $\Delta=3$) to realistic values ($\Delta=0.03$, the event rate is supressed
by factors of $\sim 100-150$ if we ignore strong mass segregation.  Thanks to
this new solution, based on more realistic physics, even for low values of
$\Delta=0.03$, we boost the rates from few tens to a few hundred per Gyr, $\sim
250$/Gyr if we consider a mass ratio of 10 between the stellar-mass black holes
and the MS stars and if we take a fractional number for stellar-mass black holes
of $f_{\bullet}=0.001$.

\subsection{Tidal separation of binaries}

Another process contributing to the creation of EMRIs has its origin in the work
of Hills \cite{Hills88}. In his work, Hills describes the possibility of finding
escaping stars which originate by this process:

\begin{quote}
\emph{``A close but Newtonian encounter between a tightly bound binary and a million solar
mass black hole causes one binary component to become bound to the black
hole and the other to be ejected at up to 4000 km/s. The discovery of even one
such hypervelocity star coming from the Galactic center would be nearly definitive
evidence for a massive black hole. The new companion of the black hole has a high
orbital velocity which increases further as its orbit shrinks by tidal dissipation. The
gravitational energy released by the orbital shrinkage of such a tidal star can be
comparable to its total nuclear energy release.''}
\end{quote}

His work, about the tidal separation of binaries by a MBH, did not have a big
impact for some 15 years until the discovery of the so-called ``hyper-velocity
stars'', stars with a velocity of $> 10^3\,\kms$, which had been predicted in
his work. Indeed, several such stars have been discovered in the last years. I
refer the reader to the work of \cite{BrownEtAl09} for a discussion of the
properties of these stars, as well as for references.

While one of the objects is ejected into the stellar system, the other binary
member can remain bound to the MBH on a rather tight orbit. If this star
happens to be a compact object, then we would have an EMRI which would be
rather ``immune'' to the problems of EMRIs caused by two-body relaxation. Since
the tidal separation happens very close to the MBH, the CO will have a shorter
apoapsis (usually only tens of times the periapsis distance) and thus,
potential tugs that lead it out of the capture orbit are reduced.  This process
was described in the work of \cite{MillerEtAl05}.  The properties of these
EMRIs are very interesting and I describe the process in this section, both
from an astrophysical point of view and the observational signature.

In Fig.(\ref{fig.BinarySplitting}) we have a schematic view of the process. A
binary which happens to fly by close enough to the central MBH will be tidally
separated because the tides acting on the pair overcome the gravity in the
binary. One of the stars is captured, meaning that it will be bound to
the MBH, after losing a little energy compared to what it had before, and the
other companion of the binary will obtain a bit more energy after the
separation, so that it will be ejected with a high velocity, as in a slingshot.
It is rather straightforward to make a toy model for the process and get the
scalings, which sheds light on the process.

\epubtkImage{.png}{
\begin{figure}[htbp]
\centerline{\includegraphics[scale=0.4,clip]{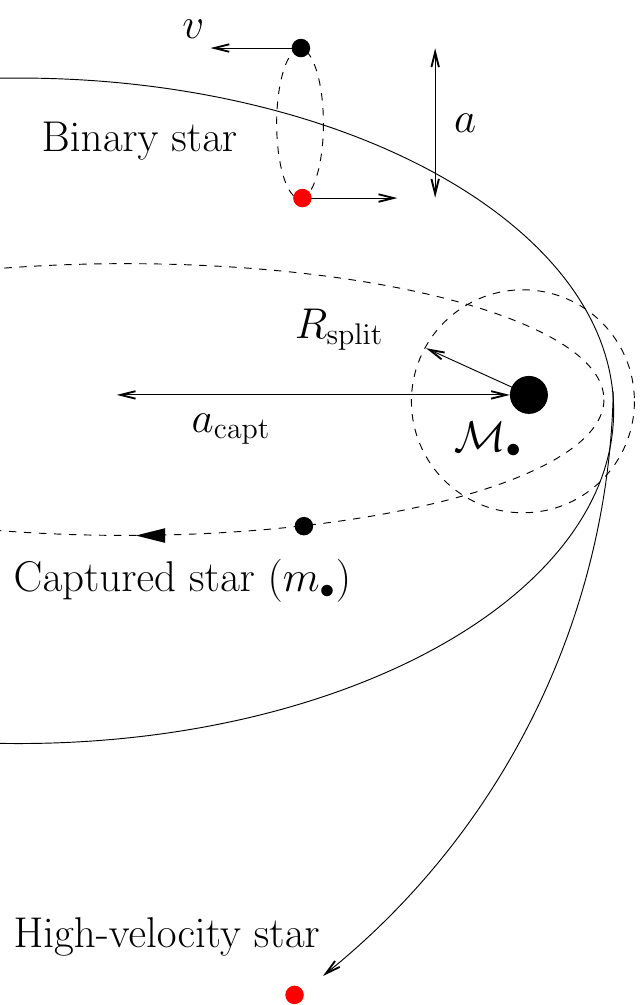}}
\caption{Schematic illustration of the process of the tidal separation of a binary by
a MBH of mass ${\cal M}_{\bullet}$. I define $R_{\rm split}$ as the radius within
which the tidal forces of the MBH overcome the binding energy of the binary}
\label{fig.BinarySplitting}
\end{figure}
}

The size of the region where this process will occur, $R_{\rm split}$, is
proportional to the size of the object, the separation of the binary (i.e.
about the semi-major axis), the
mass of the binary $m_{\rm bin}$ and the mass of the central MBH, ${\cal
M}_{\bullet}$, as it was in the case of the tidal disruption of an extended
star. Actually we can follow the analogy very closely, except in that case
we have to come closer to the MBH to have the tidal forces overcoming the
binding energy of the binary, since $a > R_{\star}$, with $R_{\star}$ the
radius of the star.

\begin{equation}
R_{\rm split} \sim a \left(\frac{{\cal M}_{\bullet}}{m_{\rm bin}}\right)^{1/3}.
\label{eq.Rsplit}
\end{equation}
The orbital velocity in the binary can be easily computed as follows,

\begin{equation}
v \sim \sqrt{\frac{G\,m_{\rm bin}}{a}}.
\label{eq.orbital_velocity_binary}
\end{equation}
I now normalise the last equation to nominal values assuming that
it is a hard binary in a galactic nucleus,

\begin{equation}
v \sim 133\,{\rm km\,s}^{-1}
\left( \frac{m_{\rm bin}}{2M_{\odot}} \right)^{1/2}
\left( \frac{a}{0.1{\rm AU}} \right)^{-1/2}.
\label{eq.}
\end{equation}

In Fig.(\ref{fig.BinarySplitting_CM_xfig}) I show a zoom-in of
Fig.(\ref{fig.BinarySplitting}), at the moment in which the binary is at the
periapsis of the MBH. We estimate now the ejection velocity.  The centre-of-mass
(CM) of the
binary has a velocity $V_{\rm CM}$ which we can easily calculate by assuming
that the encounter is parabolic

\begin{equation}
V_{\rm CM} \gtrsim \sqrt{\frac{G\,{\cal M}_{\bullet}}{R_{\rm split}}}\sim\,v
\left(\frac{{\cal M}_{\bullet}}{m_{\rm bin}}\right)^{1/3} \gg v.
\label{eq.}
\end{equation}
This allows us to estimate the ejection velocity of the slingshot star,
because the difference of energy will be

\begin{equation}
\pm \delta E \simeq V_{\rm CM}\cdot v \geq v^2 \left(\frac{{\cal
M}_{\bullet}}{m_{\rm bin}} \right)^{1/3} \simeq \frac{v_{\rm eject}^2}{2}.
\label{eq.}
\end{equation}
Then, we have that

\begin{equation}
v_{\rm eject} \gtrsim \left(\frac{{\cal M}_{\bullet}}{m_{\rm bin}}\right)^{1/6}.
\label{eq.Veject}
\end{equation}
Since we are dealing with a binary, the star of mass $m_{\bullet}$, which we
assume to be a stellar-mass black hole, will be slowed down by $v$, as in
Eq.~(\ref{eq.orbital_velocity_binary}) and the star of mass $m_{\star}$, which
can be an extended star or a compact object, will be accelerated by the same
amount. I now assume that in that moment the stars do not interact
gravitationally with each other and they only ``see'' the potential created by
the MBH. Therefore we have a simple situation with a simplified geometry that
allows us to compute the initial orbits of the two companions in the pair at
the moment of splitting.

\epubtkImage{.png}{
\begin{figure}[htbp]
\centerline{\includegraphics[scale=0.4,clip]{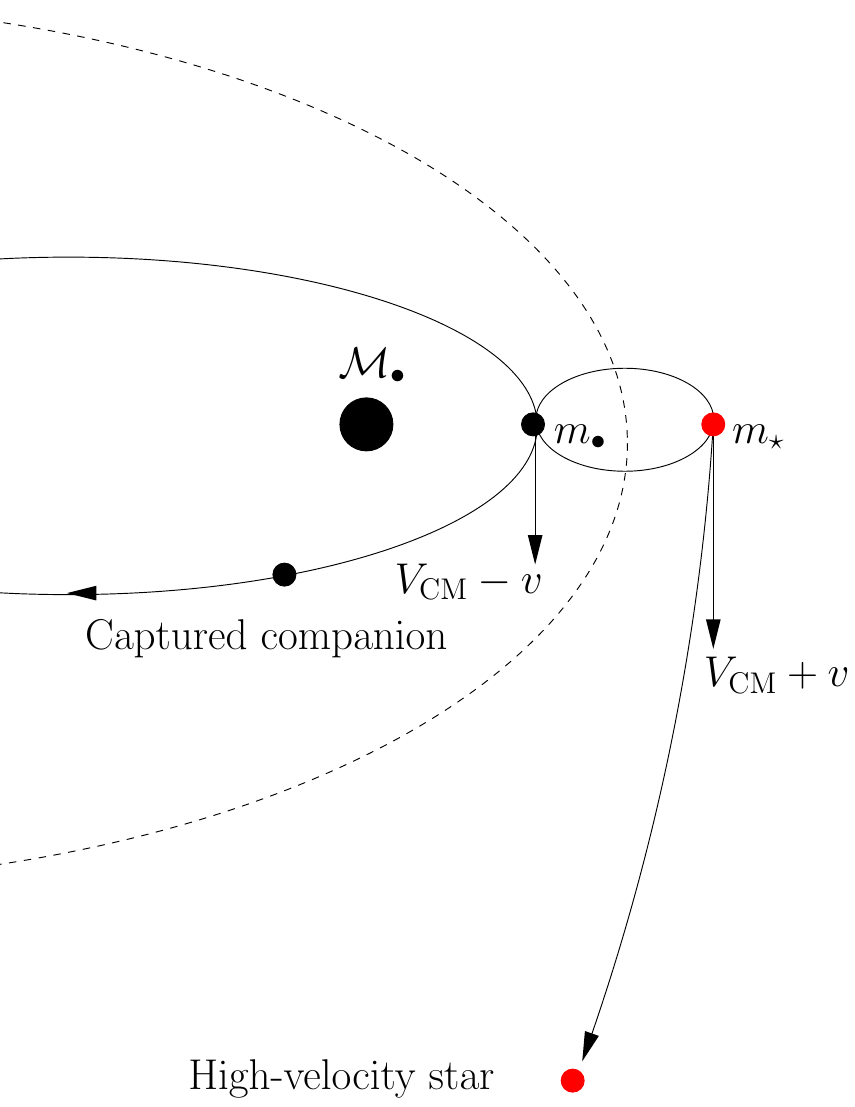}}
\caption{Zoom of Figure~\ref{fig.BinarySplitting} for the splitting of the pair.
I assume a parabolic encounter (dashed curve) for the COM of the binary and a
mass $m_{\bullet}$ and $m_{\star}$ for the stars.}
\label{fig.BinarySplitting_CM_xfig}
\end{figure}
}

Hence, the stellar-mass BH is bound to the MBH and the escaping star leaves the
system with a high velocity, which is of the order of the velocity in the
binary, typically of about $\sim 10 \kms$, multiplied by the same mass ratio as
in Eq.~(\ref{eq.Rsplit}) but to a different power, as we can see in
Eq.~(\ref{eq.Veject}).

One very interesting aspect of this particular process to produce the capture
of compact objects by MBHs is the eccentricity that the orbit has. We can
estimate it roughly by computing the semi-major axis of the bound pair after
the separation of the initial binary, $a_{\rm capt}$,

\begin{equation}
a_{\rm capt} \approx a \left(\frac{{\cal M}_{\bullet}}{m_{\rm bin}}
\right)^{2/3}.
\approx 10^4a
\label{eq.a_capt}
\end{equation}
As we can see in Figure~\ref{fig.BinarySplitting}, we can approximate the separation
radius $R_{\rm split}$ to the periapsis distance,

\begin{equation}
R_{\rm peri} = (1-e_{\rm capt})\,a_{\rm capt} \approx R_{\rm split} \sim
\left(\frac{{\cal M}_{\bullet}}{m_{\rm  bin}} \right)^{-1/3}.
\label{eq.}
\end{equation}
Hence, this kind of sources will typically have a capture eccentricity
of

\begin{equation}
e_{\rm capt} = 1-\left(\frac{{\cal M}_{\bullet}}{m_{\rm bin}} \right)^{-1/3}
\sim 0.98.
\label{eq.}
\end{equation}

Contrary to ``usual'' EMRIs, tidally-split MS stars have a low eccentricity when they
form, and possibly when they reach the bandwidth of the detector (for
convenience, we will call these tidally-split EMRIs ``TSEMRI''). This is
because no energy needs to be dissipated in order to have a capture.  As a
result, capture can occur at much larger radii than is possible in the two-body
case, as described in \cite{MillerEtAl05}. For a $10\,M_{\odot}$ object this
should be of the order $1-e_{\rm TSEMRI}\approx 0.99$. On the other hand, we have
seen that typical EMRI eccentricities when reaching the LISA bandwidth are
$1-e\approx [10^{-3},\,10^{-7}]$

In order to understand the difference in terms of detectability, we need to
introduce some definitions of the geometric model of signal analysis. We
treat the waveforms as vectors in a Hilbert space \cite{Helstrom68}, which
allows us to define the natural scalar product

\begin{equation}
 \left<h\left|s\right.\right> :=
2\int_{0}^{\infty}{df}\,\frac{ \tilde{h}(f)\tilde{s}^{*}(f) +
\tilde{h}^{*}(f)\tilde{s}(f)}{S_{n}(f)},
\label{eq.}
\end{equation}
where

\begin{equation}
\tilde{h}(f) = \int_{-\infty}^{\infty}\, dt\, h(t)e^{2\pi\imath ft}
\label{eq.}
\end{equation}

\noindent
is the Fourier transform of the time domain waveform $h(t)$. I have introduced
the $S_{n}(f)$, which is the one-sided noise spectral density of LISA, see e.g.
\cite{Thorne87,Finn92}. One can think of LISA as two detectors, so that the
signal in each of them is given by $s_{i}(t) = h_{i}(t)+n_{i}(t)$, with
$i=1,\,2$ label each detector. I adopt the assumption that the noise $n_{i}(t)$
is stationary, Gaussian, uncorrelated in each detector and characterized by the
noise spectral density $S_{n}(f)$.  Hence, we can define the signal-to-noise
ratio in each detector as

\begin{equation}
\rho_{i} = \frac{\left<h\left|s_{i}\right.\right>}{\sqrt{\left<h\left|h\right.\right>}}.
\end{equation}
Therefore, if we consider the waveform of a TSEMRI and compare it with a normal EMRI, we
can calculate the mismatch of their expected signal-to-noise ratio for LISA as

\begin{equation}
  {\cal M} := 1 -
           \frac{\left<h_{\rm TSEMRI}\left|h_{\rm EMRI}\right.\right>}
                     {\sqrt{\left<h_{\rm TSEMRI}\left|h_{\rm TSEMRI}\right.\right>
                     \left<h_{\rm EMRI}\left|h_{\rm EMRI}\right.\right>}}.
\label{eq.mismatch}
\end{equation}

For a TSEMRI and a normal EMRI starting with exactly the same orbital parameters at the
GC and coloured for LISA, I have calculated with the \textsc{LISACode}\epubtkFootnote{\url{http://www.apc.univ-paris7.fr/~petiteau/LISACode/Home.html}}
\cite{PetiteauEtAl08} that there is a missmatch of 99.9971\%.  Using the
standard definition of signal-to-noise ratio
$\rho_i=\left<h|s_i\right>/\sqrt{\left<h|h\right>}$, we have that the TSEMRI is
calculated to have an average signal-to-noise ratio of $\rho_{\rm TSMI} \sim
27637$ and the normal EMRI of $\rho_{\rm EMRI} \sim 18848$, both set to be at a
distance of 8.3 kpc.

In Figure~\ref{fig.WaveformsXYZ} I show the waveforms for an observer at
$\theta = 55$ degrees, with origin at the MBH, with a mass of
$3\cdot10^6\,M_{\odot}$ and z-axis along direction of big black hole spin.
The waveforms are from Steve Drasco and have been calculated using the kludge approximation of
\cite{GG06}.

\epubtkImage{.png}{
\begin{figure}[htbp]
\centerline{\includegraphics[width=\textwidth]{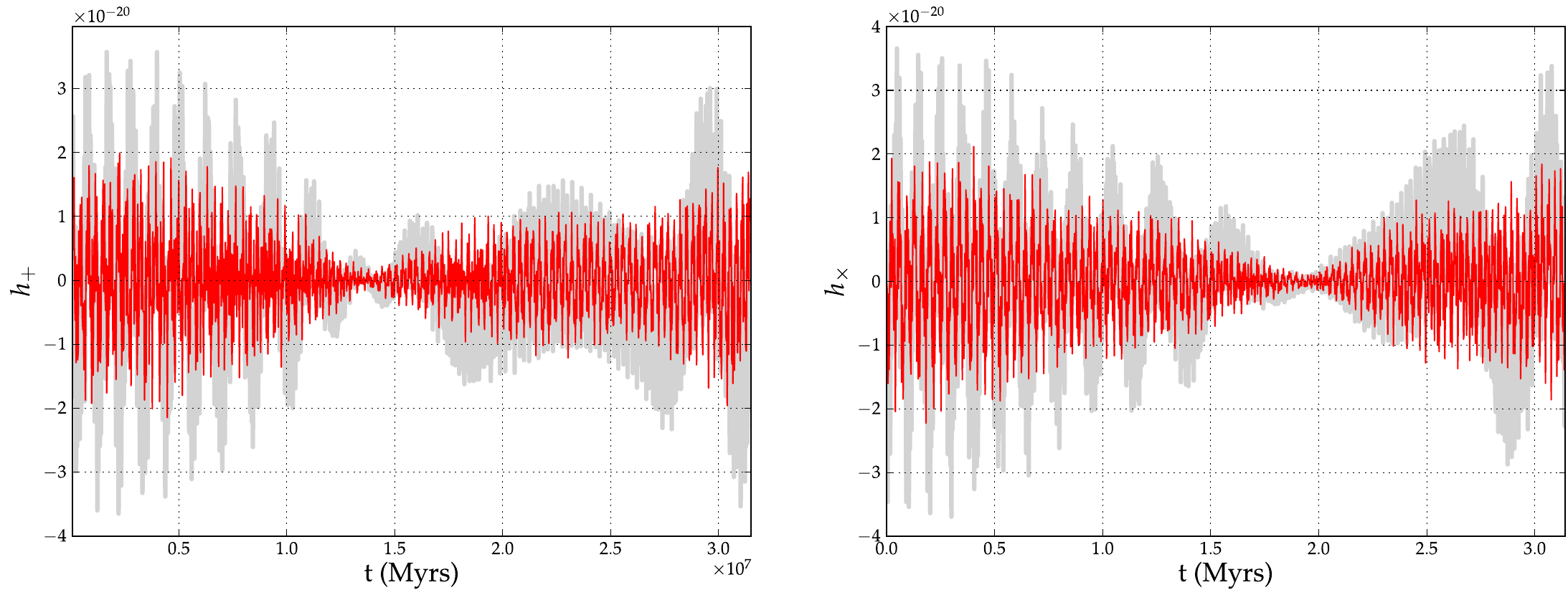}}
\caption{Waveforms of a normal EMRI and a TSEMRI (see text) superimposed for a full year of data before the
final plunge, which has been defined to happen at a periapsis of $r_{\rm
plunge} \equiv 2\times r_{\rm ISCO}$, with $r_{\rm ISCO}$ the radius of the
innermost stable circular orbit, which is $\sim 8\,M$ in this case. The mass of
the central MBH is $3\times 10^6\,M_{\odot}$, the mass of the star
$0.53\,M_{\odot}$. The spin of the MBH is set to $a = 0.5\,M$ and we
neglect the spin of the star.}
\label{fig.WaveformsXYZ}
\end{figure}
}

\epubtkImage{.png}{
\begin{figure}[htbp]
\centerline{\includegraphics[scale=0.5,clip]{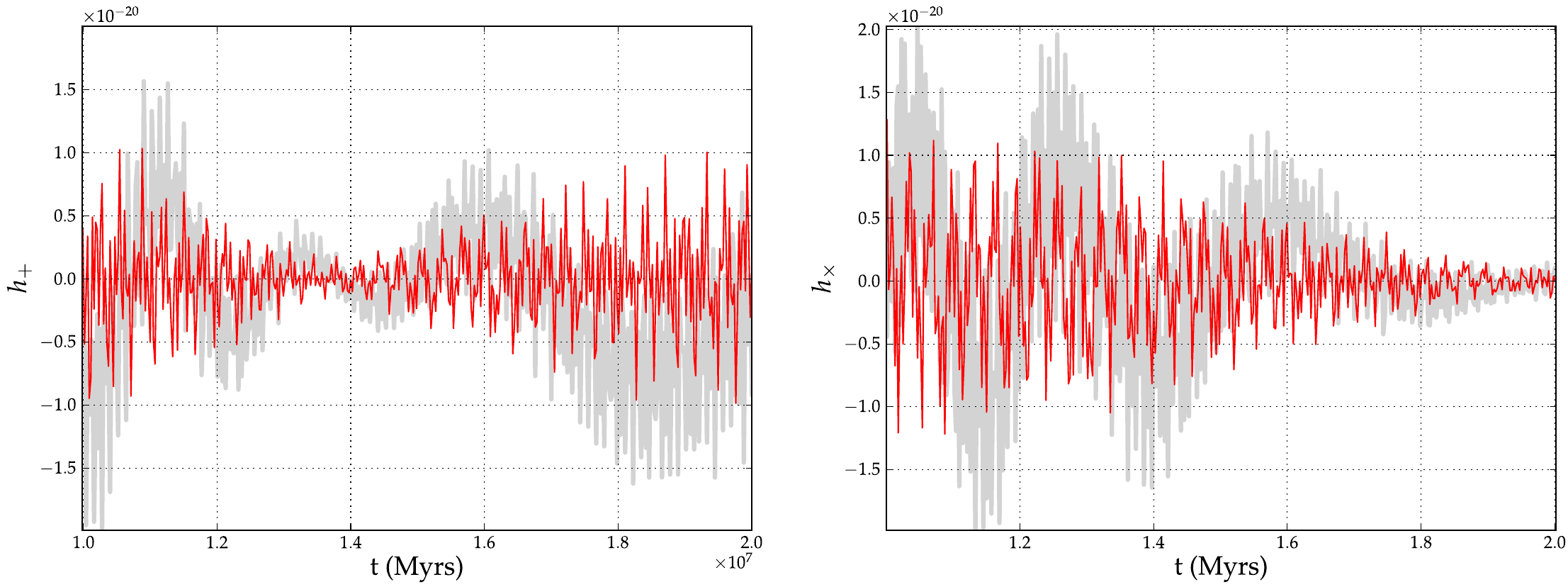}}
\caption{Zoom of Figure~\ref{fig.WaveformsXYZ}. For the two figures only one point out of
1600 is used to make the plot lighter. Still, we can see that even in the region of maximum overlap there
is a significant difference. In the computation of the mismatch function of Eq.~(\ref{eq.mismatch}), however,
the full waveform has been used.}
\label{fig.WaveformsXYZ_Zoom}
\end{figure}
}

\subsection{A barrier for captures ignored by rotating MBHs}
\label{sub.SchwBarr}

A number of authors have addressed the question of EMRI event rates in a Milky
Way-like galaxy.  The numbers differ but a common denominator to all estimates
is that the number of ``direct plunges'' is much larger than slowly decaying,
``adiabatic'' EMRIs. This is so simply because the region of the galaxy from
which potential plunges originate contains many more stars than the volume
within which we expect EMRIs, as we have seen in Section~\ref{ch.TwoBodyEMRIs}.

For a long time ``plunges'' have been considered to be irrelevant for the
purposes for which EMRIs are best. After one intense burst of radiation, the
source would be lost along with, obviously, the SBH. Some studies have looked
into that, such as the work of \cite{HopmanFreitagLarson07}, which is probably
one of the most meticulous one since it incorporates a high realism of the
physics in that regime.  However, the conclusions of the authors are that these
sources are not interesting because they could only be detected if they
originated in our own Galactic Center. Later, \cite{BerryGair2013} addressed the
possible constraints on paramters of our Milky Way's MBH if one of this bursting
sources was to be observed with LISA.

In contrast, a few years later, the work of
\cite{Amaro-SeoaneSopuertaFreitag2012} showed that since MBH are likely to be
spinning, it is actually very hard for a SBH on a plunge orbit to ``hit'' the
MBH.  They show that the majority of plunging orbits for spinning MBHs are
actually not plunging but EMRI orbits. They prove that since spin allows for
stable orbits very near the LSO in the case in which the EMRI is prograde, the
contribution of each cycle to the SNR is much bigger than each cycle of an
EMRI around a non-spinning MBH.  On the other hand, retrograde orbits ``push the
LSO outwards'' and hence, it is easier for a SBH to plunge, and the EMRI is
lost.  However, this situation is not symmetric, resulting in an effective
enhancement of the rates. These results have been also confirmed by
\cite{WillMaitra2017} using a different method based in a post-Newtonian
algorithm. In this approach these EMRI spend a lower number of cycles in the
band of the detector. However, as the authors of \cite{WillMaitra2017} state,
``(...) the PN approximation is being pushed up to or beyond its limit of
validity, so we do not wish to claim too much accuracy for our values of
$T_{\rm{plunge}}$ in Table III.''

The authors  of \cite{Amaro-SeoaneSopuertaFreitag2012} also show that vectorial
coherent relaxation is not efficient enough to turn a prograde orbit into a
retrograde one, which would be fatal for this scenario, once the evolution is
dominated by GW emission.  This result is crucial in the formation of EMRI
sources. To understand why, first we need to introduce the problem of the
so-called ``Schwarzschild barrier''.

The authors of \cite{MerrittAlexanderMikkolaWill2011} performed direct-summation
$N$-body simulations and found that EMRI event rates are severely suppressed
when introducing relativistic precession in the integrations. The precession
limits the action of torques from the stellar potential in the orbital angular
momenta.  Nevertheless, they do find some particles that do cross this barrier
(the Schwarzschild barrier, to use their nomenclature).  In
Figure~\ref{fig.Schw_Barrier} we see this scenario. This is from
\cite{MerrittAlexanderMikkolaWill2011}, and shows a Newtonian simulation in the
left panel.  The authors display the semi-major axis and eccentricity of the
two-body system consisting of one star and the MBH. In the right panel they
depict the situation with all the relativistic correction terms ``switched on''.
$a_r$ and $e_r$ are the 1PN generalisations of the semi-major axis and
eccentricity. In the upper panel the red, dotted line corresponds to the barrier,
given by their expression:

\begin{equation}
{\tilde a} = C_\mathrm{SB} \left(1-e^2\right)^{-1/3},
\end{equation}
where $C_\mathrm{SB}$ is a constant of order unity.
The blue, dash-dotted line corresponds to the GW capture.

\epubtkImage{.png}{
\begin{figure}[htbp]
\centerline{\includegraphics[width=0.8\textwidth]{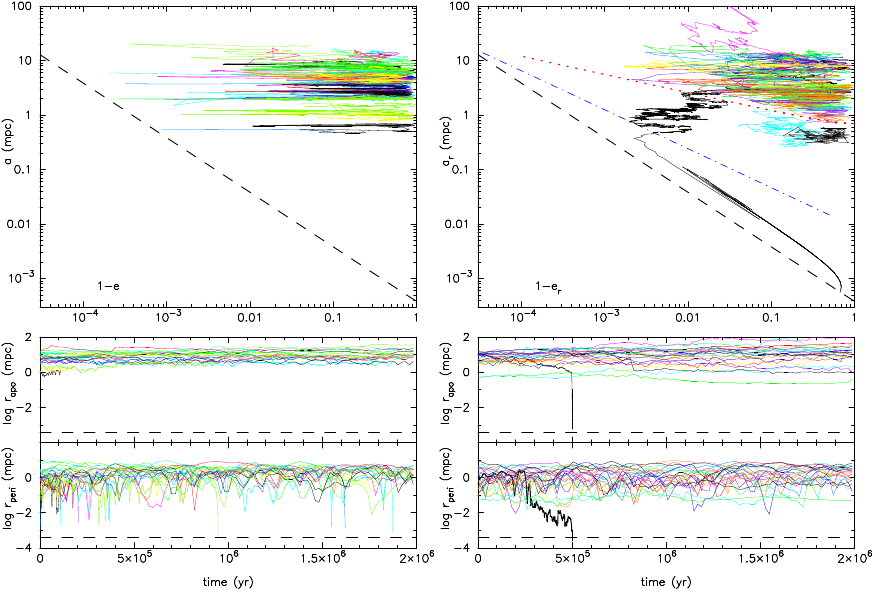}}
\caption{The Schwarzschild barrier in the work of \cite{MerrittAlexanderMikkolaWill2011}.
In the left panel we have the Newtonian case and in the right one the simulations with
relativistic corrections. We clearly see the barrier and also a particle crossing it.}
\label{fig.Schw_Barrier}
\end{figure}
}

This finding has been confirmed and quantified in the work of
\cite{BremAmaroSeoaneSopuerta2012} using a statitical sample of 2,500
direct-summation $N$-body simulations using both a post-Newtonian but also, and
for the first time, a geodesic approximation for the relativistic orbits.
However, in their work, the authors do not find a sharp transition ``barrier'',
but an area in phase space within which particles (stars) spend more time than
outside of it.

A better way of displaying this barrier is not by following a few individual
orbits, which are not representative of the phenomenon, but to depict a full
presence density map. Indeed, in Figures~\ref{fig.cusp2d} and
\ref{fig.DensityPresence}, we have the normalized presence density as a
histogram in the $(a, 1-e)$ plane for the Newtonian case,
Figure~\ref{fig.DensityPresence} (left panel) and the relativistic case (right
panel), and I give the theoretical distribution in Figure~\ref{fig.cusp2d}. In
the figures we see that on the right of the blue line there is a region within
which stars significantly spend more time than in other areas.
If we consider our specific setup, there are 3 different regions in the
$(a,1-e)$ plane where different mechanisms are efficient. In the right
region, where pericenters are large, coherent relaxation plays the dominant
role. The left border of this region is roughly given by the appearance of the
Schwarzschild precession which inhibits stellar-mass black holes from experiencing
coherent torques \cite{BremAmaroSeoaneSopuerta2012}.

\cite{Bar-OrAlexander2014} addressed this problem in terms of the adiabatic
invariance of the precession against the slowly varying random background
torques and find that this precession-induced barrier in angular momentum is
maximal for smooth noise.  The barrier is not such, nor a reflecting one. It is
an effective division of phase-space where resonant relaxation is effective,
and where it is not.

\epubtkImage{.png}{
\begin{figure}[htbp]
\centerline{\includegraphics[width=\textwidth]{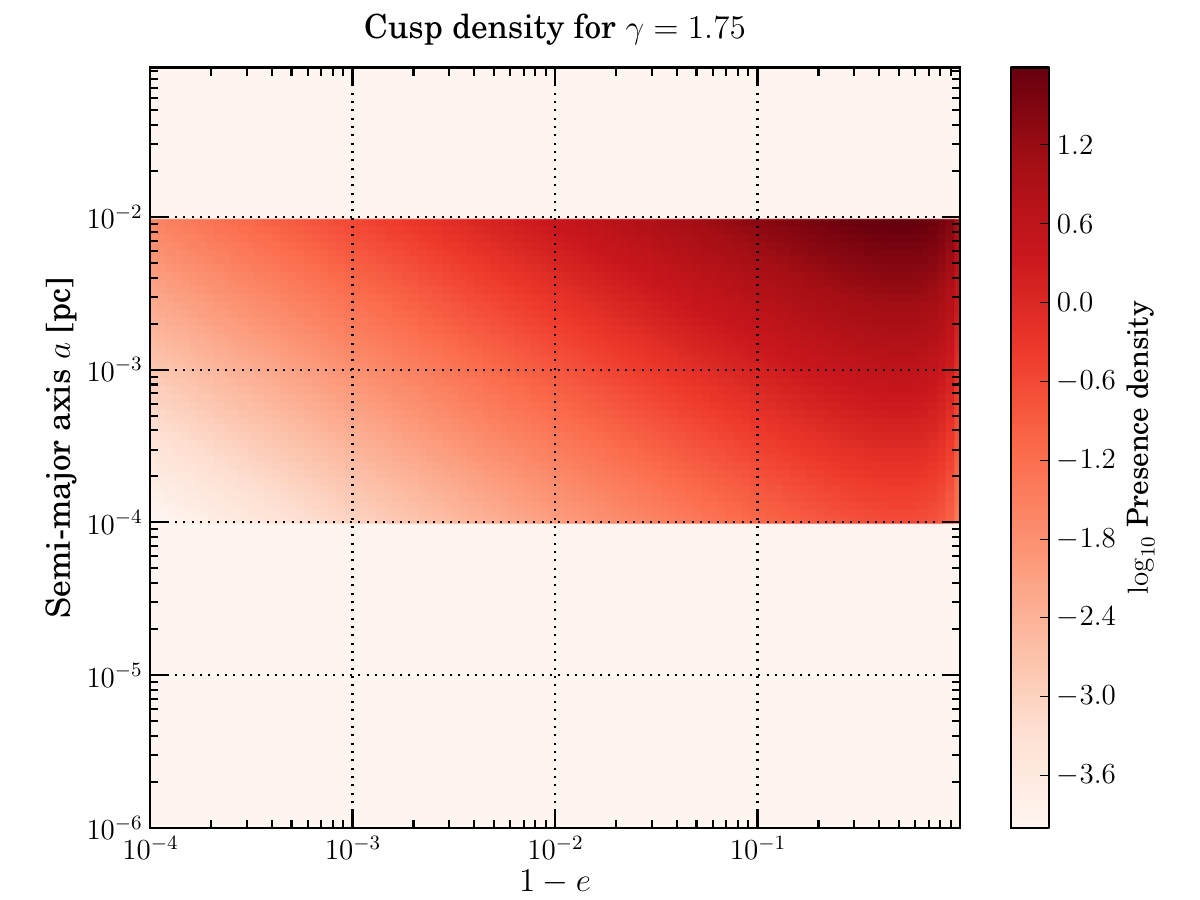}}
\caption{Theoretical distribution for the density of presence
of a cusp of power-law 1.75 around a MBH, using a truncated distribution.}
\label{fig.cusp2d}
\end{figure}
}

\epubtkImage{.png}{
\begin{figure}[htbp]
\centerline{\includegraphics[width=\textwidth]{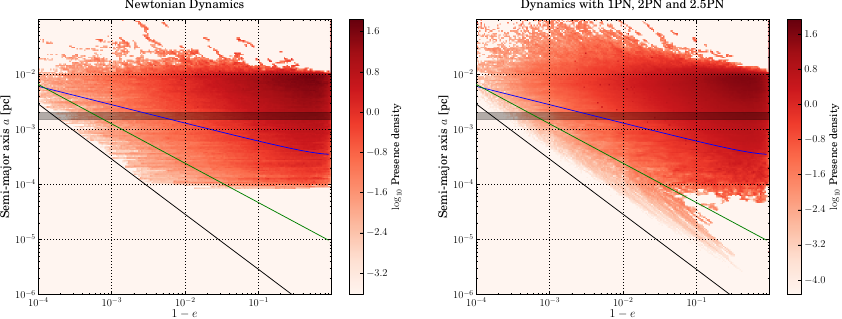}}
\caption{Integrated presence density for the Newtonian (left panel) and the
relativistic case (right panel). The lines indicate the position of
the Schwarzschild barrier with $C_{\rm SB} = 0.35$ ({\itshape blue}) and the
limit for capture onto inspiral orbits for non-resonant relaxation ({\itshape
green}).}
\label{fig.DensityPresence}
\end{figure}
}

This interesting and pioneering scenario would obviously imply a priori a
severe suppression of EMRI event rates \textit{if we relied on resonant
relaxation}.  While this is true for EMRIs originating at these distances, the
whole picture looks much more different at larger semi-major axis and
eccentricities.

We have seen in Section~\ref{sec.fundamentalEMRIs} that the small compact
object will be on a so-called ``plunging orbit'' if $e\ge e_{\rm plunge} \equiv
1-4\,R^{}_{\rm Schw}/a$.  It has been claimed a number of times by different
authors that this would result in a too short burst of gravitational radiation
which could only be detected if it was originated in our own Galactic Center
\cite{RubboEtAl2006,HopmanFreitagLarson07,YunesEtAl2008,BerryGair2013} because
one needs a coherent integration of some few thousand repeated passages
through the periapsis in the LISA bandwidth.

Therefore, such ``plunging'' objects would then be lost for the GW signal,
since they would be plunging ``directly'' through the horizon of the MBH and
only a final burst of GWs would be emitted, and such burst would be very
difficult to recover, since the very short signal would be buried in a sea of
instrumental and confusion noise and the information contained in the
signal would be practically nil.

However, \cite{Amaro-SeoaneSopuertaFreitag2013} showed that this is not true.
In Figures~\ref{fig.LSO_Spin0p4_0p7} and \ref{fig.LSO_Spin0p99_0p999} I show
plots of the location of the LSO in the plane $a$ (pc) - $(1-e)$, including the
Schwarzschild separatrix between stable and unstable orbits, $p -6 - 2e = 0$,
for both prograde and retrograde orbits and for different values of the
inclination $\iota$.  Each plot corresponds to a different value of the spin,
showing how increasing the spin makes a difference in shifting the location of
the separatrix between stable and unstable orbits, pushing prograde orbits near
$GM^{}_{\bullet}/c^{2}$ while retrograde orbits are pushed out towards
$9GM^{}_{\bullet}/c^{2}$.  The procedure I have used to build these plots is a
standard one.  Briefly, given a value of the dimensionless spin parameter
$s\equiv a^{}_{\bullet}c^{2}/(GM^{}_{\bullet})$ and a value of the eccentricity
and inclination angle $\iota$.

\epubtkImage{.png}{
\begin{figure}[htbp]
\centerline{\includegraphics[width=\textwidth]{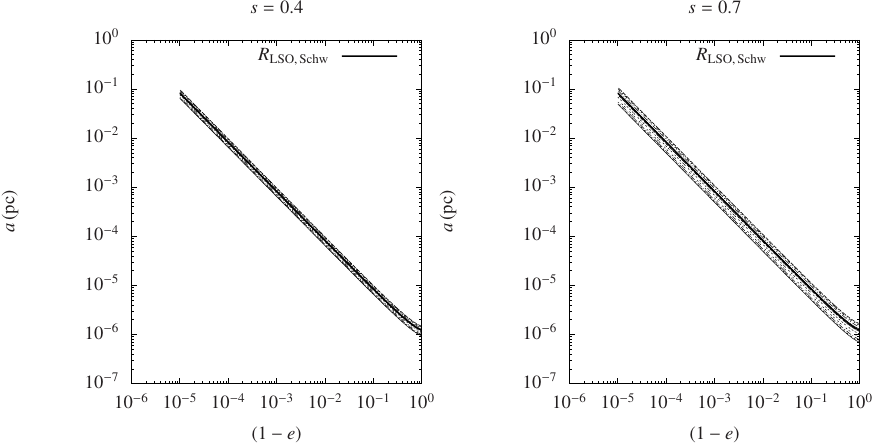}}
\caption{LSO for a MBH of mass $4\times10^4\,M^{}_{\odot}$ and a SBH of mass
$m_{\bullet}=10\,M^{}_{\odot}$ for a Kerr MBH of spin $s=0.4$ (left) and $s=0.7$
(right). The Schwarzschild separatrix is given as a solid black line. Curves
above it correspond to retrograde orbits and inclinations of
$\iota=5.72,\,22.91,\,40.10,\,57.29,\,74.48$ and $89.95^{\circ}$ starting from the
last value ($89.95^{\circ}$). In the left panel we can barely see any difference
from the different inclinations due to the low value of the spin.}
\label{fig.LSO_Spin0p4_0p7}
\end{figure}
}

\epubtkImage{.png}{
\begin{figure}[htbp]
\centerline{\includegraphics[width=\textwidth]{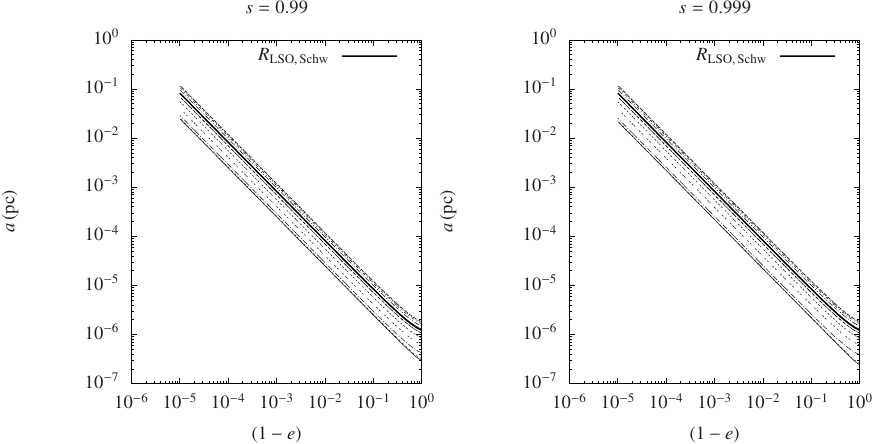}}
\caption{The same as in Figure~\ref{fig.LSO_Spin0p4_0p7} but for a spin of $s=0.99$ (left) and
$s=0.999$ (right panel). The larger the spin, the ``further away'' the Kerr LSO
gets from the Schwarzschild LSO.}
\label{fig.LSO_Spin0p99_0p999}
\end{figure}
}

In \cite{Amaro-SeoaneSopuertaFreitag2013} it was estimated that the number of cycles that
certain EMRI orbital configurations, that were thought to be plunging orbits (or
orbits with no sufficient cycles), in the case of non-spinning MBHs, can spend in
a frequency regime of $f\in [10^{-4},1]$ Hz during their last year(s) of
inspiral before plunging into the MBH.  This is important to assess how many of
these EMRIs will have sufficient Signal-to-Noise Ratio (SNR) to be detectable.
It was  found that (prograde) EMRIs that are in a ``plunge'' orbit actually spend  a
significant number of cycles, more than sufficient to be detectable with good
SNR.  The number of cycles has been associated with $N^{}_{\varphi}$ (the
number of times that the azimuthal angle $\varphi$ advances $2\pi$) which is
usual for binary systems.  However, as I have discussed above, the structure of
the waveforms from EMRIs is quite rich since they contain harmonics of three
different frequencies.  Therefore the waveforms have cycles associated with the
three frequencies $(f^{}_{r},f^{}_{\theta},f^{}_{\varphi})$ which makes them
quite complex and in principle this is good for detectability (assuming we have
the correct waveform templates). Moreover, these cycles happen just before
plunge and take place in the strong field region very near the MBH horizon.
Then, these cycles should contribute more to the SNR than cycles taking place
farther away from the MBH horizon.

The authors also estimate the impact on the event rates. Since ``direct plunges''
are actually disguised EMRIs, although with a higher eccentricity. They prove that

\begin{eqnarray}
\frac{{{a}_{\rm EMRI}^{\rm Kerr}}}{{{a}_{\rm EMRI}^{\rm Schw}}} & = & {\cal W}^{\frac{-5}{6-2\gamma}}(\iota,\,s)\\
\frac{{\dot{N}_{\rm EMRI}^{\rm Kerr}}}{{\dot{N}_{\rm EMRI}^{\rm Schw}}} & = & {\cal W}^{\frac{20\gamma-45}{12-4\gamma}} (\iota,\,s) \,.
\label{eq.NAEMRIW}
\end{eqnarray}

\noindent
Here, ${\cal W}$ is a function that depends on $\iota$, the
inclination of the EMRI and $s$, its spin\epubtkFootnote{For the derivation and some
examples of values for ${\cal W}$, I refer the reader to the work of
\cite{Amaro-SeoaneSopuertaFreitag2012}.}.  I also have assumed that the
stellar-mass black holes
distribute around the central MBH following a power-law cusp of exponent
$\gamma$, i.e., that the density profile follows $\rho \propto r^{-\gamma}$
within the region where the gravity of the MBH dominates the gravity of the
stars, with $\gamma$ ranging between 1.75 and 2 for the heavy stellar
components
\cite{Peebles72,BW76,BW77,ASEtAl04,PretoMerrittSpurzem04,AlexanderHopman09,PretoAmaroSeoane10,Amaro-SeoanePreto11}
(see \cite{Gurevich64} for an interesting first idea of this
concept)\epubtkFootnote{The authors obtained a similar solution for how electrons
distribute around a positively charged Coulomb centre.}.
For instance, for a spin of $s=0.999$ and an inclination of $\iota = 0.4\,$rad,
they estimate that ${\cal W}\sim 0.26$ and, thus, $\dot{N}_{\rm EMRI}^{\rm Kerr}
\sim 30$.

To sum up, the existence of the barrier prevents ``traditional EMRIs'' from
approaching the central MBH, but if the central MBH is spinning the rate will
be dominated by highly-eccentric extreme-mass ratio inspirals anyway, which
insolently ignore the presence of the barrier, because they are driven by chaotic
two-body relaxation.

\subsection{Extended stars EMRIs}

In this section I review the idea described in \cite{Freitag03} that MS stars
can be potential sources of GWs in our Galactic Centre.  I include this in
this section because in the whole review our standard CO is considered to
be a SBH and so, it falls into the category of ``not in the standard model''.

Indeed, a MS star can reach close enough distances to the central MBH depending
on its average density and stellar structure. For a mass of around
$0.07\,M_{\odot}$, the density of the MS star is maximum and corresponds to the
transition to a sub-stellar object \cite{CB00}. For masses smaller than $0.3
- 0.4\,M_{\odot}$, the core is totally convective and can be described with a
polytrope of index $n = 3/2$.

The articles \cite{Freitag01,Freitag03} estimated the number of single MS stars
which can become an abundant source of GWs in our GC by inspiraling into the
central MBH.  In his work, the author estimated with Monte Carlo simulations
that there must be one to a few low-mass MS stars sufficiently bound to the GC
to be discernible by LISA.  In Figure~\ref{fig.Marc2002Fig3} we show some of the
most relevant results of the investigation. Nevertheless, we note that the
assumptions made by the numerical tool are probably biasing the results to an
overestimation. These assumptions rely in the nature of the Monte Carlo code.

\epubtkImage{.png}{
\begin{figure}[htbp]
\centerline{\includegraphics[width=0.6\textwidth]{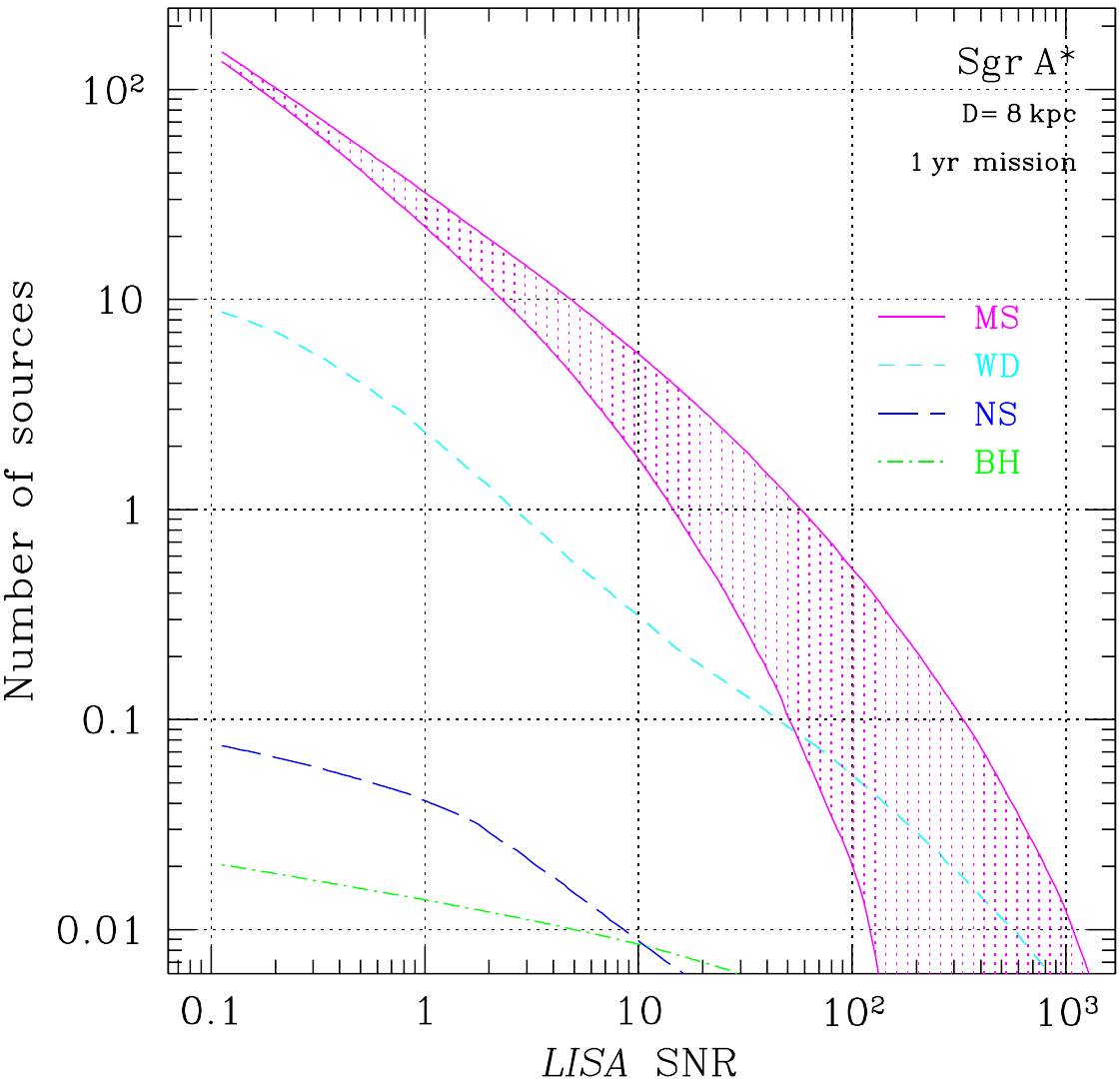}}
\caption{Number of sources at the GC depending on their
  compactness. This figure (taken from \cite{Freitag03}) displays the
  number of MSSs, WDs, NSs and stellar-mass black holes as a function of the SNR in LISA for a one year mission and assuming a distance of 8~kpc.}
\label{fig.Marc2002Fig3}
\end{figure}
}

\subsection{The butterfly effect}

An interesting effect described in the work of
\cite{Amaro-SeoaneBremCuadraArmitage2012} is the lack of determinism in an EMRI
system if a perturbing star is close enough to the binary formed by the MBH and
the SBH. One immediate question that arises is how realistic it is to assume
that we can have a second star so close to the EMRI so as to perturb it.

I estimate how likely is to have a star close enough to perturb the EMRI in a
measurable way.  For this, I take our Galactic Centre as a representative
system of the scenario that we want to analyse. If we admit that for Sgr$A^*$
half of the mass within the orbit of S-2, which has a periapsis of $6\times
10^{-4}$pc \cite{GhezEtAl08,GenzelEtAl10}, is $M_{\rm encl}/2 = \eta \times
M_{\bullet}$, with $\eta \leq 0.040$ and $M_{\bullet}$ the mass of the MBH
\cite{GillessenEtAl09}, i.e., $M_{\rm encl} = 172,000\,M_{\odot}$, and we assume
that the stars build a cusp following a power-law of the type $R^{-\gamma}$,
then we can estimate that the mass at radius $R$ is

\begin{equation}
M(R) = \int_{0}^{R} 4 \pi r^2 \rho(r) \, dr
\propto \int_0^R
r^{-\gamma+2} dr \propto R^{3-\gamma},
\end{equation}
for $\gamma<3$. Hence, the number of stars within a sphere of radius $R$ is given by
\begin{equation}
N(R) \simeq 8.6 \times 10^4 \left(\frac{R}{6\times 10^{-4} {\rm
~pc}}\right)^{3-\gamma}.
\label{eq.N_in_R}
\end{equation}

And so, the radius within we can expect to find in average a star is

\begin{equation}
R_1 \simeq 6\times 10^{-4} {\rm ~pc} \times \left(\frac{1}{8.6 \times
10^4}\right)^{\frac{1}{3-\gamma}}.
\label{eq.R1}
\end{equation}
We note, however, that the value derived for $\eta$ is not observational. As a
matter of fact, with current limitations in the observations, it is impossible
to know whether all mass enclosed by the orbit of S-2 corresponds to the MBH or
it contains also an ``extended'' component. Hence, in order to obtain $\eta$,
one has to model the system by admitting that it consists of a punctual source
(the MBH) along with a stellar component whose properties are parametrised by
following a model, not an observation.  In Figure~\ref{fig.R_1} I show the
dependence on $\gamma$ of $R_1$.  We can see that $R_1 \simeq 3 \times 10^{-7}$
pc for $\gamma = 1.5$ or $7\times 10^{-8}$ pc for $\gamma = 1.75$, see
\cite{ASEtAl04,FAK06a,PretoMerrittSpurzem04}.  These distances are of
the same order of magnitude than an EMRI which is within the bandwidth of a
LISA-like observatory. Even if this argument is based on the concept of a cusp
and, hence, it is difficult to define at such short radii, in my work with Marc Freitag
and Vassiliki Kalogera
\cite{FAK06a} we derive in our Milky Way-like G25 model some $15\,M_{\odot}$
within $3\times 10^{-4}$ pc.  It is possible that at such distances the mass
density is totally dominated by stellar-mass black holes, but the work of
\cite{FAK06a} does not allow one to resolve them for distances shorter than
0.01 pc. In this case, strong mass segregation would play a crucial
role \cite{AlexanderHopman09,PretoAmaroSeoane10,Amaro-SeoanePreto11}, since for
the kind of slopes that one can expect in the case the density is dominated by
stellar-mass black holes, the ``one-star'' radius is much shorter.

\epubtkImage{.png}{
\begin{figure}[htbp]
\centerline{\includegraphics[width=0.7\textwidth]{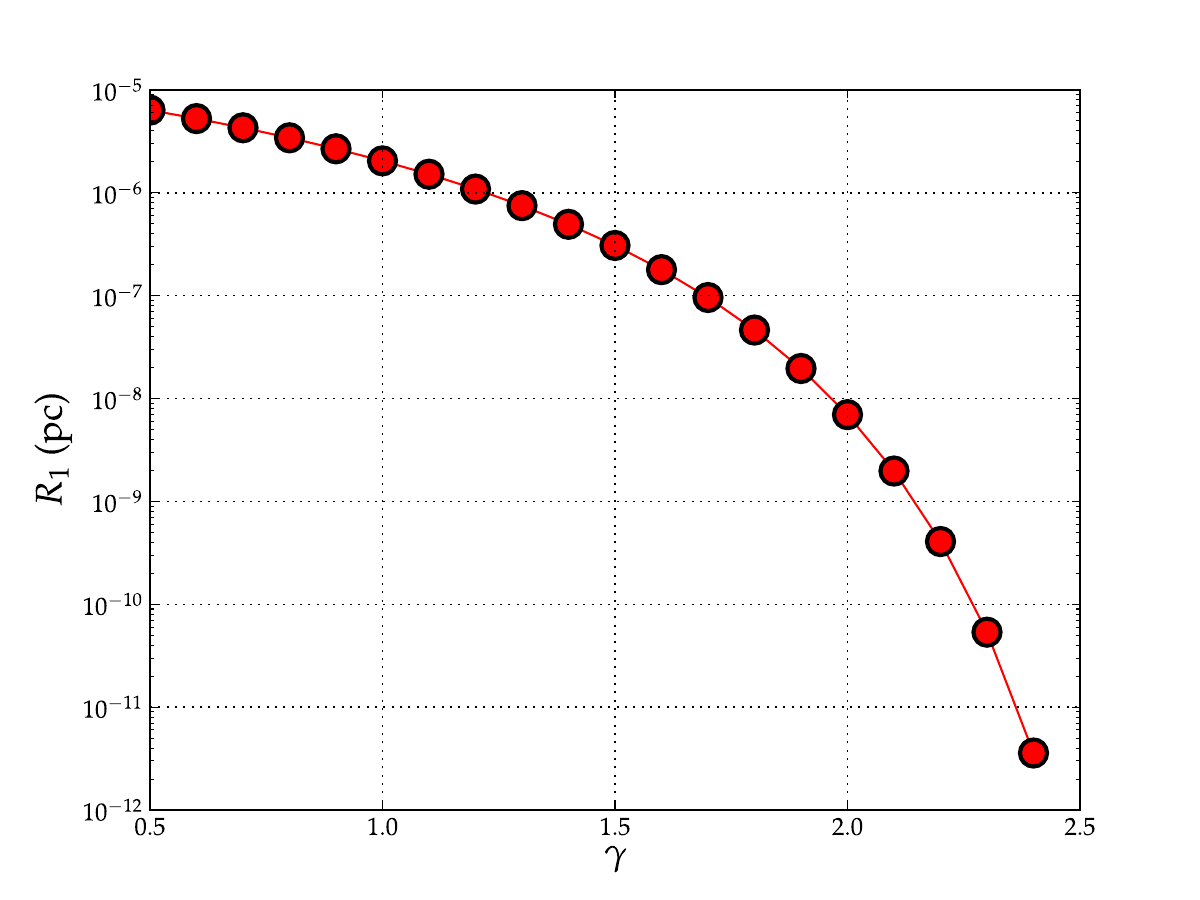}}
\caption{Evolution of the one-star radius as a function of the slope as in
Eq.~(\ref{eq.R1}).   We can see that for very mild slopes and even a core the
distances are within a millihertz gravitational wave detector similar to LISA;
i.e., of orbital periods of $10^5$ s. In this regime we expect sources of GWs. For instance,
an EMRI of $10\,M_{\odot}$ with a MBH of $4\times 10^{6}M_{\odot}$
has a semi-major axis of about $a_{\bullet}
\approx 8\times 10^{-4}$ pc and is well within the bandwidth.}
\label{fig.R_1}
\end{figure}
}

In Figure~\ref{fig.Butterfly} we have the initial setup for the fiducial case
in the work by \cite{Amaro-SeoaneBremCuadraArmitage2012}.  The mass of the
MBH is assumed to be ${M}_{\bullet} = 10^{6}\,M_{\odot}$, the initial
semi-major axis of the EMRI of $a_{\bullet,\,\rm i} \simeq 1.45\times 10^{-6}$
pc (i.e., it is well within the band of LISA) , the mass of the EMRI is
$m_{\bullet} = 10\,M_{\odot}$ (but they also successfully tested
$5$ and $1.44\,M_{\odot}$), the mass of the perturbing star is of $m_{\star} =
10\,M_{\odot}$, the initial semi-major axis of the star $a_{\star ,\,\rm i}
\simeq 4.1 \times 10^{-6}$ pc, and the initial eccentricity $e_{\star,\,\rm i}
= 0.5$ and the inclination is $i_{\bullet,\,\star} = 30^{\circ}$ at $T=0$.

\epubtkImage{.png}{
\begin{figure}[htbp]
\centerline{\includegraphics[width=0.5\textwidth]{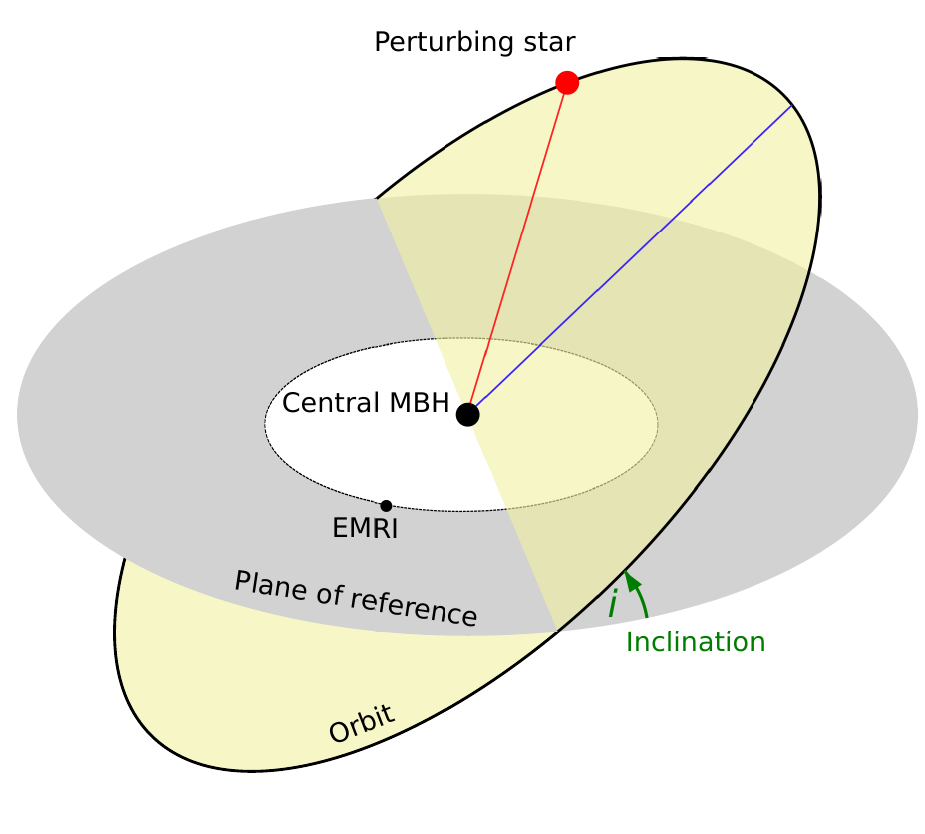}}
\caption{Description of the scenario for the butterlfy effect of the work
by \cite{Amaro-SeoaneBremCuadraArmitage2012}
(adapted from a figure by Lucas Snyder).}
\label{fig.Butterfly}
\end{figure}
}

The authors find that the interloper introduces an observable modification in
the orbit of the EMRI when using a code that uses loss of energy via
gravitational radiation at periapsis.  The interesting result, though, is that
when taking into account also the two first-order non-dissipative
post-Newtonian contributions, the orbital evolution is not deterministic.  We
do not know what the stellar distribution around a MBH is at such short
radii, but if this scenario was possible, then the detection of EMRIs would be
much more challenging than it was thought, because the waveforms developed for
detection would be of little use.  There has also been work about the role of a
massive perturber on an EMRI. I refer the reader to the work of
\cite{ChenEtAl11,YunesEtAl2011,Seto2012}.

\subsection{Role of the gas}

Another proposal is related to the presence of massive accretion discs around
MBHs.  At distances of $\sim 0.1-1$~pc from the MBH and with typical accretion
rates, these discs can be unstable to star formation
\cite{CZ99,LB03,Goodman03,GT04,ML04,Levin03,Levin06,Nayakshin06}.  If, as in
some calculations, there is a bias towards the production of massive stars in
the disc, they could evolve to become black holes, which are then dragged in
along with the disc matter.  Alternately, massive stars on orbits that cross the
disc could be captured and then evolve into black holes
\cite{SCR91,Rauch95,SubrKaras99,KS01}. Rates are highly uncertain as well as the
mass of the stellar remnants formed (which could even be IMBHs). However these
events would likely have a different signature waveform than those of the other
two classes because they should occur on co-rotating, circular orbits lying in
the equatorial plane of the spinning MBH if it has gained a significant fraction
of its mass by accreting from the disc \cite{Bardeen70,KLOP05,VMQR05}.
Moreover, there is the exciting possibility that in such a scenario the compact
object would open a gap in the disc, which could lead to an optical counterpart
to the EMRI event~\cite{Levin06}.

The work of \cite{BarausseEtAl2007} addresses the imprint on the waveform of
compact, massive tori close to the central MBH. The kludge waveforms generated
in their study were indistinguishable from pure Kerr waveforms in the regime on
which they focussed. Barausse and his collaborators later extended the study to a
non self-gravitating torus with constant specific angular momentum and found
that typically one should not expect big differences, although for a certain
region of the parameter space the hydrodynamic drag acting on the EMRI does have an
impact comparable to the radiation-reaction, so that it could, in principle, be
measurable \cite{BarausseRezzolla2008}. Later, this work was expanded in
\cite{BarausseEtAl2014}. Nevertheless, it is not clear what
the appropriate gas distribution around the MBH is in the regime of their study.
Perturbations to the SBH are likely to be negligible if accretion onto the
hole happens in a low density, radiatively inefficient flow. Such flows are
much more common than dense accretion discs, which in principle could yield
observable phase shifts during the inspiral \cite{KocsisEtAl11}, at least within
the redshift range in which we expect to observe EMRIs.

\newpage

\section{Integration of dense stellar systems and EMRIs}
\label{ch.Integration}

\subsection{Introduction}

In this section we give a summary of the current numerical approaches available
for studying stellar dynamics in systems for which relaxation is an important
factor\epubtkFootnote{A part of this section profits from \cite{Amaro-SeoaneEtAl07},
though some parts have been significantly expanded and improved.}

As of writing this article only approximate methods using a number of
simplifying assumptions have been used to estimate the rates and characteristics
of EMRIs. I review these approaches, their accomplishments and limitations.
Thanks to the rapid computational power increase and the development of new
algorithms, it is most likely that direct $N$-body techniques will soon be able
to robustly confirm or disprove these approximate results and extend them. One
of the main issues is that exceptionally long and accurate integrations are
required to account correctly for secular effects such as coherent relaxation or
Kozai oscillations. These requirements, and the extreme mass ratio pose new
challenges to developers of $N$-body codes.

We can approximately classify the different kinds of techniques employed for
studying stellar dynamics according to the dynamical regime(s) they can cope
with.  In Figure~\ref{fig.StellDynRealms} we have a classification of these
techniques. (Semi-)analytical methods are generally sufficient only to study
systems which are in dynamical equilibrium  and which are not affected by
collisional (relaxational) processes. In all other cases, including those of
importance for EMRI studies, the complications that arise if we want to extend
the analysis to more complex (realistic) situations, force us to resort to
numerical techniques.

The $N$-body codes are the most straightforward approach from a conceptual
point of view. In those, one seeks to integrate the orbital motion of $N$
particles interacting gravitationally. It is necessary to distinguish between
the \emph{direct} $N$-body approaches which are extremely accurate but slow and
the fast $N$-body approaches, which less accurate and therefore
generally deemed inadequate for studying relaxing systems because relaxation is
the cumulative effect of \emph{small} perturbations of the overall, smooth,
gravitational potential. Fast $N$-body codes are usually based on either TREE
algorithms \cite{barneshut86} or on an FFT (Fast Fourier Transform)
convolution to calculate the gravitational potential and force for each
particle \cite{superbox} or on an SCF (self-consistent-field)
\cite{Clutton-Brock73,HO92} approach.  I will not describe these numerical
techniques in this section because they have never been used to study E/IMRIs
and the approximations on which they are based make them unsuitable for an
accurate study of such systems, since relaxation plays a role of paramount
importance. Fast $N$-body algorithms can only be employed in situations in
which relaxation is not relevant or over relatively short dynamical times, such
as in studying bulk dynamics of whole galaxies.

On the other hand, if we want to study a system including both collisional
effects and dynamical equilibrium, we can employ direct $N$-body codes or use
faster approaches, like the Monte Carlo, Fokker Planck and Gas methods, which
we will describe below.  The only technique that can cope with all physical
inputs is the direct $N$-body approach, in which we make no strong assumptions
other than that gravity is Newtonian gravity (although nowadays post-Newtonian
corrections have also been incorporated, see Section~\ref{sec:directNbody}).

If we neglect capture processes driven by tidal effects, the region from which
we expect most EMRIs to come is limited to $\sim$~1--0.1~pc around the
central MBH (see subsections \ref{sub.SchwBarr}). In that zone the potential is totally spherical.  Non-spherical
structures such as triaxial bulges or stellar discs are common on scales of
100\,--\,1000~pc, and the nucleus itself may be non-spherical. For example, it
could be rotating, as a result of a merger with another nucleus \cite{MM01} or
due to dissipative interactions between the stars and a dense accretion disc
\cite{Rauch95}.

It is unclear whether this effect could enhance the replenishment of the loss
cone, see \cite{MCD91,FB02b,ASFS04,BME04a,BME04b,MerrittVasiliev11,VasilievMerritt2013},
and \cite{VasilievEtAl2014} in particular for the even more complex of binaries of
massive black holes, in the context of the ``final parsec problem''.  The problem is further
compounded, for example, by the presence of multiple stellar populations whose
spatial distributions are segregated (``mass segregation''), with more massive
stars sinking deeper into the potential well and approaching closer to the
central black hole.  Besides, two interacting stars may become gravitationally
bound (become a binary) so that during the subsequent interactions with other
stars or massive black holes they behave differently from single stars, or they may collide
into each other, then the subsequent evolution will be determined by
gas-dynamics. As these ``micro-physical'' effects are usually not incorporated into the
global modeling of the entire nuclear star clusters, considerable uncertainties
are attached to the theoretical predictions of the abundance and orbital
parameters of the stars in the relativistic regime.

Whilst assuming sphericity will probably not have any impact on the estimate of
capture rates, it is of huge relevance for ``tidal processes'', since this is
the region in which binary tidal separation and the tidal capture of giant
cores will happen. For these processes the critical radius is
beyond the influence radius of the central MBH and so triaxiality can probably
play an important role.  Due to computer power and the limitations
of simulation codes galactic nuclei have so far been modelled only as isolated
spherical clusters with purely Newtonian gravity (i.e., \cite{MCD91,FB02b}).
\cite{Vasiliev2015} used the Princeton approach to derive a new Monte Carlo
code, which presents a scheme to deal with asphericity (with other issues
remaining open), with the limitation that it assumes isotropy of background
stars population, so that it cannot model a highly flattened system with
significant rotation support.

\epubtkImage{.png}{
\begin{figure}[htbp]
\centerline{\includegraphics[scale=0.4,clip]{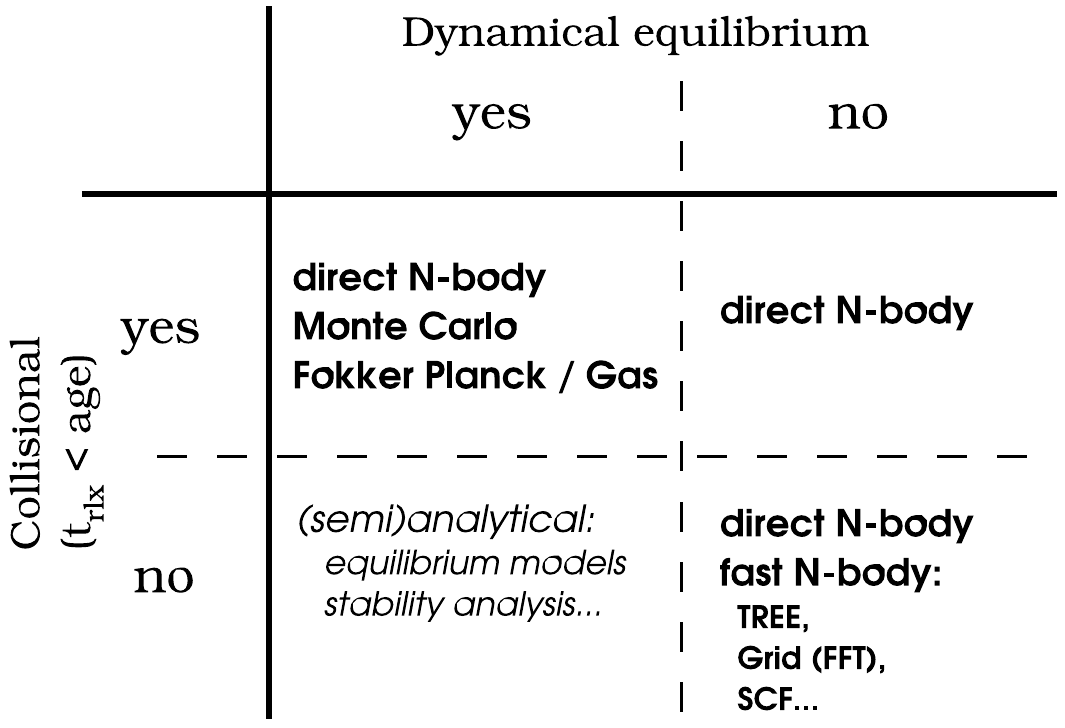}}
\caption{Possible methods to study the various realms of stellar dynamics.}
\label{fig.StellDynRealms}
\end{figure}
}

More realistic situations could only be explored with $N$-body methods or
possibly with hybrid codes (Monte Carlo combined with $N$-body, for instance).
While important approaches exist that implement small-number $N$-body
integrations in the core of Monte Carlo, see
\cite{HypkiGiersz2013,FregeauRasio2007}, these approaches typically focus on
binary scattering interactions, with less than 5 bodies. An important exception
is the work of \cite{RodriguezEtAl2015}, which can integrate
larger numbers, but it is limited to CPUs. Being based on KIRA
\cite{PortegiesZwartEtAl01}, it can in principle run on GRAPEs, a
special-purpose chip to compute gravitational forces that was used in the past
by many groups, see e.g. \cite{GRAPE6A}, but (i) it is fair to say that these
cards are obsolete, and virtually all efforts focus now on GPUs (there exists a
library that can allow a GRAPE to mimic a GPU \cite{GaburovEtAl2009}, but it is
far from trivial to do it and in any case sub-optimal), (ii) it does not
account for relativistic corrections, crucial to EMRI astrophysics, (iii) the
code requires spherical symmetry and (iv) the code does not account for a
central MBH.

\epubtkImage{.png}{
\begin{figure}[htbp]
\centerline{\includegraphics[scale=0.4,clip]{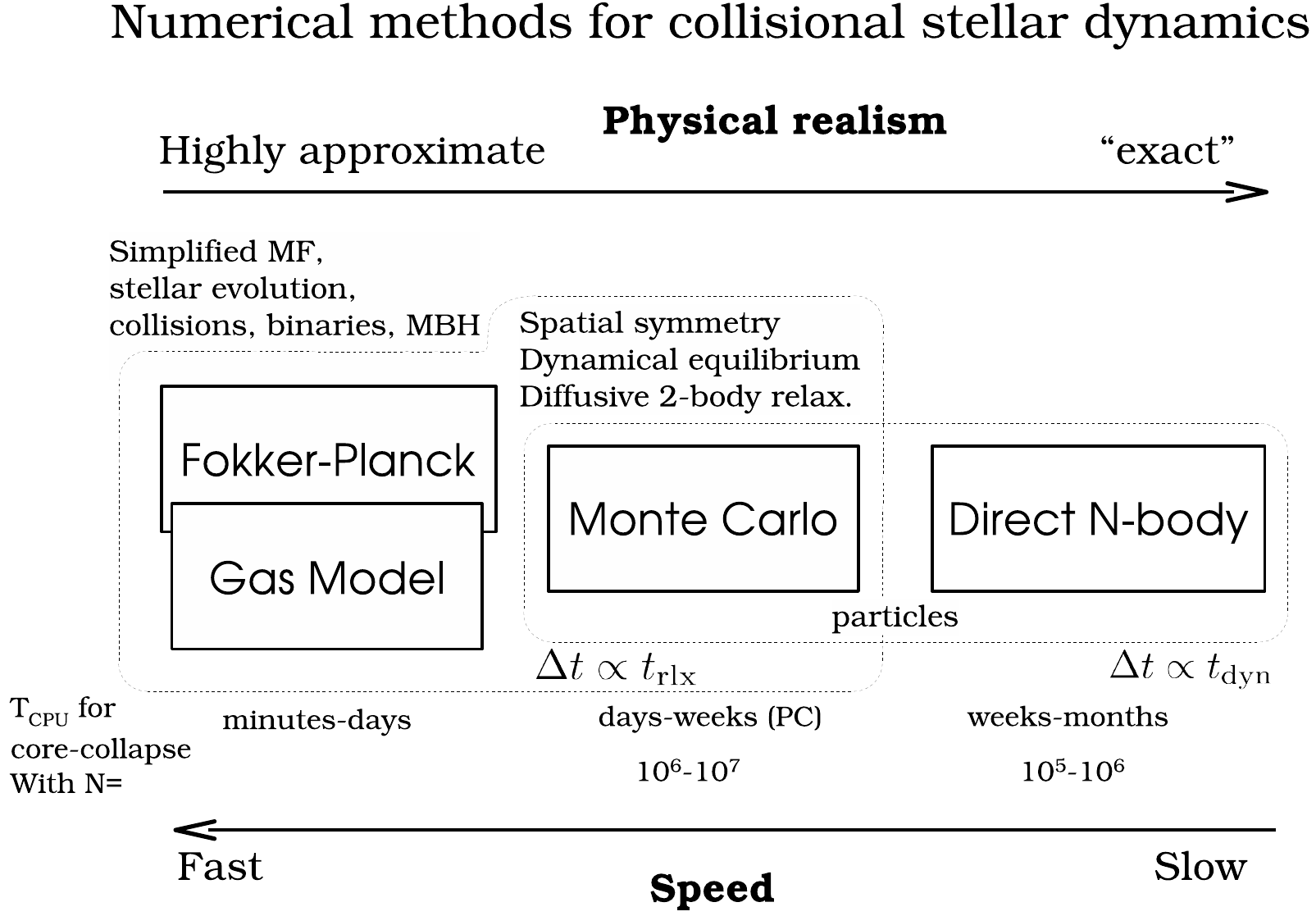}}
\caption{The various methods used to study collisional stellar dynamics.
In the case of direct $N$-body computations, the simulations require the use of either
special-purpose hardware such as the GRAPEs, Beowulf clusters for the
parallel version of Rainer Spurzem or Graphical Computing Units (GPUs).
A version of the parallel code $N$-body6++ ported to GPU architecture has
been developed, see \cite{WangEtAl2015,WangEtAl2016}, which allows us to simulate more realistic particles numbers.
}
\label{fig.CollStellNumMethods}
\end{figure}
}

In Figure~\ref{fig.CollStellNumMethods} I show a schematic illustration of
the current available codes for stellar dynamics including relaxation. The \emph{physical realism} of the codes increases from the
left to the right while the computational
speed decreases.  The two-dimensional numerical direct solutions of the
Fokker--Planck equation~\cite{Takahashi97,Takahashi96,Takahashi95} probably
require the least computational time, but these are followed closely by the
gaseous model. The idea behind it is to treat two-body relaxation as a
transport process such as in a conducting plasma \cite{HachisuEtAl78,LBE80}.
Multi-mass models have been implemented\cite{LS91,Spurzem92,GS94,ST95} and
improved for the detailed form of the conductivities by comparing to direct
$N$-body models (described below). The addition of a central accreting MBH and
a treatment for loss-cone effects was done by \cite{ASFS04} (a comprehensive
description of the code is in the appendix of the same work) for the
single-mass case, and also for a stellar mass spectrum
\cite{AmaroSeoaneThesis04}.  The advantage of these two codes is the
computational time required to perform a simulation (typically of the order of
one minute on a regular PC for a Hubble time) and since they are not
particle-based, the resolution can be envisaged as infinite, so that they are
not limited by the particle number of the system and there is practically no
numerical noise. Nevertheless, although they should be envisaged as powerful
tools to make an initial, fast exploration of the parameter space, the results
give us \emph{tendencies} of the system, rather than an accurate answer
\cite{AmaroSeoaneThesis04}. Studying the Astrophysical I/EMRI problem requires
a meticulous characterisation of the orbital parameters, so that approximate
techniques should be regarded as exploratory only~\cite{DeFreitasEtAl06}.

\subsection{The Fokker--Planck approach}

\newcommand{\DCf}[1]{\langle{#1}\rangle} 
\newcommand{\DCN}[1]{\{{#1}\}} 
\newcommand{\FCF}[1]{{\cal D}_{#1}} 

Instead of tracking the individual motion of a large number of particles, as in
$N$-body methods, one can attempt to describe a system consisting of a very
large number of stars through the \emph{1-particle phase-space distribution
function} (DF for short) $f(\vec{x},\vec{v},t)$. The best interpretation of $f$,
with the proviso that it has been properly normalised,
is as a probability density if it is normalised to 1 ---
$f(\vec{x},\vec{v},t)d^3x\,d^3v$ is the probability of finding, at time $t$,
any given particle within the volume of phase space $d^3x\,d^3v$ around the 6-D
phase-space point $(\vec{x},\vec{v})$; the average number of particles in this
volume would be ${N_{\star}}f(\vec{x},\vec{v},t)d^3x\,d^3v$, with ${N_{\star}}$
the total number of particles. If the particles move
in a common smooth potential $\Phi$, the evolution of $f$ is described by the
collisionless Boltzmann equation \cite{BT87}:

\begin{equation}
D_{t}f \equiv
\partd{f}{t}+\vec{v}\cdot\vec{\nabla}f-\vec{\nabla}\Phi\cdot\partd{f}{\vec{v}}=0.
\label{eq.CollisionlessFP}
\end{equation}

\noindent
$\Phi$ is obtained from $f$ and a possible external potential $\Phi_{\rm ext}$
(such as the one produced by a central MBH) from the Poisson equation.

In a real self-gravitating $N$-particle system the potential cannot be smooth
on small scales but has some graininess, i.e., short-term, small-scale
fluctuations, $\Phi_{\rm real}=\Phi+\Delta \Phi_{\rm grainy}$. Relaxation
describes the effects of these fluctuations on $f$. They arise because a given
particle sees the rest of the system as a collection of point masses rather
than as a smooth mass distribution. Relaxational effects, also known (somewhat
confusingly) as collisional effects, can therefore be seen as particles
influencing each other individually as opposed as to collectively.  To allow
for these effects, a \emph{collision term} has to be introduced on the right
hand side of the Boltzmann equation, in Eq.(\ref{eq.CollisionlessFP}). This
equation cannot be equated to zero if we want to take into account relaxational
effects.
The Fokker--Planck (FP) equation is derived by assuming that relaxation is due
to a large number of 2-body gravitational encounters, each of which leads to a
small deflection and occurs ``locally'', i.e., they affect the velocity of a
star without affecting its position.  This is the basis for Chandrasekhar's
theory of relaxation \cite{Chandrasekhar60,BT87,Spitzer87}.
To take care of encounters between stars, hence, we have to equate Eq.(\ref{eq.CollisionlessFP})
not to zero, but to a collision term,

\begin{equation}
D_{t}f = -\sum_{i=1}^{3} \partd{}{v_i}\left[f(\vec{x},\vec{v})\DCf{\Delta v_i}\right]
+ \frac{1}{2}\sum_{i,j=1}^{3} \frac{\partial^2}{\partial v_i\partial v_j}\left[f(\vec{x},\vec{v})\DCf{\Delta v_i \Delta v_j}\right],
\label{eq:FPEvit}
\end{equation}

\noindent
where the ``diffusion coefficient'' $\DCf{\Delta v_i}$ is the average
change in $v_i$ per unit of time due to encounters (see
\cite{RMcDJ57,BT87} for a derivation).

From Jeans' theorem \cite{Jeans15,Merritt99}, for a spherical system
in dynamical equilibrium, the DF $f$ can depend on the phase-space
coordinates $(\vec{x},\vec{v})$ only through the (specific) orbital
binding energy $E$ and angular momentum (in modulus) $J$,
\begin{equation}
f(\vec{x},\vec{v}) = F(E(\vec{x},\vec{v}),J(\vec{x},\vec{v})).
\end{equation}
In the vast majority of applications, the Fokker--Planck formalism is applied in
the two-dimensional $(E,J)$-space or, assuming isotropy, the one-dimensional
energy-space rather than the six-dimensional phase space, through the operation of
``orbit averaging'' (see \cite{Cohn79,Cohn80,Cohn85,Spitzer87} amongst others).

A standard form of the FP equation for an isotropic, spherical system is

\begin{equation}
D_t N(E)  \equiv \partd{N}{t} + \partd{N}{E}\left.\frac{dE}{dt}\right|_{\phi} =
-\partd{{\cal F}_E}{E}
\label{eq.FPiso}
\end{equation}
where
\begin{equation}
{\cal F}_E=m\FCF{E}F-\FCF{EE}\partd{F}{E}
\end{equation}
is the flux of particles in the energy
space; $\left.{dE}/{dt}\right|_{\phi}$ is the change of energy due
to the evolution of the potential $\phi$; $N(E)$ is the density of stars in
$E-$space,

\begin{equation}
N(E) = 16\pi^2 p(E) F(E)
\label{eq:FtoN}
\end{equation}

\noindent
with

\begin{equation}
p(E)=\int_0^{r_{\rm max}}r^2v\, dr.
\end{equation}

\noindent
The ``flux coefficients''  are

\begin{align}
\FCF{E}  & =  16\pi^3\lambda m_{\rm f} \int_{\phi(0)}^E dE'p(E')F_{\rm f}(E'), \\
\FCF{EE} & =  16\pi^3\lambda m_{\rm f}^2 \Big[ q(E)\int_E^0 dE'F_{\rm f}(E') +
               \int_{\phi(0)}^E dE'q(E')F_{\rm f}(E') \Big],
\label{eq:FluxCoef}
\end{align}
where $\lambda\equiv 4\pi G^2 \ln \Lambda$ and $q(E)=\frac{1}{3}\int_0^{r_{\rm max}} r^2v^3\, dr$
is the volume of phase space accessible to particles with
energies lower than energy, and $p(E)={\partial q}/{\partial E}$ \cite{Goodman83}.

We use an index ``f'' for ``field'' to distinguish the mass and DF of the
population we follow (``test-stars'') from the ``field'' objects. This
distinction does not apply to a single-component system but it is easy to
generalise to a multi-component situation by summing over components to get the
total flux coefficient
\begin{equation}
\FCF{E} = \sum_{l=1}^{N_{\rm comp}} {\FCF{E}}_{,l},\ \ \FCF{EE} = \sum_{l=1}^{N_{\rm comp}} {\FCF{EE}}_{,l},
\end{equation}
where the flux coefficient for component l can written by
replacing the subscript ``f'' by ``$l$'' in Eq.~(\ref{eq:FluxCoef}).

I now explain schematically how the FP equation is implemented numerically to
follow the evolution of star clusters.  A more detailed description can be
found in, e.g., \cite{CW90}. In the most common scheme, pioneered
by~\cite{Cohn80}, two types of steps are employed alternately, a method known
as ``operator splitting'':

\begin{enumerate}

\item \textbf{Diffusion step}. The change in the distribution function $F$ for a
discrete time-step $\Delta t$ is computed by using the FP equation
\emph{assuming the potential $\phi$ is fixed}, i.e., setting $D_t N = {\partial
N}/{\partial t} = \left.{\partial N}/{\partial t}\right|_{\rm coll}$. The FP
equation is discretized on an energy grid. The flux coefficients are computed
using the DF(s) of the previous step; this makes the equations linear in the
values of $F$ on the grid points. The finite-differencing scheme is the implicit
\cite{ChangCooper70} algorithm, based on a finite difference scheme for initial
value problems, which is first order in time and energy.

\item \textbf{Poisson step}. Now the change of potential resulting from the
modification in the DF $F$ is computed and $F$ is modified to account for the
term $\left.dE/dt\right|_\phi$, i.e., assuming $D_t N = {\partial N}/{\partial
t} + {\partial N}/{\partial E}\left.{dE}/{dt}\right|_{\phi} = 0$. This can be
done implicitly because, as long as the change in $\phi$ over $\Delta t$ is very
small, the actions of each orbit are adiabatic invariants. Hence, during the
Poisson step, the distribution function, expressed in terms of the actions, does
not change. In practice, an iterative scheme is used to compute the modified
potential, determined implicitly by the modified DF, through the Poisson
equation.  The iteration starts with the values of $\phi$, $\rho$, etc.
computed before the previous diffusion step.

\end{enumerate}

A variant of the FP equation analogous to Eq.~(\ref{eq.FPiso}) can be written which
allows for anisotropy by taking into account the dependence of $F$ on angular momentum and
including a angular momentum-flux and corresponding flux coefficients
\cite{CK78,Cohn79,Cohn85,Takahashi95,Takahashi96,Takahashi97,DCLY99}.  The
expressions for the flux coefficients are significantly longer than in the
isotropic case and I do not present them here. However, we note that in galactic
nuclei, in contrast to globular clusters, anisotropy plays a key role
because of the existence of a loss cone.

The use of the FP approach to determine the distribution of stars
around a MBH requires a few modifications. First the (Keplerian)
contribution of the MBH to the potential has to be added.
Several authors have made use of the FP or similar formalisms to study
the dynamics well within the influence radius under the assumption of a
fixed potential
\cite{BW76,BW77,LS77,CK78,HA06,HA06b,MHB06}, which is a significant
simplification. The static potential included a contribution for the
stellar nucleus in the last study \cite{MHB06} but was limited to
a Keplerian MBH potential in the other cases. The presence of the MBH
also constitutes a central sink as stars are destroyed or swallowed if
they come very close to it. This has to be implemented into FP codes
as a boundary condition. The authors of \cite{LS77} and~\cite{CK78} have developed
detailed (and rather complex) treatments of the loss cone for the
anisotropic FP formalism. It can be used in a simplified way in an
isotropic FP analysis
\cite{BW77} to obtain a good approximation to the distribution of
stars around a MBH and of the rates of consumption of stars by the
MBH. However, additional analysis is required to determine what fraction of
the swallowed stars are EMRIs and what their orbital properties are
\cite{HA05,HA06b}.

\subsection{Moment models}

\def\sr{\sigma_{\rm r}}
\def\st{\sigma_{\rm t}}
\def\s_t{\sigma_{\rm t}}

Another way to approximately solve the (collisional) Boltzmann equation is to
take velocity moments of it. The moment or order $n=0$ of the DF is the
density, the moments of order $n=1$ are bulk velocities and $n=2$ corresponds
to (anisotropic) pressures (or velocity dispersions). This is analogous to the
derivation of the Jeans equation from the collisionless Boltzmann equation
\cite{BT87} but the collision term introduces moments of order $n+1$ in the
equations for moments of order $n$.


In statistical moment models, we employ velocity moments to characterize the
local velocity distribution function. The $n$-th moment of a velocity
distribution $f(v)$ is defined as $\langle v^n \rangle = \int (v)^n\,f(v) \,\,
\mathrm{d}v$. The accuracy of these models is then limited by the order of the
highest moment included to describe the velocity distribution, as discussed in
detail in the work of \cite{SchneiderEtAl2011}.

Since each stellar dynamical process driving the evolution of a cluster has a
different impact on the local velocity distribution, this motivates us to
construct a distribution function that is able to reflect the effects of each of
these processes properly so as not to lose information that influences the
clusters evolution. The velocity distribution can be written as a series
expansion using a \emph{truncated Gauss-Hermite series}, as in the works of
\cite{Gerhard1993,vanderMarelEtAl1993} to illustrate the meaning of the first
four moments:

\begin{equation}
\label{eq:gauss_hermite}
f(v_r) \propto
\exp\left(-\frac{v_r-\bar{v}_r}{2\sigma}\right)\left[1+\sum^4_{k=3}h_k
H_k(v_r-\bar{v}_r)\right],
\end{equation}

\noindent
where $H_k$ are the Hermite polynomials (see e.g. the appendix A of \cite{vanderMarelEtAl1993}),
$v_r$ the velocity in radial direction (or the line-of-sight velocity
which is the velocity measured in direction of an observer), and $\bar{v}_r$,
$\sigma$, $h_3$ and $h_4$ are free parameters. The first moments can be related
to physical properties of the system that we are studying:

        \begin{enumerate_estret}
		\item[\emph{0th moment:}]
		The zeroth moment of a velocity distribution is 1 due to normalization.
		\item[\emph{1st moment:}]
		The first moment of a velocity distribution is the mean velocity $\bar{v}_r$ and denotes the bulk mass transport velocity.
		\item[\emph{2nd moment:}]
		The second moment of a velocity distribution is the variance $\sigma$
and is equal to the velocity dispersion. It determines the width of  $f(v_r)$
and thus the scattering of stellar velocities around the mean velocity
$\bar{v}_r$. If $f(v_r)$ is fully determined by $\bar{v}_r$ and $\sigma$ and
$h_3=h_4=0$ it is a Gaussian ({upper left pannel} in figure \ref{fig:gauss_plots}) corresponding to thermal equilibrium. Then the symmetry of the one-dimensional velocity distribution $f(v_r)$ to $\bar{v}_r$ reflects isotropy.
		\item[\emph{3rd moment:}]
		The third moment, denotes the transport of random kinetic energy and
depends on $h_3$. If the third moment of the velocity distribution does not
vanish, implying that $h_3\ne0$, then the shape of the velocity distribution is
a skewed Gaussian (figure \ref{fig:gauss_plots}, {upper right pannel}). The asymmetry indicates the direction of the energy flux, and the uneven distribution of velocities in different directions denotes anisotropy.
		\item[\emph{4th moment:}]
		The fourth moment is a measure of the excess or deficiency of
particles/stars with high velocities as compared to thermodynamical equilibrium,
and depends on the value of $h_4$. An excess of particles with high velocities
results in thicker wings of the velocity distribution and a more pointed maximum
(figure \ref{fig:gauss_plots}, {lower left pannel}). A deficiency of high
velocities causes a broader shape around the mean and thinner wings of the
velocity distribution (figure \ref{fig:gauss_plots}, {lower right pannel}).
	\end{enumerate_estret}

Third and fourth order moments therefore denote deviations from thermodynamical
equilibrium. Modeling processes that lead to the transport of random kinetic
energy in a cluster or that strongly affect the high velocity wings of the
distribution suggest the use of a model that includes 4th order moments. These
processes are, for example, the ``evaporation'' of high velocity stars from the
cluster, which reduces the number of high velocity stars. On the other hand,
binaries and a mass spectrum transfer kinetic energy
{between different} stellar
components and thereby produce high velocity stars. These high velocity stars then transfer
their excess energy to their environment in subsequent distant two-body
encounters which can lead to a transport of kinetic energy between different
regions in the GC.

\epubtkImage{.png}{
\begin{figure}[htbp]
\centerline{\includegraphics[scale=1.8,clip]{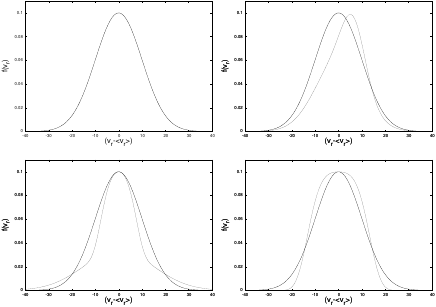}}
\caption{One-dimensional velocity distribution functions for
different cases. \emph{{Upper panels:}} From the left to the right,
I first show the Gaussian velocity distribution describing
thermodynamical equilibrium with a variance of
$\sigma=10\,\text{km}/\text{s}$. The Gaussian appears in the subsequent panels
for comparison (black). On the right, velocity distribution (grey) with a
skewness in positive $v_r$-direction indicating energy flow in $v_r$-direction.
\emph{{Lower panels}:} Two velocity distributions (grey) with an
excess and the deficit of high velocity stars, respectively, as compared to
a situation of thermodynamical equilibrium.}
\label{fig:gauss_plots}
\end{figure}
}


The so-called ``gaseous model'', is a particular case of moment models
\epubtkFootnote{
\url{http://astro-gr.org/modelling-galactic-nuclei-self-gravitating-conducting-gas-spheres/}
}.
In this
approach one assumes spherical symmetry (but not necessarily dynamical
equilibrium) and stops the infinite set of moment equations at $n=2$.  The
system is closed with the assumption that energy exchanges between stars through
2-body relaxation can be approximated by an ad hoc (local) heat conduction
prescription \cite{HachisuEtAl78,LBE80}. This reduces the study of the stellar
system to that of a self-gravitating conducting gas sphere. Multi-mass models
have been implemented \cite{LS91,Spurzem92,GS94,ST95} and the detailed forms for
the conductivities have been improved by comparing to direct $N$-body models
(described below).  The addition of a central accreting MBH and a treatment for
loss-cone effects was done in~\cite{ASFS04} for the single-mass case (a
comprehensive description of the code is in the appendix of the same work), and
in \cite{AmaroSeoaneThesis04} for a stellar mass spectrum.

The system is treated as a continuum, which is only adequate for a large number
of stars and in well populated regions of the phase space. Here I consider
spherical symmetry and single-mass stars. We handle relaxation in the
Fokker--Planck approximation, i.e., like a diffusive process determined by local
conditions. We also make use of the hydrodynamical approximation; that is to
say, only local moments of the velocity dispersion are considered, not the full
orbital structure. In particular, the effect of the two-body relaxation can be
modelled by a local heat flux equation with an appropriately tailored
conductivity.

For our description I use polar coordinates, ($r$ $\theta $, $\phi$).
The vector ${\bf v} = (v_i), i=r,\theta,\phi $ denotes the velocity in
a local Cartesian coordinate system at the spatial point
$r,\theta,\phi$. For simplicity, I will employ the notation
$u=v_{\rm r}$, $v=v_\theta$, $w=v_\phi$. The distribution function $f$, is a
function of $r$, $t$, $u$, $v^2+w^2$ only due to spherical symmetry,
and is normalised according to

\begin{equation}
\rho(r,t) = \int f(r,u,v^2+w^2,t) du\,dv\,dw.
\label{eq.rho(r,t)}
\end{equation}
Here $\rho(r,t)$ is the mass density; if $m_{\star}$ denotes the
stellar mass, we get the particle density $n=\rho/m_{\star}$.  The
Euler-Lagrange equations of motion corresponding to the Lagrange
function

\begin{equation}
{\cal L} = {1\over 2}\bigl({\dot r}^2 + r^2{\dot\theta}^2 +
         r^2 \sin^2\!\!\theta\, {\dot\phi}^2\bigr) - \Phi(r,t)
\label{eq.lagrange}
\end{equation}
are the following
\begin{eqnarray}
&{\dot u} =& - \frac{\partial \Phi}{\partial r}
+ {v^2\!+\!w^2\over r}, \nonumber \\
&{\dot v}  =& - {uv\over r} + {w^2\over r\tan\theta}, \\
\label{eq.u_v_w_dot}
&{\dot w}  =& - {uw\over r} - {vw\over r\tan\theta}. \nonumber
\end{eqnarray}
And so we get a complete local Fokker--Planck equation,

\begin{equation}
\frac{\partial f}{\partial t} +v_{\rm r}\frac{\partial f}{\partial r}+\dot{v_{\rm r}}
\frac{\partial f}{\partial v_{\rm r}}+\dot{v_{\theta}}\frac{\partial f}{\partial
v_{\theta}}+\dot{v_{\varphi}}\frac{\partial f}{\partial v_{\varphi}}=
 \left( \frac{\delta f}{\delta t} \right)_{FP}.
\label{eq.FP}
\end{equation}

In our model we do not solve the equation directly; we use a so-called
\emph{moments process}.  The moments of the velocity distribution function $f$
are defined as follows

\begin{equation}
<i,\,j,\,k> := \int^{+\infty}_{-\infty} (v_{\rm r})^{i} (v_{\theta})^{j}
(v_{\phi})^{k} \,f(r, v_{\rm r}, v_{\theta},v_{\phi},t)\,dv_{\rm r}dv_{\theta}dv_{\phi};
\label{eq.ijk}
\end{equation}

\noindent
Using the previous definition, I introduce now the following moments of the velocity distribution function,

\begin{align}
<0,0,0> & := \rho = \int f\, du\,dv\,dw,                                 \\
<1,0,0> & := {u} = \int uf\, du\,dv\,dw,                                 \\
<2,0,0> & := p_{\rm r} + \rho {u}^2 =
          \int u^2 f\, du\,dv\,dw,                                       \\
<0,2,0> & := p_\theta = \int v^2 f\, du\,dv\,dw,                         \\
<0,0,2> & := p_\phi = \int w^2 f\, du\,dv\,dw,                           \\
<3,0,0> & := F_{\rm r} + 3{u}p_{\rm r} + {u}^3 =
             \int u^3 f\, du\,dv\,dw,                                    \\
<1,2,0> & := F_\theta + {u}p_\theta =
             \int uv^2 f\, du\,dv\,dw,                                   \\
<1,0,2> & := F_\phi + {u}p_\phi =
             \int uw^2 f\, du\,dv\,dw,
\label{eq.vel_momenta}
\end{align}

\noindent
where $\rho$ is the density of stars, $u$ is the bulk velocity,
$v_{\rm r}$ and $v_{\rm t}$ are the radial and tangential flux velocities,

\begin{align}
v_{\rm r} & = {F_{\rm r} \over 3 p_{\rm r}} + u, \nonumber \\
v_{\rm t} & = {F_{\rm t} \over 2 p_{\rm t}} + u,
\label {eq.vel_energy_trans}
\end{align}

\noindent
$p_{\rm r}$ and $p_{\rm t}$ are the radial and tangential pressures, $F_{\rm r}$
is the radial and $F_{\rm t}$ the tangential kinetic energy flux
\cite{LS91}. Note that the definitions of $p_i$ and $F_i$ are such
that they are proportional to the random motion of the stars. Due to
spherical symmetry, we have $p_{\theta} = p_{\phi }=: p_{\rm t}$ and
$F_{\theta} = F_{\phi} =: F_{\rm t}/2$. By $p_{\rm r} = \rho\sigma_{\rm r}^2$ and
$p{_{\rm t}} = \rho\sigma_{\rm t}^2$ the random velocity dispersions are given,
which are closely related to observable properties in stellar
clusters.

\noindent
$F = (F_{\rm r} + F_{\rm t})/2$ is a radial flux of random kinetic energy. In
the notion of gas dynamics it is just an energy flux. Whilst for the $\theta-$
and $\phi-$ components in the set of Eqs.  (\ref{eq.vel_momenta}) are equal in
spherical symmetry, for the $r$ and $t$- quantities this is not true. In stellar
clusters the relaxation time is larger than the dynamical time and so any
possible difference between $p_{\rm r}$ and $p_{\rm t}$ may \emph{survive} many
dynamical times. We shall call such differences anisotropy. In case of {\it
weak} isotropy ($p_{\rm r}$=$p_{\rm t}$), $2F_{\rm r}$ = $3F_{\rm t}$, and thus
$v_{\rm r}$ = $v_{\rm t}$, i.e., the (radial) transport velocities of radial and
tangential random kinetic energy are equal.

The Fokker--Planck equation (\ref{eq.FP}) is multiplied by various powers of the
velocity components $u$, $v$, $w$. We get up to second order we get a set of
moment equations: A mass equation, a continuity equation, an Euler equation
(force) and radial and tangential energy equations. The system of equations is
closed by a phenomenological heat flux equation for the flux of radial and
tangential RMS ({\em root mean square}) kinetic energy, both in radial
direction. The concept is physically similar to that of \cite{LBE80}. The set of
equations is

\begin{align}
\frac{\partial{\rho}}{\partial t} & + \frac{1}{r^2}\frac{\partial}
{\partial r} (r^2u\,\rho)= 0,\nonumber \\
\frac{\partial u}{\partial t} & +u\,\frac{\partial u}{\partial r} +
{GM_{\rm r}\over r^2} +
{1\over\rho}\frac{\partial p_{\rm r}}{\partial r} + 2\,\frac{p_{\rm r} - p_{\rm t}}{\rho\, r}
= 0,\nonumber \\
\frac{\partial{p_{\rm r}}}{\partial {t}} & + \frac{1}{r^2} \frac{\partial}
{\partial r} (r^2 u \,p_{\rm r})+2 \,p_{\rm r} \frac{\partial u}{\partial r} + \frac{1}{r^2} \frac{\partial}{\partial r} (r^2 F_{\rm r})
 \frac{2F_{\rm t}}{r} = -\frac{4}{5} \frac{(2p_{\rm r}-p_{\rm t})}{\lambda_A
t_{\rm rlx}}, \nonumber \\
\frac{\partial{p_{\rm t}}}{\partial {t}} & + \frac{1}{r^2} \frac{\partial}
{\partial r} (r^2 u \,p_{\rm t})+2 \,\frac{p_{\rm t}\,u}{r}+\frac{1}{2 r^2} \frac{\partial}{\partial r}
(r^2F_{\rm t}) +
 \frac{F_{\rm t}}{r} = \frac{2}{5} \frac{(2p_{\rm r}-p_{\rm t})}{\lambda_A t_{\rm rlx}} \label{eq.set_of_eqs},
\end{align}

\noindent
where $\lambda_A$ is a numerical constant related to the time-scale of
collisional anisotropy decay. The value chosen for it has been
discussed in comparison with direct simulations performed with the
$N$--body code \cite{GS94}. The authors find that $\lambda_A=0.1$
is the physically realistic value inside the half-mass radius for all
cases of $N$, provided that close encounters and binary activity do
not carry out an important role in the system, this is, however,
inherent to systems with a large number of particles, as this is.

With the definition of the mass $M_{\rm r}$ contained in a sphere of radius
$r$

\begin{equation}
\frac{\partial M_{\rm r}}{\partial r} = 4 \pi r^2 \rho,
\label{eq.Mr}
\end{equation}
the set of equations is then equivalent to gas-dynamical
equations coupled with the equation of Poisson. To close it, we need an
independent relation, for moment equations of order $n$ contain
moments of order $n\,+\,1$. For this I use the heat conduction
closure, a phenomenological approach obtained in an analogous way to
gas dynamics. It was used for the first time by \cite{LBE80} but
restricted to isotropy. In this approximation one assumes that heat
transport is proportional to the temperature gradient,

\begin{equation}
F = -\kappa \frac{\partial T}{\partial r} = -\Lambda \frac{\partial
\sigma ^2}{\partial r}
\label{eq.temp_grad}
\end{equation}
That is the reason why such models are usually also called {\em
  conducting gas sphere models}.

It has been argued that for the classical approach
$\Lambda\propto\bar{\lambda}^2/\tau$, one has to choose the Jeans'
length $\lambda_J^2 = \sigma ^2/(4\pi G\rho)$ and the standard
Chandrasekhar local relaxation time $t_{\rm rlx}\propto \sigma
^3/\rho$ \cite{LBE80}, where $\bar{\lambda}$ is the mean free path
and $\tau$ the collisional time. In this context we obtain a
conductivity $\Lambda\propto \rho/ \sigma$. We shall consider this as
a working hypothesis. For the anisotropic model we use a mean velocity
dispersion $\sigma^2 = (\sr^2 + 2\st^2)/3$ for the temperature
gradient and assume $v_{\rm r} = v_{\rm t}$ \cite{BS86}.

\noindent
Therefore, the equations we need to close our model are

\begin{align}
v_{\rm r} & - u + \frac{\lambda}{4\pi \,G\rho\,t_{\rm rlx}} \frac{\partial
\sigma^2}{\partial r} = 0, \nonumber \\
v_{\rm r} & = v_{\rm t}.
\label{eq.closing_eqs}
\end{align}

I now introduce the interaction terms to be added to right hand of
the component equations.

\subsubsection{Equation of continuity}

I now modify the star continuity equation to include the interaction terms
\cite{LangbeinEtAl90}. The equation

\begin{equation}
\frac {\partial \rho_{\star}}{\partial {\rm t}}+ \frac{1}{{\rm r}^2} \frac{\partial}
{\partial {\rm r}}({\rm r^2} \rho_{\star} {\rm u_{\star}})=0,
\end{equation}
becomes

\begin{equation}
\frac {\partial \rho_{\star}}{\partial {\rm t}} + \frac{1}{{\rm r}^2} \frac{\partial}
{\partial {\rm r}}({\rm r^2} \rho_{\star} {\rm u_{\star}})=
 \left( \frac{\delta \rho_{\star}}
{\delta {\rm t}} \right)_{\rm coll}+\left( \frac{\delta \rho_{\star}}
{\delta {\rm t}} \right)_{\rm lc},
\end{equation}

\noindent
where the term on the right-hand side reflects the time variation of the star's density due to
stellar interactions (i.e., due to the calculation of the mean rate of gas production by
stars' {collision}s) and loss-cone (stars plunging onto the central object).

If $f(v_{\rm rel})$ is the stellar distribution of relative velocities,
then the mean rate of gas production by stellar {collision}s is

\begin{align}
\left( \frac{\delta \rho_{\star}}{\delta {\rm t}} \right)_{\rm coll} & = -\int_{|v_{\rm rel|}>\sigma_{\rm coll}} \frac{\rho_{\star}f_{\rm c}(v_{\rm
rel})}{t_{\rm coll}} f(v_{\rm rel})\,d^3v_{\rm rel},\\
\left( \frac{\delta \rho_{\star}}{\delta {\rm t}} \right)_{\rm coll} & =
-\int_{|v_{\rm rel|}>\sigma_{\rm coll}} \frac{\rho_{\star}f_{\rm c}(v_{\rm
rel})}{t_{\rm coll}} f(v_{\rm rel})d^3v_{\rm rel},
\end{align}

\noindent
where $f(v_{\rm rel})$ is a Schwarzschild - Boltzmann distribution,

\begin{equation}
f(v_{\rm rel}) =\frac{1}{2 \pi^{3/2} \sigma_{\rm r} \sigma_{\rm t}^2} \,
 {\rm exp}\,\left[ -\frac{(v_{\rm rel,r}-u_{\star})^2}{4 \sigma_{\rm r}^2}-\frac{v_{\rm rel,t}^2}
{2 \sigma_{\rm t}^2}  \right].
\end{equation}

\noindent
With regards to $f_{\rm c}$, it is the relative fraction of mass liberated per stellar
{collision} into the gaseous medium. Under certain assumptions given in the
initial work of \cite{SS66}, we can calculate it as an average
over all impact parameters resulting in $r_{\rm min}<2\,r_{\star}$ and as a function
of the relative velocity at infinity of the two colliding stars, $v_{\rm rel}$.
The authors of \cite{LangbeinEtAl90} approximate their result by

\begin{equation}
f_{\rm c}(v_{\rm rel}) = \left\{ \begin{array}{ll}
  \left(1+q_{\rm coll} \sqrt{\frac{\sigma_{\rm coll}}{v_{\rm rel}}} \right)^{-1}&
\mbox{$v_{\rm rel} > \sigma_{\rm coll}$},\\
  0 & \mbox{$v_{\rm rel} < \sigma_{\rm coll}$},
  \end{array}
   \right.
\end{equation}
with $q_{\rm coll}=100$. Hence, we have that

\begin{equation}
f_{\rm c}(v_{\rm rel}) = \left\{ \begin{array}{ll}
  0.01 & \mbox{$\sigma_{\rm coll}=v_{\rm rel}$},\\
  0 & \mbox{$\sigma_{\rm coll}>v_{\rm rel}$}.
  \end{array}
   \right.
\end{equation}

The first interaction term is

\begin{equation}
\left( \frac{\delta \rho_{\star}}{\delta {\rm t}} \right)_{\rm coll} =- \rho_{\star}
\frac{\rm f_{\rm c}}{\rm t_{\rm coll}} \left[ 1-{\rm erf} \left( \frac{\sigma_{\rm coll}}
{\sqrt {6}\sigma_{\rm r}} \right) \right]
 \left[ 1-{\rm erf} \left( \frac
{\sigma_{\rm coll}}{\sqrt {6}\sigma_{\rm t}} \right) \right]^2,
\end{equation}

\noindent
which, for simplification, we call it

\begin{equation}
\left( \frac{\delta \rho_{\star}}{\delta {\rm t}} \right)_{\rm coll}\equiv -\rho_{\star}
{\rm X_{\rm coll}}.
\end{equation}

Since the evolution of the system that we are studying can be regarded as
stationary, I introduce for each equation the ``logarithmic variables'' in
order to study the long-term evolution. On
the other hand, if the system happens to have quick changes, we should use
the ``non-logarithmic'' version of the equations. For this reason I will write
at the end of each subsection the equation in terms of the logarithmic variables.

In the case of the equation of continuity, I develop it and divide it by $\rho_{\star}$
because we are looking for the logarithm of the star density, $\partial \ln \rho_{\star}
/\partial t=(1/\rho_{\star})\partial \rho_{\star}/ \partial t$.
The result is:

\begin{equation}
\frac{\partial \ln \rho_{\star}}{\partial t}  + \frac{\partial u_{\star}}{\partial r}+
u_{\star} \frac{\partial \ln \rho_{\star}}{\partial r}+ \frac{2u_{\star}}{r}=
 \frac{1}
{\rho_{\star}}\left( \frac{\delta \rho_{\star}}{\delta {\rm t}} \right)_{\rm coll}+
\frac{1}{\rho_{\star}}\left( \frac {\delta \rho_{\star}} {\delta {\rm t}} \right)_{\rm lc}
\end{equation}

\subsubsection{Momentum balance equation}

The second equation has the following
star interaction terms:

\begin{equation}
\frac{\partial u_{\star}}{\partial t}  +u_{\star} \frac{\partial u_{\star}}{\partial r} + {GM_{\rm r}\over r^2} +
{1\over\rho_{\star}}\frac{\partial p_{\rm r}}{\partial r} +
 2\,{p_{\rm r} - p_{\rm t}\over\rho_{\star} r} =  \left( \frac{\delta
u_{\star}}{\delta t}\right)_{\rm drag}.
\label{eq.ustar_non_log}
\end{equation}

\noindent
The interaction term is due to the decelerating force that stars moving
inside the gas are subject.

\begin{equation}
\left( \frac{\delta u_{\star}}{\delta t}\right)_{\rm drag} = - X_{\rm drag}\frac{1}{\rho_{\star}}
(u_{\star}-u_{\rm g}),
\end{equation}

\noindent
where I have introduced the following definition:

\begin{equation}
X_{\rm drag} \equiv -C_{D} \frac{\pi r_{\star}^2}{m_{\star}}\rho_{\star} \rho_{\rm g} \sigma_{\rm tot},
\end{equation}

\noindent
with $\sigma_{\rm tot}^2=\sigma_{\rm r}^2+\sigma_{\rm t}^2+(u_{\star}-u_{\rm
g})^2$.
In the ``gaseous model'' I use a logarithmic expression of the equation, so that we
multiply Eq.~(\ref{eq.ustar_non_log}) by $\rho_{\star} r/p_{\rm r}$:

\begin{equation}
\frac{\rho_{\star} r}{p_{\rm r}} \left( \frac{\partial u_{\star}}{\partial t}+ u_{\star} \right)+
\frac{GM_{\rm r}}{rp_{\rm r}}\rho_{\star}+\frac{\partial \ln p_{\rm r}}
{\partial \ln r}+
 2\,\left(1-\frac{p_{\rm t}}{p_{\rm r}}\right)=-X_{\rm drag} \frac{r}{p_{\rm
r}}\left(u_{\star}-u_{\rm g}\right).
\end{equation}

\subsubsection{Radial energy equation}

Regarding the penultimate equation, the interaction terms are:

\begin{equation}
 \frac{\partial{p_{\rm r}}}{\partial {t}} + \frac{1}{r^2} \frac{\partial}{\partial r}
(r^2 u_{\star} p_{\rm r})+2 p_{\rm r} \frac{\partial u_{\star}}{\partial r}+
 \frac{4}{5} \frac{(2p_{\rm r}-p_{\rm t})} {t_{\rm rlx}} + \frac{1}{r^2} \frac{\partial}{\partial r}
(r^2 F_{\rm r})- \frac{2F_{\rm t}}{r}=
 \left( \frac{\delta p_{\rm r}}{\delta t}\right)_{\rm drag}+\left( \frac{\delta p_{\rm r}}
{\delta t}\right)_{\rm coll},
\end{equation}

\noindent
where

\begin{align}
\left( \frac{\delta p_{\rm r}}{\delta t}\right)_{\rm drag} & =-2X_{\rm drag} \sigma_{\rm r}^2, \nonumber \\
\left( \frac{\delta p_{\rm r}}{\delta t}\right)_{\rm coll} & =-X_{\rm coll}
\rho_{\star} \tilde{\sigma_{\rm r}}^2 \epsilon.
\end{align}

\noindent
In order to determine $\epsilon$, I introduce the ratio $k$ of kinetic
energy of the remaining mass after the encounter over its initial
value (before the encounter); $k$ is a measure of the inelasticity of
the {collision}: for $k=1$ we have the minimal inelasticity, just the
kinetic energy of the liberated mass fraction is dissipated, while
if $k<1$ a surplus amount of stellar kinetic energy is dissipated
during the {collision} (tidal interactions and excitation of stellar
oscillations). If we calculate the energy loss in the stellar system
per unit volume as a function of $k$, we obtain

\begin{equation}
\epsilon=f_{\rm c}^{-1}[1-k(1-f_{\rm c})].
\end{equation}

We divide by $p_{r }$ so that we get the logarithmic variable version of the equation.
We also make the following substitution:

\begin{align}
F_{\rm r} & =  3p_{\rm r}v_{\rm r}, \nonumber \\
F_{\rm t} & =  2p_{\rm t}v_{\rm t}.
\end{align}

\noindent
The resulting equation is


\begin{align}
  \frac{\partial \ln p_{\rm r}}{\partial t} & +  (u_{\star}+3v_{\rm r}) \frac{\partial \ln p_{\rm r}}{\partial r}
                                            +3 \left( \frac{\partial u_{\star}}{\partial r}+\frac{\partial v_{\rm r}}{\partial r} \right) +
\frac{2}{r}\left( u_{\star}+3v_{\rm r}-2v_{\rm t} \frac{p_{\rm t}}{p_{\rm r}}
\right) + \frac{4}{5}\frac{ 2-{p_{\rm t}}/{p_{\rm r}}}{t_{\rm rlx}} = & \nonumber \\
& \frac{1}{p_{\rm r}}
\left( \frac{\delta p_{\rm r}}{\delta t}\right)_{\rm drag}+\frac{1}{p_{\rm r}}
\left( \frac{\delta p_{\rm r}} {\delta t}\right)_{\rm coll}.
\end{align}

\subsubsection{Tangential energy equation}

To conclude the set of equations of the star component
with the interaction terms, we have the following equation:

\begin{align}
\frac{\partial{p_{\rm t}}}{\partial {t}} & + \frac{1}{r^2} \frac{\partial}{\partial r}
\left(r^2 u_{\star} p_{\rm t}\right)+ 2\, \frac{p_{\rm t}u_{\star}}{r}-
  \frac{4}{5} \frac{(2p_{\rm r}-p_{\rm t})}{t_{\rm rlx}}+
 \frac{1}{r^2} \frac{\partial}{\partial r}(r^2F_{\rm t})+\frac{2F_{\rm t}}{r}  = \nonumber \\
 &\left( \frac{\delta p_{\rm t}}{\delta t}\right)_{\rm drag}+\left( \frac{\delta p_{\rm t}}
{\delta t}\right)_{\rm coll},
\end{align}
where

\begin{align}
  \left( \frac{\delta p_{\rm t}}{\delta t}\right)_{\rm drag}&=-2X_{\rm drag} \sigma_{\rm t}^2\nonumber \\
  \left( \frac{\delta p_{\rm t}}{\delta t}\right)_{\rm coll}&=-X_{\rm coll}
\rho_{\star} \tilde{\sigma_{\rm t}}^2 \epsilon.
\end{align}

\noindent
We follow the same path like in the last case and so:

\begin{align}
\frac{\partial \ln p_{\rm t}}{\partial t} & + (u_{\star}+2v_{\rm t}) \frac{\partial \ln p_{\rm t}}{\partial r}
+\frac{\partial}{\partial r}(u_{\star}+2v_{\rm t})+
\frac{4}{r}\left(u_{\star}+2v_{\rm t}\right)- \frac{4}{5} \frac{2{p_{\rm
r}}/{p_{\rm t}}-1}{t_{\rm rlx}}=\nonumber \\
&\frac{1}{p_{\rm t}}
\left( \frac{\delta p_{\rm t}}{\delta t}\right)_{\rm drag}+
\frac{1}{p_{\rm t}} \left( \frac{\delta p_{\rm t}} {\delta t}\right)_{\rm coll}.
\end{align}

\subsection{Solving conducting, self-gravitating gas spheres}

\newcommand{\Ngrid}{\ensuremath{N_{\rm r}}}
\newcommand{\Neq}{\ensuremath{N_{\rm eq}}}
\newcommand{\qid}[1]{\ensuremath{^{(#1)}}}
\newcommand{\rid}[1]{\ensuremath{_{#1}}}
\newcommand{\xprev}{y}
\newcommand{\myH}[1]{\tilde{#1}}
\newcommand{\GrandVect}[1]{{{\cal #1}^*}}
\newcommand{\VecNeq}[1]{{\cal #1}}
\newcommand{\MatNeq}[1]{{\mathfrak #1}}
\newcommand{\itid}[1]{\ensuremath{_{[ #1]}}}

\newcommand{\DMatrix}{\ensuremath{{\blacksquare}}}
\newcommand{\CMatrix}{\ensuremath{{\square_{-}}}}
\newcommand{\EMatrix}{\ensuremath{{\square_{+}}}}
\newcommand{\CEMatrix}{\ensuremath{{\square_{\pm}}}}
\newcommand{\DerivMatrix}{\ensuremath{{\partial\GrandVect{F}}/{\partial \GrandVect{X}}}}
\newcommand{\DerivMatrixDisp}{\ensuremath{\frac{\partial\GrandVect{F}}{\partial \GrandVect{X}}}}

In this subsection, I explain briefly how the gaseous model is solved.
The algorithm used is a partially implicit
Newton-Raphson-Henyey iterative scheme, see \cite{HenyeyEtAl59,KW94}, their section 11.2.

Putting aside the bounding conditions, the set of equations to be
solved are Eq.(\ref{eq.set_of_eqs}) to Eq.(\ref{eq.closing_eqs}). In
practice, however, the equations are rewritten using the logarithm of
all positive quantities as dependant functions. As explained in
\cite{GS94}, this greatly improves energy conservation. Formally, one
may write this system as follows

\begin{align}
 \frac{\partial x\qid{i}}{\partial t} +  f\qid{i}\left(
\left\{x\qid{j},\frac{\partial x\qid{j}}{\partial r}\right\}_{j=1}^{\Neq}
\right) =0, & \mbox{\ \ for\ } i=1\dotso 4 \nonumber \\
f\qid{i}\left(
\left\{x\qid{j},\frac{\partial x\qid{j}}{\partial r}\right\}_{j=1}^{\Neq}
\right) =0, & \mbox{\ \ for\ } i=5\dotso \Neq\,,
\label{eq:difeq_set}
\end{align}

\noindent
where the $x\qid{i}$ are the local quantities defining the state of the cluster, i.e.

\begin{equation}
\underline{x} \equiv \left\{x\qid{1}, x\qid{2},\dotso x\qid{\Neq}\right\}
 \equiv \{ \log\rho,\, u,\, \log p_{\rm r},\, \log p_{\rm\, t},\,
 \log M_{\rm r},\, v_{\rm r}-u,\, v_{\rm t}-u \},
\end{equation}

\noindent
with $\Neq=7$ in the present application.

To be solved numerically, this set of coupled partial differential
equations have to be discretized according to time and radius. Let us
first consider time stepping. Let $\Delta t$ be the time step. Assume
we know the solution $\underline{x}(t-\Delta t)$ at time $t-\Delta t$
and want to compute $\underline{x}(t)$. For the sake of numerical
stability, a partially implicit scheme is used. I adopt the shorthand
notations $x\qid{i} \equiv x\qid{i}(t)$ and $\xprev\qid{i} \equiv x\qid{i}(t-\Delta t)$. Time
derivation is replaced by finite differences,
\begin{equation}
\frac{\partial x\qid{i}}{\partial t} \rightarrow \Delta t^{-1}(x\qid{i}-\xprev\qid{i}).
\label{eq:time_discr}
\end{equation}
In the terms $f\qid{i}$, I replace the $x\qid{j}$ by $\myH{x}\qid{j}$
which are intermediate values between $\xprev\qid{j}$ and $x\qid{j}$,
$\myH{x}\qid{j} =
\zeta x\qid{j} + (1-\zeta)\xprev\qid{j}$, with $\zeta= 0.55$ for
stability purposes
\cite{GS94}.

Spatial discretisation is done by defining all quantities (at a given
time) on a radial mesh, $\{r\rid{1}, r\rid{2},\dotso r\rid{N_{\rm
r}}\}$ with $r\rid{1}=0$ and $r\rid{\Ngrid}=r_{\rm max}$. A staggered
mesh is implemented. While values of $r$, $u$, $v_{\rm t}$, $v_{\rm r}$
and $M_{\rm r}$
are defined at the boundaries of the mesh cells, $\rho$, $p_{\rm t}$ and
$p_{\rm r}$ are defined at the centre of each cell. When the value of a ``boundary'' quantity
is needed at the centre of a cell, or vice-versa, one does a simple
averaging, i.e., $\hat{b}\rid{k} = (b\rid{k-1}+b\rid{k})/2$,
$\hat{c}\rid{k} = (c\rid{k}+c\rid{k+1})/2$. Let us adopt
the notation $x\qid{j}\rid{k}$ for the value of $x\qid{j}$ at position
$r\rid{k}$ (or $\hat{r}\rid{k}$) and $\Delta r\rid{k}\equiv
r\rid{k}-r\rid{k-1}$. Then, radial derivatives in the terms $f\qid{i}$
are approximated by finite differences,

\begin{equation}
\frac{\partial x\qid{j}}{\partial r}
\rightarrow
\frac{\myH{{x}}\qid{j}\rid{k}-\myH{{x}}\qid{j}\rid{k-1}}{\Delta r\rid{k}},
\label{eq:rad_discr1}
\end{equation}

\noindent
if the derivative has to be evaluated at a point where $x\rid{k}$ is
defined (centre or border of a cell), or

\noindent
\begin{equation}
\frac{\partial x\qid{j}}{\partial r}
\rightarrow
\frac{\hat{\myH{{x}}}\qid{j}\rid{k}-\hat{\myH{{x}}}\qid{j}\rid{k-1}}{\Delta r\rid{k}} =
\frac{\myH{x}\qid{j}\rid{k+1}-\myH{x}\qid{j}\rid{k-1}}{2\Delta r\rid{k}},
\label{eq:rad_discr2}
\end{equation}

\noindent
otherwise. As an exception I use upstream differencing in
$\partial{u}/\partial{r}$ for the second equation in Eq.(\ref{eq.set_of_eqs}),
i.e., the difference quotient is displaced by half
a mesh point upstream to improve stability.

By making the substitutions for ${\partial x\qid{j}}/{\partial t}$ and
${\partial x\qid{j}}/{\partial r}$ in the set of differential
equations
(\ref{eq:difeq_set}), one obtains, at each mesh point $r\rid{k}$, a set
of $\Neq$ non-linear algebraic equations linking the new values to be
determined, $\underline{x}\rid{k-1}$ and $\underline{x}\rid{k}$, to
the ``old'' ones, $\underline{\xprev}\rid{k-1}$ and
$\underline{\xprev}\rid{k}$, which are known,
\begin{equation}
\begin{split}
 {\cal F}\qid{i}\rid{k}\left(\underline{x}\rid{k-1},\underline{x}\rid{k} |
\underline{\xprev}\rid{k-1},\underline{\xprev}\rid{k}\right)=0\\
 \ i=1\dotso \Neq, \ k=1\dotso \Ngrid.
\end{split}
\end{equation}

Note that the structure of the equations is the same at all mesh
points, except for $k=1$ and $k=\Ngrid$. In particular, terms with index $k-1$ do
not appear in ${\cal F}\qid{i}\rid{1}$. Also, one has to keep in mind that
only the $\underline{x}\rid{k-1}$ and $\underline{x}\rid{k}$ are
unknown; the $\underline{\xprev}\rid{k-1}$ and
$\underline{\xprev}\rid{k}$ play the role of fixed parameters in these
equations (as do the $\Delta r\rid{k}$). If one defines a $(\Neq\times
\Ngrid)$-dimension vector $\GrandVect{X}$ whose component
$\Neq(k-1)+i$ is $x\qid{i}\rid{k}$, one can write the system of $\Neq\times
\Ngrid$ equations as $\GrandVect{F}(\GrandVect{X})=0$, i.e.
\begin{equation}
\GrandVect{F}(\GrandVect{X}) \equiv \left( \begin{array}{c}
{\cal F}\qid{1}\rid{1} \\
{\cal F}\qid{2}\rid{1} \\
\vdots \\
{\cal F}\qid{\Neq}\rid{1} \\
{\cal F}\qid{1}\rid{2} \\
\vdots \\
{\cal F}\qid{\Neq}\rid{2} \\
\vdots \\
{\cal F}\qid{1}\rid{\Ngrid} \\
\vdots \\
{\cal F}\qid{\Neq}\rid{\Ngrid} \\
\end{array}\right) =  \left( \begin{array}{c}
0 \\
\vdots \\
0 \\
\end{array}\right),
\label{eq:discreq_set}
\end{equation}

\noindent
where I have defined

\begin{equation}
\GrandVect{X} \equiv \left( \begin{array}{c}
x\qid{1}\rid{1} \\
x\qid{2}\rid{1} \\
\vdots \\
x\qid{\Neq}\rid{1} \\
x\qid{1}\rid{2} \\
\vdots \\
x\qid{\Neq}\rid{2} \\
\vdots \\
x\qid{1}\rid{\Ngrid} \\
\vdots \\
x\qid{\Neq}\rid{\Ngrid} \\
\end{array}\right).
\end{equation}

The system is solved iteratively using a Newton-Raphson scheme. If
$\GrandVect{X}\itid{m}$ is the approximation to the solution of
Eq.~(\ref{eq:discreq_set}) after iteration $m$, with
$\GrandVect{F}\itid{m} \equiv \GrandVect{F}(\GrandVect{X}\itid{m})\ne 0$,
the solution is refined through the relation

\begin{equation}
\GrandVect{X}\itid{m+1} = \GrandVect{X}\itid{m} -
\left(\DerivMatrixDisp\right)^{-1}
\GrandVect{F}\itid{m},
\label{eq:NewtRaphsonStep}
\end{equation}

where $(\DerivMatrix)^{-1}$ is
the inverse of the matrix of derivatives. The latter, of dimension
$(\Neq\, \Ngrid)\times(\Neq\, \Ngrid)$, has the following structure

\begin{equation}
\setcounter{MaxMatrixCols}{20}
\frac{\partial \GrandVect{F}}{\partial \GrandVect{X}} =
\begin{pmatrix}
\DMatrix & \EMatrix \\
\CMatrix & \DMatrix & \EMatrix \\
 & \CMatrix & \DMatrix & \EMatrix \\
 & & \ddots & \ddots\\
 & & \CMatrix\rid{k} & \DMatrix\rid{k} & \EMatrix\rid{k} \\
 & & & \ddots & \ddots\\
 & & & \CMatrix & \DMatrix & \EMatrix \\
 & & & & \CMatrix & \DMatrix \\
\end{pmatrix}.
\end{equation}
\setcounter{MaxMatrixCols}{10}
In this diagram, each square is a $\Neq\times \Neq$ sub-matrix. For $2\le k
\le \Ngrid-1$, the lines ${\Neq}k-6$ to ${\Neq}k$ of $\DerivMatrix$ are composed
of a group of 3 such ${\Neq}\times {\Neq}$ matrices, $\CMatrix\rid{k},
\DMatrix\rid{k}, \EMatrix\rid{k}$ that span columns ${\Neq}k-13$ to
${\Neq}k+{\Neq}$, while the rest is composed of zeros,

\begin{equation}
\begin{aligned}
\DMatrix\rid{k} &= \begin{pmatrix}
\frac{\partial {\cal F}\qid{1}\rid{k}}{\partial x\qid{1}\rid{k}} & \frac{\partial {\cal F}\qid{1}\rid{k}}{\partial x\qid{2}\rid{k}} & \cdots & \frac{\partial {\cal F}\qid{1}\rid{k}}{\partial x\qid{{\Neq}}\rid{k}} \\
\vdots & & & \vdots \\
\frac{\partial {\cal F}\qid{{\Neq}}\rid{k}}{\partial x\qid{1}\rid{k}} & \frac{\partial {\cal F}\qid{{\Neq}}\rid{k}}{\partial x\qid{2}\rid{k}} & \cdots & \frac{\partial {\cal F}\qid{{\Neq}}\rid{k}}{\partial x\qid{{\Neq}}\rid{k}} \\
\end{pmatrix},\\
\CEMatrix\rid{k} &= \begin{pmatrix}
\frac{\partial {\cal F}\qid{1}\rid{k}}{\partial x\qid{1}\rid{k\pm 1}} & \frac{\partial {\cal F}\qid{1}\rid{k}}{\partial x\qid{2}\rid{k\pm 1}} & \cdots & \frac{\partial {\cal F}\qid{1}\rid{k}}{\partial x\qid{{\Neq}}\rid{k\pm 1}} \\
\vdots & & & \vdots \\
\frac{\partial {\cal F}\qid{{\Neq}}\rid{k}}{\partial x\qid{1}\rid{k\pm 1}} & \frac{\partial {\cal F}\qid{{\Neq}}\rid{k}}{\partial x\qid{2}\rid{k\pm 1}} & \cdots & \frac{\partial {\cal F}\qid{{\Neq}}\rid{k}}{\partial x\qid{{\Neq}}\rid{k\pm 1}} \\
\end{pmatrix}.\\
\end{aligned}
\end{equation}
We can see this more explicitly in the two big matrix expressions of ${\partial \GrandVect{F}}/{\partial \GrandVect{X}}$,

\setcounter{MaxMatrixCols}{20}
\epubtkImage{.png}{
\begin{figure*}[htb]
\begin{equation}
\frac{\partial \GrandVect{F}}{\partial \GrandVect{X}} =
\begin{pmatrix}
\frac{\partial {\cal F}^{1}\rid{1}}{\partial x^{1}\rid{1}} & \cdots & \frac{\partial {\cal F}^{1}\rid{1}}{\partial x^{N_{\rm eq}}\rid{1}} & \frac{\partial {\cal F}^{1}\rid{1}}{\partial x^{1}\rid{2}} & \cdots & \frac{\partial {\cal F}^{1}\rid{1}}{\partial x^{N_{\rm eq}}\rid{2}} & 0 & \hdotsfor{4} & 0\\
\vdots & & & & & \vdots &\vdots & & & & &\vdots \\
\frac{\partial {\cal F}^{N_{\rm eq}}\rid{1}}{\partial x^{1}\rid{1}} & \cdots & \frac{\partial {\cal F}^{N_{\rm eq}}\rid{1}}{\partial x^{N_{\rm eq}}\rid{1}} & \frac{\partial {\cal F}^{N_{\rm eq}}\rid{1}}{\partial x^{1}\rid{2}} & \cdots & \frac{\partial {\cal F}^{N_{\rm eq}}\rid{1}}{\partial x^{N_{\rm eq}}\rid{2}} & 0 & \hdotsfor{4} & 0\\
\frac{\partial {\cal F}^{1}\rid{2}}{\partial x^{1}\rid{1}} & \cdots & \frac{\partial {\cal F}^{1}\rid{2}}{\partial x^{N_{\rm eq}}\rid{1}} & \frac{\partial {\cal F}^{1}\rid{2}}{\partial x^{1}\rid{2}} & \cdots & \frac{\partial {\cal F}^{1}\rid{2}}{\partial x^{N_{\rm eq}}\rid{2}} & \frac{\partial {\cal F}^{1}\rid{2}}{\partial x^{1}\rid{3}} & \cdots & \frac{\partial {\cal F}^{1}\rid{2}}{\partial x^{N_{\rm eq}}\rid{3}} & 0 & \hdotsfor{1} & 0\\
\vdots & & & & &  &  & & \vdots & \vdots & & \vdots \\
\frac{\partial {\cal F}^{N_{\rm eq}}\rid{2}}{\partial x^{1}\rid{1}} & \cdots & \frac{\partial {\cal F}^{N_{\rm eq}}\rid{2}}{\partial x^{N_{\rm eq}}\rid{1}} & \frac{\partial {\cal F}^{N_{\rm eq}}\rid{2}}{\partial x^{1}\rid{2}} & \cdots & \frac{\partial {\cal F}^{N_{\rm eq}}\rid{2}}{\partial x^{N_{\rm eq}}\rid{2}} & \frac{\partial {\cal F}^{N_{\rm eq}}\rid{2}}{\partial x^{1}\rid{3}} & \cdots & \frac{\partial {\cal F}^{1}\rid{2}}{\partial x^{N_{\rm eq}}\rid{3}} & 0 & \hdotsfor{1} & 0\\
\hdotsfor{12} \\
\hdotsfor{12} \\
0 & \hdotsfor{4} & 0 & \frac{\partial {\cal F}^{1}\rid{N_{\rm r}}}{\partial x^{1}\rid{N_{\rm r}-1}} & \cdots & \frac{\partial {\cal F}^{1}\rid{N_{\rm r}}}{\partial x^{N_{\rm eq}}\rid{N_{\rm r}-1}} & \frac{\partial {\cal F}^{1}\rid{N_{\rm r}}}{\partial x^{1}\rid{N_{\rm r}}} & \cdots & \frac{\partial {\cal F}^{1}\rid{N_{\rm r}}}{\partial x^{N_{\rm eq}}\rid{N_{\rm r}}} \\
\vdots & & & & & \vdots & \vdots & & & & & \vdots \\
0 & \hdotsfor{4} & 0 & \frac{\partial {\cal F}^{N_{\rm eq}}\rid{N_{\rm r}}}{\partial x^{1}\rid{N_{\rm r}-1}} & \cdots & \frac{\partial {\cal F}^{N_{\rm eq}}\rid{N_{\rm r}}}{\partial x^{N_{\rm eq}}\rid{N_{\rm r}-1}} & \frac{\partial {\cal F}^{N_{\rm eq}}\rid{N_{\rm r}}}{\partial x^{1}\rid{N_{\rm r}}} & \cdots & \frac{\partial {\cal F}^{N_{\rm eq}}\rid{N_{\rm r}}}{\partial x^{N_{\rm eq}}\rid{N_{\rm r}}} \\
\end{pmatrix},
\label{eq.matriu_totxa1}
\end{equation}
\end{figure*}
}
\setcounter{MaxMatrixCols}{10}

\newcommand{\DiagMatrix}{\begin{pspicture}(0,0)(0.65,0.8)
\psframe[fillstyle=vlines,hatchangle=0,hatchcolor=black,hatchsep=1.4pt,hatchwidth=0.5pt](0,0)(0.8,0.8)
\end{pspicture}}

\newcommand{\AfterDiagMatrix}{\begin{pspicture}(0,0)(0.65,0.8)
\psframe[fillstyle=vlines,hatchangle=45,hatchcolor=black,hatchsep=2.8pt,hatchwidth=0.3pt](0,0)(0.8,0.8)
\end{pspicture}}

\newcommand{\BeforeDiagMatrix}{\begin{pspicture}(0,0)(0.65,0.8)
\psframe[fillstyle=vlines,hatchangle=-45,hatchcolor=black,hatchsep=2.8pt,hatchwidth=0.3pt](0,0)(0.8,0.8)
\end{pspicture}}

\setcounter{MaxMatrixCols}{10}

\epubtkImage{.png}{
\begin{figure}[htbp]
\centerline{\includegraphics[width=0.57\textwidth]{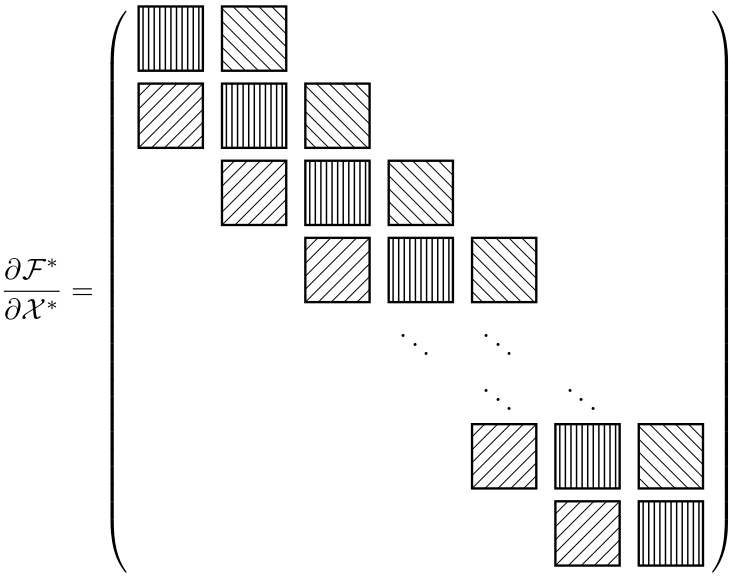}}
\end{figure}
}

The Heyney method is a way to take advantage of the special structure
of the matrix $\DerivMatrix$ to solve system (\ref{eq:NewtRaphsonStep})
efficiently, with the number of operations scaling like ${\cal O}(N_{\rm
r})$ rather than ${\cal O}(\Ngrid^3)$ as would be the case if one
uses a general-purpose matrix inversion scheme\epubtkFootnote{Memory usage
is also reduced, scaling like ${\cal O}(\Ngrid)$ rather than ${\cal
O}(\Ngrid^2)$.}. Setting
$\GrandVect{B}\equiv-\GrandVect{F}\itid{m}$ and $\GrandVect{W}\equiv
\GrandVect{X}\itid{m+1}-\GrandVect{X}\itid{m}$,
Eq.~(\ref{eq:NewtRaphsonStep}) is equivalent to
\begin{equation}
  \left(\DerivMatrixDisp\right) \GrandVect{W} = \GrandVect{B},
\end{equation}

\noindent
where $\GrandVect{W}$ is the unknown vector. I further decompose vectors
$\GrandVect{W}$ and $\GrandVect{B}$ into {\Neq}--dimensional sub-vectors,
each one representing the values at a given mesh point,

\begin{equation}
\GrandVect{W} = \begin{pmatrix}
\VecNeq{W}\rid{1} \\
\VecNeq{W}\rid{2} \\
\vdots \\
\VecNeq{W}\rid{k} \\
\vdots \\
\VecNeq{W}\rid{\Ngrid} \\
\end{pmatrix}.
\label{eq:MtimesWequalB}
\end{equation}
Then, the system (\ref{eq:MtimesWequalB}) can be written as a set of
coupled {\Neq}--dimensional vector equations,
\begin{equation}
\addtolength{\arraycolsep}{-3.5pt}
\begin{array}{lclclcl}
                                            &
&\DMatrix\rid{1}\VecNeq{W}\rid{1}& + &\EMatrix\rid{1}\VecNeq{W}\rid{2}  & = &\VecNeq{B}\rid{1},\\
\CMatrix\rid{k}\VecNeq{W}\rid{k-1}          & +
&\DMatrix\rid{k}\VecNeq{W}\rid{k}& + &\EMatrix\rid{k}\VecNeq{W}\rid{k+1}& = &\VecNeq{B}\rid{k},\\
\CMatrix\rid{\Ngrid}\VecNeq{W}\rid{\Ngrid-1}& + &\DMatrix\rid{\Ngrid}\VecNeq{W}\rid{\Ngrid}&&                           & = &\VecNeq{B}\rid{\Ngrid}.
\end{array}
\addtolength{\arraycolsep}{3.5pt}
\end{equation}
The algorithm operates in two steps. First, going from $k=1$ to
$\Ngrid$, one defines recursively a sequence of {\Neq}--vectors
$\VecNeq{V}\rid{k}$ and $({\Neq}\times {\Neq})$--matrices
$\MatNeq{M}\rid{k}$ through
\begin{equation}
\begin{aligned}
\VecNeq{V}\rid{1} &= \left(\DMatrix\rid{1}\right)^{-1} \VecNeq{B}\rid{1},\\
\MatNeq{M}\rid{1} &= \left(\DMatrix\rid{1}\right)^{-1} \EMatrix\rid{1},\\
\VecNeq{V}\rid{k} &=
\left(\DMatrix\rid{k}-\CMatrix\rid{k}\MatNeq{M}\rid{k-1}\right)^{-1}\left(\VecNeq{B}\rid{k}-\CMatrix\rid{k}\VecNeq{V}\rid{k-1}\right),\\
\MatNeq{M}\rid{k} &= \left(\DMatrix\rid{k}-\CMatrix\rid{k}\MatNeq{M}\rid{k-1}\right)^{-1} \EMatrix\rid{k},~ 2\le k \le \Ngrid.
\end{aligned}
\end{equation}
$\MatNeq{M}\rid{\Ngrid}$ is not defined. In the second step, the
values of the unknown $\VecNeq{V}\rid{k}$ are computed, climbing back
from $k=\Ngrid$ to $1$, with
\begin{equation}
\begin{aligned}
 \VecNeq{W}\rid{\Ngrid} &= \VecNeq{V}\rid{\Ngrid},\\
 \VecNeq{W}\rid{k} &= \VecNeq{V}\rid{k} - \MatNeq{M}\rid{k}\VecNeq{W}\rid{k+1},~ 1\le k \le \Ngrid-1.
\end{aligned}
\end{equation}
Note that, with this algorithm, only $({\Neq}\times {\Neq})$ matrices
have to be inverted. I use Gauss elimination for this purpose because
this venerable technique proves to be robust enough to properly deal
with the kind of badly conditioned matrices that often appear in this
application.

The initial model for the Newton-Raphson algorithm is given by the
structure of the cluster at the previous time,
$\GrandVect{X}\itid{0}(t)=\GrandVect{X}(t-\Delta t)$ One iterates
until the following convergence criteria are met. Let us set $\delta
x\qid{i}\rid{k} \equiv
\left.x\qid{i}\rid{k}\right|\itid{m+1}-\left.x\qid{i}\rid{k}\right|\itid{m}$. Then,
the condition for logarithmic quantities is
\begin{equation}
 \max_{i=1...\Neq} \frac{1}{\Ngrid} \sum_{k=1...\Ngrid}
 \left(\delta x\qid{i}\rid{k}\right)^2  < \varepsilon_1,
\end{equation}
with $\varepsilon_1=10^{-6}$. For velocities ($u$, $v_{\rm r}-u$,
$v_{\rm t}-u$), one checks
\begin{equation}
\max_{i=1...\Neq} \frac{1}{\Ngrid} \sum_{k=1...\Ngrid}
\left(\frac{ \delta x\qid{i}\rid{k} }{ x\qid{i}\rid{k}+\varepsilon_1 w\rid{k} }\right)^2  < \varepsilon_2,
\end{equation}
with $\varepsilon_2=10^{-3}$ and $w\rid{k}=r\rid{k}(4\pi
G\rho\rid{k})^{1/2}$. Generally, two iterations are sufficient to
reach convergence.

\subsection{The Local Approximation}

There are two alternative methods for further simplification of FP or moment models. One is the
orbit average, which uses the fact that that any distribution function, being a
steady state solution of the collisionless Boltzmann equation, can be
expressed as a function of the constants of motion of an individual
particle (Jeans' theorem). For the sake of simplicity, it is assumed
that all orbits in the system are regular, as it is the case for
example in a spherically symmetric potential; thus the distribution
function $f$ now only depends maximally on three independent integrals
of motion (strong Jeans' theorem). Let us transform the Fokker--Planck
equation to a new set of variables, which comprise the constants of
motion instead of the velocities $v_i$.  Since in a spherically
symmetric system the distribution only depends on energy and the
modulus of the angular momentum vector, the number of independent
coordinates in this example can be reduced from six to two, and all
terms in the transformed equation containing derivatives of other
variables than energy and angular momentum vanish (in particular those containing
derivatives of the spatial coordinates $x_i$). Integrating the
remaining parts of the Fokker--Planck equation over the spatial
coordinates is called orbit averaging, because in our present example
(a spherical system) it would be an integration over accessible
coordinate space for a given energy and angular momentum (which is a spherical shell
between $r_{\rm min}(E,\,J)$ and $r_{\rm max}(E,\,J)$, the minimum and
maximum radius for stars with energy $E$ and angular momentum $J$).
Such volume integration is, since $f$ does not depend anymore on
$x_i$ carried over to the diffusion coefficients $D$, which become
orbit-averaged diffusion coefficients.

Orbit-averaged Fokker--Planck models effectively deal with the diffusion of orbits
according to the changes of their constants of motion, taking into account the
potential and the orbital structure of the system in a self-consistent way.
However, they are not free of any problems or approximations.  They require
checks and tests, for example by comparisons with other methods, like the one
described in the following.  We treat relaxation like the addition of a big
non-correlated number of two-body encounters. Close encounters are rare and
thus I suppose that each encounter produces a very small deflection angle.
Thence, relaxation can be regarded as a diffusion process\epubtkFootnote{Anyhow, it
  has been argued that rare deflections with a large angle may play a important
role in the vicinity of a BH \cite{LT80}.}.

A typical two-body encounter in a large stellar system takes place in
a volume whose linear dimensions are small compared to other typical
radii of the system (total system dimension, or scaling radii of
changes in density or velocity dispersion). Consequently, it is
assumed that an encounter only changes the velocity, not the position
of a particle.  Thenceforth, encounters do not produce any changes
${\Delta \bf x}$, so all related terms in the Fokker--Planck equation
vanish. However, the local approximation goes even further and assumes
that the entire cumulative effect of all encounters on a test particle
can approximately be calculated as if the particle were surrounded by
a very big homogeneous system with the local distribution function
(density, velocity dispersions) everywhere. We are left with a
Fokker--Planck equation containing only derivatives with respect to the
velocity variables, but still depending on the spatial coordinates (a
local Fokker--Planck equation).

In practical astrophysical applications, the diffusion coefficients
occurring in the Fokker--Planck equation are not directly calculated,
containing the probability $\Psi$ for a velocity change
${\Delta \bf v}$ from an initial velocity ${\bf v}$. Since $D(\Delta
v_i)$, and $D(\Delta v_i\Delta v_j)$ have dimensions of velocity
(change) per time unit, and squared velocity (change) per time unit,
respectively, one calculates such velocity changes in a more direct
way, considering a test star moving in a homogeneous sea of field
stars. Let the test star have a velocity ${\bf v}$ and consider an
encounter with a field star of velocity ${\bf v}_f$. The result of the
encounter (i.e., velocity changes $\Delta v_i$ of the test star) is
completely determined by the impact parameter $p$ and the relative
velocity at infinity $v_{\rm rel} = \vert{\bf v} - {\bf v}_f\vert $;
thus by an integration of the type
\begin{equation}
\langle \Delta {\dot v}_i\rangle_p = 2\pi\int (\Delta v_i)
 \,v_{\rm rel}\, n_f\, p\, dp ,
\end{equation}
the rate of change of the test star
velocity due to encounters with $v_{\rm rel}$, in the field of stars with
particle density $n_f$,
averaged over all relevant impact parameters is computed. The
integration is normally carried out from $p_0$ (impact
parameter for $90^{\circ}$ deflection) until $R$, which is some
maximum linear dimension of the system under consideration. Such
integration generates
in subsequent equations the Coulomb logarithm $\ln \Lambda $;
as we have seen previously, it can be well approximated by
$\ln (0.11 N)$, where $N$ is the total particle number.
The diffusion
coefficient finally is
\begin{equation}
D(\Delta v_i) = \int \langle\Delta {\dot v}_i\rangle_p
                 f({\bf v}_f) d^3\,{\bf v}_f,
\end{equation}
where $f({\bf v}_f)$ is the velocity distribution of the field stars.
In an equal mass system, $f({\bf v}_f)$ should be equal to the distribution
function of the test stars occurring in the
Fokker--Planck equation for self-consistency.
In the case of a multi-mass system, however, $f({\bf v}_f)$ could be
different from the test-star distribution, if the diffusion coefficient
arising from encounters between two different species of stars is
to be calculated.
The diffusion coefficients are (for an exact procedure see \cite{BinneyTremaine08}):
\begin{align}
D(\Delta v_i)    = & 4\pi G^2 m_f \ln\Lambda \frac{\partial{}}{\partial{v_i}}h({\bf v}), \nonumber \\
D(\Delta v_iv_j) = & 4\pi G^2 m_f \ln\Lambda \frac{{\partial^2}}{ \partial v_i\partial v_j} g({\bf v}),
\end{align}
where $h({\bf v})$ and $g({\bf v})$ are given by the Rosenbluth potentials
\cite{RosenbluthEtAl57},

\begin{align}
  h({\bf v}) &= (m+m_f)
 \int {f({\bf v}_f)\over\vert{\bf v}-{\bf v}_f\vert}
 d^3\!{\bf v}_f,  \nonumber \\
 g({\bf v}) &= m_f \int f({\bf v}_f) \vert{\bf v}-{\bf v}_f\vert
 d^3\!{\bf v}_f \ .
\end{align}

With these results we can finally write down the local Fokker--Planck
equation in its standard form
for the Cartesian coordinate system of the $v_i$:

\begin{equation}
   \left(\frac{\delta f}{\delta t} \right)_{\rm enc} = -4\pi G^2 m_f \ln\Lambda
                                                   \Biggl[
     \sum_{i=1}^3 \frac{\partial{}}{\partial {v_i}}\Bigl(
        f({\bf v}) \frac{\partial{h}}{\partial {v_i}} \Bigr)
         + {1\over 2} \sum_{i,j=1}^3
       \frac{{\partial^2}}{\partial v_i\partial v_j}\Bigl(
        f({\bf v})
       {\partial^2 g \over\partial v_i\partial v_j}
       \Bigr)\Biggr]
\end{equation}

Note that in \cite{RosenbluthEtAl57} the above equation
is given in a covariant notation, which allows for a straightforward
transformation into other curvilinear coordinate systems.

Before going ahead the question is raised, why such approximation
can be reasonable, regarding the long-range gravitational force,
and the impossibility to shield gravitational forces as in the
case of Coulomb forces in a plasma by opposite charges. The key is
that logarithmic intervals in impact parameter $p$ contribute equally
to the mean square velocity change of a test particle,
provided $p\gg p_0$
(see, e.g., \cite{Spitzer87}, section 2.1).
On one side, the lower limit of impact parameters ($p_0$, the
$90^o$ deflection angle impact parameter) is small compared to
the mean interparticle distance $d$ but, on the other side,
$D$ is a typical radius connected with a change in density or
velocity dispersions (e.g., the scale height in a disc of a galaxy),
and $R$ is the maximum total dimension of the system.

Let us assume $D=100\,d$, and $R=100\,D$. In that case
the volume of the spherical shell with radius between $D$ and $R$
is $10^6$ times larger than the volume of the shell defined by
the radii $d$ and $D$. Nevertheless the contribution of both shells
to diffusion coefficients or the relaxation time is approximately
equal. This is a heuristic illustration of why the {local approximation}
is not so bad; the reason is that there are a lot
more encounters with particles in the outer, larger shell, but the
effect is exactly compensated by the larger deflection angle for
encounters happening with particles from the inner shell.
If we are in the core or in the plane of a galactic disc the density
would fall off further out, so the actual error will be smaller than
outlined in the above example. By the same reasoning one can
see, however, that the {local approximation} for a particle in a
low-density region, which suffers from relaxation by a nearby
density concentration, is prone to failure.

These simple examples should illustrate that under certain
conditions the {local approximation} is a priori not bad.
On the other hand, it is obvious from
our previous arguments that, if we are interested in relaxation effects
on particles in a low-density environment, whose orbit occasionally
passes distant, high-density regions, the {local approximation} could
be completely wrong. One might think here, for example, of stars on
radially elongated orbits in the halo of globular clusters or
of stars, globular clusters, or other objects
as massive black holes, on spherical orbits in the galactic
halo, passing the galactic disc. In these situations an orbit-averaged
treatment seems much more appropriate.

\subsection{Monte-Carlo codes}

The Monte-Carlo (MC) numerical scheme is intermediate in realism and numerical
efficiency between Fokker--Planck or moment/gas approaches, which are very fast
but based on a significantly idealised description of the stellar system, and
direct $N$-body codes, which treat (Newtonian) gravity in an essentially
assumption-free way but are extremely demanding in terms of computing time. The
MC scheme was first introduced by H\'enon to follow the relaxational evolution
of globular clusters \cite{Henon71a,Henon71b,Henon73,Henon75}. To my knowledge
there exist three independent codes in active development and use that are
based on H\'enon's ideas. The first is the one written by M.~Giersz (see
\cite{Giersz06}), which implements many of the developments first introduced by
Stodo{\l}kiewicz \cite{Stodol82,Stodol86}.  The second code is the one written
by K.~Joshi, (Cluster Monte Carlo, \textsc{MCM}), see \cite{JRPZ00,JNR01} and
greatly improved and extended by A.~G\"urkan and J.~Fregeau (see for instance
\cite{FregeauEtAl03,GFR04,FGR06,GFR06} and \cite{PattabiramanEtAl2013}
describing the latest parallel version). Finally, M.~Freitag developed an MC
code specifically aimed at the study of galactic nuclei containing a central
MBH \cite{FB01a,FB02b,FASK06}
\epubtkFootnote{\url{http://astro-gr.org/monte-carlo-simulations-for-stellar-dynamics/}}.
The description of the method given here is based on this particular
implementation.

The MC technique assumes that the cluster is spherically
symmetric\epubtkFootnote{But see the work of \cite{Vasiliev2015}, which has
developed a MC code base on
the Princeton approach. The code features a scheme to deal with asphericity, with the limitation that
it assumes isotropy of the population of background stars, so that it cannot
model a highly flattened system with significant rotation support.} and
represents it as a set of particles, each of which may be considered as a
homogeneous spherical shell of stars sharing the same orbital and stellar
properties. The number of particles may be lower than the number of stars in
the simulated cluster but the number of stars per particle has to be the same
for each particle.  Another important assumption is that the system is always
in dynamical equilibrium so that orbital time scales need not be resolved and
the natural time-step is a fraction of the relaxation (or collision) time.
Instead of being determined by integration of its orbit, the position of a
particle (i.e., the radius $R$ of the shell) is picked up at random, with a
probability density for $R$ that reflects the time spent at that radius:
$\mathrm{d}P/\mathrm{d}R\propto 1/V_\mathrm{r}(R)$ where $V_\mathrm{r}$ is the
radial velocity. The Freitag scheme adopts time steps that are a small
fraction $f$ of the local relaxation (or collision) time: $\delta t(R) \simeq
f_{\delta t} \left(\trlx^{-1} + \tcoll^{-1}\right)^{-1}$.  Consequently the
central parts of the cluster, where evolution is faster, are updated much more
frequently than the outer parts. At each step, a pair of neighbouring particles
is selected randomly with probability $P_\mathrm{selec} \propto 1/\delta t(R)$.
This ensures that a particle stays for an \emph{average} time $\delta t(R)$ at
$R$ before being updated.

Relaxation is treated as a diffusive process, using the classical Chandrasekhar
theory on which FP codes are also based. The long-term effects on orbits of the
departure of the gravitational field from a smooth stationary potential are
assumed to arise from a large number of uncorrelated, small angle, hyperbolic
2-body encounters. If a star of mass $M_1$ travels with relative velocity
$v_\mathrm{rel}$ through a homogeneous field of stars of mass $M_2$ with number
density $n$ for a time $\delta t$, then in the centre-of-mass reference frame,
its trajectory will be deflected by an angle $\theta_{\delta t}$ with average
values

\begin{align}
\nonumber \langle \theta_{\delta t} \rangle &= 0, \mbox{\ \ and}\\
\langle \theta^2_{\delta t}
\rangle &= 8\pi \ln\Lambda \, G^2 n \left(M_1+M_2\right)^2 \delta t,
\label{eq.thetaMC}
\end{align}
where $G$ is the gravitational constant and $\ln\Lambda\simeq 10-15$
is the Coulomb logarithm. In the MC code, at each step, the velocities
of the particles of the selected pair are modified by a
hyperbolic encounter with deflection angle
$\theta_\mathrm{eff}=\sqrt{\langle \theta^2_{\delta t} \rangle}$. The
particles are then put at random positions on the slightly modified
orbits. As a given particle will be selected many times, at various
positions on its orbit, the MC scheme will integrate the
effect of relaxation over the particle's orbit and over all possible
field particles. Proper averaging is ensured if the time steps are
sufficiently short for the orbit to be modified significantly only
after a large number of effective encounters. The energy is trivially
conserved to machine accuracy in such a scheme because the same
deflection angle $\theta_\mathrm{eff}$ is applied to both particles in
an interacting pair. Only the direction of the relative velocity
vector is changed by $\theta_\mathrm{eff}$.

Using a binary tree structure which allows quick determination and updating of
the potential created by the particles, the self gravity of the stellar cluster
is included accurately. This potential is not completely smooth because the
particles are infinitesimally thin spherical shells whose radii change
discontinuously. Test computations have been used to verify that the
additional, unwanted, relaxation is negligible provided the number of particles
is larger than a few tens of thousands.

Although H\'enon's method is based on the assumption than all departures from
the smooth potential can be treated as 2-body small angle scatterings, it is
flexible enough to incorporate more realism. The dynamical effect of binaries
(i.e., the dominant 3- and 4-body processes), which may be important in the
evolution of globular clusters, have been included in various MC codes through
the use of approximate analytical cross-sections \cite{Stodol86,GS00,RFJ01}.
The works of \cite{FGR06,GFR06,HypkiGiersz2013} introduced a much more
realistic treatment of binaries by on-the-fly, explicit integrations of the 3-
or 4-body interactions, a brute force approach that is necessary to deal with
the full diversity of unequal-mass binary interactions. This approach was
pioneered by \cite{GS03} in a hybrid code where binaries are followed as MC
particles while single stars are treated as a gaseous component.  In
particular, the code \textsc{MCM} of \cite{JRPZ00,JNR01} has been later further
developed to integrate larger numbers of particles than earlier attempts with
the integrator of the work of Fregeau \cite{FregeauEtAl2004}, named
\textsc{RAPID}, see \cite{RodriguezEtAl2015,FregeauRasio2007}, but it is
limited to CPUs, and the code does not account for a central MBH in its current
status.

The few 2-body encounters that lead to large angle ($> \pi/10$, say)
deflections are usually neglected. In globular clusters, these ``kicks'' have a
negligible imprint on the overall dynamics \cite{Henon75,Goodman83} but it has
been suggested that they lead to a high ejection rate from the density cusp
around a central (I)MBH \cite{LT80}. Kicks can be introduced in the MC code,
where they are treated in a way similar to collisions, with a cross section
$\pi b_\mathrm{l.a.}^2$, where
$b_\mathrm{l.a.}=f_{\mathrm{l.a.}}G(M_1+M_2)v_\mathrm{rel}^{-2}$.
$f_{\mathrm{l.a.}}$ is a numerical factor to distinguish between kicks and
``normal'' small angle scatterings (impact parameter $> b_\mathrm{l.a.}$).
However, simulations seem to indicate that such kicks have little
influence on the evolution of a stellar cusp around a MBH \cite{FASK06}.

The MC code is much faster than a direct $N$-body integration: a simulation of
a Milky-Way-type galactic nucleus represented by $10^7$ particles requires
between a few days and a few weeks of computation on a single CPU. Furthermore,
with the proper scaling with the number of stars, the number of stars
represented is independent of the number of particles. A high particle number
is obviously desirable for robust statistics, particularly when it comes to
rare events such as star-MBH interactions. In contrast, because they treat
gravitational (Newtonian) interactions on a elementary level, without relying
on any theory about their collective and/or long-term effects, the results of
direct $N$-body codes can generally be applied only to systems with a number of
stars equal to the number of particles used.

\subsection{Applications of Monte-Carlo and Fokker--Planck
simulations to the EMRI problem}
\label{sec.EMRIStatMethods}

MC and FP codes are only appropriate for studying how collisional effects
(principally relaxation) affect spherical systems in dynamical equilibrium.
These assumptions are probably valid within the radius of influence of MBHs
with masses in the LISA range. Indeed, assuming naively that the Sgr~A*
cluster at the centre of our Galaxy is typical (as far as the total stellar
mass and density is concerned) and that one can scale to other galactic nuclei
using the $M - \sigma$ relation in the form $\sigma = \sigma_{\rm MW}
(\MBH/3.6\times 10^6\,\Msun)^{1/\beta}$ with $\beta\approx4-5$
\cite{FM00,TremaineEtAl02}, one can estimate the relaxation time at the radius
of influence to be $\trlx(R_{\rm infl}) \approx 25\times 10^9\,{\rm
yr}\,(\MBH/3.6\times 10^6\,\Msun)^{(2-3/\beta)}$.

Although observations suggest a large spread amongst the values of the
relaxation time at the influence radius of MBHs with similar masses (see, e.g.,
Figure~4 of \cite{MHB06}), most galactic nuclei hosting MBHs less massive than a
few $10^6\,\Msun$ are probably relaxed and amenable to MC or FP treatment. Even
if the age of the system is significantly smaller than its relaxation time,
such approaches are valid as long as the nucleus is in dynamical equilibrium,
with a smooth, spherical distribution of matter. In such conditions,
relaxational processes are still controlling the EMRI rate, no matter how long
the relaxation time is, but one cannot assume a steady-state rate of diffusion
of stars onto orbits with small periapsis, as is often done in FP codes (see
the discussion in \cite{MM03}, in the different context of the evolution of
binary MBHs).

The H\'enon-type MC scheme of \cite{FB02b} has been used
to
determine the structure of galactic nuclei \cite{FB02b,FASK06}.
Predictions for the distribution of stars around a MBH have also been obtained
by solving some form of the Fokker--Planck equation
\cite{BW77,MCD91,HA06,HA06b,MHB06} or using the gaseous model
\cite{AmaroSeoaneThesis04,ASFS04}.  These methods have proved useful to
determine how relaxation, collisions, large-angle scatterings, MBH growth,
etc., shape the distribution of stars around the MBH, which is an obvious
prerequisite for the determination of the rate and characteristics of EMRIs. Of
particular importance is the inward segregation of stellar BHs as they lose
energy to lighter objects. This effect, combined with the fact that stellar BHs
produce GWs with higher amplitude than lower-mass stars, explains why they are
expected to dominate the EMRI detection rate \cite{SR97,HA06b}. An advantage of
the MC approach is that it can easily and realistically include a continuous
stellar mass spectrum and extra physical ingredients. However, the first point
might not be critical here as MC results suggest that, for models where all the
stars were born $\sim$~10~Gyr ago, the pattern of mass segregation can be well
approximated by a population of two components only, one representing the
stellar BHs and the other representing all other (lighter) objects
\cite{FASK06}.  Furthermore, the uncertainties are certainly dominated by our
lack of knowledge about where and when stellar formation takes place in
galactic nuclei, what the masses of the stars which form might be, and what
type of compact remnants they become.

The most recent FP results concerning mass segregation were obtained under the
assumptions of a fixed potential and an isotropic velocity dispersion, with the
effects of (standard or resonant) relaxation being averaged over angular momentum at a given
energy. The MC code includes the self-gravity of the cluster so the simulated
region can extend past the radius of influence, allowing a more natural outer
boundary condition. We note that one has to impose a steeper density drop-off
at large radii than what is observed to limit the number of particles to a
reasonable value while keeping a good resolution in the region of influence.
The MC code naturally allows anisotropy and implicitly follows relaxation in
both energy and angular momentum. Anisotropic FP codes for spherical self-gravitating systems
exist \cite{Takahashi96,Takahashi97,DCLY99} but, to our knowledge, none are
currently in use that also include a central MBH. Unique amongst all stellar
dynamical codes based on the Chandrasekhar theory of relaxation is {\sc Fopax},
a FP code which assumes axial rather than spherical symmetry, thus permitting
the study of clusters and nuclei with significant global rotation
(see~\cite{FSK06} and references therein) and which has been adapted
to include a central MBH \cite{FiestasThesis06}.

Determining the EMRI rates and characteristics is a harder challenge for
statistical stellar dynamics codes because these events are intrinsically rare
and critically sensitive to rather fine details of the stellar dynamics around
a MBH. As I explained previously, the main difficulty,
in comparison with, for example, tidal disruptions, is that EMRIs are not
``one-passage'' events but  must be gradual.  The first estimate of EMRI rates
was performed by \cite{HB95}. Assuming a static cusp profile,
they followed the evolution of the orbits of test-particles subject to GW
emission, Equations~(\ref{aevol}) and (\ref{eccevol}), and 2-body relaxation
introduced by random perturbations of the energy and angular momentum according
to pre-computed ``diffusion coefficients''. Hopman \& Alexander \cite{HA05}
have used a refined version of this ``single-particle Monte-Carlo method'', as
well as the Fokker--Planck equation, to make a more detailed analysis. It was
found that no more than $\sim10\%$ of the compact objects swallowed by the MBH
are EMRIs, while the rest are direct plunges.

Determination of EMRI rates and characteristics were also attempted with
Freitag's MC code \cite{Freitag01,Freitag03,Freitag03b}. Despite its present
limitations, this approach might serve to inspire future, more accurate,
computations and is therefore worth describing in some detail. The MC code does
not include GW emission explicitly (or any other relativistic effects). At the
end of each step in which two particles have experienced an encounter (to
simulate 2-body relaxation), each particle is tested for entry into the
``radiation-dominated'' regime, defined by Eq.~(\ref{eq.EMRIcond}) (with $C_{\rm
EMRI}=1$). A complication arises because the time step $\delta t$ used in the
MC code is a fraction $f_{\delta t}=10^{-3}-10^{-2}$ of the local relaxation
time $\trlx(R)$, which is generally much larger than the critical timescale
defined by the equality $\tGW(e,a) = C_{\rm EMRI}\, (1-e)\trlx$. In other
words, the effective diffusion angle $\theta_{\rm eff}$ is generally much
larger than the opening angle of the ``radiation cone'',
$\tilde\theta\equiv(1-\tilde{e})^{1/2}$. So that the entry of the particle into
the radiation cone (corresponding to a possible EMRI) is not missed, it is
assumed that, over $\delta t$, the energy of a given particle does not change.
Hence, each time it comes back to a given distance from the centre, its
velocity vector has the same modulus but relaxation makes its direction execute
a random walk with an individual step per orbital period of $\theta_{\rm orb} =
\theta_{\rm eff} (P_{\rm orb}/\delta t)^{1/2}$. Entry into the unstable or
radiation cone is tested at each of these sub-steps. If the particle is found
on a plunge or radiation-dominated orbit, it is immediately removed from the
simulation and its mass is added to the MBH.

Unfortunately, in addition to this approximate way of treating relaxation on
small time scales, there are a few reasons why the results of these simulations
may be only indicative. One is the way $\trlx$ is estimated, using the
coefficient in front of $\delta t$ in Eq.~(\ref{eq.thetaMC}), i.e., an estimate
based on the neighbouring particle. Even if it is correct on average, this
estimate is affected by a very high level of statistical noise and its value
can be far too long in some cases (e.g., when the relative velocity between the
particles in the pair is much larger than the local velocity dispersion). This
could lead one to conclude erroneously that a star has reached the
radiation-dominated regime and will become an EMRI. To improve on this one
could base the $\trlx$ estimate on more than one point on the orbit and on more
than one ``field-particle'' ({the number of stars within a distance of
$10^{-2}$\,pc of Sgr~A* is probably larger than 1000, so $\trlx$ is a
well-defined quantity even at such small scales}). Another limitation is that
GW emission is not included in the orbital evolution, which forces one to
assume an abrupt transition when $\tau_{\rm GW} = (1-e)\trlx$. Hopman \&
Alexander \cite{HA05} have also shown that a value of $C_{\rm EMRI}$ as small
as $10^{-3}$ might be required to be sure the EMRI will be successful.
Furthermore, the MC simulations carried out so far suffer from relatively poor
resolution, with each particle having the statistical weight of a few tens of
stars. To improve this one would need to limit the simulation to a smaller
volume (such as the influence region) or develop a parallel implementation of
the MC code to use $\sim 10^8$ particles.

\subsection{Direct-summation $N$-body codes}
\label{sec:directNbody}

We finally consider the direct $N$-body
approach~\cite{Aarseth99,Aarseth03,PortegiesZwartEtAl01}.  This is the most
expensive method because it involves integrating all gravitational forces for
all particles at every time step, without making any a priori assumptions about
the system. The $N$-body codes use the improved Hermite integration scheme as
described in~\cite{Aarseth99,Aarseth03}, which requires computation of not
only the accelerations but also of their time derivatives.  Since these approaches
integrate Newton's equations directly, all Newtonian gravitational effects are
included naturally. More relevant for this subject is that the family of the
direct $N$-body codes of Aarseth also includes versions in which both \emph{KS
regularisation} and \emph{chain regularisation} are employed, so that when
particles are tightly bound or their separation becomes too small during a
hyperbolic encounter, the system is regularised (as described first
in~\cite{KS65,Aarseth03}) to prevent dangerous small individual time steps.
This means that we can accurately follow and resolve individual orbits in the
system. Other schemes which make use of a softening in the gravitational forces
(i.e., $1/(r^2+\epsilon^2)$ instead of $1/r^2$, where $\epsilon$ is the
softening parameter) cannot be employed because $\epsilon$ can induce
unacceptable errors in the calculations.  The $N$-body codes scale as
$N_{\star}^2$, or $\Delta t \propto t_{\rm dyn}$, which means that even with
special-purpose hardware, a simulation can take of the order of weeks if not
months. This hardware is the GRAPE (short for GRAvity PipE), a family of
hardware which acts as a Newtonian force accelerator.  For instance, a GRAPE-6A
PCI card has a peak performance of 130 Gflop, roughly equivalent to 100 single
PCs \cite{GRAPE6A}.  It is possible to parallelise basic versions of the
direct $N$-body codes (without including regularisation schemes) on clusters of PCs,
each equipped with one GRAPE-6A PCI card. This leads to efficiencies greater
than 50\% and speeds in excess of 2 TFlops and thus the possibility of
simulating up to $N_{\star} = 2\cdot 10^6$ stars~\cite{HarfstEtAl06}.
Nevertheless, when we consider the situation relevant to an EMRI, in which mass
ratios are large and we need to follow thousands of orbits, the Hermite
integrator is not suitable and problems show up even in the Newtonian regime.
Aarseth et al. \cite{Aarseth06,Aarseth03} summarise different methods developed
to cope with large systems with one or more massive bodies. The problem becomes
even more difficult when including relativistic corrections to the forces when
the stellar-mass black hole approaches the central MBH, because extremely small
time-scales are involved in the integration. Progress is being made in this
direction with a developed time-transformed leapfrog method
\cite{MA02} (for a description of the leapfrog integrator see~\cite{MM06})
and the even more promising wheel-spoke regularisation, which was developed to
handle situations in which a very massive object is surrounded by strongly
bound particles, precisely the situation for EMRIs \cite{Zare74,Aarseth03}.
Additionally, one must include post-Newtonian corrections in the direct
$N$-body code because secular effects such as Kozai or resonant relaxation may be
smoothed out significantly by relativistic precession and thus have an impact
on the number of captures, see e.g. \cite{MerrittEtAl11}.

\subsubsection{Relativistic corrections: The post-Newtonian approach}

Direct $N$-body have been modified to take into account the role of relativity.
The first inclusion of relativistic corrections at 1PN, 2PN (periapsis shifts)
and 2.5PN (energy loss in the form of gravitational wave emission) in an
$N$-body code was presented in my work of \cite{KupiEtAl06}. Later, in my work
of \cite{BremAmaro-SeoaneSpurzem2014}, we presented the first implementation of
the effect of spin in mergers in a direct-summation code, NBODY6. We employ
non-spinning post-Newtonian (PN) corrections to the Newtonian accelerations up
to 3.5 PN order as well as the spin-orbit coupling up to next-to-lowest order
and the lowest order spin-spin coupling.

In the work of \cite{KupiEtAl06}, we included perturbations in the \emph{KS
regularisation} scheme, so that the forces (actually the accelerations) were
modified by

\begin{equation}
{F}  = \underbrace{{F}_0}_{\rm Newtonian}
+\underbrace{\underbrace{c^{-2}{F}_2}_{1{\rm PN}} +
\underbrace{c^{-4}{F}_4}_{2{\rm PN}}}_{\rm periapsis~shift} +
\underbrace{\underbrace{c^{-5}{F}_5}_{2.5{\rm PN}}}_{\rm GW} +
\mathcal{O}(c^{-6})
\label{eq.F_PN}
\end{equation}

\noindent
These corrections are valid for two isolated bodies and shall thus only be
applied to the Newtonian acceleration in the case of strong, relativistic
pair-interactions where the perturbation by third bodies is sufficiently small.
Because of this, one should restrict the implementation of PN terms to
regularised KS pairs. Note that formally the perturbation force in the KS
formalism does not need to be small compared to the two-body force, see the
work of \cite{Mikkola1997}. If the internal KS time step is properly adjusted,
the method will work even for relativistic terms becoming comparable to the
Newtonian force component.  For this reason I also choose the center-of-mass
frame, which is equivalent to the centre-of-mass Hamiltonian in the ADM
(Arnowit, Deser and Misner) formalism, see the work of \cite{BlanchetIyer03},
and not the formulation in the general frame.

These KS pairs are only formed when the interaction between two bodies becomes
strong enough so that the pair, as mentioned, has to be regularised. During the
KS regularisation the relative motion of the companions is still far from
relativistic. Hence, only a small, relativistic subset of all regularised KS
pairs will need post-Newtonian corrections.

In the centre-of-mass frame,

\begin{equation}
\frac{d \bm{v}}{dt}=-\frac{m}{r^2}\Big[(1+{\cal A})\,\bm{n} + {\cal B}\,\bm{v}
\Big]+ \bm{C}_{\rm 1.5,SO} + \bm{C}_{\rm 2,SS} + \bm{C}_{\rm 2.5,SO} \,,
\label{eq.Blanchet}
\end{equation}

\noindent
where the relative separation of the binary components is $x^i=y_1^i-y_2^i$,
$r=|{\bf x}|$ and $n^i={x^i}/{r}$; ${\cal A}$ and ${\cal B}$ are given by the
expressions (3.10a) and (3.10b) of \cite{BlanchetIyer03}.
The spin terms $\bm{C}_{\rm N}$, where ${\rm N}$ denotes the PN order, are
taken from the work \cite{FayeEtAl2006} and \cite{TagoshiEtAl01}.  SO stands for
spin-orbit and SS for spin-spin coupling.

We can organize the different terms in the following form, using forces per
unit mass, $f^i_{g}$, i.e. accelerations:

\begin{align}
f^{i}_{g} = &- \frac{GM}{r^{2}}n^{i} + \frac{GM}{r^{2}}\Bigl\{ \left({\cal A'}^{}_{\rm 1PN} +
               {\cal A'}^{}_{\rm 2PN}\right)n^{i} + \frac{\bm{n v}}{c}
               \left({\cal B'}^{}_{\rm 1PN} + {\cal B'}^{}_{\rm 2PN} \right)\frac{v^{i}}{c} +
               \frac{\bm{n v}}{c}{\cal A'}^{}_{\rm 2.5PN}\;n^{i} +\nonumber \\[1mm]
            & {\cal B'}^{}_{\rm 2.5PN}\frac{v^{i}}{c}\Bigr\}\,,
\label{PN_force}
\end{align}

\noindent
where here $M$ is the two-body total mass.
I list here the PN coefficients for $m_\star \ne 0$, see, e.g. the work of \cite{Blanchet2006}, in particular Eq.~(131):
\begin{align}
\label{eq.PNbegin}
{\cal A'}^{}_{\rm 1PN} = & \frac{3}{2}\nu\left(\frac{\bm{n v}}{c}\right)^{2} -
                                (1+3\nu)\frac{v^{2}}{c^{2}}  + \left(4+2\nu\right)\frac{R^{}_{g}}{r}\,,\\[1mm]
{\cal A'}^{}_{\rm 2PN} = & -\frac{15}{8}\nu\left(1+3\nu\right)\left(\frac{\bm{n v}}{c}\right)^{4}
                           +\nu\left(3-4\nu\right)\left[\frac{3}{2}\left(\frac{\bm{n v}}{c}\right)^{2}
                           -\frac{v^{2}}{c^{2}}\right]\frac{v^{2}}{c^{2}} \nonumber \\[1mm]
                       + & \frac{R^{}_{g}}{r} \Bigl\{ 2\left(1+\frac{25}{2}\nu+\nu^{2}\right)\left(\frac{\bm{n v}}{c}\right)^{2}
                           +\nu\left(\frac{13}{2}-2\nu\right) \frac{v^{2}}{c^{2}}\Bigr\}
                       - \left(9+\frac{87}{4}\nu\right)\frac{R^{2}_{g}}{r^2}\,, \\[1mm]
{\cal A'}^{}_{\rm 2.5PN} = &  \frac{24}{5}\frac{R^{}_{g}}{r}\frac{v^2}{c^{2}} + \frac{136}{15}\nu\left(\frac{R^{}_{g}}{r}\right)^{2}\,,\\[1mm]
{\cal B'}^{}_{\rm 1PN} = & 4-2\nu\,,\\[1mm]
{\cal B'}^{}_{\rm 2PN} = & -\frac{3}{2}\nu\left(3 +2\nu\right)\left(\frac{\bm{n v}}{c}\right)^{2} + \nu\left(\frac{15}{2}+2\nu\right) \frac{v^2}{c^{2}}
                         -\left(2+\frac{41 \nu}{2} +4 \nu^2\right)\frac{R^{}_{g}}{r}\,,\\[1mm]
{\cal B'}^{}_{\rm 2.5PN} = & -\frac{24}{5}\nu\left(\frac{R^{}_{g}}{r}\right)^{2} -\frac{8}{5}\nu\frac{R^{}_{g}}{r}\frac{v^{2}}{c^{2}}\,.
\label{eq.PNend}
\end{align}
where $\nu$ is the symmetric mass ratio, $\nu = m_\star M_\bullet/M^2$, with $m_\star$ the mass of the stellar-mass black hole,
and $R^{}_{g} = GM/c^{2}$.
One can verify that the coefficients in Eq.~(\ref{eq.carlos1}) to Eq.~(\ref{eq.carlos2})
agree with Eq.~(\ref{eq.PNbegin}) to (\ref{eq.PNend}) for $\nu = 0$.

Whilst the gauge choice was not a problem for the system studied in my work
of~\cite{KupiEtAl06}, since we were interested in the global dynamical
evolution, for the EMRI problem the centre-of-mass frame (located at the origin
of the coordinates) must be employed. The integration cannot be extended to
velocities higher than $\sim$~0.3~c, because at these velocities the
post-Newtonian formalism can no longer be applied accurately. This means that
we cannot reach the final coalescence of the stellar BH with the MBH, but this
is not a big issue, because this part of the evolution does not contribute
significantly to the SNR of the GW signal.  We note that it will not be
possible to include in $N$-body codes all the $\PN$ corrections that are
required for accurate modelling of the phase evolution of the EMRI during the
last few years before plunge. However, the $N$-body codes are not required in
that regime, since the system is then decoupled from the rest of the stellar
cluster.

The expressions for the accelerations are:

\begin{align}
\bm{a}_2  & = \frac{Gm_2}{r^2} \Biggl \{
                          \bm{n}\left[-v_1^2-2v_2^2+4\bm{v_1 v_2}+\frac{3}{2}(\bm{n v_2})^2+
                          5\left(\frac{Gm_1}{r}\right)+4\left(\frac{Gm_2}{r}\right)\right]+ \nonumber \\
               & (\bm{v}_1-\bm{v}_2) \left[4\bm{n v_1}-3\bm{n v_2}\right] \Biggr \}
\label{eq.a2}
\end{align}


\begingroup
\allowdisplaybreaks
\begin{align}
\bm{a}_4 = & \frac{Gm_2}{r^2} \Biggl\{ \bm{n} \Biggl[-2v_2^4+4v_2^2(\bm{v_1 v_2})-2(\bm{v_1 v_2})^2 +
                  \frac{3}{2}v_1^2(\bm{n v_2})^2+ \frac{9}{2}v_2^2(\bm{n v_2})^2-6(\bm{v_1 v_2})(\bm{n v_2})^2 -                         \nonumber \\
               & \frac{15}{8}(\bm{n v_2})^4+ \left( \frac{Gm_1}{r} \right) \Bigl(-\frac{15}{4}v_1^2+ \frac{5}{4}v_2^2-
                    \frac{5}{2}\bm{v_1 v_2} +\frac{39}{2}(\bm{n v_1})^2-39(\bm{n v_1})(\bm{n v_2})+\frac{17}{2}(\bm{n v_2})^2 \Bigr)+          \nonumber \\
               & \left(\frac{Gm_2}{r}\right)(4v_2^2-8\bm{v_1 v_2}+2(\bm{n v_1})^2 -4(\bm{n v_1})(\bm{n v_2})-6(\bm{n v_2})^2) \Biggr]+         \nonumber \\
               &     (\bm{v}_1-\bm{v}_2) \Biggl[ v^2_1(\bm{n v_2})+ 4v_2^2(\bm{n v_1})-5v_2^2(\bm{n v_2})
                     -4(\bm{v_1 v_2})(\bm{n v_1})+ 4(\bm{v_1 v_2})(\bm{n v_2})-                                                          \nonumber \\
               & 6(\bm{n v_1})(\bm{n v_2})^2 +\frac{9}{2}(\bm{n v_2})^3+\left( \frac{Gm_1}{r} \right)
                     \left(-\frac{63}{4}\bm{n v_1}+\frac{55}{4}\bm{n v_2} \right) +                                          \nonumber \\
               & \left( \frac{Gm_2}{r} \right) \left(-2\bm{n v_1}-2\bm{n v_2} \right) \Biggr] \Biggr\}+
                    \frac{G^3m_2}{r^4}\bm{n} \left[-\frac{57}{4}m_1^2-9m_2^2-\frac{69}{2}m_1m_2\right],
\label{eq.a4}
\end{align}
\endgroup

\begin{align}
\bm{a}_5 = & \frac{4}{5}\frac{G^2m_1m_2}{r^3}\Biggl\{\left(\bm{v}_1- \bm{v}_2\right)\left[-\left(\bm{v}_1-
                 \bm{v}_2\right)^2+2\left(\frac{Gm_1}{r}\right) -8\left(\frac{Gm_2}{r}\right)\right] + \nonumber \\
               & \bm{n}(\bm{n v_1}-\bm{n v_2})\left[3(\bm{v}_1-\bm{v}_2)^2- 6\left(\frac{Gm_1}{r} \right)+
                 \frac{52}{3}\left(\frac{Gm_2}{r}\right)\right]\Biggr\}.
\label{eq.a5}
\end{align}

\noindent
The basis of direct {\sc Nbody4} and {\sc Nbody6}++ codes relies on an improved
Hermit integrator scheme of the works \cite{Makino1992, Aarseth1999} for which
we need not only the accelerations but also their time derivative, given by

\begin{equation}
\bm{\dot{a}}_0= -Gm_2\left(\frac{\bm{v}_1-\bm{v}_2}{r^3}+3\frac{\bm{n}}{r^3}(\bm{n v_1}-\bm{n v_2})\right)
\end{equation}

\begingroup
\allowdisplaybreaks
\begin{align}
\bm{\dot{a}}_2 & = Gm_2 \mbox{\Huge\{} -\Biggl[(\bm{v}_1-\bm{v}_2)\frac{v_1^2}{r^3}+
             2\bm{n}\frac{v_1a_1}{r^2}+ 3\bm{n}\frac{v_1^2(\bm{n v_2}-\bm{n v_1})}{r^3}\Biggr]-        \nonumber \\
     &   2\Biggl[(\bm{v}_1-\bm{v}_2)\frac{v_2^2}{r^3}+2\bm{n}\frac{\bm{v_2 a_2}}{r^2}+
             3\bm{n}\frac{v_2^2(\bm{n v_2}-\bm{n v_1})}{r^3}\Biggr] +                                      \nonumber \\
     &   4\Big[(\bm{v}_1-\bm{v}_2)\frac{\bm{v_1 v_2}}{r^3}+\bm{n}\frac{\bm{a_1 v_2}+\bm{a_2 v_1}}{r^2}+
             3\bm{n}\frac{\bm{v_1 v_2}(\bm{n v_2}-\bm{n v_1})}{r^3}\Big]+                                        \nonumber \\
     &\frac{3}{2}\Biggl[(\bm{v}_1-\bm{v}_2)\frac{(\bm{n v_2})^2}{r^3}+
             2\bm{n}(\bm{n v_2})\frac{r(\bm{n a_2})+\bm{v_1 v_2}-v_2^2}{r^3} +
             5\bm{n}\frac{(\bm{n v_2})^2(\bm{n v_2}-\bm{n v_1})}{r^3}\Biggr]+                                    \nonumber \\
     & G\Biggl[\frac{\bm{v}_1-\bm{v}_2}{r^4}+4\bm{n}\frac{\bm{n v_2}-\bm{n v_1}}{r^4}\Biggr]
             (5m_1+4m_2)+4\frac{\bm{n v_1}}{r^2}(\bm{a}_1-\bm{a}_2)+
             3\frac{\bm{n v_2}}{r^2}(\bm{a}_2-\bm{a}_1) +                                        \nonumber \\
     & 4\frac{v_1^2-\bm{v_1 v_2}+r(\bm{n a_1})+3(\bm{n v_2}-\bm{n v_1})\bm{n v_1}}{r^3}(\bm{v}_1-\bm{v}_2)+              \nonumber \\
     & 3\frac{\bm{v_1 v_2}-v_2^2+r(\bm{n a_2})+3(\bm{n v_2}-\bm{n v_1})\bm{n v_2}}{r^3}(\bm{v}_2-\bm{v}_1) \mbox{\Huge\}}
\end{align}
\endgroup

\begingroup
\allowdisplaybreaks
\begin{align}
\bm{\dot{a}}_4 & = Gm_2 \mbox{\Huge\{} -2\left[(\bm{v}_1-\bm{v}_2)\frac{v_2^4}{r^3}+
             \bm{n}\frac{4v_2^2(a_2v_2)}{r^2}+3\bm{n}\frac{v_2^4(\bm{n v_2}-\bm{n v_1})}{r^3}\right]+  \nonumber \\
     &   4\left[(\bm{v}_1-\bm{v}_2)\frac{v_2^2\bm{v_1 v_2}}{r^3}+
             2\bm{n}\frac{(\bm{v_2 a_2})(\bm{v_1 v_2})}{r^2}+\bm{n}\frac{v_2^2(\bm{a_1 v_2}+v_1a_2)}{r^2}+
             3\bm{n}\frac{v_2^2(\bm{v_1 v_2})(\bm{n v_2}-\bm{n v_1})}{r^3}\right]-                               \nonumber \\
     &   2\left[(\bm{v}_1-\bm{v}_2)\frac{(v_2v_1)^2}{r^3}+
             2\bm{n}\frac{(\bm{v_1 v_2})(\bm{a_1 v_2}+\bm{a_2 v_1})}{r^2}+
             3\bm{n}\frac{(\bm{v_1 v_2})^2(\bm{n v_2}-\bm{n v_1})}{r^3}\right]+                                  \nonumber \\
     &    \frac{3}{2}\Biggl[(\bm{v}_1-\bm{v}_2)\frac{v_1^2(\bm{n v_2})^2}{r^3}+
             2\bm{n}\frac{v_1a_1(\bm{n v_2})^2}{r^2}+2\bm{n}\frac{v_1^2(\bm{n v_2})}{r^2}
             \left(\bm{n a_2}+\frac{\bm{v_1 v_2}-v_2^2}{r}\right)+5\bm{n}
             \frac{v_1^2(\bm{n v_2})^2}{r^3}(\bm{n v_2}-\bm{n v_1}) \Biggr] +                                        \nonumber \\
     &    \frac{9}{2} \left[ (\bm{v}_1-\bm{v}_2)\frac{v_2^2(\bm{n v_2})^2}{r^3}+
             2\bm{n}\frac{\bm{v_2 a_2}(\bm{n v_2})^2}{r^2}+2\bm{n}\frac{v_2^2(\bm{n v_2})}{r^2}
             \left(\bm{n a_2}+\frac{\bm{v_1 v_2}-v_2^2}{r}\right)+
             5\bm{n}\frac{v_2^2(\bm{n v_2})^2}{r^3}(\bm{n v_2}-\bm{n v_1})\right] -                              \nonumber \\
     & 6 \Biggl[ (\bm{v}_1-\bm{v}_2)\frac{\bm{v_1 v_2}(\bm{n v_2})^2}{r^3}+
             \bm{n} \frac{(\bm{a_1 v_2}+v_1a_2)(\bm{n v_2})^2}{r^2} +                                      \nonumber \\
     & 2\bm{n}\frac{\bm{v_1 v_2}(\bm{n v_2})}{r^2} \left(\bm{n a_2}+\frac{\bm{v_1 v_2}-v_2^2}{r}\right)+
             5\bm{n}\frac{\bm{v_1 v_2}(\bm{n v_2})^2(\bm{n v_2}-\bm{n v_1})}{r^3} \Biggr] -                            \nonumber \\
     &  \frac{15}{8}\left[(\bm{v}_1-\bm{v}_2)\frac{(\bm{n v_2})^4}{r^3}+
             4\bm{n}\frac{(\bm{n v_2})^3}{r^2}\left(\bm{n a_2}+\frac{\bm{v_1 v_2}-v_2^2}{r}\right)+
             7\bm{n}\frac{(\bm{n v_2})^4}{r^3}(\bm{n v_2}-\bm{n v_1})\right] +                                   \nonumber \\
     &  Gm_1  \Bigg\langle  
                   -\frac{15}{4}\left[(\bm{v}_1-\bm{v}_2)\frac{v_1^2}{r^4}+
                    2\bm{n}\frac{v_1a_1}{r^3}+
                    4\bm{n}\frac{v_1^2(\bm{n v_2}-\bm{n v_1})}{r^4}\right] +                               \nonumber \\
     &  \frac{5}{4}\left[(\bm{v}_1-\bm{v}_2)\frac{v_2^2}{r^4}+
                    2\bm{n}\frac{\bm{v_2 a_2}}{r^3}+ 4\bm{n}
                    \frac{v_2^2(\bm{n v_2}-\bm{n v_1})}{r^4}\right]-                                           \nonumber \\
     &  \frac{5}{2}\left[(\bm{v}_1-\bm{v}_2)\frac{\bm{v_1 v_2}}{r^4}+
                     \bm{n}\frac{\bm{a_1 v_2}+v_1a_2}{r^3}+4\bm{n}
                     \frac{\bm{v_1 v_2}(\bm{n v_2}-\bm{n v_1})}{r^4}\right] +                                        \nonumber \\
     &  \frac{39}{2}\left[(\bm{v}_1-\bm{v}_2)\frac{(\bm{n v_1})^2}{r^4}+
                     2\bm{n}\frac{\bm{n v_1}}{r^3}\left(\bm{n a_1}+\frac{v_1^2-\bm{v_1 v_2}}{r}\right)+
                     6\bm{n}\frac{(\bm{n v_1})^2(\bm{n v_2}-\bm{n v_1})}{r^4}\right] -                           \nonumber \\
     &  39 \Biggl[(\bm{v}_1-\bm{v}_2)\frac{(\bm{n v_1})(\bm{n v_2})}{r^4}+
                     \frac{\bm{n}}{r^3}\left((\bm{n v_1})(\bm{n a_2})+(\bm{n v_2})(\bm{n a_1})+
                     \frac{\bm{n v_1}(\bm{v_1 v_2}-v_2^2)}{r}+\frac{\bm{n v_2}(v_1^2-\bm{v_1 v_2})}{r}\right) +            \nonumber \\
     & 6\bm{n}\frac{(\bm{n v_1})(\bm{n v_2})}{r^4}(\bm{n v_2}-\bm{n v_1})\Biggr] +\frac{17}{2}\left[(\bm{v}_1-
                     \bm{v}_2)\frac{(\bm{n v_2})^2}{r^4}+2\bm{n}\frac{\bm{n v_2}}{r^3}
                     \left(\bm{n a_2}+\frac{\bm{v_1 v_2}-v_2^2}{r}\right)+6\bm{n}
                     \frac{(\bm{n v_2})^2(\bm{n v_2}-\bm{n v_1})}{r^4}\right]
             \Bigg \rangle  +                                                                      \nonumber \\ 
     & Gm_2 \Bigg\langle   
                 4\left[(\bm{v}_1-\bm{v}_2)\frac{v_2^2}{r^4}+2\bm{n}
                 \frac{\bm{v_2 a_2}}{r^3}+4\bm{n}\frac{v_2^2(\bm{n v_2}-\bm{n v_1})}{r^4}\right]-                \nonumber \\
     &           8\left[(\bm{v}_1-\bm{v}_2)\frac{\bm{v_1 v_2}}{r^4}+\bm{n}
                 \frac{\bm{a_1 v_2}+v_1a_2}{r^3}+4\bm{n}\frac{\bm{v_1 v_2}(\bm{n v_2}-\bm{n v_1})}{r^4}\right]+        \nonumber \\
     & 2\left[(\bm{v}_1-\bm{v}_2)\frac{(\bm{n v_1})^2}{r^4}+2\bm{n}
                 \frac{\bm{n v_1}}{r^3}\left\{\bm{n a_1}+\frac{v_1^2-\bm{v_1 v_2}}{r}\right\}+
                 6\bm{n}\frac{(\bm{n v_1})^2(\bm{n v_2}-\bm{n v_1})}{r^4}\right] -                               \nonumber \\
     & 4 \Biggl[ (\bm{v}_1-\bm{v}_2)\frac{(\bm{n v_1})(\bm{n v_2})}{r^4}+
                 \frac{\bm{n}}{r^3}\left\{(\bm{n v_1})(\bm{n a_2})+(\bm{n v_2})(\bm{n a_1})+
                 \frac{\bm{n v_1}(\bm{v_1 v_2}-v_2^2)}{r}+\frac{\bm{n v_2}(v_1^2-\bm{v_1 v_2})}{r}\right\}+                \nonumber \\
     & 6\bm{n}\frac{(\bm{n v_1})(\bm{n v_2})}{r^4}(\bm{n v_2}-\bm{n v_1}) \Biggr] -
                 6 \left[(\bm{v}_1-\bm{v}_2)\frac{(\bm{n v_2})^2}{r^4}+2\bm{n}
                 \frac{\bm{n v_2}}{r^3}\left\{\bm{n a_2}+\frac{\bm{v_1 v_2}-v_2^2}{r}\right\}+
                 6\bm{n}\frac{(\bm{n v_2})^2(\bm{n v_2}-\bm{n v_1})}{r^4}\right]
        { \Bigg \rangle } +                                                                        \nonumber \\ 
     & (\bm{a}_1-\bm{a}_2)\frac{v_1^2(\bm{n v_2})}{r^2}+(\bm{v}_1-\bm{v}_2)
                 \left\{\frac{2(v_1a_1)(\bm{n v_2})}{r^2}+\frac{v_1^2}{r^2}
                 \left(\bm{n a_2}+\frac{\bm{v_1 v_2}-v_2^2}{r}+3\frac{\bm{n v_2}}{r}(\bm{n v_2}-\bm{n v_1})\right) \right\} +    \nonumber \\
     & 4(\bm{a}_1-\bm{a}_2)\frac{v_2^2(\bm{n v_1})}{r^2}+4(\bm{v}_1-\bm{v}_2)
                 \left\{\frac{2(\bm{v_2 a_2})(\bm{n v_1})}{r^2}+\frac{v_2^2}{r^2}
                 \left(\bm{n a_1}+\frac{v_1^2-\bm{v_1 v_2}}{r}+3\frac{\bm{n v_1}}{r}(\bm{n v_2}-\bm{n v_1})\right) \right\} -    \nonumber \\
     & 5(\bm{a}_1-\bm{a}_2)\frac{v_2^2(\bm{n v_2})}{r^2}-5(\bm{v}_1-\bm{v}_2)
                 \left\{\frac{2(\bm{v_2 a_2})(\bm{n v_2})}{r^2}+\frac{v_2^2}{r^2}
                 \left(\bm{n a_2}+\frac{\bm{v_1 v_2}-v_2^2}{r}+3\frac{\bm{n v_2}}{r}(\bm{n v_2}-\bm{n v_1})\right) \right\} -    \nonumber \\
     & 4(\bm{a}_1-\bm{a}_2)\frac{(\bm{v_1 v_2})(\bm{n v_1})}{r^2}-4(\bm{v}_1-\bm{v}_2)
                 \Biggl\{ \frac{(\bm{a_1 v_2}+v_1a_2)(\bm{n v_1})}{r^2}+                                       \nonumber \\
     & \frac{\bm{v_1 v_2}}{r^2} \left( \bm{n a_1}+\frac{v_1^2-\bm{v_1 v_2}}{r}+3\frac{(\bm{v_1 v_2})(\bm{n v_1})}{r}(\bm{n v_2}-\bm{n v_1})
                 \right) \Biggr\} +                                                                \nonumber \\
     & 4(\bm{a}_1-\bm{a}_2)\frac{(\bm{v_1 v_2})(\bm{n v_2})}{r^2}+4(\bm{v}_1-\bm{v}_2)
                 \Biggl\{\frac{(\bm{a_1 v_2}+v_1a_2)(\bm{n v_2})}{r^2}+                                        \nonumber \\
     & \frac{\bm{v_1 v_2}}{r^2} \left( \bm{n a_2}+\frac{\bm{v_1 v_2}-v_2^2}{r}+
                 3\frac{(\bm{v_1 v_2})(\bm{n v_2})}{r}(\bm{n v_2}-\bm{n v_1}) \right) \Biggr\} -6(\bm{a}_1
                 -\bm{a}_2)\frac{(\bm{n v_1})(\bm{n v_2})^2}{r^2}-                                         \nonumber \\
     & 6(\bm{v}_1-\bm{v}_2)\Biggl\{\frac{(\bm{n v_2})^2}{r^2}
                 \left(\bm{n a_1}+\frac{v_1^2-\bm{v_1 v_2}}{r}\right)+\frac{2(\bm{n v_1})(\bm{n v_2})}{r^2}
                 \left(\bm{n a_2}+\frac{\bm{v_1 v_2}-v_2^2}{r}\right) +
                 5\frac{(\bm{n v_1})(\bm{n v_2})^2}{r^3}(\bm{n v_2}-\bm{n v_1})\Biggr\}\ +                                  \nonumber \\
     & \frac{9}{2}(\bm{a}_1-\bm{a}_2)\frac{(\bm{n v_2})^3}{r^2}+
                 \frac{9}{2}(\bm{v}_1-\bm{v}_2)\left\{\frac{3(\bm{n v_2})^2}{r^2}
                 \left(\bm{n a_2}+\frac{\bm{v_1 v_2}-v_2^2}{r}\right)+
                 5\frac{(\bm{n v_2})^3}{r^3}(\bm{n v_2}-\bm{n v_1})\right\}+                                          \nonumber \\
     & G \Biggl \langle 
                 \left[(\bm{a}_1-\bm{a}_2)\frac{\bm{n v_1}}{r^3}+
                 (\bm{v}_1-\bm{v}_2)\left\{\frac{\bm{n a_1}}{r^3}+
                 \frac{v_1^2-\bm{v_1 v_2}}{r^4}+4\frac{\bm{n v_1}}{r^4}(\bm{n v_2}-\bm{n v_1})\right\}\right]
                 \left[-\frac{63}{4}m_1-2m_2\right] +                                              \nonumber \\
     & \left[(\bm{a}_1-\bm{a}_2)\frac{\bm{n v_2}}{r^3}+(\bm{v}_1-\bm{v}_2)
                 \left\{\frac{\bm{n a_2}}{r^3}+\frac{\bm{v_1 v_2}-v_2^2}{r^4}+4\frac{\bm{n v_2}}{r^4}(\bm{n v_2}-\bm{n v_1})
                 \right\}\right]\times\left[\frac{55}{4}m_1-2m_2\right]
                 \Biggr \rangle 
                 \mbox{\Huge\}} +                                                                  \nonumber \\
     & G^3 m_2\left(-\frac{57}{4}m_1^2-9m_2^2-\frac{69}{2}m_1m_2\right)
                 \left(\frac{\bm{v}_1-\bm{v}_2}{r^5}+5\bm{n}\frac{\bm{n v_2}-\bm{n v_1}}{r^5}\right)
\end{align}
\endgroup

\begingroup
\allowdisplaybreaks
\begin{align}
              \bm{\dot{a}}_5 & = \frac{4}{5}G^2m_1m_2\mbox{\Huge\{}-\frac{(\bm{a}_1-\bm{a}_2)
                                          (\bm{v}_1-\bm{v}_2)^2}{r^3}-2\frac{\bm{v}_1-
                                          \bm{v}_2}{r^3}(v_1a_1+\bm{v_2 a_2}-v_2a_1-v_1a_2)+                         \nonumber \\
                                 & 6\frac{\bm{v}_1-\bm{v}_2}{r^4}(\bm{v}_1-\bm{v}_2)^2(\bm{n v_1}-\bm{n v_2})+
                                          G(2m_1-8m_2)\left[\frac{\bm{a}_1-\bm{a}_2}{r^4}+
                                          4\frac{\bm{v}_1-\bm{v}_2}{r^5}(\bm{n v_2}-\bm{n v_1})\right]+                \nonumber \\
                                 & 3 {\Biggl[}\bm{n}\frac{(\bm{v}_1-\bm{v}_2)^2}{r^3}
                                          \left(\bm{n a_1}-\bm{n a_2}+\frac{v_1^2+v_2^2-2\bm{v_1 v_2}}{r}\right)+2\bm{n}
                                          \frac{\bm{n v_1}-\bm{n v_2}}{r^3}(v_1a_1+\bm{v_2 a_2}-v_2a_1-v_1a_2)-                      \nonumber \\
                                 & 5\bm{n}\frac{(\bm{v}_1-\bm{v}_2)^2(\bm{n v_1}-\bm{n v_2})^2}{r^4} {\Biggr]} +
                                          G\left(\frac{52}{3}m_2-6m_1\right){\Biggl[}(\bm{v}_1-\bm{v}_2)
                                          \frac{\bm{n v_1}-\bm{n v_2}}{r^5}+                                                   \nonumber \\
                                 & \frac{\bm{n}}{r^4}\left(\bm{n a_1}-\bm{n a_2}+ \frac{v_1^2-2\bm{v_1 v_2}+v_2^2}{r}\right)-
                                   6\bm{n}\frac{(\bm{n v_2}-\bm{n v_1})^2}{r^5} {\Biggr]} \mbox{\Huge\}}
\label{eq.adot5}
\end{align}
\endgroup

\noindent
In the last expressions, I have used different kinds of brackets, including
angle brackets for legibility reasons.
G{\'a}bor Kupi and I derived in 2005 independently the equations for our common
work or \cite{KupiEtAl06} and compared the results with the output of a computer
algebra system by defining a function that was the substraction of our
derivations and the numerical output. The result turned out to be smaller than
$10^{-18}$. On the other hand, the terms have been now tested by a number of
independent works that adapted them in their codes. Also, we did in
\cite{BremAmaro-SeoaneSpurzem2014} a series of tests that compared the
terms with the semi-Keplerian approximation of \cite{Peters64}.

In addition to the effects on the acceleration, the spin of compact objects of
Eq.(\ref{eq.Blanchet}) undergoes precession in relativistic two-body
interactions. This is also taken into account by integrating the spin
precession equations

\begin{align}
 \frac{d \bm{S}}{d t} &= \frac{1}{c^2} \bm{U}_{\rm 1,SO} + \frac{1}{c^3} \bm{U}_{\rm 1.5,SS} + \frac{1}{c^4} \bm{U}_{\rm 2,SO}, \\
 \frac{d \bm{\Sigma}}{d t} &= \frac{1}{c^2} \bm{V}_{\rm 1,SO} + \frac{1}{c^3} \bm{V}_{\rm 1.5,SS} + \frac{1}{c^4} \bm{V}_{\rm 2,SO}, \\
 \bm{S} &= \bm{S}_1 + \bm{S}_2, \\
 \bm{\Sigma} &= m\left(\frac{\bm{S}_2}{m_2}-\frac{\bm{S}_1}{m_1}\right).
\end{align}
$\bm{S}$ and $\bm{\Sigma}$ describe the spin state of the pair.
The individual terms for $\bm{U}_N$ and $\bm{V}_N$, where $N$ denotes the PN order, can be found in the works of
\cite{FayeEtAl2006} and \cite{BuonannoEtAl03}.

In Fig.~(\ref{fig.PNintegration}) I show the evolution of a binary that formed
in an $N-$body simulation with the code developed in our work of
\cite{KupiEtAl06}. The binary corresponded to two stellar-mass black holes
with a mass ratio of 10 and I am only using the PN terms for perihelion shift and
gravitational radiation loss.

\epubtkImage{.png}{
\begin{figure}[htbp]
\centerline{\includegraphics[width=\textwidth]{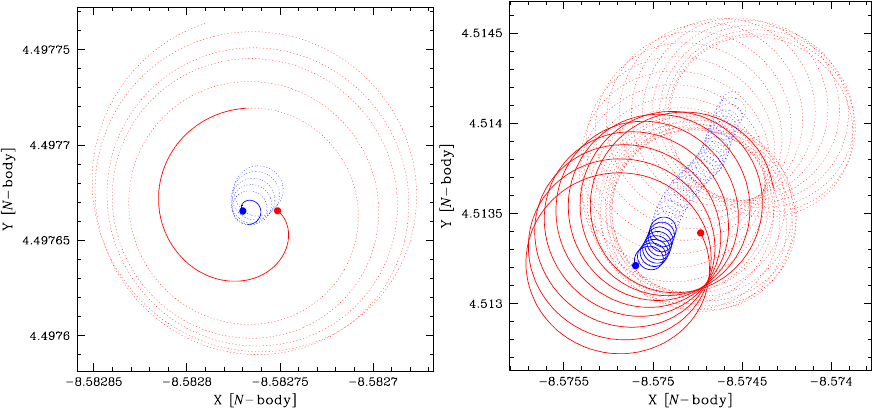}}
\caption{
Projection in the x--y plane of wo moments in the orbital evolution of a binary.
The length units are given in $N-$body units
For this example, the binary has a mass ratio of 10, and is integrated with $N-$body6
including the post-Newtonian treatment described in the work of \cite{KupiEtAl06}.
The expressions corresponding to the relativistic corrections for the accelerations and
their time derivatives are given in Eq.~(\ref{eq.a2}) to Eq.~(\ref{eq.adot5}).
For this particular simulation I did, an $N-$body unit of lenght corresponds to $0.21$ pc
(see section \ref{sec.NbodyUnits}).
}
\label{fig.PNintegration}
\end{figure}
}

\subsubsection{Relativistic corrections: A geodesic solver}

In the paper \cite{BremAmaro-SeoaneSopuerta2014} we presented, for the first
time, a geodesic approximation for the relativistic orbits in an $N-$body code.
I show in this section the geodesic equations of motion in a form that is
suitable to be included in an $N-$body code that uses a Newtonian-type
formulation of the equations of motion (initially presented in the appendix of
paper \cite{BremAmaro-SeoaneSopuerta2014}).  Also, so as to be able to compare
results with post-Newtonian approach, I show the geodesic equations using
harmonic coordinates for Schwarzschild, which are compatible with the harmonic
gauge condition of post-Newtonian theory.

Since we are integrating stars, we need to consider the geodesics for
massive particles (i.e. timelike geodesics). Given our system of spacetime coordinates
$\{x^\mu\} = \{t,x^i\}$ ($\mu\,,\nu\,,\dots = 0-3$;\; $i\,, j\,, \ldots = 1-3$),
a geodesic will be given by $\{x^\mu(\tau)\}$, where $\tau$ denotes the particle's proper time.
The components of the velocity vector are defined as
\begin{equation}
u^\mu = \frac{dx^\mu(\tau)}{d\tau} \,.
\end{equation}
This four-velocity vector satisfies:
\begin{equation}
g^{}_{\mu\nu}u^{\mu}u^{\nu} = - c^{2} \,, \label{unorm}
\end{equation}
where $g^{}_{\mu\nu}$ is the Schwarzschild metric in our coordinate system and $c$ denotes the speed of light.
Since we are interested in geodesics, the velocity vector must satisfy the
following equation of motion, see e.g. the book \cite{MisnerThorneWheeler1973}.
\begin{equation}
u^\nu\nabla^{}_\nu u^\mu = 0\,, \label{geqs}
\end{equation}
where $\nabla^{}_\mu$ denotes the canonical covariant derivative associated
with the spacetime metric $g_{\mu\nu}$.  Expanding this equation we have
\begin{equation}
\frac{du^\rho}{d\tau} + \Gamma^{\rho}_{\mu\nu}u^\mu u^\nu = 0\,, \label{egeqs}
\end{equation}
being $\Gamma^\rho_{\mu\nu}$ the Christoffel symbols associated with the
spacetime metric $g^{}_{\mu\nu}$.  They are given in terms of the metric by:
\begin{equation}
\Gamma^{\mu}_{\alpha\beta} = \frac{1}{2} g^{\mu\nu}\left(
\frac{\partial g^{}_{\alpha\nu}}{\partial x^{\beta}} +
\frac{\partial g^{}_{\beta\nu}}{\partial x^{\alpha}} -
\frac{\partial g^{}_{\alpha\beta}}{\partial x^{\rho}} \right)\,.
\label{christoffel-symbols}
\end{equation}
Using the splitting of time and space we can write the velocity vector as follows:
\begin{equation}
\vec{\bm u} = u^t \frac{\partial}{\partial t} + u^i\frac{\partial}{\partial x^i}\,,
\end{equation}
where $\{u^t,u^i\}$ are the velocity components in the $\{t,x^i\}$ coordinate
system:
\begin{equation}
u^t = \frac{\partial t(\tau)}{\partial \tau}\,,~~~~~
u^i = \frac{\partial x^i(\tau)}{\partial \tau} \,.
\end{equation}
Therefore, on the trajectory of the particle we can write
\begin{equation}
u^i = \frac{dx^i(t)}{dt}\frac{\partial t}{\partial \tau} = v^i u^t \equiv \Gamma v^{i}\,,
\end{equation}
where $v^i$ are the spatial components of the velocity
\begin{equation}
v^i = \frac{dx^i(t)}{dt}\,,
\end{equation}
and $\Gamma$ is the general relativistic version of the special relativistic gamma
factor, which is given in terms of the components of the spatial
velocity and the metric tensor as:
\begin{equation}
\Gamma^{2} = -\frac{c^{2}}{g^{}_{tt} + 2g^{}_{ti}v^i + g^{}_{ij}v^iv^j}\,.
\label{gamma-factor}
\end{equation}
which, in the weak-field limit ($g^{}_{tt}\approx -c^{2}\,$, $g^{}_{ti}\approx
0\,$, $g^{}_{ij}\approx \delta^{}_{ij}\,$), has the usual
expression:
\begin{equation}
\Gamma^{2} \approx \frac{1}{1-\frac{v^{2}}{c^{2}}}\,, \qquad (v^{2}\equiv \delta^{}_{ij}v^{i}v^{j})\,.
\end{equation}

At this point, we can now adopt a Newtonian point of view by looking at
the geodesic equations for the six quantities: $\{x^i(t),v^i(t)\}$,
that is, for the spatial coordinates and spatial velocity components.
They can be written as:
\begin{align}
\frac{dx^i}{dt} &= v^i\,, \label{geq1}  \\
\frac{dv^i}{dt} &= \,f^i_{g} \,, \label{geq2}
\end{align}
where, as we have mentioned before, the forces, $f^i_{g}$, are actually forces per unit mass, i.e.
accelerations, since they should not depend on the mass of the body (according
to the equivalence principle).  Moreover, these specific forces depend
on the spacetime metric (and its first derivatives) and on $v^i$.
We can write them as
\begin{equation}
\nonumber f^i_{g} =
v^i\,\Gamma^t_{tt}-\Gamma^i_{tt} + 2\left(v^i\,\Gamma^t_{tj}-\Gamma^i_{tj}\right)v^j
+ \left(v^i\,\Gamma^t_{jk}-\Gamma^i_{jk}\right)v^jv^k\,. \label{force}
\end{equation}
Given initial conditions $\{x^i_o,v^i_o\}$ equations (\ref{geq1},\ref{geq2})
have a unique solution $\{x^i(t),v^i(t)\}\,$.
Note that the $c^{2}$ factor dividing
the forces, when going to the right-hand side of the equation (multiplying the
Christoffel symbols) will cancel the $c^{2}$ factor in the denominator of $r^{}_{g}\,$
[see expressions in Eqs.~(\ref{gamma-t-tt})-(\ref{gamma-i-jk})].

Since up to now the development has been quite general, let us now consider the case of
a non-spinning (Schwarzschild) MBH black hole of mass $M_{\bullet}$.  The metric components,
in harmonic coordinates, can be written in the following form:
\begin{align}
g^{}_{tt} & =  - \frac{1-\frac{r^{}_{g}}{r}}{1+\frac{r^{}_{g}}{r}}\, c^{2} \,, \\[1mm]
g^{}_{ti} & = 0\,, \\[1mm]
g^{}_{ij} & =  \frac{1+\frac{r^{}_{g}}{r}}{1-\frac{r^{}_{g}}{r}}\, n^{}_{i}n^{}_{j}
+\left(1+\frac{r^{}_{g}}{r}\right)^{2}\left(\delta^{}_{ij} - n^{}_{i}n^{}_{j}\right) \,,
\end{align}
where
\begin{equation}
r = \sqrt{\delta^{}_{ij}\,x^{i}x^{j}}\,, \qquad
n^{i} = \frac{x^{i}}{r}\,,\qquad
r^{}_{g} = \frac{GM_{\bullet}}{c^{2}}\,.
\end{equation}
From here, the components of the inverse metric are:
\begin{align}
g^{tt} & =  - \frac{1+\frac{r^{}_{g}}{r}}{1-\frac{r^{}_{g}}{r}}\,\frac{1}{c^{2}} \,, \\[1mm]
g^{ti} & = 0\,, \\[1mm]
g^{ij} & =  \frac{1-\frac{r^{}_{g}}{r}}{1+\frac{r^{}_{g}}{r}}\, n^{i}n^{j}
+\frac{1}{\left(1+\frac{r^{}_{g}}{r}\right)^{2}}\left(\delta^{ij} - n^{i}n^{j}\right) \,,
\end{align}
where $x^{}_{i}=\delta^{}_{ij}\,x^{j}$ and $n^{}_{i} =\delta^{}_{ij}\,n^{j}$.

To determine the forces we need to compute the Christoffel symbols.
From their definition~(\ref{christoffel-symbols}) we find the following result
\begin{align}
\Gamma^{t}_{tt} = & 0 \,, \label{gamma-t-tt} \\[2mm]
\Gamma^{t}_{ti} = & \frac{r^{}_{g}}{r^{2}}\frac{n^{}_{i}}{1-\left(\frac{r^{}_{g}}{r}\right)^{2}} \,,
\label{gamma-t-ti} \\[2mm]
\Gamma^{t}_{ij} = & 0 \,, \label{gamma-t-ij} \\[2mm]
\Gamma^{i}_{tt} = & \frac{r^{}_{g}}{r^{2}}\frac{1-\frac{r^{}_{g}}{r}}{\left(1+\frac{r^{}_{g}}{r}\right)^{3}}\,n^{i}\,c^{2}
\,, \label{gamma-i-tt} \\[2mm]
\Gamma^{i}_{tj} = & 0\,, \label{gamma-i-tj} \\[2mm]
 \Gamma^{i}_{jk} = & \frac{r^{}_{g}}{r^{2}}\frac{1}{1+\frac{r^{}_{g}}{r}} \left[ \left(1+\frac{r^{}_{g}}{r}\right)\,n^{i}
\left(\delta^{}_{jk}-n^{}_{j}n^{}_{k}\right)
- \frac{n^{i}n^{}_{j}n^{}_{k}}{1-\frac{r^{}_{g}}{r}}
-2 n^{}_{(j}\left(\delta^{i}_{k)}- n^{i}n^{}_{k)} \right) \right]  \label{gamma-i-jk}\,.
\end{align}
And this determines completely the geodesic equations of motion in Eqs.~(\ref{geq1}) and~(\ref{geq2}).

Finally, we can make a post-Newtonian expansion of the equations of motion.  That is, an expansion
for $r^{}_{g}/r \ll 1\,$, and $v/c \ll 1\,$.  In our case, the expression for the {\em force}
simplifies to [see Eq.~(\ref{force}) and Eqs.~(\ref{gamma-t-tt})-(\ref{gamma-i-jk})]:
\begin{equation}
f^i_{g} = -\Gamma^i_{tt} + 2\,v^i\,\Gamma^t_{tj}v^j
-\Gamma^i_{jk}v^jv^k\,. \label{force2}
\end{equation}

\noindent
Expanding this we get:

\begin{align}
f^{i}_{g} = & -\frac{r^{}_{g}c^{2}}{r^{2}}\left[ 1 - 4\frac{r^{}_{g}}{r} + 9\left(\frac{r^{}_{g}}{r}\right)^{2}
- 16\left(\frac{r^{}_{g}}{r}\right)^{3} \right]\,n^{i} + 2\,\frac{r^{}_{g}c^{2}}{r^{2}} \left[ 1+\left(\frac{r^{}_{g}}{r}\right)^{2}\right] \left(\frac{n^{}_{j}v^{j}}{c}\right) \frac{v^{i}}{c}- \nonumber \\[2mm]
& \frac{r^{}_{g}c^{2}}{r^{2}} \Bigl\{ \,n^{i}\left(\delta^{}_{jk}-n^{}_{j}n^{}_{k}\right)
- \left[ 1 + \left(\frac{r^{}_{g}}{r}\right)^{2} \right] n^{i}n^{}_{j}n^{}_{k}
- 2 \left[ 1 - \frac{r^{}_{g}}{r} + \left(\frac{r^{}_{g}}{r}\right)^{2} - \left(\frac{r^{}_{g}}{r}\right)^{3} \right]\nonumber \\
& n^{}_{(j}\left(\delta^{i}_{k)}- n^{i}n^{}_{k)} \right) \Bigr\} \times \frac{v^{j}}{c}\frac{v^{k}}{c}\,,
\end{align}

\noindent
where the first two rows correspond to the first two terms in Eq.~(\ref{force2}).  We have expanded in Taylor
series the functions of $r^{}_{g}/r$ up to order $(r^{}_{g}/r)^{4}\,$.  We can now collect the terms and
we find the following expression, which is valid to order 2PN [see Eq.~(\ref{PN_force}) below]:

\begin{equation}
f^{i}_{g} = - \frac{GM_{\bullet}}{r^{2}}n^{i}
+ \frac{GM_{\bullet}}{r^{2}}\Bigl\{ \left({\cal A}^{}_{\rm 1PN} + {\cal A}^{}_{\rm 2PN}\right)n^{i}
+ \frac{\bm{n\cdot v}}{c} \left({\cal B}^{}_{\rm 1PN} + {\cal B}^{}_{\rm 2PN}\right)\frac{v^{i}}{c}\Bigr\}\,,
\label{geodesic-PN_force}
\end{equation}

\noindent
where
\begin{align}
\frac{\bm{n\cdot v}}{c} = &\frac{\bm{x}}{cr}\frac{d\bm{x}}{dt} = \frac{1}{2cr}\frac{d\bm{x}^{2}}{dt} = \frac{1}{2cr}\frac{dr}{dt}
= \frac{\dot{r}}{c} \,, \nonumber \\
v^{2} = &\bm{v\cdot v} = \delta^{}_{ij}v^{i}v^{j}\,,
\end{align}
and
\begin{align}
\label{eq.carlos1}
{\cal A}^{}_{\rm 1PN}  = & 4\frac{r^{}_{g}}{r} - \frac{v^{2}}{c^{2}}\,, \\[1mm]
{\cal A}^{}_{\rm 2PN}  = & -9\left(\frac{r^{}_{g}}{r}\right)^{2} + 2 \left(\frac{\bm{n\cdot v}}{c}\right)^{2}\frac{r^{}_{g}}{r}\,, \\[1mm]
{\cal B}^{}_{\rm 1PN}  = & 4\,, \\[1mm]
{\cal B}^{}_{\rm 2PN}  = & -2\frac{r^{}_{g}}{r}\,.
\label{eq.carlos2}
\end{align}

\subsubsection{$N-$body units and conversion}
\label{sec.NbodyUnits}
In $N-$body simulations we use the so-called $N-$body units, as defined in e.g.
the book \cite{HeggieHut03}, although they were introduced in the work of
\cite{HM86}.  In these units, the total mass of the system $M$ and $G$ are set
to unity, $M = G = 1$. Hence, to convert length $r$, mass $m$, time $t$ and
velocity $v$ from $N-$body (``Nbody'') to physical units (``phys''), we need to
multiply them by a conversion factor ``conv'':

\begin{align}
r_{\rm phys} & =r_{\rm conv}\cdot r_{_{\rm Nbody}}\nonumber \\
m_{\rm phys} & =m_{\rm conv}\cdot m_{_{\rm Nbody}}\nonumber \\
t_{\rm phys} & =t_{\rm conv}\cdot t_{_{\rm Nbody}}\nonumber \\
v_{\rm phys} & =v_{\rm conv}\cdot v_{_{\rm Nbody}}.\nonumber
\end{align}

\noindent
We usually fix $r_{\rm conv}$ by deciding the size of the system,
and $m_{\rm conv}$ is fixed to the average mass of a star in the system,
so that

\begin{align}
t_{\rm conv}=\sqrt{\frac{r_{\rm conv}^{3}}{G\,m_{\rm conv}}}\nonumber \\
v_{\rm conv}=\sqrt{\frac{G\,m_{\rm conv}}{r_{\rm conv}}}.\nonumber
\end{align}

\newpage

\section{A comment on the use of the Ancient Greek word barathron and
bathron to refer to a black hole}
\label{sec.barathron}

In the literature related to the main subject of this review I have now and
again found the expression peribathron to refer to the concept of
periapsis. Some authors use it in the case in which one has a black hole
instead of a simple star (for which we have periastron).  In this last
section I will try to convince the reader (from the Latin {\em cum vincere},
i.e. win with her or him the truth) that this is not correct. Nevertheless, if
one has an uncontrollable urge to be pedantic, I would instead recommend the
term apo- or peribarathron. In the next lines I explain why.

From the Dictionnaire grec français, A. Bailly. Ed. Hachette, édition n$^{\rm
o}$44, p. 347 we can find that

\vspace{0.3cm}

\begin{quotation}
\noindent \textgreek{βάραθρον}, -\textgreek{ου}; \textgreek{τό} {\bf I} trou profond, d’où: {\bf 1} abîme, gouffre,
ARSTT.  Probl. 26, 28; part.  à Athènes, barathre, gouffre où l’on précipitait
les condamnés (cf. à Sparte \textgreek{καιάδας}, DT.  7, 133; PLAT. Gorg. 516 e; AR. Nub.
1450 // 2 fig. ruine, perte, DÈM. 101, 1; d’où cause de ruine ou de perte, ``un
vrai gouffre'', en parl. d’une femme, THÉOPH. (ATH. 587 f) // {\bf II} ornement
de femme, AR. fr. 309, 8 // {\bf III} c. \textgreek{βράθυ} HDT. l.  c. v. ce mot (R.
*gwera-/ gwre/o\footnote{Unfortunately I was unable to reproduce the
Indo-European root with the appropriate diacritical marks, due to limitations
of typography, so I have approximated it.  }, cf. \textgreek{ἔβρων}, lat. uoro, etc.).
\end{quotation}

\vspace{0.3cm}

\noindent
In p. 340, we have the definition of \textgreek{βάθρον}

\begin{quotation}
\noindent
\textgreek{βάθρον}, -\textgreek{ου}; \textgreek{τό} surface servant de fondement, d’où: {\bf 1} base HDT. 1, 183;
part.  piédestal d’une statue, HDT. 5, 85; ESCH. Pers. 811; XÉN. Eq. 1, 1 //
{\bf 2} degré, marche, HDT. 7, 23; SOPH. O. C. 1591; part. degré d’échelle,
échelon, EUR. Ph. 1179; HDN. 4, 2, 9; fig. \textgreek{κινδύνου} \textgreek{βάθρα}, EUR. Cycl. 352, les
approches (litt. les degrés d’un danger)// {\bf 3} banc, siège SOPH. O. R. 142;
O. C. 101; PLAT. Prot. 315 c; DÉM. 313, 12, etc.; \textgreek{βάθρον} (\textgreek{Δίκμς}) SOPH. Ant.
854, le trône (de la Justice)// {\bf 4} en gén. fondement solide, d’où sol
(d’une maison, SOPH. Aj. 860; d’un pays, SOPH. Aj. 135; Phil. 1000); ou
fondations, assises (de la terre, SOPH. O. C. 1662; d’une ville, EUR. Suppl.
1198); fig. fondement, support, PD. O. 13, 6; \textgreek{ἐν} \textgreek{βάθροις} \textgreek{εἶναι} EUR. Tr. 47,
être ferme sur de solides fondements; \textgreek{ἐκ} \textgreek{βάθρον} DH. 8, 1; LUC. de fond en
comble (\textgreek{βαίνω} -\textgreek{θρον}; \textgreek{βαθρός})
\end{quotation}

\vspace{0.3cm}

The Greek-English Lexicon, Liddell-Scott, gives similar definitions. The short version
can be consulted at the website of the Perseus Project.

The barathron (as it has been transliterated e.g. by García Gual in the
Odyssey, is described as an abyss, the pit of Tartarus, even though the name
itself has never been employed exclusively -- there is no ``Barathron'' par
excellence. Here are the oldest eferences to it in the Homeric texts
(reproduced below). By the way in the Epic and in the Ionic dialect of
Herodotus, one finds not \textgreek{βάραθρον}, but the dialect variety
\textgreek{βέρεθρον}, as Bailly indicates.

\vspace{0.3cm}
\begin{quotation}
Iliad 8, 14 (Zeus very angry)

\noindent
``Hear me,'' said he, ``gods and goddesses, that I
may speak even as I am minded. Let none of you neither
goddess nor god try to cross me, but obey me every one of you
that I may bring this matter to an end. If I see anyone acting
apart and helping either Trojans or Danaans, he shall be beaten
inordinately ere he come back again to Olympus; or I will hurl
him down into dark Tartarus far into the {\bf deepest pit} under the
earth, where the gates are iron and the floor bronze, as far
beneath Hades as heaven is high above the earth, that you may
learn how much the mightiest I am among you.''
\end{quotation}

\vspace{0.3cm}
\begin{quotation}
Odyssey 12, 94 (Spanish, by García Gual)

\noindent
La otra ruta se abre entre dos promontorios. La cima
de uno de ellos se clava en el cielo anchuroso(...)
Tenebrosa caverna se abre a mitad de su altura
orientada a las sombras de ocaso y del Érebo (...)
Ni el más hábil arquero podría desde el fondo del barco
con su flecha alcanzar la oquedad de la cueva en que Escila
vive haciendo sentir desde allí sus terribles aullidos (...)
La mitad de su cuerpo se esconde en la cóncava gruta;
las cabezas, empero, por fuera del {\bf báratro} horrible
van mirando hacie el pie de la escarpa y exploran su presa (...)
\end{quotation}

Regarding this dialect variation (\textgreek{βέρεθρον}), we should note that it has been
attested in a shorter or contracted form through syncope: \textgreek{βέθρον}. This is what
may lead to the mistake of using ``bathron'' instead of the correct form
``barathron''. No dictionary mentions a potential form \textgreek{βάθρον} with the
appropriate meaning, which might have existed under the same rule as \textgreek{βέθρον},
i.e. as syncope form of the Attic variant \textgreek{βάραθρον}.

Chantraine, in his book La formation des noms en grec ancien, states that
barathron is a word with religious meaning. This should have been formed from
an Indo-European root meaning ``to devour'' plus the suffix -thron (-\textgreek{θρον}),
which expresses an instrument, tool or place, much like the suffixes -tron and
-terion. Its meaning is therefore something akin to ``place of devouring'',
``instrument or means of devouring'', ``tool of devouring''. English ``devour''
comes through French from the Latin form voro, from the same Indo-European
root.

Bathron, on the other hand, seems to be formed from the same root as the verb
\textgreek{βαίνω}, which expresses the action of walking, lean the feet on the floor or in
general to do any kind of movement on the floor. To this root one must add the
same suffix of before and, so, we have a ``place, tool or instrument to lean
on''.

In conclusion, I maintain the words ``bathron'' and ``barathron'' are are
nearly antonyms. ``Barathron'' expresses what one needs to express in the
context of a black hole; ``bathron'' can be envisaged as the opposite. What's
more, and what's worse, in modern Greek, a ``bathron'' is a cesspit.

The Barathron was in ancient Greece a cliff in Athens, below which lay an
inaccessible or invisible place, where criminals were thrown to their deaths.
This is attested in the works of Aristophanes, Plato and Herodotus (7,133), and
compiled by Bailly. It is not clear whether this Barathron was a common name or
a proper noun, namely a specific cliff. One should look in the ``Clouds'' of
Aristophanes.

According to Bailly, a cliff similar to the Barathron of Athens existed with
the name ``kaiadas''.  So we can choose: peribarathron or perikaiadas.

Nevertheless, I would suggest something different. If you do not fancy
periapsis or apoapsis, you could always make up a new word based on a modern
language, such as English.  In that case I would advocate for the very nice
term my friend Dave J. Vanecek coined in an e-mail exchange: ``periholion'',
from the English word ``hole''.

\newpage

%
%

\section{Acknowledgements}

I first of all thank my children, Ant{\'o}n and Natalia, for being the nicest
children in the world, and for their daily present of smiles during the long
preparation of this work, in spite of my spastic and even spasmodic changes of
mood as the article progressed (or retrogressed).

I am particularly thankful to one of my best and closest friends, and also one
of the brightest minds in this field, Tal Alexander. Tal has given to me not
just the present of his friendship, humour, company and countless hours of
smiles and laugh, which have made my life more beautiful, but also his wit and
sharpness in virtually any subject we discussed, in science or any other
intellectual topic. He is a reference to me as human being and scientist. Many
of the ideas I had would have never happened if it had not been for my
discussions with him.

It is a pleasure for me to show my most sincere gratitude to Francine Leeuwin
and Marc Dewi Freitag. The many discussions contributed enormously to the
writing up of this review. More importantingly, they ``also'' contributed
enormously to my \emph{human} formation during my freshman years of PhD student
in Heidelberg. I am also thankful to Bernard Schutz, Carlos F. Sopuerta, Xian
Chen, Steve Drasco, Rainer Spurzem, Rainer Sch{\"o}del, Simos Konstantinidis,
Miguel Preto, and Cole Miller both for discussions and their friendship (and
waveforms, in the case of Steve). I am indebted to Emily Davidson for her
titanic work of checking my English, but also to Emma Robinson, Jon
Gair, Melissa and Taka Tanaka.  Part of this work has been finished at the
Max-Planck Institut for Gravitational Physics (Albert Einstein Institute), Am
M{\"u}hlenberg 1, D-14476 Potsdam.

\newpage


\begin{thebibliography}{100}

\bibitem{Aarseth99}
Aarseth, S.J., ``From NBODY1 to NBODY6: The Growth of an Industry'', {\em Publ.
  Astron. Soc. Japan}, {\bf 111}, 1333--1346 (1999).
  {\small[\href{http://adsabs.harvard.edu/abs/1999PASP..111.1333A}{ADS}]}.

\bibitem{Aarseth1999}
Aarseth, S.J., ``From NBODY1 to NBODY6: The Growth of an Industry'', {\em Publ.
  Astron. Soc. Japan}, {\bf 111}, 1333--1346 (1999).
  {\small[\href{http://adsabs.harvard.edu/abs/1999PASP..111.1333A}{ADS}]}.

\bibitem{Aarseth03}
Aarseth, S.J., ``Black hole binary dynamics'', {\em Astrophys. Space Sci.},
  {\bf 285}, 367--372 (2003).
  {\small[\href{http://adsabs.harvard.edu/abs/2003Ap&SS.285..367A}{ADS}]}.

\bibitem{Aarseth06}
Aarseth, S.J., ``$N$-body codes'', in {\em Highlights of Astronomy 14},
  Modelling Dense Stellar Systems, 26th meeting of the IAU, Joint Discussion
  14, 22-23 August 2006, Prague, Czech Republic, JD14, \#1, Proceedings of the
  IAU, 14. Cambridge University Press, (2006).
  {\small[\href{http://dx.doi.org/10.1017/S1743921307011210}{DOI}]},
  {\small[\href{http://adsabs.harvard.edu/abs/2006IAUJD..14E...1A}{ADS}]}.

\bibitem{Alexander07}
Alexander, T., ``Stellar Relaxation Processes Near the Galactic Massive Black
  Hole'', {\em arXiv}, e-print, (2007).
  {\small[\href{http://adsabs.harvard.edu/abs/2007arXiv0708.0688A}{ADS}]},
  {\small[\href{http://arxiv.org/abs/0708.0688}{{arXiv:0708.0688}}]}.

\bibitem{AH03}
Alexander, T.  and Hopman, C., ``Orbital inspiral into a massive black hole in
  a galactic center'', {\em Astrophys. J. Lett.}, {\bf 590}, L29--L32 (2003).

\bibitem{AlexanderHopman09}
Alexander, T.  and Hopman, C., ``Strong mass segregation around a massive black
  hole'', {\em Astrophys. J.}, {\bf 697}, 1861--1869 (2009).
  {\small[\href{http://dx.doi.org/10.1088/0004-637X/697/2/1861}{DOI}]},
  {\small[\href{http://adsabs.harvard.edu/abs/2009ApJ...697.1861A}{ADS}]}.

\bibitem{AmaroSeoaneThesis04}
Amaro-Seoane, P., {\em Dynamics of dense gas-star systems. Black holes and
  their precursors}, Ph.D. thesis, (Heidelberg University, 2004).
  {\small[\href{http://adsabs.harvard.edu/abs/2004PhDT.........2A}{ADS}]}Online
  version: \newline\url{http://www.ub.uni-heidelberg.de/archiv/4826}.

\bibitem{Amaro-SeoaneBarrancoBernalRezzolla10}
Amaro-Seoane, P., Barranco, J., Bernal, A.  and Rezzolla, L., ``Constraining
  scalar fields with stellar kinematics and collisional dark matter'', {\em J.
  Cosmol. Astropart. Phys.}, {\bf 11}, 2 (2010).
  {\small[\href{http://dx.doi.org/10.1088/1475-7516/2010/11/002}{DOI}]},
  {\small[\href{http://adsabs.harvard.edu/abs/2010JCAP...11..002A}{ADS}]},
  {\small[\href{http://arxiv.org/abs/1009.0019}{{arXiv:1009.0019
  {\small[astro-ph.CO]}}}]}.

\bibitem{Amaro-SeoaneBremCuadraArmitage2012}
Amaro-Seoane, P., Brem, P., Cuadra, J.  and Armitage, P.J., ``The Butterfly
  Effect in the Extreme-mass Ratio Inspiral Problem'', {\bf 744}, L20 (2012).
  {\small[\href{http://dx.doi.org/10.1088/2041-8205/744/2/L20}{DOI}]},
  {\small[\href{http://adsabs.harvard.edu/abs/2012ApJ...744L..20A}{ADS}]},
  {\small[\href{http://arxiv.org/abs/1108.5174}{{arXiv:1108.5174
  {\small[astro-ph.CO]}}}]}.

\bibitem{Amaro-SeoaneChen2014}
{Amaro-Seoane}, P.  and {Chen}, X., ``{The Fragmenting Past of the Disk at the
  Galactic Center: The Culprit for the Missing Red Giants}'', {\bf 781}, L18
  (January 2014).
  {\small[\href{http://dx.doi.org/10.1088/2041-8205/781/1/L18}{DOI}]},
  {\small[\href{http://adsabs.harvard.edu/abs/2014ApJ...781L..18A}{ADS}]},
  {\small[\href{http://arxiv.org/abs/1310.0458}{{arXiv:1310.0458
  {\small[astro-ph.CO]}}}]}.

\bibitem{ASF06}
Amaro-Seoane, P.  and Freitag, M., ``Intermediate-Mass Black Holes in Colliding
  Clusters: Implications for Lower Frequency Gravitational-Wave Astronomy'',
  {\em Astrophys. J. Lett.}, {\bf 653}, L53--L56 (2006).
  {\small[\href{http://dx.doi.org/10.1086/510405}{DOI}]},
  {\small[\href{http://adsabs.harvard.edu/abs/2006ApJ...653L..53A}{ADS}]},
  {\small[\href{http://arxiv.org/abs/astro-ph/0610478}{{astro-ph/0610478}}]}.

\bibitem{ASFS04}
Amaro-Seoane, P., Freitag, M.  and Spurzem, R., ``Accretion of stars on to a
  massive black hole: a realistic diffusion model and numerical studies'', {\em
  Mon. Not. R. Astron. Soc.}, {\bf 352}, 655--672 (2004).
  {\small[\href{http://adsabs.harvard.edu/abs/2004MNRAS.352..655A}{ADS}]}.

\bibitem{ASEtAl04}
Amaro-Seoane, P., Freitag, M.  and Spurzem, R., ``Accretion of stars on to a
  massive black hole: A realistic diffusion model and numerical studies'', {\em
  Mon. Not. R. Astron. Soc.} (2004).
  {\small[\href{http://adsabs.harvard.edu/abs/2004astro.ph..1163A}{ADS}]},
  {\small[\href{http://arxiv.org/abs/astro-ph/0401163}{{astro-ph/0401163}}]}.

\bibitem{Amaro-SeoaneEtAl07}
Amaro-Seoane, P., Gair, J.R., Freitag, M., Miller, M.C., Mandel, I., Cutler,
  C.J.  and Babak, S., ``Intermediate and Extreme Mass-Ratio Inspirals --
  Astrophysics, Science Applications and Detection using LISA'', {\em Class.
  Quantum Grav.}, {\bf 24}, R113--R170 (2007).
  {\small[\href{http://dx.doi.org/10.1088/0264-9381/24/17/R01}{DOI}]},
  {\small[\href{http://adsabs.harvard.edu/abs/2007CQGra..24..113A}{ADS}]},
  {\small[\href{http://arxiv.org/abs/arXiv:astro-ph/0703495}{{arXiv:astro-ph/0703495}}]}.

\bibitem{Amaro-SeoanePreto11}
Amaro-Seoane, P.  and Preto, M., ``The impact of realistic models of mass
  segregation on the event rate of extreme-mass ratio inspirals and cusp
  re-growth'', {\em Class. Quantum Grav.}, {\bf 28}, 094017 (2011).
  {\small[\href{http://dx.doi.org/10.1088/0264-9381/28/9/094017}{DOI}]},
  {\small[\href{http://adsabs.harvard.edu/abs/2011CQGra..28i4017A}{ADS}]},
  {\small[\href{http://arxiv.org/abs/1010.5781}{{arXiv:1010.5781
  {\small[astro-ph.CO]}}}]}.

\bibitem{Amaro-SeoaneSopuertaFreitag2012}
Amaro-Seoane, P., Sopuerta, C.F.  and Freitag, M., ``The role of the
  supermassive black hole spin in the estimation of the EMRI event rate'', {\em
  arXiv}, e-print, (2012).
  {\small[\href{http://arxiv.org/abs/1205.4713}{{arXiv:1205.4713}}]}.

\bibitem{Amaro-SeoaneSopuertaFreitag2013}
Amaro-Seoane, P., Sopuerta, C.F.  and Freitag, M.D., ``The role of the
  supermassive black hole spin in the estimation of the EMRI event rate'', {\em
  Mon. Not. R. Astron. Soc.}, {\bf 429}, 3155--3165 (2013).
  {\small[\href{http://dx.doi.org/10.1093/mnras/sts572}{DOI}]},
  {\small[\href{http://adsabs.harvard.edu/abs/2013MNRAS.429.3155A}{ADS}]},
  {\small[\href{http://arxiv.org/abs/1205.4713}{{arXiv:1205.4713
  {\small[astro-ph.CO]}}}]}.

\bibitem{AS01}
Amaro-Seoane, P.  and Spurzem, R., ``The loss-cone problem in dense nuclei'',
  {\em Mon. Not. R. Astron. Soc.}, {\bf 327}, 995--1003 (2001).
  {\small[\href{http://adsabs.harvard.edu/abs/2001MNRAS.327..995A}{ADS}]}.

\bibitem{Amaro-SeoaneEtAl2012}
Amaro-Seoane, P. {et~al.}, ``eLISA: Astrophysics and cosmology in the
  millihertz regime'', {\em arXiv}, e-print, (2012).
  {\small[\href{http://adsabs.harvard.edu/abs/2012arXiv1201.3621A}{ADS}]},
  {\small[\href{http://arxiv.org/abs/1201.3621}{{arXiv:1201.3621
  {\small[astro-ph.CO]}}}]}.

\bibitem{Amaro-SeoaneEtAl2012b}
Amaro-Seoane, P. {et~al.}, ``Low-frequency gravitational-wave science with
  eLISA/NGO'', {\em arXiv}, e-print, (2012).
  {\small[\href{http://adsabs.harvard.edu/abs/2012arXiv1202.0839A}{ADS}]},
  {\small[\href{http://arxiv.org/abs/1202.0839}{{arXiv:1202.0839
  {\small[gr-qc]}}}]}.

\bibitem{Amaro-SeoaneEtAl2017}
{Amaro-Seoane}, P. {et~al.}, ``{Laser Interferometer Space Antenna}'', {\em
  ArXiv e-prints} (February 2017).
  {\small[\href{http://adsabs.harvard.edu/abs/2017arXiv170200786A}{ADS}]},
  {\small[\href{http://arxiv.org/abs/1702.00786}{{arXiv:1702.00786
  {\small[astro-ph.IM]}}}]}.

\bibitem{ArmanoEtAl2016}
Armano, M. {et~al.}, ``Sub-Femto-$g$ Free Fall for Space-Based Gravitational
  Wave Observatories: LISA Pathfinder Results'', {\em Phys. Rev. Lett.}, {\bf
  116}, 231101 (Jun 2016).
  {\small[\href{http://dx.doi.org/10.1103/PhysRevLett.116.231101}{DOI}]}URL:
  \newline\url{http://link.aps.org/doi/10.1103/PhysRevLett.116.231101}.

\bibitem{ArmanoEtAl2018}
Armano, M. {et~al.}, ``Beyond the Required LISA Free-Fall Performance: New LISA
  Pathfinder Results down to $20{\,\mu}$ Hz'', {\em Phys. Rev. Lett.}, {\bf
  120}, 061101 (Feb 2018).
  {\small[\href{http://dx.doi.org/10.1103/PhysRevLett.120.061101}{DOI}]}URL:
  \newline\url{https://link.aps.org/doi/10.1103/PhysRevLett.120.061101}.

\bibitem{AKO75}
Arons, J., Kulsrud, R.M.  and Ostriker, J.P., ``A multiple pulsar model for
  quasi-stellar objects and active galactic nuclei'', {\em Astrophys. J.}, {\bf
  198}, 687--705 (1975).
  {\small[\href{http://adsabs.harvard.edu/abs/1975ApJ...198..687A}{ADS}]}.

\bibitem{BabakEtAl10}
Babak, S. {et~al.} (Challenge 3 participants), ``The Mock LISA Data Challenges:
  from challenge 3 to challenge 4'', {\em Class. Quantum Grav.}, {\bf 27},
  084009 (2010).
  {\small[\href{http://dx.doi.org/10.1088/0264-9381/27/8/084009}{DOI}]},
  {\small[\href{http://adsabs.harvard.edu/abs/2010CQGra..27h4009B}{ADS}]},
  {\small[\href{http://arxiv.org/abs/0912.0548}{{arXiv:0912.0548
  {\small[gr-qc]}}}]}.

\bibitem{BabakEtAl2017}
{Babak}, S. {et~al.}, ``{Science with the space-based interferometer LISA. V:
  Extreme mass-ratio inspirals}'', {\em ArXiv e-prints} (March 2017).
  {\small[\href{http://adsabs.harvard.edu/abs/2017arXiv170309722B}{ADS}]},
  {\small[\href{http://arxiv.org/abs/1703.09722}{{arXiv:1703.09722
  {\small[gr-qc]}}}]}.

\bibitem{BW76}
Bahcall, J.N.  and Wolf, R.A., ``Star distribution around a massive black hole
  in a globular cluster'', {\em Astrophys. J.}, {\bf 209}, 214--232 (1976).
  {\small[\href{http://adsabs.harvard.edu/abs/1976ApJ...209..214B}{ADS}]}.

\bibitem{BW77}
Bahcall, J.N.  and Wolf, R.A., ``The star distribution around a massive black
  hole in a globular cluster. {II} Unequal star masses'', {\em Astrophys. J.},
  {\bf 216}, 883--907 (1977).
  {\small[\href{http://adsabs.harvard.edu/abs/1977ApJ...216..883B}{ADS}]}.

\bibitem{Bar-OrAlexander2014}
{Bar-Or}, B.  and {Alexander}, T., ``{The statistical mechanics of relativistic
  orbits around a massive black hole}'', {\em Classical and Quantum Gravity},
  {\bf 31}(24), 244003 (December 2014).
  {\small[\href{http://dx.doi.org/10.1088/0264-9381/31/24/244003}{DOI}]},
  {\small[\href{http://adsabs.harvard.edu/abs/2014CQGra..31x4003B}{ADS}]},
  {\small[\href{http://arxiv.org/abs/1404.0351}{{arXiv:1404.0351}}]}.

\bibitem{BC04}
Barack, L.  and Cutler, C., ``LISA capture sources: Approximate waveforms,
  signal-to-noise ratios, and parameter estimation accuracy'', {\em Phys. Rev.
  D}, {\bf 69}, 082005 (2004).
  {\small[\href{http://adsabs.harvard.edu/abs/2004PhRvD..69h2005B}{ADS}]},
  {\small[\href{http://arxiv.org/abs/gr-qc/0310125}{{gr-qc/0310125}}]}.

\bibitem{BarausseEtAl2014}
{Barausse}, E., {Cardoso}, V.  and {Pani}, P., ``{Can environmental effects
  spoil precision gravitational-wave astrophysics?}'', {\em Ph. Rv. D}, {\bf
  89}(10), 104059 (May 2014).
  {\small[\href{http://dx.doi.org/10.1103/PhysRevD.89.104059}{DOI}]},
  {\small[\href{http://adsabs.harvard.edu/abs/2014PhRvD..89j4059B}{ADS}]},
  {\small[\href{http://arxiv.org/abs/1404.7149}{{arXiv:1404.7149
  {\small[gr-qc]}}}]}.

\bibitem{BarausseRezzolla2008}
Barausse, E.  and Rezzolla, L., ``Influence of the hydrodynamic drag from an
  accretion torus on extreme mass-ratio inspirals'', {\em Phys. Rev. D}, {\bf
  77} (2008).
  {\small[\href{http://dx.doi.org/10.1103/PhysRevD.77.104027}{DOI}]},
  {\small[\href{http://adsabs.harvard.edu/abs/2008PhRvD..77j4027B}{ADS}]},
  {\small[\href{http://arxiv.org/abs/0711.4558}{{arXiv:0711.4558
  {\small[gr-qc]}}}]}.

\bibitem{BarausseEtAl2007}
Barausse, E., Rezzolla, L., Petroff, D.  and Ansorg, M., ``Gravitational waves
  from extreme mass ratio inspirals in nonpure Kerr spacetimes'', {\em Phys.
  Rev. D}, {\bf 75} (2007).
  {\small[\href{http://dx.doi.org/10.1103/PhysRevD.75.064026}{DOI}]},
  {\small[\href{http://adsabs.harvard.edu/abs/2007PhRvD..75f4026B}{ADS}]},
  {\small[\href{http://arxiv.org/abs/arXiv:gr-qc/0612123}{{arXiv:gr-qc/0612123}}]}.

\bibitem{Bardeen70}
Bardeen, J.M., ``Kerr Metric Black Holes'', {\em Nature}, {\bf 226}, 64--65
  (1970). {\small[\href{http://dx.doi.org/10.1038/226064a0}{DOI}]},
  {\small[\href{http://adsabs.harvard.edu/abs/1970Natur.226...64B}{ADS}]}.

\bibitem{barneshut86}
Barnes, J.  and Hut, P., ``A hierarchical O(N log N) force-calculation
  algorithm'', {\em Nature}, {\bf 324}, 446--449 (1986).
  {\small[\href{http://dx.doi.org/10.1038/324446a0}{DOI}]},
  {\small[\href{http://adsabs.harvard.edu/abs/1986Natur.324..446B}{ADS}]}.

\bibitem{BartkoEtAl10}
Bartko, H. {et~al.}, ``An Extremely Top-Heavy Initial Mass Function in the
  Galactic Center Stellar Disks'', {\em Astrophys. J.}, {\bf 708}, 834--840
  (2010). {\small[\href{http://dx.doi.org/10.1088/0004-637X/708/1/834}{DOI}]},
  {\small[\href{http://adsabs.harvard.edu/abs/2010ApJ...708..834B}{ADS}]},
  {\small[\href{http://arxiv.org/abs/0908.2177}{{arXiv:0908.2177}}]}.

\bibitem{BaumgardtEtAl2017}
{Baumgardt}, H., {Amaro-Seoane}, P.  and {Sch{\"o}del}, R., ``{The distribution
  of stars around the Milky Way's black hole III: Comparison with
  simulations}'', {\em ArXiv e-prints} (January 2017).
  {\small[\href{http://adsabs.harvard.edu/abs/2017arXiv170103818B}{ADS}]},
  {\small[\href{http://arxiv.org/abs/1701.03818}{{arXiv:1701.03818}}]}.

\bibitem{BME04a}
Baumgardt, H., Makino, J.  and Ebisuzaki, T., ``Massive Black Holes in Star
  Clusters. {I}. Equal-Mass Clusters'', {\em Astrophys. J.}, {\bf 613},
  1133--1142 (2004).
  {\small[\href{http://adsabs.harvard.edu/abs/2004ApJ...613.1133B}{ADS}]}.

\bibitem{BME04b}
Baumgardt, H., Makino, J.  and Ebisuzaki, T., ``Massive Black Holes in Star
  Clusters. {II}. Realistic Cluster Models'', {\em Astrophys. J.}, {\bf 613},
  1143--1156 (2004).
  {\small[\href{http://adsabs.harvard.edu/abs/2004ApJ...613.1143B}{ADS}]}.

\bibitem{Begelman10}
Begelman, M.C., ``Evolution of supermassive stars as a pathway to black hole
  formation'', {\em Mon. Not. R. Astron. Soc.}, {\bf 402}, 673--681 (2010).
  {\small[\href{http://dx.doi.org/10.1111/j.1365-2966.2009.15916.x}{DOI}]},
  {\small[\href{http://adsabs.harvard.edu/abs/2010MNRAS.402..673B}{ADS}]},
  {\small[\href{http://arxiv.org/abs/0910.4398}{{arXiv:0910.4398}}]}.

\bibitem{BenderEtAl05}
Bender, P.L., Armitage, P.J., Begelman, M.C.  and Pema, R., ``Massive Black
  Hole Formation and Growth'', {\em White Paper submitted to the NASA SEU
  Roadmap Committee} (2005).

\bibitem{BH97}
Bender, P.L.  and Hils, D., ``Confusion noise level due to galactic and
  extragalactic binaries'', {\em Class. Quantum Grav.}, {\bf 14}, 1439--1444
  (1997).
  {\small[\href{http://adsabs.harvard.edu/abs/1997CQGra..14.1439B}{ADS}]}.

\bibitem{BenderKormendyEtAl2005}
Bender, R. {et~al.}, ``HST STIS Spectroscopy of the Triple Nucleus of M31: Two
  Nested Disks in Keplerian Rotation around a Supermassive Black Hole'', {\em
  Astrophys. J.}, {\bf 631}, 280--300 (2005).
  {\small[\href{http://dx.doi.org/10.1086/432434}{DOI}]},
  {\small[\href{http://adsabs.harvard.edu/abs/2005ApJ...631..280B}{ADS}]},
  {\small[\href{http://arxiv.org/abs/arXiv:astro-ph/0509839}{{arXiv:astro-ph/0509839}}]}.

\bibitem{BerryGair2013}
Berry, C.P.L.  and Gair, J.R., ``Observing the Galaxy's massive black hole with
  gravitational wave bursts'', {\em Mon. Not. R. Astron. Soc.}, {\bf 429},
  589--612 (2013).
  {\small[\href{http://dx.doi.org/10.1093/mnras/sts360}{DOI}]},
  {\small[\href{http://adsabs.harvard.edu/abs/2013MNRAS.429..589B}{ADS}]},
  {\small[\href{http://arxiv.org/abs/1210.2778}{{arXiv:1210.2778
  {\small[astro-ph.HE]}}}]}.

\bibitem{BS86}
Bettwieser, E.  and Spurzem, R., ``Anisotropy in stellar dynamics'', {\bf 161},
  102--112 (1986).
  {\small[\href{http://adsabs.harvard.edu/abs/1986A&A...161..102B}{ADS}]}.

\bibitem{BT87}
Binney, J.  and Tremaine, S., {\em Galactic Dynamics}, (Princeton University
  Press, 1987).
  {\small[\href{http://adsabs.harvard.edu/abs/1987gady.book.....B}{ADS}]}.

\bibitem{BinneyTremaine08}
Binney, J.  and Tremaine, S., {\em Galactic Dynamics: Second Edition},
  (Princeton University Press, Princeton, NJ, 2008), 2nd edition.
  {\small[\href{http://adsabs.harvard.edu/abs/2008gady.book.....B}{ADS}]},
  {\small[\href{http://books.google.com/books?id=6mF4CKxlbLsC}{Google Books}]}.

\bibitem{Blanchet2006}
{Blanchet}, L., ``{Gravitational Radiation from Post-Newtonian Sources and
  Inspiralling Compact Binaries}'', {\em Living Reviews in Relativity}, {\bf
  9}, 4 (June 2006).
  {\small[\href{http://dx.doi.org/10.12942/lrr-2006-4}{DOI}]},
  {\small[\href{http://adsabs.harvard.edu/abs/2006LRR.....9....4B}{ADS}]}.

\bibitem{BlanchetIyer03}
Blanchet, L.  and Iyer, B.R., ``Third post-Newtonian dynamics of compact
  binaries: equations of motion in the centre-of-mass frame'', {\em Class.
  Quantum Grav.}, {\bf 20}, 755--776 (2003).
  {\small[\href{http://adsabs.harvard.edu/abs/2003CQGra..20..755B}{ADS}]},
  {\small[\href{http://arxiv.org/abs/gr-qc/0209089}{{gr-qc/0209089}}]}.

\bibitem{BregmanAlexander09}
Bregman, M.  and Alexander, T., ``Accretion Disk Warping by Resonant
  Relaxation: The Case of Maser Disk NGC 4258'', {\bf 700}, L192--L195 (2009).
  {\small[\href{http://dx.doi.org/10.1088/0004-637X/700/2/L192}{DOI}]},
  {\small[\href{http://adsabs.harvard.edu/abs/2009ApJ...700L.192B}{ADS}]},
  {\small[\href{http://arxiv.org/abs/0903.2051}{{arXiv:0903.2051
  {\small[astro-ph.GA]}}}]}.

\bibitem{BremAmaroSeoaneSopuerta2012}
Brem, P., Amaro-Seoane, P.  and Sopuerta, C.F., ``Blocking low-eccentricity
  EMRIs: A statistical direct-summation $N$-body study of the Schwarzschild
  barrier'', {\em arXiv}, e-print, (2012).
  {\small[\href{http://adsabs.harvard.edu/abs/2012arXiv1211.5601B}{ADS}]},
  {\small[\href{http://arxiv.org/abs/1211.5601}{{arXiv:1211.5601
  {\small[astro-ph.CO]}}}]}.

\bibitem{BremAmaro-SeoaneSopuerta2014}
{Brem}, P., {Amaro-Seoane}, P.  and {Sopuerta}, C.~F., ``{Blocking
  low-eccentricity EMRIs: a statistical direct-summation N-body study of the
  Schwarzschild barrier}'', {\bf 437}, 1259--1267 (January 2014).
  {\small[\href{http://dx.doi.org/10.1093/mnras/stt1948}{DOI}]},
  {\small[\href{http://adsabs.harvard.edu/abs/2014MNRAS.437.1259B}{ADS}]},
  {\small[\href{http://arxiv.org/abs/1211.5601}{{arXiv:1211.5601}}]}.

\bibitem{BremAmaro-SeoaneSpurzem2014}
{Brem}, P., {Amaro-Seoane}, P.  and {Spurzem}, R., ``{Relativistic mergers of
  compact binaries in clusters: the fingerprint of the spin}'', {\bf 434},
  2999--3007 (October 2013).
  {\small[\href{http://dx.doi.org/10.1093/mnras/stt1220}{DOI}]},
  {\small[\href{http://adsabs.harvard.edu/abs/2013MNRAS.434.2999B}{ADS}]},
  {\small[\href{http://arxiv.org/abs/1302.3135}{{arXiv:1302.3135}}]}.

\bibitem{BrownEtAl09}
Brown, W.R., Geller, M.J., Kenyon, S.J.  and Bromley, B.C., ``The Anisotropic
  Spatial Distribution of Hypervelocity Stars'', {\bf 690}, L69--L71 (2009).
  {\small[\href{http://dx.doi.org/10.1088/0004-637X/690/1/L69}{DOI}]},
  {\small[\href{http://adsabs.harvard.edu/abs/2009ApJ...690L..69B}{ADS}]},
  {\small[\href{http://arxiv.org/abs/0811.0612}{{arXiv:0811.0612}}]}.

\bibitem{BuchholzEtAl09}
Buchholz, R.M., Sch{\"o}del, R.  and Eckart, A., ``Composition of the galactic
  center star cluster. Population analysis from adaptive optics narrow band
  spectral energy distributions'', {\em aa}, {\bf 499}, 483--501 (2009).
  {\small[\href{http://dx.doi.org/10.1051/0004-6361/200811497}{DOI}]},
  {\small[\href{http://adsabs.harvard.edu/abs/2009A&A...499..483B}{ADS}]},
  {\small[\href{http://arxiv.org/abs/0903.2135}{{arXiv:0903.2135}}]}.

\bibitem{BuonannoEtAl03}
{Buonanno}, A., {Chen}, Y.  and {Vallisneri}, M., ``{Detecting gravitational
  waves from precessing binaries of spinning compact objects: Adiabatic
  limit}'', {\em Ph.Rv. D}, {\bf 67}(10), 104025 (May 2003).
  {\small[\href{http://dx.doi.org/10.1103/PhysRevD.67.104025}{DOI}]},
  {\small[\href{http://adsabs.harvard.edu/abs/2003PhRvD..67j4025B}{ADS}]},
  {\small[\href{http://arxiv.org/abs/gr-qc/0211087}{{gr-qc/0211087}}]}.

\bibitem{CB00}
Chabrier, G.  and Baraffe, I., ``Theory of Low-Mass Stars and Substellar
  Objects'', {\em Annu. Rev. Astron. Astrophys.}, {\bf 38}, 337--377 (2000).
  {\small[\href{http://adsabs.harvard.edu/abs/2000ARA&A..38..337C}{ADS}]},
  {\small[\href{http://arxiv.org/abs/astro-ph/0006383}{{astro-ph/0006383}}]}.

\bibitem{Chandra42}
Chandrasekhar, S., ``Principles of stellar dynamics'', {\em Physical Sciences
  Data} (1942).
  {\small[\href{http://adsabs.harvard.edu/abs/1942psd..book.....C}{ADS}]}.

\bibitem{Chandrasekhar60}
Chandrasekhar, S., {\em Principles of stellar dynamics}, (New York: Dover,
  1960, Enlarged ed., 1960).
  {\small[\href{http://adsabs.harvard.edu/abs/1960psd..book.....C}{ADS}]}.

\bibitem{ChangCooper70}
Chang, J.S.  and Cooper, G., ``A practical Difference Scheme for Fokker-Planck
  Equations'', {\em J. Comp. Phys.}, {\bf 6}, 1--16 (1970).

\bibitem{ChenAmaro-Seoane2014a}
{Chen}, X.  and {Amaro-Seoane}, P., ``{A Rapidly Evolving Region in the
  Galactic Center: Why S-stars Thermalize and More Massive Stars are
  Missing}'', {\bf 786}, L14 (May 2014).
  {\small[\href{http://dx.doi.org/10.1088/2041-8205/786/2/L14}{DOI}]},
  {\small[\href{http://adsabs.harvard.edu/abs/2014ApJ...786L..14C}{ADS}]},
  {\small[\href{http://arxiv.org/abs/1401.6456}{{arXiv:1401.6456}}]}.

\bibitem{ChenEtAl11}
Chen, X., Sesana, A., Madau, P.  and Liu, F.K., ``Tidal Stellar Disruptions by
  Massive Black Hole Pairs. II. Decaying Binaries'', {\em Astrophys. J.}, {\bf
  729}, 13 (2011).
  {\small[\href{http://dx.doi.org/10.1088/0004-637X/729/1/13}{DOI}]},
  {\small[\href{http://adsabs.harvard.edu/abs/2011ApJ...729...13C}{ADS}]},
  {\small[\href{http://arxiv.org/abs/1012.4466}{{arXiv:1012.4466
  {\small[astro-ph.GA]}}}]}.

\bibitem{CW90}
Chernoff, D.F.  and Weinberg, M.D., ``Evolution of globular clusters in the
  Galaxy'', {\em Astrophys. J.}, {\bf 351}, 121--156 (1990).
  {\small[\href{http://adsabs.harvard.edu/abs/1990ApJ...351..121C}{ADS}]}.

\bibitem{Clutton-Brock73}
Clutton-Brock, M., ``The Gravitational Field of Three Dimensional Galaxies'',
  {\em Astrophysics and Space Science}, {\bf 23}, 55--69 (1973).
  {\small[\href{http://dx.doi.org/10.1007/BF00647652}{DOI}]},
  {\small[\href{http://adsabs.harvard.edu/abs/1973Ap&SS..23...55C}{ADS}]}.

\bibitem{Cohn79}
Cohn, H., ``Numerical integration of the Fokker-Planck equation and the
  evolution of star clusters'', {\em Astrophys. J.}, {\bf 234}, 1036--1053
  (1979).
  {\small[\href{http://adsabs.harvard.edu/abs/1979ApJ...234.1036C}{ADS}]}.

\bibitem{Cohn80}
Cohn, H., ``Late core collapse in star clusters and the gravothermal
  instability'', {\em Astrophys. J.}, {\bf 242}, 765--771 (1980).
  {\small[\href{http://adsabs.harvard.edu/abs/1980ApJ...242..765C}{ADS}]}.

\bibitem{Cohn85}
Cohn, H., ``Direct Fokker-Planck calculations'', in {\em IAU Symp. 113:
  Dynamics of Star Clusters}, pp. 161--177, (1985).
  {\small[\href{http://adsabs.harvard.edu/abs/1985IAUS..113..161C}{ADS}]}.

\bibitem{CK78}
Cohn, H.  and Kulsrud, R.M., ``The stellar distribution around a black hole -
  Numerical integration of the Fokker-Planck equation'', {\em Astrophys. J.},
  {\bf 226}, 1087--1108 (1978).
  {\small[\href{http://adsabs.harvard.edu/abs/1978ApJ...226.1087C}{ADS}]}.

\bibitem{CZ99}
Collin, S.  and Zahn, J.P., ``Accretion Disks and Star Formation'', in Terzian,
  Y., Khachikian, E.  and Weedman, D., eds., {\em IAU Symp. 194: Activity in
  Galaxies and Related Phenomena}, 194, p. 246, (1999).
  {\small[\href{http://adsabs.harvard.edu/abs/1999IAUS..194..246C}{ADS}]}.

\bibitem{CH06}
Cutler, C.  and Harms, J., ``Big Bang Observer and the neutron-star-binary
  subtraction problem'', {\em Phys. Rev. D}, {\bf 73}, 042001 (2006).
  {\small[\href{http://dx.doi.org/10.1103/PhysRevD.73.042001}{DOI}]},
  {\small[\href{http://adsabs.harvard.edu/abs/2006PhRvD..73d2001C}{ADS}]},
  {\small[\href{http://arxiv.org/abs/gr-qc/0511092}{{gr-qc/0511092}}]}.

\bibitem{CKP94}
Cutler, C., Kennefick, D.  and Poisson, E., ``Gravitational radiation reaction
  for bound motion around a Schwarzschild black hole'', {\em Phys. Rev. D},
  {\bf 50}, 3816--3835 (1994).
  {\small[\href{http://dx.doi.org/10.1103/PhysRevD.50.3816}{DOI}]},
  {\small[\href{http://adsabs.harvard.edu/abs/1994PhRvD..50.3816C}{ADS}]}.

\bibitem{Danzmann00}
Danzmann, K., ``LISA Mission Overview'', {\em Advances in Space Research}, {\bf
  25}, 1129--1136 (2000).
  {\small[\href{http://adsabs.harvard.edu/abs/2000AdSpR..25.1129D}{ADS}]}.

\bibitem{DDC87a}
David, L.P., Durisen, R.H.  and Cohn, H.N., ``The evolution of active galactic
  nuclei. I - Models without stellar evolution'', {\em Astrophys. J.}, {\bf
  313}, 556--575 (1987).
  {\small[\href{http://adsabs.harvard.edu/abs/1987ApJ...313..556D}{ADS}]}.

\bibitem{DDC87b}
David, L.P., Durisen, R.H.  and Cohn, H.N., ``The evolution of active galactic
  nuclei. {II} - Models with stellar evolution'', {\em Astrophys. J.}, {\bf
  316}, 505--516 (1987).
  {\small[\href{http://adsabs.harvard.edu/abs/1987ApJ...316..505D}{ADS}]}.

\bibitem{DeFreitasEtAl06}
de~Freitas~Pacheco, J.A., Filloux, C.  and Regimbau, T., ``Capture rates of
  compact objects by supermassive black holes'', {\em Phys. Rev. D}, {\bf 74},
  023001 (2006).
  {\small[\href{http://dx.doi.org/10.1103/PhysRevD.74.023001}{DOI}]},
  {\small[\href{http://adsabs.harvard.edu/abs/2006PhRvD..74b3001D}{ADS}]},
  {\small[\href{http://arxiv.org/abs/astro-ph/0606427}{{astro-ph/0606427}}]}.

\bibitem{DoEtAl09}
Do, T., Ghez, A.M., Morris, M.R., Lu, J.R., Matthews, K., Yelda, S.  and
  Larkin, J., ``High Angular Resolution Integral-Field Spectroscopy of the
  Galaxy's Nuclear Cluster: A Missing Stellar Cusp?'', {\em Astrophys. J.},
  {\bf 703}, 1323--1337 (2009).
  {\small[\href{http://dx.doi.org/10.1088/0004-637X/703/2/1323}{DOI}]},
  {\small[\href{http://adsabs.harvard.edu/abs/2009ApJ...703.1323D}{ADS}]},
  {\small[\href{http://arxiv.org/abs/0908.0311}{{arXiv:0908.0311}}]}.

\bibitem{DCLY99}
Drukier, G.A., Cohn, H.N., Lugger, P.M.  and Yong, H., ``Anisotropic
  Fokker-Planck Models for the Evolution of Globular Star Clusters: The
  Core-Halo Connection'', {\em Astrophys. J.}, {\bf 518}, 233--245 (1999).
  {\small[\href{http://adsabs.harvard.edu/abs/1999ApJ...518..233D}{ADS}]}.

\bibitem{DS83}
Duncan, M.~J.  and Shapiro, S.~L., ``Monte Carlo simulations of the evolution
  of galactic nuclei containing massive, central black holes'', {\em Astrophys.
  J.}, {\bf 268}, 565--581 (1983).
  {\small[\href{http://adsabs.harvard.edu/abs/1983ApJ...268..565D}{ADS}]}.

\bibitem{EilonEtAl09}
Eilon, E., Kupi, G.  and Alexander, T., ``The Efficiency of Resonant Relaxation
  Around a Massive Black Hole'', {\em Astrophys. J.}, {\bf 698}, 641--647
  (2009). {\small[\href{http://dx.doi.org/10.1088/0004-637X/698/1/641}{DOI}]},
  {\small[\href{http://adsabs.harvard.edu/abs/2009ApJ...698..641E}{ADS}]},
  {\small[\href{http://arxiv.org/abs/0807.1430}{{arXiv:0807.1430}}]}.

\bibitem{Amaro-SeoaneEtAl2013}
{eLISA Consortium} {et~al.}, ``{The Gravitational Universe}'', {\em ArXiv
  e-prints} (May 2013).
  {\small[\href{http://adsabs.harvard.edu/abs/2013arXiv1305.5720E}{ADS}]},
  {\small[\href{http://arxiv.org/abs/1305.5720}{{arXiv:1305.5720
  {\small[astro-ph.CO]}}}]}.

\bibitem{ElsonEtAl87}
Elson, R., Hut, P.  and Inagaki, S., ``Dynamical evolution of globular
  clusters'', {\em Annu. Rev. Astron. Astrophys.}, {\bf 25}, 565--601 (1987).
  {\small[\href{http://adsabs.harvard.edu/abs/1987ARA&A..25..565E}{ADS}]}.

\bibitem{FayeEtAl2006}
{Faye}, G., {Blanchet}, L.  and {Buonanno}, A., ``{Higher-order spin effects in
  the dynamics of compact binaries. I. Equations of motion}'', {\em Ph.Rv. D},
  {\bf 74}(10), 104033 (November 2006).
  {\small[\href{http://dx.doi.org/10.1103/PhysRevD.74.104033}{DOI}]},
  {\small[\href{http://adsabs.harvard.edu/abs/2006PhRvD..74j4033F}{ADS}]},
  {\small[\href{http://arxiv.org/abs/gr-qc/0605139}{{gr-qc/0605139}}]}.

\bibitem{superbox}
Fellhauer, M., Kroupa, P., Baumgardt, H., Bien, R., Boily, C.M., Spurzem, R.
  and Wassmer, N., ``SUPERBOX - an efficient code for collisionless galactic
  dynamics'', {\em Nature}, {\bf 5}, 305--326 (2000).
  {\small[\href{http://dx.doi.org/10.1016/S1384-1076(00)00032-4}{DOI}]},
  {\small[\href{http://adsabs.harvard.edu/abs/2000NewA....5..305F}{ADS}]},
  {\small[\href{http://arxiv.org/abs/arXiv:astro-ph/0007226}{{arXiv:astro-ph/0007226}}]}.

\bibitem{FF04}
Ferrarese, L.  and Ford, H., ``Supermassive Black Holes in Galactic Nuclei:
  Past, Present and Future Research'', {\em Space Sci. Rev.}, {\bf 116},
  523--624 (2005).
  {\small[\href{http://dx.doi.org/10.1007/s11214-005-3947-6}{DOI}]},
  {\small[\href{http://adsabs.harvard.edu/abs/2005SSRv..116..523F}{ADS}]}.

\bibitem{FM00}
Ferrarese, L.  and Merritt, D., ``A Fundamental Relation between Supermassive
  Black Holes and Their Host Galaxies'', {\em Astrophys. J. Lett.}, {\bf 539},
  L9--L12 (2000).
  {\small[\href{http://adsabs.harvard.edu/abs/2000ApJ...539L...9F}{ADS}]}.

\bibitem{FPPMWJ01}
Ferrarese, L., Pogge, R.W., Peterson, B.M., Merritt, D., Wandel, A.  and
  Joseph, C.L., ``Supermassive Black Holes in Active Galactic Nuclei. {I}. The
  Consistency of Black Hole Masses in Quiescent and Active Galaxies'', {\em
  Astrophys. J. Lett.}, {\bf 555}, L79--L82 (2001).
  {\small[\href{http://adsabs.harvard.edu/abs/2001ApJ...555L..79F}{ADS}]}.

\bibitem{FiestasThesis06}
Fiestas, J., {\em Dynamical evolution of rotating globular clusters with
  embedded black holes}, Ph.D. thesis, (Heidelberg University, 2006).
  {\small[\href{http://adsabs.harvard.edu/abs/2006PhDT.........2F}{ADS}]}.

\bibitem{FSK06}
Fiestas, J., Spurzem, R.  and Kim, E., ``2D Fokker-Planck models of rotating
  clusters'', {\em Mon. Not. R. Astron. Soc.}, {\bf 373}, 677--686 (2006).
  {\small[\href{http://dx.doi.org/10.1111/j.1365-2966.2006.11036.x}{DOI}]},
  {\small[\href{http://adsabs.harvard.edu/abs/2006MNRAS.tmp.1193F}{ADS}]},
  {\small[\href{http://arxiv.org/abs/astro-ph/0609056}{{astro-ph/0609056}}]}.

\bibitem{Finn92}
Finn, L.S., ``Detection, measurement, and gravitational radiation'', {\em prd},
  {\bf 46}, 5236--5249 (1992).
  {\small[\href{http://dx.doi.org/10.1103/PhysRevD.46.5236}{DOI}]},
  {\small[\href{http://adsabs.harvard.edu/abs/1992PhRvD..46.5236F}{ADS}]},
  {\small[\href{http://arxiv.org/abs/arXiv:gr-qc/9209010}{{arXiv:gr-qc/9209010}}]}.

\bibitem{FT00}
Finn, L.S.  and Thorne, K.S., ``Gravitational waves from a compact star in a
  circular, inspiral orbit, in the equatorial plane of a massive, spinning
  black hole, as observed by LISA'', {\em Phys. Rev. D}, {\bf 62}, 124021
  (2000).
  {\small[\href{http://adsabs.harvard.edu/abs/2000PhRvD..62l4021F}{ADS}]}.

\bibitem{FR76}
Frank, J.  and Rees, M.J., ``Effects of massive central black holes on dense
  stellar systems'', {\em Mon. Not. R. Astron. Soc.}, {\bf 176}, 633--647
  (1976).
  {\small[\href{http://adsabs.harvard.edu/abs/1976MNRAS.176..633F}{ADS}]}.

\bibitem{FregeauEtAl03}
Fregeau, J.M., G{\"u}rkan, M.A., Joshi, K.J.  and Rasio, F.A., ``Monte Carlo
  Simulations of Globular Cluster Evolution. III. Primordial Binary
  Interactions'', {\em Astrophys. J.}, {\bf 593}, 772--787 (2003).
  {\small[\href{http://adsabs.harvard.edu/abs/2003ApJ...593..772F}{ADS}]}.

\bibitem{FGR06}
Fregeau, J.M., G{\"u}rkan, M.A.  and Rasio, F.A., ``Star Cluster Evolution with
  Primordial Binaries'', {\em arXiv}, e-print, (2005).
  {\small[\href{http://arxiv.org/abs/astro-ph/0512032}{{arXiv:astro-ph/0512032}}]}.

\bibitem{FregeauEtAl2004}
{Fregeau}, J.~M., {Cheung}, P., {Portegies Zwart}, S.~F.  and {Rasio}, F.~A.,
  ``{Stellar collisions during binary-binary and binary-single star
  interactions}'', {\bf 352}, 1--19 (July 2004).
  {\small[\href{http://dx.doi.org/10.1111/j.1365-2966.2004.07914.x}{DOI}]},
  {\small[\href{http://adsabs.harvard.edu/abs/2004MNRAS.352....1F}{ADS}]},
  {\small[\href{http://arxiv.org/abs/astro-ph/0401004}{{astro-ph/0401004}}]}.

\bibitem{FregeauRasio2007}
{Fregeau}, J.~M.  and {Rasio}, F.~A., ``{Monte Carlo Simulations of Globular
  Cluster Evolution. IV. Direct Integration of Strong Interactions}'', {\bf
  658}, 1047--1061 (April 2007).
  {\small[\href{http://dx.doi.org/10.1086/511809}{DOI}]},
  {\small[\href{http://adsabs.harvard.edu/abs/2007ApJ...658.1047F}{ADS}]},
  {\small[\href{http://arxiv.org/abs/astro-ph/0608261}{{astro-ph/0608261}}]}.

\bibitem{Freitag01}
Freitag, M., ``Monte Carlo cluster simulations to determine the rate of compact
  star inspiralling to a central galactic black hole'', {\em Class. Quantum
  Grav.}, {\bf 18}, 4033--4038 (2001).
  {\small[\href{http://adsabs.harvard.edu/abs/2001CQGra..18.4033F}{ADS}]}.

\bibitem{Freitag03b}
Freitag, M., ``Captures of stars by a massive black hole: Investigations in
  numerical stellar dynamics'', in Centrella, J.M., ed., {\em The Astrophysics
  of Gravitational Wave Sources}, AIP Conference Proceedings, 686, pp.
  109--112. American Institute of Physics, (2003).

\bibitem{Freitag03}
Freitag, M., ``Gravitational waves from stars orbiting the Sagittarius A* black
  hole'', {\em Astrophys. J. Lett.}, {\bf 583}, L21--L24 (2003).
  {\small[\href{http://arxiv.org/abs/astro-ph/0211209}{{astro-ph/0211209}}]}.

\bibitem{FAK06b}
Freitag, M., Amaro-Seoane, P.  and Kalogera, V., ``Models of mass segregation
  at the Galactic Centre'', {\em Journal of Physics Conference Series}, {\bf
  54}, 252--258 (2006).
  {\small[\href{http://dx.doi.org/10.1088/1742-6596/54/1/040}{DOI}]},
  {\small[\href{http://adsabs.harvard.edu/abs/2006JPhCS..54..252F}{ADS}]},
  {\small[\href{http://arxiv.org/abs/arXiv:astro-ph/0607001}{{arXiv:astro-ph/0607001}}]}.

\bibitem{FASK06}
Freitag, M., Amaro-Seoane, P.  and Kalogera, V., ``Stellar Remnants in Galactic
  Nuclei: Mass Segregation'', {\em Astrophys. J.}, {\bf 649}, 91--117 (2006).
  {\small[\href{http://dx.doi.org/10.1086/506193}{DOI}]},
  {\small[\href{http://adsabs.harvard.edu/abs/2006ApJ...649...91F}{ADS}]},
  {\small[\href{http://arxiv.org/abs/astro-ph/0603280}{{astro-ph/0603280}}]}.

\bibitem{FAK06a}
Freitag, M., Amaro-Seoane, P.  and Kalogera, V., ``Stellar Remnants in Galactic
  Nuclei: Mass Segregation'', {\em Astrophys. J.}, {\bf 649}, 91--117 (2006).
  {\small[\href{http://dx.doi.org/10.1086/506193}{DOI}]},
  {\small[\href{http://adsabs.harvard.edu/abs/2006ApJ...649...91F}{ADS}]},
  {\small[\href{http://arxiv.org/abs/arXiv:astro-ph/0603280}{{arXiv:astro-ph/0603280}}]}.

\bibitem{FB01a}
Freitag, M.  and Benz, W., ``A New {M}onte {C}arlo Code for Star Cluster
  Simulations: {I}.Relaxation'', {\em Astron. Astrophys.}, {\bf 375}, 711--738
  (2001).
  {\small[\href{http://www.edpsciences.org/articles/aa/abs/2001/32/aa1125/aa1125.html}{ADS}]}.

\bibitem{FB02b}
Freitag, M.  and Benz, W., ``A New {M}onte {C}arlo Code for Star Cluster
  Simulations: {II}.Central Black Hole and Stellar Collisions'', {\em Astron.
  Astrophys.}, {\bf 394}, 345--374 (2002).
  {\small[\href{http://www.edpsciences.com/articles/aa/abs/2002/40/aa2593/aa2593.html}{ADS}]}.

\bibitem{FB05}
Freitag, M.  and Benz, W., ``A comprehensive set of simulations of
  high-velocity collisions between main-sequence stars'', {\em Mon. Not. R.
  Astron. Soc.}, {\bf 358}, 1133--1158 (2005).
  {\small[\href{http://adsabs.harvard.edu/abs/2005MNRAS.358.1133F}{ADS}]},
  {\small[\href{http://arxiv.org/abs/astro-ph/0403621}{{astro-ph/0403621}}]}.

\bibitem{GRAPE6A}
{Fukushige}, T., {Makino}, J.  and {Kawai}, A., ``{GRAPE-6A: A Single-Card
  GRAPE-6 for Parallel PC-GRAPE Cluster Systems}'', {\em Publ. Astron. Soc.
  Japan}, {\bf 57}, 1009--1021 (December 2005).
  {\small[\href{http://dx.doi.org/10.1093/pasj/57.6.1009}{DOI}]},
  {\small[\href{http://adsabs.harvard.edu/abs/2005PASJ...57.1009F}{ADS}]},
  {\small[\href{http://arxiv.org/abs/astro-ph/0504407}{{astro-ph/0504407}}]}.

\bibitem{GaburovEtAl2009}
{Gaburov}, E., {Harfst}, S.  and {Portegies Zwart}, S., ``{SAPPORO: A way to
  turn your graphics cards into a GRAPE-6}'', {\em New Astronomy}, {\bf 14},
  630--637 (October 2009).
  {\small[\href{http://dx.doi.org/10.1016/j.newast.2009.03.002}{DOI}]},
  {\small[\href{http://adsabs.harvard.edu/abs/2009NewA...14..630G}{ADS}]},
  {\small[\href{http://arxiv.org/abs/0902.4463}{{arXiv:0902.4463
  {\small[astro-ph.IM]}}}]}.

\bibitem{GG06}
Gair, J.R.  and Glampedakis, K., ``Improved approximate inspirals of test
  bodies into Kerr black holes'', {\em Phys. Rev. D}, {\bf 73}, 064037 (2006).
  {\small[\href{http://dx.doi.org/10.1103/PhysRevD.73.064037}{DOI}]},
  {\small[\href{http://adsabs.harvard.edu/abs/2006PhRvD..73f4037G}{ADS}]},
  {\small[\href{http://arxiv.org/abs/arXiv:gr-qc/0510129}{{arXiv:gr-qc/0510129}}]}.

\bibitem{Gair2009}
{Gair}, J.~R., ``{Probing black holes at low redshift using LISA EMRI
  observations}'', {\em Classical and Quantum Gravity}, {\bf 26}(9), 094034
  (May 2009).
  {\small[\href{http://dx.doi.org/10.1088/0264-9381/26/9/094034}{DOI}]},
  {\small[\href{http://adsabs.harvard.edu/abs/2009CQGra..26i4034G}{ADS}]},
  {\small[\href{http://arxiv.org/abs/0811.0188}{{arXiv:0811.0188
  {\small[gr-qc]}}}]}.

\bibitem{GallegoEtAl2017}
{Gallego-Cano}, E., {Sch{\"o}del}, R., {Dong}, H., {Nogueras-Lara}, F.,
  {Gallego-Calvente}, A.~T., {Amaro-Seoane}, P.  and {Baumgardt}, H., ``{The
  distribution of old stars around the Milky Way's central black hole I: Star
  counts}'', {\em ArXiv e-prints} (January 2017).
  {\small[\href{http://adsabs.harvard.edu/abs/2017arXiv170103816G}{ADS}]},
  {\small[\href{http://arxiv.org/abs/1701.03816}{{arXiv:1701.03816}}]}.

\bibitem{GRH02}
Gebhardt, K., Rich, R.M.  and Ho, L.C., ``A $20000 M_\odot$ Black Hole in the
  Stellar Cluster G1'', {\em Astrophys. J. Lett.}, {\bf 578}, L41--L45 (2002).
  {\small[\href{http://adsabs.harvard.edu/abs/2002ApJ...578L..41G}{ADS}]}.

\bibitem{GebhardtEtAl01}
Gebhardt, K. {et~al.}, ``M33: A Galaxy with No Supermassive Black Hole'', {\em
  AJ}, {\bf 122}, 2469--2476 (2001).
  {\small[\href{http://cdsaas.u-strasbg.fr:2001/AJ/journal/issues/v122n5/201284/201284.html}{ADS}]}.

\bibitem{GenzelEtAl10}
Genzel, R., Eisenhauer, F.  and Gillessen, S., ``The Galactic Center massive
  black hole and nuclear star cluster'', {\em Reviews of Modern Physics}, {\bf
  82}, 3121--3195 (2010).
  {\small[\href{http://dx.doi.org/10.1103/RevModPhys.82.3121}{DOI}]},
  {\small[\href{http://adsabs.harvard.edu/abs/2010RvMP...82.3121G}{ADS}]},
  {\small[\href{http://arxiv.org/abs/1006.0064}{{arXiv:1006.0064
  {\small[astro-ph.GA]}}}]}.

\bibitem{Gerhard1993}
{Gerhard}, O.~E., ``{Line-of-sight velocity profiles in spherical galaxies:
  breaking the degeneracy between anisotropy and mass.}'', {\bf 265}, 213
  (November 1993).
  {\small[\href{http://dx.doi.org/10.1093/mnras/265.1.213}{DOI}]},
  {\small[\href{http://adsabs.harvard.edu/abs/1993MNRAS.265..213G}{ADS}]}.

\bibitem{GerssenEtAl02}
Gerssen, J., van~der Marel, R.P., Gebhardt, K., Guhathakurta, P., Peterson,
  R.C.  and Pryor, C., ``Hubble Space Telescope Evidence for an
  Intermediate-Mass Black Hole in the Globular Cluster M15. {II}. Kinematic
  Analysis and Dynamical Modeling'', {\em AJ}, {\bf 124}, 3270--3288 (2002).
  {\small[\href{http://adsabs.harvard.edu/abs/2002AJ....124.3270G}{ADS}]}.

\bibitem{GezariEtAl03}
Gezari, S., Halpern, J.P., Komossa, S., Grupe, D.  and Leighly, K.M.,
  ``Follow-Up Hubble Space Telescope/Space Telescope Imaging Spectroscopy of
  Three Candidate Tidal Disruption Events'', {\em Astrophys. J.}, {\bf 592},
  42--51 (2003). {\small[\href{http://dx.doi.org/10.1086/375553}{DOI}]},
  {\small[\href{http://adsabs.harvard.edu/abs/2003ApJ...592...42G}{ADS}]}.

\bibitem{GhezEtAl05}
Ghez, A.M., Salim, S., Hornstein, S.D., Tanner, A., Lu, J.R., Morris, M.,
  Becklin, E.E.  and Duch{\^ e}ne, G., ``Stellar Orbits around the Galactic
  Center Black Hole'', {\em Astrophys. J.}, {\bf 620}, 744--757 (2005).
  {\small[\href{http://adsabs.harvard.edu/abs/2005ApJ...620..744G}{ADS}]}.

\bibitem{GhezEtAl03b}
Ghez, A.M. {et~al.}, ``The First Measurement of Spectral Lines in a
  Short-Period Star Bound to the Galaxy's Central Black Hole: A Paradox of
  Youth'', {\em Astrophys. J. Lett.}, {\bf 586}, L127--L131 (2003).
  {\small[\href{http://adsabs.harvard.edu/abs/2003ApJ...586L.127G}{ADS}]}.

\bibitem{GhezEtAl08}
Ghez, A.M. {et~al.}, ``Measuring Distance and Properties of the Milky Way's
  Central Supermassive Black Hole with Stellar Orbits'', {\em Astrophys. J.},
  {\bf 689}, 1044--1062 (2008).
  {\small[\href{http://dx.doi.org/10.1086/592738}{DOI}]},
  {\small[\href{http://adsabs.harvard.edu/abs/2008ApJ...689.1044G}{ADS}]},
  {\small[\href{http://arxiv.org/abs/0808.2870}{{arXiv:0808.2870}}]}.

\bibitem{Giersz06}
Giersz, M., ``Monte Carlo simulations of star clusters - III. A million-body
  star cluster'', {\em Mon. Not. R. Astron. Soc.}, {\bf 371}, 484--494 (2006).
  {\small[\href{http://dx.doi.org/10.1111/j.1365-2966.2006.10693.x}{DOI}]},
  {\small[\href{http://adsabs.harvard.edu/abs/2006MNRAS.371..484G}{ADS}]},
  {\small[\href{http://arxiv.org/abs/astro-ph/0512606}{{astro-ph/0512606}}]}.

\bibitem{GH94b}
Giersz, M.  and Heggie, D.C., ``Statistics of N-Body Simulations - Part Two -
  Equal Masses after Core Collapse'', {\em Mon. Not. R. Astron. Soc.}, {\bf
  270}, 298 (1994).
  {\small[\href{http://adsabs.harvard.edu/abs/1994MNRAS.270..298G}{ADS}]}.

\bibitem{GH96}
Giersz, M.  and Heggie, D.C., ``Statistics of $N$-body simulations - {III}.
  Unequal masses'', {\em Mon. Not. R. Astron. Soc.}, {\bf 279}, 1037--1056
  (1996).
  {\small[\href{http://adsabs.harvard.edu/abs/1996MNRAS.279.1037G}{ADS}]}.

\bibitem{GS94}
Giersz, M.  and Spurzem, R., ``Comparing direct $N$-body integration with
  anisotropic gaseous models of star clusters'', {\em Mon. Not. R. Astron.
  Soc.}, {\bf 269}, 241 (1994).
  {\small[\href{http://adsabs.harvard.edu/abs/1994MNRAS.269..241G}{ADS}]}.

\bibitem{GS00}
Giersz, M.  and Spurzem, R., ``A stochastic {M}onte {C}arlo approach to model
  real star cluster evolution - {II}. Self-consistent models and primordial
  binaries'', {\em Mon. Not. R. Astron. Soc.}, {\bf 317}, 581 (2000).
  {\small[\href{http://adsabs.harvard.edu/abs/2000MNRAS.317..581G}{ADS}]}.

\bibitem{GS03}
Giersz, M.  and Spurzem, R., ``A stochastic {M}onte {C}arlo approach to
  modelling real star cluster evolution - {III}. Direct integration of three-
  and four-body interactions'', {\em Mon. Not. R. Astron. Soc.}, {\bf 343},
  781--795 (2003).
  {\small[\href{http://adsabs.harvard.edu/abs/2003MNRAS.343..781G}{ADS}]}.

\bibitem{GillessenEtAl09}
Gillessen, S., Eisenhauer, F., Trippe, S., Alexander, T., Genzel, R., Martins,
  F.  and Ott, T., ``Monitoring Stellar Orbits Around the Massive Black Hole in
  the Galactic Center'', {\em Astrophys. J.}, {\bf 692}, 1075--1109 (2009).
  {\small[\href{http://dx.doi.org/10.1088/0004-637X/692/2/1075}{DOI}]},
  {\small[\href{http://adsabs.harvard.edu/abs/2009ApJ...692.1075G}{ADS}]},
  {\small[\href{http://arxiv.org/abs/0810.4674}{{arXiv:0810.4674}}]}.

\bibitem{GO64}
Ginzburg, V.  and Ozernoy, L.M., {\em Sov. Phys. JETP}, {\bf 20}, 489 (1964).

\bibitem{GongEtAl11}
Gong, X. {et~al.}, ``A scientific case study of an advanced LISA mission'',
  {\em Class. Quantum Grav.}, {\bf 28}, 094012 (2011).
  {\small[\href{http://dx.doi.org/10.1088/0264-9381/28/9/094012}{DOI}]},
  {\small[\href{http://adsabs.harvard.edu/abs/2011CQGra..28i4012G}{ADS}]}.

\bibitem{GongEtAl2015}
{Gong}, X. {et~al.}, ``{Descope of the ALIA mission}'', {\em Journal of Physics
  Conference Series}, {\bf 610}(1), 012011 (May 2015).
  {\small[\href{http://dx.doi.org/10.1088/1742-6596/610/1/012011}{DOI}]},
  {\small[\href{http://adsabs.harvard.edu/abs/2015JPhCS.610a2011G}{ADS}]},
  {\small[\href{http://arxiv.org/abs/1410.7296}{{arXiv:1410.7296
  {\small[gr-qc]}}}]}.

\bibitem{Goodman83}
Goodman, J., ``Core collapse with strong encounters'', {\em Astrophys. J.},
  {\bf 270}, 700--710 (1983).
  {\small[\href{http://adsabs.harvard.edu/abs/1983ApJ...270..700G}{ADS}]}.

\bibitem{Goodman03}
Goodman, J., ``Self-gravity and quasi-stellar object discs'', {\em Mon. Not. R.
  Astron. Soc.}, {\bf 339}, 937--948 (2003).
  {\small[\href{http://dx.doi.org/10.1046/j.1365-8711.2003.06241.x}{DOI}]},
  {\small[\href{http://adsabs.harvard.edu/abs/2003MNRAS.339..937G}{ADS}]},
  {\small[\href{http://arxiv.org/abs/astro-ph/0201001}{{astro-ph/0201001}}]}.

\bibitem{GT04}
Goodman, J.  and Tan, J.C., ``Supermassive Stars in Quasar Disks'', {\em
  Astrophys. J.}, {\bf 608}, 108--118 (2004).
  {\small[\href{http://dx.doi.org/10.1086/386360}{DOI}]},
  {\small[\href{http://adsabs.harvard.edu/abs/2004ApJ...608..108G}{ADS}]},
  {\small[\href{http://arxiv.org/abs/astro-ph/0307361}{{astro-ph/0307361}}]}.

\bibitem{GualandrisMerritt2012}
Gualandris, A.  and Merritt, D., ``Long-term Evolution of Massive Black Hole
  Binaries. IV. Mergers of Galaxies with Collisionally Relaxed Nuclei'', {\em
  Astrophys. J.}, {\bf 744}, 74 (2012).
  {\small[\href{http://dx.doi.org/10.1088/0004-637X/744/1/74}{DOI}]},
  {\small[\href{http://adsabs.harvard.edu/abs/2012ApJ...744...74G}{ADS}]},
  {\small[\href{http://arxiv.org/abs/1107.4095}{{arXiv:1107.4095
  {\small[astro-ph.GA]}}}]}.

\bibitem{GuelketinEtAl09}
{G{\"u}ltekin}, K. {et~al.}, ``{The M-{$\sigma$} and M-L Relations in Galactic
  Bulges, and Determinations of Their Intrinsic Scatter}'', {\bf 698}, 198--221
  (June 2009).
  {\small[\href{http://dx.doi.org/10.1088/0004-637X/698/1/198}{DOI}]},
  {\small[\href{http://adsabs.harvard.edu/abs/2009ApJ...698..198G}{ADS}]},
  {\small[\href{http://arxiv.org/abs/0903.4897}{{arXiv:0903.4897
  {\small[astro-ph.GA]}}}]}.

\bibitem{Gurevich64}
Gurevich, A.V., ``Instability of the disturbed zone in the vicinity of a
  charged body in plasma'', {\em Geomag. Aeronom.}, {\bf 4}, 247--255 (1964).

\bibitem{GFR06}
G{\"u}rkan, M.A., Fregeau, J.M.  and Rasio, F.A., ``Massive Black Hole Binaries
  from Collisional Runaways'', {\em Astrophys. J. Lett.}, {\bf 640}, L39--L42
  (2006). {\small[\href{http://dx.doi.org/10.1086/503295}{DOI}]},
  {\small[\href{http://adsabs.harvard.edu/abs/2006ApJ...640L..39G}{ADS}]},
  {\small[\href{http://arxiv.org/abs/astro-ph/0512642}{{astro-ph/0512642}}]}.

\bibitem{GFR04}
G{\"u}rkan, M.A., Freitag, M.  and Rasio, F.A., ``Formation of Massive Black
  Holes in Dense Star Clusters. {I}. Mass Segregation and Core Collapse'', {\em
  Astrophys. J.}, {\bf 604} (2004).
  {\small[\href{http://adsabs.harvard.edu/abs/2003astro.ph..8449G}{ADS}]},
  {\small[\href{http://arxiv.org/abs/astro-ph/0308449}{{astro-ph/0308449}}]}.

\bibitem{HachisuEtAl78}
Hachisu, I., Nakada, Y., Nomoto, K.  and Sugimoto, D., ``Post-collapse
  evolution of a gaseous cluster model'', {\em Prog. Theor. Phys.}, {\bf 393},
  60 (1978).

\bibitem{Hara78}
Hara, T., ``Evolution of a super-massive star in a dense stellar system'', {\em
  Prog. Theor. Phys.}, {\bf 60}, 711--723 (1978).

\bibitem{HarfstEtAl06}
Harfst, S., Gualandris, A., Merritt, D., Spurzem, R., Portegies~Zwart, S.  and
  Berczik, P., ``Performance Analysis of Direct N-Body Algorithms on
  Special-Purpose Supercomputers'', e-print, (2006).
  {\small[\href{http://adsabs.harvard.edu/abs/2006astro.ph..8125H}{ADS}]},
  {\small[\href{http://arxiv.org/abs/astro-ph/0608125}{{astro-ph/0608125}}]}.

\bibitem{HeggieHut03}
{Heggie}, D.  and {Hut}, P., {\em {The Gravitational Million-Body Problem: A
  Multidisciplinary Approach to Star Cluster Dynamics, by Douglas Heggie and
  Piet Hut.~ Cambridge University Press, 2003, 372 pp.}}, (2003).
  {\small[\href{http://adsabs.harvard.edu/abs/2003gmbp.book.....H}{ADS}]}.

\bibitem{HM86}
{Heggie}, D.~C.  and {Mathieu}, R.~D., ``Standardised Units and Time Scales'',
  in {Hut}, P.  and {McMillan}, S. L.~W., eds., {\em The Use of Supercomputers
  in Stellar Dynamics}, p. 233. Springer-Verlag, (1986).

\bibitem{Helstrom68}
Helstrom, C.W., {\em Statistical Theory of Signal Detection}, (London, 1968).

\bibitem{Henon71b}
H{\'e}non, M., ``The {M}onte {C}arlo Method'', {\em Ap\&SS}, {\bf 14}, 151--167
  (1971). {\small[\href{http://dx.doi.org/10.1007/BF00649201}{DOI}]},
  {\small[\href{http://adsabs.harvard.edu/abs/1971Ap&SS..14..151H}{ADS}]}.

\bibitem{Henon71a}
H{\'e}non, M., ``Monte Carlo Models of Star Clusters'', {\em Astrophys. Space
  Sci.}, {\bf 13}, 284--299 (1971).
  {\small[\href{http://dx.doi.org/10.1007/BF00649159}{DOI}]},
  {\small[\href{http://adsabs.harvard.edu/abs/1971Ap&SS..13..284H}{ADS}]}.

\bibitem{Henon73}
H{\'{e}}non, M., ``Collisional dynamics of spherical stellar systems'', in
  Martinet, L.  and Mayor, M., eds., {\em Dynamical structure and evolution of
  stellar systems, Lectures of the 3rd Advanced Course of the Swiss Society for
  Astronomy and Astrophysics (SSAA)}, pp. 183--260, (1973).
  {\small[\href{http://adsabs.harvard.edu/abs/1973dses.conf.....C}{ADS}]}.

\bibitem{Henon75}
H{\'e}non, M., ``Two Recent Developments Concerning the {M}onte {C}arlo
  Method'', in Hayli, A., ed., {\em IAU Symp. 69: Dynamics of Stellar Systems},
  p. 133, (1975).
  {\small[\href{http://adsabs.harvard.edu/abs/1975IAUS...69..133H}{ADS}]}.

\bibitem{HenyeyEtAl59}
Henyey, L.G., Wilets, L., B{\" o}hm, K.H., Lelevier, R.  and Levee, R.D., ``A
  Method for Atomic Computation of Stellar Evolution'', {\em Astrophys. J.},
  {\bf 129}, 628 (1959).
  {\small[\href{http://adsabs.harvard.edu/abs/1959ApJ...129..628H}{ADS}]}.

\bibitem{HO92}
Hernquist, L.  and Ostriker, J.P., ``A self-consistent field method for
  galactic dynamics'', {\em Astrophys. J.}, {\bf 386}, 375--397 (1992).
  {\small[\href{http://adsabs.harvard.edu/abs/1992ApJ...386..375H}{ADS}]}.

\bibitem{Herrnstein99}
Herrnstein, J.R. {et~al.}, ``A geometric distance to the galaxy NGC 4258 from
  orbital motions in a nuclear gas disk.'', {\em Nature}, {\bf 400}, 539--541
  (1999).
  {\small[\href{http://adsabs.harvard.edu/abs/1999Natur.400..539H}{ADS}]}.

\bibitem{HH98}
Hillenbrand, L.A.  and Hartmann, L.W., ``A Preliminary Study of the Orion
  Nebula Cluster Structure and Dynamics'', {\em Astrophys. J.}, {\bf 492}, 540
  (1998).
  {\small[\href{http://adsabs.harvard.edu/abs/1998ApJ...492..540H}{ADS}]}.

\bibitem{Hills75}
Hills, J.G., ``Possible power source of Seyfert galaxies and QSOs'', {\em
  Nature}, {\bf 254}, 295--298 (1975).
  {\small[\href{http://adsabs.harvard.edu/abs/1975Natur.254..295H}{ADS}]}.

\bibitem{Hills88}
Hills, J.G., ``Hyper-velocity and tidal stars from binaries disrupted by a
  massive Galactic black hole'', {\em Nature}, {\bf 331}, 687--689 (1988).
  {\small[\href{http://adsabs.harvard.edu/abs/1988Natur.331..687H}{ADS}]}.

\bibitem{HB95}
Hils, D.  and Bender, P.L., ``Gradual approach to coalescence for compact stars
  orbiting massive black holes'', {\em Astrophys. J. Lett.}, {\bf 445}, L7--L10
  (1995).
  {\small[\href{http://adsabs.harvard.edu/abs/1995ApJ...445L...7H}{ADS}]}.

\bibitem{Holley-BockelmannEtAl2001}
Holley-Bockelmann, K., Mihos, J.C., Sigurdsson, S.  and Hernquist, L., ``Models
  of Cuspy Triaxial Galaxies'', {\em Astrophys. J.}, {\bf 549}, 862--870
  (2001). {\small[\href{http://dx.doi.org/10.1086/319453}{DOI}]},
  {\small[\href{http://adsabs.harvard.edu/abs/2001ApJ...549..862H}{ADS}]},
  {\small[\href{http://arxiv.org/abs/arXiv:astro-ph/0011504}{{arXiv:astro-ph/0011504}}]}.

\bibitem{Holley-BockelmannEtAl2002}
Holley-Bockelmann, K., Mihos, J.C., Sigurdsson, S., Hernquist, L.  and Norman,
  C., ``The Evolution of Cuspy Triaxial Galaxies Harboring Central Black
  Holes'', {\em Astrophys. J.}, {\bf 567}, 817--827 (2002).
  {\small[\href{http://dx.doi.org/10.1086/338683}{DOI}]},
  {\small[\href{http://adsabs.harvard.edu/abs/2002ApJ...567..817H}{ADS}]},
  {\small[\href{http://arxiv.org/abs/arXiv:astro-ph/0111029}{{arXiv:astro-ph/0111029}}]}.

\bibitem{HA05}
Hopman, C.  and Alexander, T., ``The Orbital Statistics of Stellar Inspiral and
  Relaxation near a Massive Black Hole: Characterizing Gravitational Wave
  Sources'', {\em Astrophys. J.}, {\bf 629}, 362--372 (2005).
  {\small[\href{http://dx.doi.org/10.1086/431475}{DOI}]},
  {\small[\href{http://adsabs.harvard.edu/abs/2005ApJ...629..362H}{ADS}]}.

\bibitem{HopmanAlexander05}
Hopman, C.  and Alexander, T., ``The Orbital Statistics of Stellar Inspiral and
  Relaxation near a Massive Black Hole: Characterizing Gravitational Wave
  Sources'', {\em Astrophys. J.}, {\bf 629}, 362--372 (2005).
  {\small[\href{http://dx.doi.org/10.1086/431475}{DOI}]},
  {\small[\href{http://adsabs.harvard.edu/abs/2005ApJ...629..362H}{ADS}]}.

\bibitem{HA06b}
Hopman, C.  and Alexander, T., ``The Effect of Mass Segregation on
  Gravitational Wave Sources near Massive Black Holes'', {\em Astrophys. J.
  Lett.}, {\bf 645}, L133--L136 (2006).
  {\small[\href{http://dx.doi.org/10.1086/506273}{DOI}]},
  {\small[\href{http://adsabs.harvard.edu/abs/2006ApJ...645L.133H}{ADS}]},
  {\small[\href{http://arxiv.org/abs/astro-ph/0603324}{{astro-ph/0603324}}]}.

\bibitem{HopmanAlexander06}
Hopman, C.  and Alexander, T., ``Resonant Relaxation near a Massive Black Hole:
  The Stellar Distribution and Gravitational Wave Sources'', {\em Astrophys.
  J.}, {\bf 645}, 1152--1163 (2006).
  {\small[\href{http://dx.doi.org/10.1086/504400}{DOI}]},
  {\small[\href{http://adsabs.harvard.edu/abs/2006ApJ...645.1152H}{ADS}]},
  {\small[\href{http://arxiv.org/abs/arXiv:astro-ph/0601161}{{arXiv:astro-ph/0601161}}]}.

\bibitem{HA06}
Hopman, C.  and Alexander, T., ``Resonant Relaxation near a Massive Black Hole:
  The Stellar Distribution and Gravitational Wave Sources'', {\em Astrophys.
  J.}, {\bf 645}, 1152--1163 (2006).
  {\small[\href{http://dx.doi.org/10.1086/504400}{DOI}]},
  {\small[\href{http://adsabs.harvard.edu/abs/2006ApJ...645.1152H}{ADS}]},
  {\small[\href{http://arxiv.org/abs/astro-ph/0601161}{{astro-ph/0601161}}]}.

\bibitem{HopmanFreitagLarson07}
Hopman, C., Freitag, M.  and Larson, S.L., ``Gravitational wave bursts from the
  Galactic massive black hole'', {\em Mon. Not. R. Astron. Soc.}, {\bf 378},
  129--136 (2007).
  {\small[\href{http://dx.doi.org/10.1111/j.1365-2966.2007.11758.x}{DOI}]},
  {\small[\href{http://adsabs.harvard.edu/abs/2007MNRAS.378..129H}{ADS}]},
  {\small[\href{http://arxiv.org/abs/arXiv:astro-ph/0612337}{{arXiv:astro-ph/0612337}}]}.

\bibitem{HuangEtAl2017}
{Huang}, S. {et~al.}, ``{Gravitational wave detection in space--a new window in
  astronomy}'', {\em Scientia Sinica Physica, Mechanica {\&} Astronomica}, {\bf
  47}(1), 010404 (January 2017).
  {\small[\href{http://dx.doi.org/10.1360/SSPMA2016-00438}{DOI}]},
  {\small[\href{http://adsabs.harvard.edu/abs/2017SSPMA..47a0404H}{ADS}]}.

\bibitem{HypkiGiersz2013}
{Hypki}, A.  and {Giersz}, M., ``{MOCCA code for star cluster simulations - I.
  Blue stragglers, first results}'', {\bf 429}, 1221--1243 (February 2013).
  {\small[\href{http://dx.doi.org/10.1093/mnras/sts415}{DOI}]},
  {\small[\href{http://adsabs.harvard.edu/abs/2013MNRAS.429.1221H}{ADS}]},
  {\small[\href{http://arxiv.org/abs/1207.6700}{{arXiv:1207.6700
  {\small[astro-ph.GA]}}}]}.

\bibitem{IS85}
Inagaki, S.  and Saslaw, W.C., ``Equipartition in multicomponent gravitational
  systems'', {\em Astrophys. J.}, {\bf 292}, 339--347 (1985).
  {\small[\href{http://adsabs.harvard.edu/abs/1985ApJ...292..339I}{ADS}]}.

\bibitem{IW84}
Inagaki, S.  and Wiyanto, P., ``On equipartition of kinetic energies in
  two-component star clusters'', {\em Publ. Astron. Soc. Japan}, {\bf 36},
  391--402 (1984).
  {\small[\href{http://adsabs.harvard.edu/abs/1984PASJ...36..391I}{ADS}]}.

\bibitem{Ivanov02}
Ivanov, P.B., ``On the formation rate of close binaries consisting of a
  super-massive black hole and a white dwarf'', {\em Mon. Not. R. Astron.
  Soc.}, {\bf 336}, 373--381 (2002).
  {\small[\href{http://adsabs.harvard.edu/abs/2002MNRAS.336..373I}{ADS}]}.

\bibitem{Jeans15}
Jeans, J.H., ``On the theory of star-streaming and the structure of the
  universe'', {\em Mon. Not. R. Astron. Soc.}, {\bf 76}, 70--84 (1915).
  {\small[\href{http://adsabs.harvard.edu/abs/1915MNRAS..76...70J}{ADS}]}.

\bibitem{JNR01}
Joshi, K.J., Nave, C.P.  and Rasio, F.A., ``Monte Carlo Simulations of Globular
  Cluster Evolution. II. Mass Spectra, Stellar Evolution, and Lifetimes in the
  Galaxy'', {\em Astrophys. J.}, {\bf 550}, 691--702 (2001).
  {\small[\href{http://adsabs.harvard.edu/abs/2001ApJ...550..691J}{ADS}]}.

\bibitem{JRPZ00}
Joshi, K.J., Rasio, F.A.  and Portegies~Zwart, S., ``Monte Carlo Simulations of
  Globular Cluster Evolution. I. Method and Test Calculations'', {\em
  Astrophys. J.}, {\bf 540}, 969--982 (2000).
  {\small[\href{http://adsabs.harvard.edu/abs/2000ApJ...540..969J}{ADS}]}.

\bibitem{KS01}
Karas, V.  and {\v S}ubr, L., ``Orbital decay of satellites crossing an
  accretion disc'', {\em Astron. Astrophys.}, {\bf 376}, 686--696 (2001).
  {\small[\href{http://dx.doi.org/10.1051/0004-6361:20011009}{DOI}]},
  {\small[\href{http://adsabs.harvard.edu/abs/2001A&A...376..686K}{ADS}]},
  {\small[\href{http://arxiv.org/abs/astro-ph/0107232}{{astro-ph/0107232}}]}.

\bibitem{KhalEtAl07}
Khalisi, E., Amaro-Seoane, P.  and Spurzem, R., ``A comprehensive NBODY study
  of mass segregation in star clusters: energy equipartition and escape'', {\em
  Mon. Not. R. Astron. Soc.}, {\bf 374}, 703--720 (2007).
  {\small[\href{http://dx.doi.org/10.1111/j.1365-2966.2006.11184.x}{DOI}]},
  {\small[\href{http://adsabs.harvard.edu/abs/2007MNRAS.374..703K}{ADS}]},
  {\small[\href{http://arxiv.org/abs/arXiv:astro-ph/0602570}{{arXiv:astro-ph/0602570}}]}.

\bibitem{KLG98}
Kim, S.S., Lee, H.M.  and Goodman, J., ``Two-Component Fokker-Planck Models for
  the Evolution of Isolated Globular Clusters'', {\em Astrophys. J.}, {\bf
  495}, 786 (1998).
  {\small[\href{http://adsabs.harvard.edu/abs/1998ApJ...495..786K}{ADS}]}.

\bibitem{KLOP05}
King, A.R., Lubow, S.H., Ogilvie, G.I.  and Pringle, J.E., ``Aligning spinning
  black holes and accretion discs'', {\em Mon. Not. R. Astron. Soc.}, {\bf
  363}, 49--56 (2005).
  {\small[\href{http://dx.doi.org/10.1111/j.1365-2966.2005.09378.x}{DOI}]},
  {\small[\href{http://adsabs.harvard.edu/abs/2005MNRAS.363...49K}{ADS}]},
  {\small[\href{http://arxiv.org/abs/arXiv:astro-ph/0507098}{{arXiv:astro-ph/0507098}}]}.

\bibitem{KW94}
Kippenhahn, R.  and Weigert, A., {\em Stellar Structure and Evolution},
  (Springer-Verlag Berlin Heidelberg, 1994).
  {\small[\href{http://adsabs.harvard.edu/abs/1994sse..book.....K}{ADS}]}.

\bibitem{KocsisEtAl11}
Kocsis, B., Yunes, N.  and Loeb, A., ``Observable signatures of extreme
  mass-ratio inspiral black hole binaries embedded in thin accretion disks'',
  {\em Phys. Rev. D}, {\bf 84}, 024032 (2011).
  {\small[\href{http://dx.doi.org/10.1103/PhysRevD.84.024032}{DOI}]},
  {\small[\href{http://adsabs.harvard.edu/abs/2011PhRvD..84b4032K}{ADS}]},
  {\small[\href{http://arxiv.org/abs/1104.2322}{{arXiv:1104.2322
  {\small[astro-ph.GA]}}}]}.

\bibitem{KomossaEtAl2004}
Komossa, S., Halpern, J., Schartel, N., Hasinger, G., Santos-Lleo, M.  and
  Predehl, P., ``A Huge Drop in the X-Ray Luminosity of the Nonactive Galaxy RX
  J1242.6-1119A, and the First Postflare Spectrum: Testing the Tidal Disruption
  Scenario'', {\bf 603}, L17--L20 (2004).
  {\small[\href{http://dx.doi.org/10.1086/382046}{DOI}]},
  {\small[\href{http://adsabs.harvard.edu/abs/2004ApJ...603L..17K}{ADS}]},
  {\small[\href{http://arxiv.org/abs/arXiv:astro-ph/0402468}{{arXiv:astro-ph/0402468}}]}.

\bibitem{KongEtAl09}
Kong, A.K.H., Heinke, C.O., Di~Stefano, R., Barmby, P., Lewin, W.H.G.  and
  Primini, F.A., ``X-Ray Localization of the Intermediate-Mass Black Hole in
  the Globular Cluster G1 with Chandra'', {\em arXiv}, e-print, (2009).
  {\small[\href{http://adsabs.harvard.edu/abs/2009arXiv0910.3944K}{ADS}]},
  {\small[\href{http://arxiv.org/abs/0910.3944}{{arXiv:0910.3944}}]}.

\bibitem{Kormendy03}
Kormendy, J., ``The Stellar-Dynamical Search for Supermassive Black Holes in
  Galactic Nuclei'', in Ho, L., ed., {\em Coevolution of Black Holes and
  Galaxies, Carnegie Observatories, Pasadena}, (2003).
  {\small[\href{http://arxiv.org/abs/astro-ph/0306353}{{astro-ph/0306353}}]}.

\bibitem{Kormendy04}
Kormendy, J., ``The Stellar-Dynamical Search for Supermassive Black Holes in
  Galactic Nuclei'', {\em Coevolution of Black Holes and Galaxies}, 1 (2004).
  {\small[\href{http://adsabs.harvard.edu/abs/2004cbhg.symp....1K}{ADS}]},
  {\small[\href{http://arxiv.org/abs/arXiv:astro-ph/0306353}{{arXiv:astro-ph/0306353}}]}.

\bibitem{KormendyHo2013}
{Kormendy}, J.  and {Ho}, L.~C., ``{Coevolution (Or Not) of Supermassive Black
  Holes and Host Galaxies}'', {\bf 51}, 511--653 (August 2013).
  {\small[\href{http://dx.doi.org/10.1146/annurev-astro-082708-101811}{DOI}]},
  {\small[\href{http://adsabs.harvard.edu/abs/2013ARA%26A..51..511K}{ADS}]},
  {\small[\href{http://arxiv.org/abs/1304.7762}{{arXiv:1304.7762
  {\small[astro-ph.CO]}}}]}.

\bibitem{Krolik1999}
{Krolik}, J.~H., {\em {Active galactic nuclei : from the central black hole to
  the galactic environment}}, (1999).
  {\small[\href{http://adsabs.harvard.edu/abs/1999agnc.book.....K}{ADS}]}.

\bibitem{Kroupa01}
Kroupa, P., ``On the variation of the initial mass function'', {\em Mon. Not.
  R. Astron. Soc.}, {\bf 322}, 231--246 (2001).
  {\small[\href{http://adsabs.harvard.edu/abs/2001MNRAS.322..231K}{ADS}]}.

\bibitem{KroupaEtAl93}
Kroupa, P., Tout, C.A.  and Gilmore, G., ``The distribution of low-mass stars
  in the Galactic disc'', {\em Mon. Not. R. Astron. Soc.}, {\bf 262}, 545--587
  (1993).
  {\small[\href{http://adsabs.harvard.edu/abs/1993MNRAS.262..545K}{ADS}]}.

\bibitem{KupiEtAl06}
Kupi, G., Amaro-Seoane, P.  and Spurzem, R., ``Dynamics of compact objects
  clusters: A post-Newtonian study'', {\em arXiv}, e-print, (2006).
  {\small[\href{http://arxiv.org/abs/astro-ph/0602125}{{arXiv:astro-ph/0602125}}]}.

\bibitem{KS65}
Kustaanheimo, P.E.  and Stiefel, E.L., ``Perturbation theory of Kepler motion
  based on spinor regularization'', {\em J. Reine Angew. Math.}, {\bf 218},
  204--219 (1965).

\bibitem{LangbeinEtAl90}
Langbein, T., Fricke, K.J., Spurzem, R.  and Yorke, H.W., ``Interactions
  between stars and gas in galactic nuclei'', {\em aap}, {\bf 227}, 333--341
  (1990).
  {\small[\href{http://adsabs.harvard.edu/abs/1990A&A...227..333L}{ADS}]}.

\bibitem{Larson03}
Larson, S.L., ``Online Sensitivity Curve Generator'',  (2003)URL:
  \newline\url{http://www.srl.caltech.edu/~shane/sensitivity/}.

\bibitem{LHH00}
Larson, S.L., Hiscock, W.A.  and Hellings, R.W., ``Sensitivity curves for
  spaceborne gravitational wave interferometers'', {\em Phys. Rev. D}, {\bf
  62}, 062001 (2000).
  {\small[\href{http://adsabs.harvard.edu/abs/2000PhRvD..62f2001L}{ADS}]}.

\bibitem{Lauer98}
Lauer, T.R., Faber, S.M., Ajhar, E.A., Grillmair, C.J.  and Scowen, P.A., ``M32
  +/- 1'', {\em AJ}, {\bf 116}, 2263--2286 (1998).
  {\small[\href{http://adsabs.harvard.edu/abs/1998AJ....116.2263L}{ADS}]}.

\bibitem{Levin03}
Levin, Y., ``Formation of massive stars and black holes in self-gravitating AGN
  discs, and gravitational waves in LISA band'', {\em arXiv}, e-print, (2003).
  {\small[\href{http://arxiv.org/abs/astro-ph/0307084}{{arXiv:astro-ph/0307084}}]}.

\bibitem{Levin06}
Levin, Y., ``Starbursts near supermassive black holes: young stars in the
  Galactic Center, and gravitational waves in LISA band'', {\em arXiv},
  e-print, (2006).
  {\small[\href{http://adsabs.harvard.edu/abs/2006astro.ph..3583L}{ADS}]},
  {\small[\href{http://arxiv.org/abs/astro-ph/0603583}{{arXiv:astro-ph/0603583}}]}.

\bibitem{LB03}
Levin, Y.  and Beloborodov, A.M., ``Stellar Disk in the Galactic Center: A
  Remnant of a Dense Accretion Disk?'', {\em Astrophys. J. Lett.}, {\bf 590},
  L33--L36 (2003).
  {\small[\href{http://adsabs.harvard.edu/abs/2003ApJ...590L..33L}{ADS}]}.

\bibitem{LF78}
Lightman, A.P.  and Fall, S.M., ``An approximate theory for the core collapse
  of two-component gravitating systems'', {\em Astrophys. J.}, {\bf 221},
  567--579 (1978).
  {\small[\href{http://adsabs.harvard.edu/abs/1978ApJ...221..567L}{ADS}]}.

\bibitem{LS77}
Lightman, A.P.  and Shapiro, S.L., ``The distribution and consumption rate of
  stars around a massive, collapsed object'', {\em Astrophys. J.}, {\bf 211},
  244--262 (1977).
  {\small[\href{http://adsabs.harvard.edu/abs/1977ApJ...211..244L}{ADS}]}.

\bibitem{LT80}
Lin, D.N.~C.  and Tremaine, S., ``A reinvestigation of the standard model for
  the dynamics of a massive black hole in a globular cluster'', {\em Astrophys.
  J.}, {\bf 242}, 789--798 (1980).
  {\small[\href{http://adsabs.harvard.edu/abs/1980ApJ...242..789L}{ADS}]}.

\bibitem{LS91}
Louis, P.D.  and Spurzem, R., ``Anisotropic gaseous models for the evolution of
  star clusters'', {\em Mon. Not. R. Astron. Soc.}, {\bf 251}, 408--426 (1991).
  {\small[\href{http://adsabs.harvard.edu/abs/1991MNRAS.251..408L}{ADS}]}.

\bibitem{LyndenBell67}
Lynden-Bell, D., ``Statistical mechanics of violent relaxation in stellar
  systems'', {\em Mon. Not. R. Astron. Soc.}, {\bf 136}, 101 (1967).
  {\small[\href{http://adsabs.harvard.edu/abs/1967MNRAS.136..101L}{ADS}]}.

\bibitem{LB69}
Lynden-Bell, D., ``Galactic Nuclei as Collapsed Old Quasars'', {\em Nature},
  {\bf 223}, 690 (1969).
  {\small[\href{http://adsabs.harvard.edu/abs/1969Natur.223..690L}{ADS}]}.

\bibitem{LBE80}
Lynden-Bell, D.  and Eggleton, P.P., ``On the consequences of the gravothermal
  catastrophe'', {\em Mon. Not. R. Astron. Soc.}, {\bf 191}, 483--498 (1980).
  {\small[\href{http://adsabs.harvard.edu/abs/1980MNRAS.191..483L}{ADS}]}.

\bibitem{LR71}
Lynden-Bell, D.  and Rees, M.J., ``On quasars, dust and the galactic centre'',
  {\em Mon. Not. R. Astron. Soc.}, {\bf 152}, 461 (1971).
  {\small[\href{http://adsabs.harvard.edu/abs/1971MNRAS.152..461L}{ADS}]}.

\bibitem{Lynden-BellWood1968}
{Lynden-Bell}, D.  and {Wood}, R., ``{The gravo-thermal catastrophe in
  isothermal spheres and the onset of red-giant structure for stellar
  systems}'', {\bf 138}, 495 (1968).
  {\small[\href{http://dx.doi.org/10.1093/mnras/138.4.495}{DOI}]},
  {\small[\href{http://adsabs.harvard.edu/abs/1968MNRAS.138..495L}{ADS}]}.

\bibitem{MT99}
Magorrian, J.  and Tremaine, S., ``Rates of tidal disruption of stars by
  massive central black holes'', {\em Mon. Not. R. Astron. Soc.}, {\bf 309},
  447--460 (1999).
  {\small[\href{http://adsabs.harvard.edu/abs/1999MNRAS.309..447M}{ADS}]}.

\bibitem{Makino96}
Makino, J., ``Postcollapse Evolution of Globular Clusters'', {\em Astrophys.
  J.}, {\bf 471}, 796 (1996).
  {\small[\href{http://adsabs.harvard.edu/abs/1996ApJ...471..796M}{ADS}]}.

\bibitem{Makino1992}
{Makino}, J.  and {Aarseth}, S.~J., ``{On a Hermite integrator with Ahmad-Cohen
  scheme for gravitational many-body problems}'', {\em Publ. Astron. Soc.
  Japan}, {\bf 44}, 141--151 (April 1992).
  {\small[\href{http://adsabs.harvard.edu/abs/1992PASJ...44..141M}{ADS}]}.

\bibitem{MS79}
Marchant, A.B.  and Shapiro, S.L., ``Star clusters containing massive, central
  black holes. {II} - Self-consistent potentials'', {\em Astrophys. J.}, {\bf
  234}, 317--328 (1979).
  {\small[\href{http://adsabs.harvard.edu/abs/1979ApJ...234..317M}{ADS}]}.

\bibitem{MS80}
Marchant, A.B.  and Shapiro, S.L., ``Star clusters containing massive, central
  black holes. {III} - Evolution calculations'', {\em Astrophys. J.}, {\bf
  239}, 685--704 (1980).
  {\small[\href{http://adsabs.harvard.edu/abs/1980ApJ...239..685M}{ADS}]}.

\bibitem{McMS94}
McCaughrean, M.J.  and Stauffer, J.R., ``High resolution near-infrared imaging
  of the trapezium: A stellar census'', {\em AJ}, {\bf 108}, 1382--1397 (1994).
  {\small[\href{http://adsabs.harvard.edu/abs/1994AJ....108.1382M}{ADS}]}.

\bibitem{Merritt99}
Merritt, D., ``Elliptical Galaxy Dynamics'', {\em Publ. Astron. Soc. Japan},
  {\bf 111}, 129--168 (1999).
  {\small[\href{http://dx.doi.org/10.1086/316307}{DOI}]},
  {\small[\href{http://adsabs.harvard.edu/abs/1999PASP..111..129M}{ADS}]},
  {\small[\href{http://arxiv.org/abs/arXiv:astro-ph/9810371}{{arXiv:astro-ph/9810371}}]}.

\bibitem{Merritt06}
Merritt, D., ``Dynamics of galaxy cores and supermassive black holes'', {\em
  Reports of Progress in Physics}, {\bf 69}, 2513--2579 (2006).
  {\small[\href{http://adsabs.harvard.edu/abs/2006RPPh...69.2513M}{ADS}]},
  {\small[\href{http://arxiv.org/abs/astro-ph/0605070}{{astro-ph/0605070}}]}.

\bibitem{Merritt2010}
Merritt, D., ``The Distribution of Stars and Stellar Remnants at the Galactic
  Center'', {\em Astrophys. J.}, {\bf 718}, 739--761 (2010).
  {\small[\href{http://dx.doi.org/10.1088/0004-637X/718/2/739}{DOI}]},
  {\small[\href{http://adsabs.harvard.edu/abs/2010ApJ...718..739M}{ADS}]},
  {\small[\href{http://arxiv.org/abs/0909.1318}{{arXiv:0909.1318
  {\small[astro-ph.GA]}}}]}.

\bibitem{MerrittAlexanderMikkolaWill2011}
Merritt, D., Alexander, T., Mikkola, S.  and Will, C.M., ``Stellar dynamics of
  extreme-mass-ratio inspirals'', {\em Phys. Rev. D}, {\bf 84}, 044024 (2011).
  {\small[\href{http://dx.doi.org/10.1103/PhysRevD.84.044024}{DOI}]},
  {\small[\href{http://adsabs.harvard.edu/abs/2011PhRvD..84d4024M}{ADS}]},
  {\small[\href{http://arxiv.org/abs/1102.3180}{{arXiv:1102.3180
  {\small[astro-ph.CO]}}}]}.

\bibitem{MerrittEtAl11}
Merritt, D., Alexander, T., Mikkola, S.  and Will, C., ``Stellar Dynamics of
  Extreme-Mass-Ratio Inspirals'', {\em arXiv}, e-print, (2011).
  {\small[\href{http://adsabs.harvard.edu/abs/2011arXiv1102.3180M}{ADS}]},
  {\small[\href{http://arxiv.org/abs/1102.3180}{{arXiv:1102.3180
  {\small[astro-ph.CO]}}}]}.

\bibitem{MFJ01}
Merritt, D., Ferrarese, L.  and Joseph, C.L., ``No Supermassive Black Hole in
  M33?'', {\em Science}, {\bf 293}, 1116--1119 (2001).
  {\small[\href{http://adsabs.harvard.edu/abs/2001Sci...293.1116M}{ADS}]}.

\bibitem{MHB06}
Merritt, D., Harfst, S.  and Bertone, G., ``Collisionally Regenerated Dark
  Matter Structures in Galactic Nuclei'', {\em arXiv}, e-print, (2006).
  {\small[\href{http://adsabs.harvard.edu/abs/2006astro.ph.10425M}{ADS}]},
  {\small[\href{http://arxiv.org/abs/astro-ph/0610425}{{arXiv:astro-ph/0610425}}]}.

\bibitem{MP04}
Merritt, D.  and Poon, M.Y., ``Chaotic Loss Cones and Black Hole Fueling'',
  {\em Astrophys. J.}, {\bf 606}, 788--798 (2004).
  {\small[\href{http://adsabs.harvard.edu/abs/2004ApJ...606..788M}{ADS}]}.

\bibitem{MerrittVasiliev11}
Merritt, D.  and Vasiliev, E., ``Orbits Around Black Holes in Triaxial
  Nuclei'', {\em Astrophys. J.}, {\bf 726}, 61 (2011).
  {\small[\href{http://dx.doi.org/10.1088/0004-637X/726/2/61}{DOI}]},
  {\small[\href{http://adsabs.harvard.edu/abs/2011ApJ...726...61M}{ADS}]},
  {\small[\href{http://arxiv.org/abs/1005.0040}{{arXiv:1005.0040
  {\small[astro-ph.GA]}}}]}.

\bibitem{Michell1784}
Michell, J., ``On the Means of Discovering the Distance, Magnitude, \&c. of the
  Fixed Stars...'', {\em Philos. Trans. R. Soc. London}, {\bf 74}, 35--57
  (1784). {\small[\href{http://dx.doi.org/10.1098/rstl.1784.0008}{DOI}]},
  {\small[\href{http://adsabs.harvard.edu/abs/1784RSPT...74...35M}{ADS}]}.

\bibitem{Mikkola1997}
{Mikkola}, S., ``{Numerical Treatment of Small Stellar Systems with
  Binaries}'', {\em Celestial Mechanics and Dynamical Astronomy}, {\bf 68},
  87--104 (May 1997).
  {\small[\href{http://dx.doi.org/10.1023/A:1008291715719}{DOI}]},
  {\small[\href{http://adsabs.harvard.edu/abs/1997CeMDA..68...87M}{ADS}]}.

\bibitem{MA02}
Mikkola, S.  and Aarseth, S., ``A Time-Transformed Leapfrog Scheme'', {\em
  Celest. Mech. Dyn. Astron.}, {\bf 84}, 343--354 (2002).
  {\small[\href{http://adsabs.harvard.edu/abs/2002CeMDA..84..343M}{ADS}]}.

\bibitem{MM06}
Mikkola, S.  and Merritt, D., ``Algorithmic regularization with
  velocity-dependent forces'', {\em Mon. Not. R. Astron. Soc.}, {\bf 372},
  219--223 (2006).
  {\small[\href{http://dx.doi.org/10.1111/j.1365-2966.2006.10854.x}{DOI}]},
  {\small[\href{http://adsabs.harvard.edu/abs/2006MNRAS.372..219M}{ADS}]},
  {\small[\href{http://arxiv.org/abs/astro-ph/0605054}{{astro-ph/0605054}}]}.

\bibitem{MC03}
Miller, M.C.  and Colbert, E.J.M., ``Intermediate-Mass Black Holes'', {\em
  International Journal of Modern Physics D}, {\bf 13}, 1--64 (2004).
  {\small[\href{http://dx.doi.org/10.1142/S0218271804004426}{DOI}]},
  {\small[\href{http://adsabs.harvard.edu/abs/2004IJMPD..13....1M}{ADS}]},
  {\small[\href{http://arxiv.org/abs/astro-ph/0308402}{{astro-ph/0308402}}]}.

\bibitem{MFHL05}
Miller, M.C., Freitag, M., Hamilton, D.P.  and Lauburg, V.M., ``Binary
  Encounters with Supermassive Black Holes: Zero-Eccentricity LISA Events'',
  {\em Astrophys. J. Lett.}, {\bf 631}, L117--L120 (2005).
  {\small[\href{http://dx.doi.org/10.1086/497335}{DOI}]},
  {\small[\href{http://adsabs.harvard.edu/abs/2005ApJ...631L.117M}{ADS}]}.

\bibitem{MillerEtAl05}
Miller, M.C., Freitag, M., Hamilton, D.P.  and Lauburg, V.M., ``Binary
  Encounters with Supermassive Black Holes: Zero-Eccentricity LISA Events'',
  {\bf 631}, L117--L120 (2005).
  {\small[\href{http://dx.doi.org/10.1086/497335}{DOI}]},
  {\small[\href{http://adsabs.harvard.edu/abs/2005ApJ...631L.117M}{ADS}]}.

\bibitem{ML04}
Milosavljevi{\'c}, M.  and Loeb, A., ``The Link between Warm Molecular Disks in
  Maser Nuclei and Star Formation near the Black Hole at the Galactic Center'',
  {\em Astrophys. J. Lett.}, {\bf 604}, L45--L48 (2004).
  {\small[\href{http://dx.doi.org/10.1086/383467}{DOI}]},
  {\small[\href{http://adsabs.harvard.edu/abs/2004ApJ...604L..45M}{ADS}]},
  {\small[\href{http://arxiv.org/abs/astro-ph/0401221}{{astro-ph/0401221}}]}.

\bibitem{MM01}
Milosavljevi{\'{c}}, M.  and Merritt, D., ``Formation of Galactic Nuclei'',
  {\em Astrophys. J.}, {\bf 563}, 34--62 (2001).
  {\small[\href{http://adsabs.harvard.edu/abs/2001ApJ...563...34M}{ADS}]},
  {\small[\href{http://arxiv.org/abs/astro-ph/0103350}{{astro-ph/0103350}}]}.

\bibitem{MM03}
Milosavljevi{\'{c}}, M.  and Merritt, D., ``Long-Term Evolution of Massive
  Black Hole Binaries'', {\em Astrophys. J.}, {\bf 596}, 860--878 (2003).
  {\small[\href{http://adsabs.harvard.edu/abs/2003ApJ...596..860M}{ADS}]}.

\bibitem{MEG00}
Miralda-Escud{\'e}, J.  and Gould, A., ``A Cluster of Black Holes at the
  Galactic Center'', {\em Astrophys. J.}, {\bf 545}, 847--853 (2000).
  {\small[\href{http://adsabs.harvard.edu/abs/2000ApJ...545..847M}{ADS}]},
  {\small[\href{http://arxiv.org/abs/astro-ph/0003269}{{astro-ph/0003269}}]}.

\bibitem{MisnerThorneWheeler1973}
{Misner}, C.~W., {Thorne}, K.~S.  and {Wheeler}, J.~A., {\em {Gravitation}},
  (1973).
  {\small[\href{http://adsabs.harvard.edu/abs/1973grav.book.....M}{ADS}]}.

\bibitem{Miyoshi95}
Miyoshi, M., Moran, J., Herrnstein, J., Greenhill, L., Nakai, N., Diamond, P.
  and Inoue, M., ``Evidence for a Black-Hole from High Rotation Velocities in a
  Sub-Parsec Region of NGC4258'', {\em Nature}, {\bf 373}, 127 (1995).
  {\small[\href{http://adsabs.harvard.edu/abs/1995Natur.373..127M}{ADS}]}.

\bibitem{MontgomeryEtAl2009}
{Montgomery}, C., {Orchiston}, W.  and {Whittingham}, I., ``{Michell, Laplace
  and the origin of the black hole concept}'', {\em Journal of Astronomical
  History and Heritage}, {\bf 12}, 90--96 (July 2009).
  {\small[\href{http://adsabs.harvard.edu/abs/2009JAHH...12...90M}{ADS}]}.

\bibitem{MoranEtAl99}
Moran, J.M., Greenhill, L.J.  and Herrnstein, J.R., ``Observational Evidence
  for Massive Black Holes in the Centers of Active Galaxies'', {\em J.
  Astrophys. Astron.}, {\bf 20}, 165 (1999).
  {\small[\href{http://adsabs.harvard.edu/abs/1999JApA...20..165M}{ADS}]}.

\bibitem{MouawadEtAl05}
Mouawad, N., Eckart, A., Pfalzner, S., Sch{\" o}del, R., Moultaka, J.  and
  Spurzem, R., ``Weighing the cusp at the Galactic Centre'', {\em Astron.
  Nachr.}, {\bf 326}, 83--95 (2005).
  {\small[\href{http://dx.doi.org/10.1002/ansa.200410351}{DOI}]},
  {\small[\href{http://adsabs.harvard.edu/abs/2005AN....326...83M}{ADS}]}.

\bibitem{MCD91}
Murphy, B.W., Cohn, H.N.  and Durisen, R.H., ``Dynamical and luminosity
  evolution of active galactic nuclei - Models with a mass spectrum'', {\em
  Astrophys. J.}, {\bf 370}, 60--77 (1991).
  {\small[\href{http://adsabs.harvard.edu/abs/1991ApJ...370...60M}{ADS}]}.

\bibitem{Nayakshin06}
Nayakshin, S., ``Massive stars in subparsec rings around galactic centres'',
  {\em Mon. Not. R. Astron. Soc.}, {\bf 372}, 143--150 (2006).
  {\small[\href{http://dx.doi.org/10.1111/j.1365-2966.2006.10772.x}{DOI}]},
  {\small[\href{http://adsabs.harvard.edu/abs/2006MNRAS.372..143N}{ADS}]},
  {\small[\href{http://arxiv.org/abs/astro-ph/0512255}{{astro-ph/0512255}}]}.

\bibitem{Ostriker00}
Ostriker, J.P., ``Collisional Dark Matter and the Origin of Massive Black
  Holes'', {\em Phys. Rev. Lett.}, {\bf 84}, 5258--5260 (2000).
  {\small[\href{http://dx.doi.org/10.1103/PhysRevLett.84.5258}{DOI}]},
  {\small[\href{http://adsabs.harvard.edu/abs/2000PhRvL..84.5258O}{ADS}]},
  {\small[\href{http://arxiv.org/abs/arXiv:astro-ph/9912548}{{arXiv:astro-ph/9912548}}]}.

\bibitem{PattabiramanEtAl2013}
{Pattabiraman}, B., {Umbreit}, S., {Liao}, W.-k., {Choudhary}, A., {Kalogera},
  V., {Memik}, G.  and {Rasio}, F.~A., ``{A Parallel Monte Carlo Code for
  Simulating Collisional N-body Systems}'', {\bf 204}, 15 (February 2013).
  {\small[\href{http://dx.doi.org/10.1088/0067-0049/204/2/15}{DOI}]},
  {\small[\href{http://adsabs.harvard.edu/abs/2013ApJS..204...15P}{ADS}]},
  {\small[\href{http://arxiv.org/abs/1206.5878}{{arXiv:1206.5878
  {\small[astro-ph.IM]}}}]}.

\bibitem{Peebles72}
Peebles, P.J.E., ``Star Distribution Near a Collapsed Object'', {\em Astrophys.
  J.}, {\bf 178}, 371--376 (1972).
  {\small[\href{http://adsabs.harvard.edu/abs/1972ApJ...178..371P}{ADS}]}.

\bibitem{Peters64}
Peters, P.C., ``Gravitational Radiation and the Motion of Two Point Masses'',
  {\em Phys. Rev.}, {\bf 136}, 1224--1232 (1964).
  {\small[\href{http://adsabs.harvard.edu/abs/1964PhRv..136.1224P}{ADS}]}.

\bibitem{PM63}
Peters, P.C.  and Mathews, J., ``Gravitational Radiation from Point Masses in a
  Keplerian Orbit'', {\em Phys. Rev.}, {\bf 131}, 435--440 (1963).
  {\small[\href{http://adsabs.harvard.edu/abs/1963PhRv..131..435P}{ADS}]}.

\bibitem{PetiteauEtAl08}
Petiteau, A., Auger, G., Halloin, H., Jeannin, O., Plagnol, E., Pireaux, S.,
  Regimbau, T.  and Vinet, J.-Y., ``LISACode: A scientific simulator of LISA'',
  {\em Phys. Rev. D}, {\bf 77}, 023002 (2008).
  {\small[\href{http://dx.doi.org/10.1103/PhysRevD.77.023002}{DOI}]},
  {\small[\href{http://adsabs.harvard.edu/abs/2008PhRvD..77b3002P}{ADS}]},
  {\small[\href{http://arxiv.org/abs/0802.2023}{{arXiv:0802.2023}}]}.

\bibitem{Phinney02}
Phinney, E.S., {\em LISA Science Requirements}, (2002).

\bibitem{Phinney1989}
{Phinney}, E.~S., ``{Manifestations of a Massive Black Hole in the Galactic
  Center}'', in {Morris}, M., ed., {\em The Center of the Galaxy}, IAU
  Symposium, 136, p. 543, (1989).
  {\small[\href{http://adsabs.harvard.edu/abs/1989IAUS..136..543P}{ADS}]}.

\bibitem{PinkneyEtAl03}
Pinkney, J. {et~al.}, ``Kinematics of 10 Early-Type Galaxies from Hubble Space
  Telescope and Ground-based Spectroscopy'', {\em Astrophys. J.}, {\bf 596},
  903--929 (2003).
  {\small[\href{http://adsabs.harvard.edu/abs/2003ApJ...596..903P}{ADS}]}.

\bibitem{PoonMerritt01}
Poon, M.Y.  and Merritt, D., ``Orbital Structure of Triaxial Black Hole
  Nuclei'', {\em Astrophys. J.}, {\bf 549}, 192--204 (2001).
  {\small[\href{http://dx.doi.org/10.1086/319060}{DOI}]},
  {\small[\href{http://adsabs.harvard.edu/abs/2001ApJ...549..192P}{ADS}]}.

\bibitem{PortegiesZwartEtAl06}
Portegies~Zwart, S.F., Baumgardt, H., McMillan, S.L.W., Makino, J., Hut, P.
  and Ebisuzaki, T., ``The Ecology of Star Clusters and Intermediate-Mass Black
  Holes in the Galactic Bulge'', {\em Astrophys. J.}, {\bf 641}, 319--326
  (2006). {\small[\href{http://dx.doi.org/10.1086/500361}{DOI}]},
  {\small[\href{http://adsabs.harvard.edu/abs/2006ApJ...641..319P}{ADS}]},
  {\small[\href{http://arxiv.org/abs/arXiv:astro-ph/0511397}{{arXiv:astro-ph/0511397}}]}.

\bibitem{PZMM00}
Portegies~Zwart, S.F.  and McMillan, S.L.W., ``Black Hole Mergers in the
  Universe'', {\bf 528}, L17--L20 (2000).
  {\small[\href{http://adsabs.harvard.edu/abs/2000ApJ...528L..17P}{ADS}]}.

\bibitem{PortegiesZwartEtAl01}
Portegies~Zwart, S.F., McMillan, S.L.W., Hut, P.  and Makino, J., ``Star
  cluster ecology - IV. Dissection of an open star cluster: photometry'', {\em
  Mon. Not. R. Astron. Soc.}, {\bf 321}, 199--226 (2001).
  {\small[\href{http://adsabs.harvard.edu/abs/2001MNRAS.321..199P}{ADS}]},
  {\small[\href{http://arxiv.org/abs/arXiv:astro-ph/0005248}{{arXiv:astro-ph/0005248}}]}.

\bibitem{Preto2010}
{Preto}, M., ``{Gravitational Waves Notes, Issue \#3 : ``Stellar cusps in
  galactic nuclei - How stars distribute around a massive black hole''}'', {\em
  ArXiv e-prints} (May 2010).
  {\small[\href{http://adsabs.harvard.edu/abs/2010arXiv1005.4048A}{ADS}]},
  {\small[\href{http://arxiv.org/abs/1005.4048}{{arXiv:1005.4048
  {\small[astro-ph.CO]}}}]}.

\bibitem{PretoAmaroSeoane10}
Preto, M.  and Amaro-Seoane, P., ``On Strong Mass Segregation Around a Massive
  Black Hole: Implications for Lower-Frequency Gravitational-Wave
  Astrophysics'', {\bf 708}, L42--L46 (2010).
  {\small[\href{http://dx.doi.org/10.1088/2041-8205/708/1/L42}{DOI}]},
  {\small[\href{http://adsabs.harvard.edu/abs/2010ApJ...708L..42P}{ADS}]},
  {\small[\href{http://arxiv.org/abs/0910.3206}{{arXiv:0910.3206}}]}.

\bibitem{PMS04}
Preto, M., Merritt, D.  and Spurzem, R., ``$N$-Body Growth of a Bahcall-Wolf
  Cusp around a Black Hole'', {\em Astrophys. J. Lett.}, {\bf 613}, L109--L112
  (2004).
  {\small[\href{http://adsabs.harvard.edu/abs/2004ApJ...613L.109P}{ADS}]}.

\bibitem{PretoMerrittSpurzem04}
Preto, M., Merritt, D.  and Spurzem, R., ``N-Body Growth of a Bahcall-Wolf Cusp
  around a Black Hole'', {\bf 613}, L109--112 (2004).
  {\small[\href{http://dx.doi.org/10.1086/425139}{DOI}]},
  {\small[\href{http://adsabs.harvard.edu/abs/2004ApJ...613L.109P}{ADS}]}.

\bibitem{Quinlan96}
Quinlan, G.D., ``The time-scale for core collapse in spherical star clusters'',
  {\em New Astronomy}, {\bf 1}, 255--270 (1996).
  {\small[\href{http://adsabs.harvard.edu/abs/1996NewA....1..255Q}{ADS}]}.

\bibitem{QS90}
Quinlan, G.D.  and Shapiro, S.L., ``The dynamical evolution of dense star
  clusters in galactic nuclei'', {\em Astrophys. J.}, {\bf 356}, 483--500
  (1990).
  {\small[\href{http://adsabs.harvard.edu/abs/1990ApJ...356..483Q}{ADS}]}.

\bibitem{RM98}
Raboud, D.  and Mermilliod, J.-C., ``Evolution of mass segregation in open
  clusters: some observational evidences'', {\em Astron. Astrophys.}, {\bf
  333}, 897--909 (1998).
  {\small[\href{http://adsabs.harvard.edu/abs/1998A&A...333..897R}{ADS}]}.

\bibitem{RFJ01}
Rasio, F.A., Fregeau, J.M.  and Joshi, K.J., ``Binaries and Globular Cluster
  Dynamics'', in {\em The influence of binaries on stellar population studies},
  Astrophysics and Space Science Library, 264, p. 387. Kluwer Academic
  Publishers, (2001).
  {\small[\href{http://adsabs.harvard.edu/abs/2001ibsp.conf..387R}{ADS}]}.

\bibitem{Rauch95}
Rauch, K.P., ``Dynamical evolution of star clusters around a rotating black
  hole with an accretion disc'', {\em Mon. Not. R. Astron. Soc.}, {\bf 275},
  628--640 (1995).
  {\small[\href{http://adsabs.harvard.edu/abs/1995MNRAS.275..628R}{ADS}]}.

\bibitem{RI98}
Rauch, K.P.  and Ingalls, B., ``Resonant tidal disruption in galactic nuclei'',
  {\em Mon. Not. R. Astron. Soc.}, {\bf 299}, 1231--1241 (1998).
  {\small[\href{http://adsabs.harvard.edu/abs/1998MNRAS.299.1231R}{ADS}]}.

\bibitem{RT96}
Rauch, K.P.  and Tremaine, S., ``Resonant relaxation in stellar systems'', {\em
  New Astronomy}, {\bf 1}, 149--170 (1996).
  {\small[\href{http://adsabs.harvard.edu/abs/1996NewA....1..149R}{ADS}]}.

\bibitem{Rees84}
Rees, M.J., ``Black Hole Models for Active Galactic Nuclei'', {\em Annu. Rev.
  Astron. Astrophys.}, {\bf 22}, 471--506 (1984).
  {\small[\href{http://adsabs.harvard.edu/abs/1984ARA&A..22..471R}{ADS}]}.

\bibitem{Rees88}
Rees, M.J., ``Tidal disruption of stars by black holes of 10 to the 6th-10 to
  the 8th solar masses in nearby galaxies'', {\em Nature}, {\bf 333}, 523--528
  (1988).
  {\small[\href{http://adsabs.harvard.edu/abs/1988Natur.333..523R}{ADS}]}.

\bibitem{Rees1988}
{Rees}, M.~J., ``{Tidal disruption of stars by black holes of 10 to the 6th-10
  to the 8th solar masses in nearby galaxies}'', {\bf 333}, 523--528 (June
  1988). {\small[\href{http://dx.doi.org/10.1038/333523a0}{DOI}]},
  {\small[\href{http://adsabs.harvard.edu/abs/1988Natur.333..523R}{ADS}]}.

\bibitem{RodriguezEtAl2015}
{Rodriguez}, C.~L., {Pattabiraman}, B., {Chatterjee}, S., {Choudhary}, A.,
  {Liao}, W.-k., {Morscher}, M.  and {Rasio}, F.~A., ``{A New Hybrid Technique
  for Modeling Dense Star Clusters}'', {\em ArXiv e-prints} (November 2015).
  {\small[\href{http://adsabs.harvard.edu/abs/2015arXiv151100695R}{ADS}]},
  {\small[\href{http://arxiv.org/abs/1511.00695}{{arXiv:1511.00695
  {\small[astro-ph.IM]}}}]}.

\bibitem{RMcDJ57}
Rosenbluth, M.N., MacDonald, W.M.  and Judd, D.L., ``Fokker-Planck Equation for
  an Inverse-Square Force'', {\em Phys. Rev.}, {\bf 107}, 1--6 (1957).
  {\small[\href{http://dx.doi.org/10.1103/PhysRev.107.1}{DOI}]},
  {\small[\href{http://adsabs.harvard.edu/abs/1957PhRv..107....1R}{ADS}]}.

\bibitem{RosenbluthEtAl57}
Rosenbluth, M.N., MacDonald, W.M.  and Judd, D.L., ``Fokker-Planck Equation for
  an Inverse-Square Force'', {\em Phys. Rev.}, {\bf 107}, 1--6 (1957).
  {\small[\href{http://dx.doi.org/10.1103/PhysRev.107.1}{DOI}]},
  {\small[\href{http://adsabs.harvard.edu/abs/1957PhRv..107....1R}{ADS}]}.

\bibitem{RHBF06}
Rubbo, L.J., Holley-Bockelmann, K.  and Finn, L.S., ``Event Rate for Extreme
  Mass Ratio Burst Signals in the Laser Interferometer Space Antenna Band'',
  {\em Astrophys. J. Lett.}, {\bf 649}, L25--L28 (2006).
  {\small[\href{http://dx.doi.org/10.1086/508326}{DOI}]},
  {\small[\href{http://adsabs.harvard.edu/abs/2006ApJ...649L..25R}{ADS}]}.

\bibitem{RubboEtAl2006}
Rubbo, L.J., Holley-Bockelmann, K.  and Finn, L.S., ``Event Rate for Extreme
  Mass Ratio Burst Signals in the Laser Interferometer Space Antenna Band'',
  {\em Astrophys. J. Lett.}, {\bf 649}, L25--L28 (2006).
  {\small[\href{http://dx.doi.org/10.1086/508326}{DOI}]},
  {\small[\href{http://adsabs.harvard.edu/abs/2006ApJ...649L..25R}{ADS}]}.

\bibitem{Salpeter55}
Salpeter, E.E., ``The Luminosity Function and Stellar Evolution'', {\em
  Astrophys. J.}, {\bf 121}, 161 (1955).
  {\small[\href{http://adsabs.harvard.edu/abs/1955ApJ...121..161S}{ADS}]}.

\bibitem{Schaffer79}
Schaffer, S., ``John Mitchell and Black Holes'', {\em Journal for the History
  of Astronomy}, {\bf 10}, 42 (1979).
  {\small[\href{http://adsabs.harvard.edu/abs/1979JHA....10...42S}{ADS}]}.

\bibitem{SchneiderEtAl2011}
{Schneider}, J., {Amaro-Seoane}, P.  and {Spurzem}, R., ``{Higher-order moment
  models of dense stellar systems: applications to the modelling of the stellar
  velocity distribution function}'', {\bf 410}, 432--454 (January 2011).
  {\small[\href{http://dx.doi.org/10.1111/j.1365-2966.2010.17454.x}{DOI}]},
  {\small[\href{http://adsabs.harvard.edu/abs/2011MNRAS.410..432S}{ADS}]},
  {\small[\href{http://arxiv.org/abs/1006.1365}{{arXiv:1006.1365}}]}.

\bibitem{SchoedelEtAl2014}
{Sch{\"o}del}, R., {Feldmeier}, A., {Kunneriath}, D., {Stolovy}, S.,
  {Neumayer}, N., {Amaro-Seoane}, P.  and {Nishiyama}, S., ``{Surface
  brightness profile of the Milky Way's nuclear star cluster}'', {\bf 566}, A47
  (June 2014).
  {\small[\href{http://dx.doi.org/10.1051/0004-6361/201423481}{DOI}]},
  {\small[\href{http://adsabs.harvard.edu/abs/2014A%26A...566A..47S}{ADS}]},
  {\small[\href{http://arxiv.org/abs/1403.6657}{{arXiv:1403.6657}}]}.

\bibitem{SchoedelEtAl2017}
{Sch{\"o}del}, R., {Gallego-Cano}, E., {Dong}, H., {Nogueras-Lara}, F.,
  {Gallego-Calvente}, A.~T., {Amaro-Seoane}, P.  and {Baumgardt}, H., ``{The
  distribution of stars around the Milky Way's central black hole II: Diffuse
  light from sub-giants and dwarfs}'', {\em ArXiv e-prints} (January 2017).
  {\small[\href{http://adsabs.harvard.edu/abs/2017arXiv170103817S}{ADS}]},
  {\small[\href{http://arxiv.org/abs/1701.03817}{{arXiv:1701.03817}}]}.

\bibitem{SchoedelEtAl03}
Sch{\" o}del, R., Ott, T., Genzel, R., Eckart, A., Mouawad, N.  and Alexander,
  T., ``Stellar Dynamics in the Central Arcsecond of Our Galaxy'', {\em
  Astrophys. J.}, {\bf 596}, 1015--1034 (2003).
  {\small[\href{http://adsabs.harvard.edu/abs/2003ApJ...596.1015S}{ADS}]}.

\bibitem{Schoedel02}
Sch{\" o}del, R. {et~al.}, ``A star in a 15.2-year orbit around the
  supermassive black hole at the centre of the Milky Way'', {\em Nature}, {\bf
  419}, 694--696 (2002).
  {\small[\href{http://adsabs.harvard.edu/abs/2002Natur.419..694S}{ADS}]}.

\bibitem{Seto2012}
Seto, N., ``Relativistic resonant relations between massive black hole binary
  and extreme mass ratio inspiral'', {\em Phys. Rev. D}, {\bf 85}, 064037
  (2012). {\small[\href{http://dx.doi.org/10.1103/PhysRevD.85.064037}{DOI}]},
  {\small[\href{http://adsabs.harvard.edu/abs/2012PhRvD..85f4037S}{ADS}]},
  {\small[\href{http://arxiv.org/abs/1202.4761}{{arXiv:1202.4761
  {\small[astro-ph.CO]}}}]}.

\bibitem{SM78}
Shapiro, S.L.  and Marchant, A.B., ``Star clusters containing massive, central
  black holes - {M}onte {C}arlo simulations in two-dimensional phase space'',
  {\em Astrophys. J.}, {\bf 225}, 603--624 (1978).
  {\small[\href{http://adsabs.harvard.edu/abs/1978ApJ...225..603S}{ADS}]}.

\bibitem{ST79}
Shapiro, S.L.  and Teukolsky, S.A., ``Gravitational collapse of supermassive
  stars to black holes - Numerical solution of the Einstein equations'', {\bf
  234}, L177--L181 (1979).
  {\small[\href{http://dx.doi.org/10.1086/183134}{DOI}]},
  {\small[\href{http://adsabs.harvard.edu/abs/1979ApJ...234L.177S}{ADS}]}.

\bibitem{ST85}
Shapiro, S.L.  and Teukolsky, S.A., ``The collapse of dense star clusters to
  supermassive black holes - The origin of quasars and AGNs'', {\em Astrophys.
  J. Lett.}, {\bf 292}, L41--L44 (1985).
  {\small[\href{http://adsabs.harvard.edu/abs/1985ApJ...292L..41S}{ADS}]}.

\bibitem{SR97}
Sigurdsson, S.  and Rees, M.J., ``Capture of stellar mass compact objects by
  massive black holes in galactic cusps'', {\em Mon. Not. R. Astron. Soc.},
  {\bf 284}, 318--326 (1997).
  {\small[\href{http://adsabs.harvard.edu/abs/1997MNRAS.284..318S}{ADS}]}.

\bibitem{Soltan82}
So{\l}tan, A., ``Masses of quasars'', {\em Mon. Not. R. Astron. Soc.}, {\bf
  200}, 115--122 (1982).
  {\small[\href{http://adsabs.harvard.edu/abs/1982MNRAS.200..115S}{ADS}]}.

\bibitem{SH71a}
Spitzer, L.~J.  and Hart, M.~H., ``Random Gravitational Encounters and the
  Evolution of Spherical Systems. I. Method'', {\em Astrophys. J.}, {\bf 164},
  399 (1971).
  {\small[\href{http://adsabs.harvard.edu/abs/1971ApJ...164..399S}{ADS}]}.

\bibitem{Spitzer69}
Spitzer~Jr, L., ``Equipartition and the Formation of Compact Nuclei in
  Spherical Stellar Systems'', {\em Astrophys. J. Lett.}, {\bf 158}, L139
  (1969).
  {\small[\href{http://adsabs.harvard.edu/abs/1969ApJ...158L.139S}{ADS}]}.

\bibitem{Spitzer87}
Spitzer~Jr, L., {\em Dynamical evolution of globular clusters}, (Princeton
  University Press, Princeton, NJ, 1987).
  {\small[\href{http://adsabs.harvard.edu/abs/1987degc.book.....S}{ADS}]}.

\bibitem{SH71b}
Spitzer~Jr, L.  and Hart, M.H., ``Random Gravitational Encounters and the
  Evolution of Spherical Systems. II. Models'', {\em Astrophys. J.}, {\bf 166},
  483 (1971).
  {\small[\href{http://adsabs.harvard.edu/abs/1971ApJ...166..483S}{ADS}]}.

\bibitem{SS66}
Spitzer~Jr, L.  and Saslaw, W.C., ``On the Evolution of Galactic Nuclei'', {\em
  Astrophys. J.}, {\bf 143}, 400 (1966).
  {\small[\href{http://adsabs.harvard.edu/abs/1966ApJ...143..400S}{ADS}]}.

\bibitem{SS75a}
Spitzer~Jr, L.  and Shull, J.M., ``Random gravitational encounters and the
  evolution of spherical systems. VI. Plummer's model'', {\em Astrophys. J.},
  {\bf 200}, 339--342 (1975).
  {\small[\href{http://adsabs.harvard.edu/abs/1975ApJ...200..339S}{ADS}]}.

\bibitem{Springel2005}
{Springel}, V., ``{The cosmological simulation code GADGET-2}'', {\bf 364},
  1105--1134 (December 2005).
  {\small[\href{http://dx.doi.org/10.1111/j.1365-2966.2005.09655.x}{DOI}]},
  {\small[\href{http://adsabs.harvard.edu/abs/2005MNRAS.364.1105S}{ADS}]},
  {\small[\href{http://arxiv.org/abs/astro-ph/0505010}{{astro-ph/0505010}}]}.

\bibitem{Spurzem92}
Spurzem, R., ``Evolution of Stars and Gas in Galactic Nuclei'', in {\em Rev.
  Modern Astron.}, 5, pp. 161--173, (1992).
  {\small[\href{http://adsabs.harvard.edu/abs/1992RvMA....5..161S}{ADS}]}.

\bibitem{SA96}
Spurzem, R.  and Aarseth, S.J., ``Direct collisional simulation of 10000
  particles past core collapse'', {\em Mon. Not. R. Astron. Soc.}, {\bf 282},
  19 (1996).
  {\small[\href{http://adsabs.harvard.edu/abs/1996MNRAS.282...19S}{ADS}]}.

\bibitem{ST95}
Spurzem, R.  and Takahashi, K., ``Comparison between Fokker-Planck and gaseous
  models of star clusters in the multi-mass case revisited'', {\em Mon. Not. R.
  Astron. Soc.}, {\bf 272}, 772--784 (1995).
  {\small[\href{http://adsabs.harvard.edu/abs/1995MNRAS.272..772S}{ADS}]}.

\bibitem{Stodol82}
Stodo{\l}kiewicz, J.S., ``Dynamical evolution of globular clusters. I'', {\em
  Acta Astron.}, {\bf 32}, 63--91 (1982).
  {\small[\href{http://adsabs.harvard.edu/abs/1982AcA....32...63S}{ADS}]}.

\bibitem{Stodol86}
Stodo{\l}kiewicz, J.S., ``Dynamical evolution of globular clusters. {II} -
  Binaries Method'', {\em Acta Astron.}, {\bf 36}, 19--41 (1986).
  {\small[\href{http://adsabs.harvard.edu/abs/1986AcA....36...19S}{ADS}]}.

\bibitem{SubrKaras99}
{\v S}ubr, L.  and Karas, V., ``An orbiter crossing an accretion disc'', {\em
  Astron. Astrophys.}, {\bf 352}, 452--458 (1999).
  {\small[\href{http://adsabs.harvard.edu/abs/1999A&A...352..452S}{ADS}]}.

\bibitem{SCR91}
Syer, D., Clarke, C.J.  and Rees, M.J., ``Star-disc interactions near a massive
  black hole'', {\em Mon. Not. R. Astron. Soc.}, {\bf 250}, 505--512 (1991).
  {\small[\href{http://adsabs.harvard.edu/abs/1991MNRAS.250..505S}{ADS}]}.

\bibitem{SU99}
Syer, D.  and Ulmer, A., ``Tidal disruption rates of stars in observed
  galaxies'', {\em Mon. Not. R. Astron. Soc.}, {\bf 306}, 35--42 (1999).
  {\small[\href{http://adsabs.harvard.edu/abs/1999MNRAS.306...35S}{ADS}]}.

\bibitem{TagoshiEtAl01}
{Tagoshi}, H., {Ohashi}, A.  and {Owen}, B.~J., ``{Gravitational field and
  equations of motion of spinning compact binaries to 2.5 post-Newtonian
  order}'', {\em Ph.Rv. D}, {\bf 63}(4), 044006 (February 2001).
  {\small[\href{http://dx.doi.org/10.1103/PhysRevD.63.044006}{DOI}]},
  {\small[\href{http://adsabs.harvard.edu/abs/2001PhRvD..63d4006T}{ADS}]},
  {\small[\href{http://arxiv.org/abs/gr-qc/0010014}{{gr-qc/0010014}}]}.

\bibitem{Takahashi93}
Takahashi, K., ``A new powerful method for solving the orbit-averaged
  Fokker-Planck equation regarding stellar dynamics'', {\em Publ. Astron. Soc.
  Japan}, {\bf 45}, 233--242 (1993).
  {\small[\href{http://adsabs.harvard.edu/abs/1993PASJ...45..233T}{ADS}]}.

\bibitem{Takahashi95}
Takahashi, K., ``Fokker-Planck Models of Star Clusters with Anisotropic
  Velocity Distributions {I}. Pre-Collapse Evolution'', {\em Publ. Astron. Soc.
  Japan}, {\bf 47}, 561--573 (1995).
  {\small[\href{http://adsabs.harvard.edu/abs/1995PASJ...47..561T}{ADS}]}.

\bibitem{Takahashi96}
Takahashi, K., ``Fokker-Planck Models of Star Clusters with Anisotropic
  Velocity Distributions {II}. Post-Collapse Evolution'', {\em Publ. Astron.
  Soc. Japan}, {\bf 48}, 691--700 (1996).
  {\small[\href{http://adsabs.harvard.edu/abs/1996PASJ...48..691T}{ADS}]}.

\bibitem{Takahashi97}
Takahashi, K., ``Fokker-Planck Models of Star Clusters with Anisotropic
  Velocity Distributions {III}. Multi-Mass Clusters'', {\em Publ. Astron. Soc.
  Japan}, {\bf 49}, 547--560 (1997).
  {\small[\href{http://adsabs.harvard.edu/abs/1997PASJ...49..547T}{ADS}]}.

\bibitem{Terlevich89}
Terlevich, R., ``Active galactic nuclei without black-holes'', in {\em
  Evolutionary Phenomena in Galaxies}, pp. 149--158, (1989).
  {\small[\href{http://adsabs.harvard.edu/abs/1989epg..conf..149T}{ADS}]}.

\bibitem{Thorne87}
Thorne, K.S., ``Gravitational radiation'', in Hawking, S.W.  and Israel, W.,
  eds., {\em Three Hundred Years of Gravitation}, pp. 330--458, (Cambridge
  University Press, Cambridge; New York, 1987).
  {\small[\href{http://adsabs.harvard.edu/abs/1987thyg.book..330T}{ADS}]},
  {\small[\href{http://books.google.com/books?id=Vq787qC5PWQC&pg=PA330}{Google
  Books}]}.

\bibitem{TremaineEtAl02}
Tremaine, S. {et~al.}, ``The Slope of the Black Hole Mass versus Velocity
  Dispersion Correlation'', {\em Astrophys. J.}, {\bf 574}, 740--753 (2002).
  {\small[\href{http://adsabs.harvard.edu/abs/2002ApJ...574..740T}{ADS}]}.

\bibitem{vanderMarelEtAl1993}
{van der Marel}, R.~P.  and {Franx}, M., ``{A new method for the identification
  of non-Gaussian line profiles in elliptical galaxies}'', {\bf 407}, 525--539
  (April 1993). {\small[\href{http://dx.doi.org/10.1086/172534}{DOI}]},
  {\small[\href{http://adsabs.harvard.edu/abs/1993ApJ...407..525V}{ADS}]}.

\bibitem{Vasiliev2015}
{Vasiliev}, E., ``{A new Monte Carlo method for dynamical evolution of
  non-spherical stellar systems}'', {\bf 446}, 3150--3161 (January 2015).
  {\small[\href{http://dx.doi.org/10.1093/mnras/stu2360}{DOI}]},
  {\small[\href{http://adsabs.harvard.edu/abs/2015MNRAS.446.3150V}{ADS}]},
  {\small[\href{http://arxiv.org/abs/1411.1757}{{arXiv:1411.1757}}]}.

\bibitem{VasilievEtAl2014}
{Vasiliev}, E., {Antonini}, F.  and {Merritt}, D., ``{The Final-parsec Problem
  in Nonspherical Galaxies Revisited}'', {\bf 785}, 163 (April 2014).
  {\small[\href{http://dx.doi.org/10.1088/0004-637X/785/2/163}{DOI}]},
  {\small[\href{http://adsabs.harvard.edu/abs/2014ApJ...785..163V}{ADS}]},
  {\small[\href{http://arxiv.org/abs/1311.1167}{{arXiv:1311.1167}}]}.

\bibitem{VasilievMerritt2013}
Vasiliev, E.  and Merritt, D., ``The loss cone problem in axisymmetric
  nuclei'', {\em arXiv}, e-print, (2013).
  {\small[\href{http://adsabs.harvard.edu/abs/2013arXiv1301.3150V}{ADS}]},
  {\small[\href{http://arxiv.org/abs/1301.3150}{{arXiv:1301.3150
  {\small[astro-ph.GA]}}}]}.

\bibitem{VMQR05}
Volonteri, M., Madau, P., Quataert, E.  and Rees, M.J., ``The Distribution and
  Cosmic Evolution of Massive Black Hole Spins'', {\em Astrophys. J.}, {\bf
  620}, 69--77 (2005). {\small[\href{http://dx.doi.org/10.1086/426858}{DOI}]},
  {\small[\href{http://adsabs.harvard.edu/abs/2005ApJ...620...69V}{ADS}]},
  {\small[\href{http://arxiv.org/abs/arXiv:astro-ph/0410342}{{arXiv:astro-ph/0410342}}]}.

\bibitem{WM04}
Wang, J.  and Merritt, D., ``Revised Rates of Stellar Disruption in Galactic
  Nuclei'', {\em Astrophys. J.}, {\bf 600}, 149--161 (2004).
  {\small[\href{http://adsabs.harvard.edu/abs/2004ApJ...600..149W}{ADS}]}.

\bibitem{WangEtAl2015}
{Wang}, L., {Spurzem}, R., {Aarseth}, S., {Nitadori}, K., {Berczik}, P.,
  {Kouwenhoven}, M.~B.~N.  and {Naab}, T., ``{NBODY6++GPU: ready for the
  gravitational million-body problem}'', {\em mn}, {\bf 450}, 4070--4080 (July
  2015). {\small[\href{http://dx.doi.org/10.1093/mnras/stv817}{DOI}]},
  {\small[\href{http://adsabs.harvard.edu/abs/2015MNRAS.450.4070W}{ADS}]},
  {\small[\href{http://arxiv.org/abs/1504.03687}{{arXiv:1504.03687
  {\small[astro-ph.IM]}}}]}.

\bibitem{WangEtAl2016}
{Wang}, L. {et~al.}, ``{The DRAGON simulations: globular cluster evolution with
  a million stars}'', {\em mn}, {\bf 458}, 1450--1465 (May 2016).
  {\small[\href{http://dx.doi.org/10.1093/mnras/stw274}{DOI}]},
  {\small[\href{http://adsabs.harvard.edu/abs/2016MNRAS.458.1450W}{ADS}]},
  {\small[\href{http://arxiv.org/abs/1602.00759}{{arXiv:1602.00759
  {\small[astro-ph.SR]}}}]}.

\bibitem{WJR00}
Watters, W.A., Joshi, K.J.  and Rasio, F.A., ``Thermal and Dynamical
  Equilibrium in Two-Component Star Clusters'', {\em Astrophys. J.}, {\bf 539},
  331--341 (2000).
  {\small[\href{http://adsabs.harvard.edu/abs/2000ApJ...539..331W}{ADS}]}.

\bibitem{WeinbergEtAl05}
Weinberg, N.N., Milosavljevi{\'c}, M.  and Ghez, A.M., ``Stellar Dynamics at
  the Galactic Center with an Extremely Large Telescope'', {\em Astrophys. J.},
  {\bf 622}, 878--891 (2005).
  {\small[\href{http://dx.doi.org/10.1086/428079}{DOI}]},
  {\small[\href{http://adsabs.harvard.edu/abs/2005ApJ...622..878W}{ADS}]},
  {\small[\href{http://arxiv.org/abs/arXiv:astro-ph/0404407}{{arXiv:astro-ph/0404407}}]}.

\bibitem{WillMaitra2017}
{Will}, C.~M.  and {Maitra}, M., ``{Relativistic orbits around spinning
  supermassive black holes: Secular evolution to 4.5 post-Newtonian order}'',
  {\em Ph. Rv. D}, {\bf 95}(6), 064003 (March 2017).
  {\small[\href{http://dx.doi.org/10.1103/PhysRevD.95.064003}{DOI}]},
  {\small[\href{http://adsabs.harvard.edu/abs/2017PhRvD..95f4003W}{ADS}]},
  {\small[\href{http://arxiv.org/abs/1611.06931}{{arXiv:1611.06931
  {\small[gr-qc]}}}]}.

\bibitem{YT02}
{Yu}, Q.  and {Tremaine}, S., ``{Observational constraints on growth of massive
  black holes}'', {\bf 335}, 965--976 (October 2002).
  {\small[\href{http://adsabs.harvard.edu/cgi-bin/nph-bib_query?bibcode=2002MNRAS.335..965Y&db_key=AST}{ADS}]}.

\bibitem{YunesEtAl2011}
Yunes, N., Miller, M.C.  and Thornburg, J., ``Effect of massive perturbers on
  extreme mass-ratio inspiral waveforms'', {\em Phys. Rev. D}, {\bf 83}, 044030
  (2011). {\small[\href{http://dx.doi.org/10.1103/PhysRevD.83.044030}{DOI}]},
  {\small[\href{http://adsabs.harvard.edu/abs/2011PhRvD..83d4030Y}{ADS}]},
  {\small[\href{http://arxiv.org/abs/1010.1721}{{arXiv:1010.1721
  {\small[astro-ph.GA]}}}]}.

\bibitem{YunesEtAl2008}
Yunes, N., Sopuerta, C.F., Rubbo, L.J.  and Holley-Bockelmann, K.,
  ``Relativistic Effects in Extreme Mass Ratio Gravitational Wave Bursts'',
  {\em Astrophys. J.}, {\bf 675}, 604--613 (2008).
  {\small[\href{http://dx.doi.org/10.1086/525839}{DOI}]},
  {\small[\href{http://adsabs.harvard.edu/abs/2008ApJ...675..604Y}{ADS}]},
  {\small[\href{http://arxiv.org/abs/0704.2612}{{arXiv:0704.2612}}]}.

\bibitem{Zare74}
Zare, K., ``A regularization of multiple encounters in gravitational n-body
  problems'', {\em Celestial Mechanics}, {\bf 10}, 207--215 (1974).
  {\small[\href{http://adsabs.harvard.edu/abs/1974CeMec..10..207Z}{ADS}]}.

\bibitem{ZHR02}
Zhao, H.-S., Haehnelt, M.G.  and Rees, M.J., ``Feeding black holes at galactic
  centres by capture from isothermal cusps'', {\em New Astronomy}, {\bf 7},
  385--394 (2002).
  {\small[\href{http://adsabs.harvard.edu/abs/2002NewA....7..385Z}{ADS}]}.

\end{thebibliography}

\end{document}